\documentclass[12pt]{msuthesis_98}
\usepackage{amssymb,amsmath}
\usepackage{doublespace}
\usepackage[dvips]{graphicx}
\newcommand{\daveseqn}{\addtocounter{equation}{1}}
\newcommand{\davetag}[1]{\tag{\theequation #1}}
\newcommand\rj{$\lbrace r_j \rbrace_{i =1 , \ldots , N}\,$}
\newlength{\bagfigheight}
\newlength{\HBTfigwidth}
\newcommand{\expectval}[1]{\left< #1 \right>}
\newcommand{\ME}[3]{\left< #1 \right| #2 \left| #3 \right>}
\newcommand{\OverLap}[2]{\left< \left. #1\right| #2 \right>} 
\newcommand{\bra}[1]{\left< #1 \right|}
\newcommand{\ket}[1]{\left| #1 \right>}
\newcommand{\Oper}{\hat{\cal O}}

\newcommand{\dirslash}[1]{\not{\!#1}}
\newcommand{\leftpartial}{\stackrel{\leftarrow}{\partial}}
\newcommand{\rightpartial}{\stackrel{\rightarrow}{\partial}}
\newcommand{\bothpartial}{\stackrel{\leftrightarrow}{\partial}}

\newcommand{\grad}{\vec{\nabla}}
\newcommand{\dn}[2]{{d}^{#1} #2 \:}
\newcommand{\dnpi}[2]{ \frac{ {d}^{#1} #2 }{ {(2\pi)}^{#1} } \:}
\newcommand{\dnpiinvar}[2]{ \frac{ {d}^{#1} #2 }{2 \left|{#2}_{0}\right|
                                    {(2\pi)}^{#1} } \:}
\newcommand{\intc}{\int_{\cal C}}
\newcommand{\deltaftn}[2]{ \delta^{#1} \! \left( #2 \right) }
\newcommand{\twopideltaftn}[2] {(2\pi)^{#1}\deltaftn{#1}{#2}}
\newcommand{\thetaftn}[1]{\theta \! \left( #1 \right)}
\newcommand{\twoargftn}[3]{{\cal #1} \!\left( #2 , #3 \right) }

\newcommand{\sgn}[1]{\operatorname{sgn}\left( #1 \right)}
\newcommand{\BesselJ}[2]{J_{#1}\!\left( #2 \right)}
\newcommand{\Order}[1]{{\cal O}\left( #1 \right)}
\newcommand{\Smatrix}[1]{ {\left| S_{\rm #1} \right| }^2 }
\newcommand{\ReactRate}[3]{{\cal W}_{\rm #1}\left( #2,#3 \right)}
\newcommand{\ReactRateonearg}[1]{{\cal W}_{\rm #1}}
\newcommand{\Jcurrent}[5]{{J}^{#1 #2}_{\rm #5}\!\left( #3,#4 \right)} 
\newcommand{\Jcurrentdn}[4]{{J}_{#1 #2}\!\left( #3,#4 \right)} 
\newcommand{\Afield}[4]{{A}_{#1 #2}\!\left( #3,#4 \right)}
\newcommand{\Afieldup}[4]{{A}^{#1 #2}\!\left( #3,#4 \right)}
\newcommand{\Afieldname}[5]{{A}_{#1 #2}^{#5}\!\left( #3,#4 \right)}
\newcommand{\Afieldupname}[5]{{A}^{#1 #2}_{#5}\!\left( #3,#4 \right)}
\newcommand{\EPdist}{ \frac{ d n_{\gamma} \!\left( x , q \right) }
                          { \dn{3}{x} \dn{3}{q} d q^2} }
\newcommand{\EEdist}{ \frac{ d n_{e^{-}} \!\left( y , p \right) }
                          { \dn{3}{y} \dn{3}{p} d p^2} }
\newcommand{\density}[3]{{#1} \left( #2,#3 \right)}
\newcommand{\denslabel}[4]{{#1}_{#4} \left( #2,#3 \right)}
\newcommand{\denslabelstar}[4]{{#1}^{*}_{#4} \left( #2,#3 \right)}
\newcommand{\ggtrless}[2]{\density{g^{\gtrless}}{#1}{#2}}
\newcommand{\ggtr}[2]{\density{g^{>}}{#1}{#2}}
\newcommand{\gless}[2]{\density{g^{<}}{#1}{#2}}
\newcommand{\dgtrless}[4]{\density{d^{\gtrless}_{#1 #2}}{#3}{#4}}
\newcommand{\dgtrlessup}[4]{\density{d^{\gtrless #1 #2}}{#3}{#4}}

\newcommand{\dless}[4]{\density{d^{<}_{#1 #2}}{#3}{#4}}
\newcommand{\sgtrless}[4]{\density{s^{\gtrless}_{#1 #2}}{#3}{#4}}
\newcommand{\sgtr}[4]{\density{s^{>}_{#1 #2}}{#3}{#4}}
\newcommand{\sless}[4]{\density{s^{<}_{#1 #2}}{#3}{#4}}
\newcommand{\pol}[2]{\epsilon^{#1} \!\left( #2 \right) }
\newcommand{\polstar}[2]{\epsilon^{* #1} \!\left( #2 \right) }
\newcommand{\poldn}[2]{\epsilon_{#1} \!\left( #2 \right) }
\newcommand{\polstardn}[2]{\epsilon^{*}_{#1} \!\left( #2 \right) }
\newcommand{\Dprop}[6]{{D}_{#1 #2 #3 #4}\!\left( #5,#6 \right)}
\newcommand{\Gprop}[2]{{G}\!\left( #1,#2 \right)}
\newcommand{\Sprop}[6]{{S}_{#1 #2 #3 #4}\!\left( #5,#6 \right)}

\newcommand{\Dmisc}[7]{{D}^{#7}_{#1 #2 #3 #4}\!\left( #5,#6 \right)}
\newcommand{\Dmiscup}[7]{{D}^{#7 #1 #2 #3 #4}\!\left( #5,#6 \right)}
\newcommand{\Gmisc}[3]{{G}^{#3}\!\left( #1,#2 \right)}
\newcommand{\Smisc}[7]{{S}^{#7}_{#1 #2 #3 #4}\!\left( #5,#6 \right)}

\newcommand{\Dplus}[6]{\Dmisc{#1}{#2}{#3}{#4}{#5}{#6}{+}}
\newcommand{\Gplus}[2]{\Gmisc{#1}{#2}{+}}
\newcommand{\Splus}[6]{\Smisc{#1}{#2}{#3}{#4}{#5}{#6}{+}}
\newcommand{\Dmin}[6]{\Dmisc{#1}{#2}{#3}{#4}{#5}{#6}{-}}
\newcommand{\Gmin}[2]{\Gmisc{#1}{#2}{-}}
\newcommand{\Smin}[6]{\Smisc{#1}{#2}{#3}{#4}{#5}{#6}{-}}
\newcommand{\Dgreat}[6]{\Dmisc{#1}{#2}{#3}{#4}{#5}{#6}{>}}
\newcommand{\Ggreat}[2]{\Gmisc{#1}{#2}{>}}
\newcommand{\Sgreat}[6]{\Smisc{#1}{#2}{#3}{#4}{#5}{#6}{>}}
\newcommand{\Dless}[6]{\Dmisc{#1}{#2}{#3}{#4}{#5}{#6}{<}}
\newcommand{\Gless}[2]{\Gmisc{#1}{#2}{<}}
\newcommand{\Sless}[6]{\Smisc{#1}{#2}{#3}{#4}{#5}{#6}{<}}
\newcommand{\Dcaus}[6]{\Dmisc{#1}{#2}{#3}{#4}{#5}{#6}{\rm c}}
\newcommand{\Gcaus}[2]{\Gmisc{#1}{#2}{\rm c}}
\newcommand{\Scaus}[6]{\Smisc{#1}{#2}{#3}{#4}{#5}{#6}{\rm c}}

\newcommand{\Gacaus}[2]{\Gmisc{#1}{#2}{\rm a}}

\newcommand{\Dgl}[4]{\Dmisc{}{}{#1}{#2}{#3}{#4}{\gtrless}}
\newcommand{\Dglup}[4]{\Dmiscup{}{}{#1}{#2}{#3}{#4}{\gtrless}}
\newcommand{\Sgl}[4]{\Smisc{}{}{#1}{#2}{#3}{#4}{\gtrless}}
\newcommand{\Ggl}[2]{\Gmisc{#1}{#2}{\gtrless}}

\newcommand{\Slg}[4]{\Smisc{}{}{#1}{#2}{#3}{#4}{\lessgtr}}
\newcommand{\Glg}[2]{\Gmisc{#1}{#2}{\lessgtr}}
\newcommand{\Dpm}[4]{\Dmisc{}{}{#1}{#2}{#3}{#4}{\pm}}
\newcommand{\Spm}[4]{\Smisc{}{}{#1}{#2}{#3}{#4}{\pm}}
\newcommand{\Gpm}[2]{\Gmisc{#1}{#2}{\pm}}
\newcommand{\Dnon}[6]{\Dmisc{#1}{#2}{#3}{#4}{#5}{#6}{0}}
\newcommand{\Gnon}[2]{\Gmisc{#1}{#2}{0}}
\newcommand{\Snon}[6]{\Smisc{#1}{#2}{#3}{#4}{#5}{#6}{0}}
\begin{document}
\begin{frontmatter}
   \title{ACCESSING THE SPACE--TIME DEVELOPMENT OF HEAVY--ION COLLISIONS WITH THEORY AND EXPERIMENT}
   \author{David Alan Brown}
   \dept{Department of Physics and Astronomy}
   \submityear{1998}
   \titlepage
   \abstract{
In this thesis we discuss ways to access the space-time
development of heavy-ion reactions using both theory and experiment.
From the theoretical side, we discuss modeling ultra-relativistic, 
parton-dominated, heavy-ion reactions.  This discussion is broken into a 
discussion of transport-like models for massless particles and a discussion 
of the parton model in phase-space.  From the experimental side, we
discuss using intensity interferometry to image the relative distribution
of emission points.

Transport models may offer a way to understand the space-time development 
of ultra-relativistic, parton-dominated, heavy-ion reactions at RHIC and the 
LHC.  
Two key approximations needed to derive semi-classical transport equations, the 
Quasi-Particle and Quasi-Classical approximations, may not be valid for partons.
Using QED, we outline a
derivation of a transport-like theory which does not rely on these two 
approximations.  This theory rests on the
phase-space Generalized Fluctuation-Dissipation Theorem. 
This theorem and the phase-space particle self-energies 
give a set of coupled phase-space evolution equations.  We illustrate how 
these evolution equations can be used perturbatively or to derive 
semi-classical transport equations.  

To connect the parton phase-space densities to the 
experimentally measured Parton Distribution Functions, the parton model must
be translated into phase-space.  Within QED, we study whether two key 
components of the parton model,
factorization and evolution, can be formulated 
in phase-space.  We rewrite the QED analog of the parton model, 
the Weizs\"acker-Williams Approximation, in terms of phase-space 
quantities, demonstrating factorization in phase-space.  
Evolution of the parton densities is equivalent to summing a class of ladder 
diagrams.  We study a simplified QED version of these ladders while studying 
the phase-space photon and electron densities surrounding a classical point 
charge.  We find that the densities take the form given in the phase-space 
Generalized Fluctuation-Dissipation Theorem.  
We use the tools developed here to discuss the shape of a nucleon's parton 
cloud.

\thispagestyle{empty}
We can access the space-time development of a heavy-ion reaction directly 
by imaging the source function from particle correlation functions.
We discuss several methods to perform this 
inversion.  We concentrate on one such method, the Optimized 
Discretization method, where the source resolution depends on the relative
particle separation and is adjusted to available data and their errors.  
This method can be supplemented using known constraints on the source.
We test the inversion methods by restoring simulated pp sources.
From the restored sources, one can extract the average freeze-out phase-space 
density, entropy at freeze-out and the amount of the source that lie
outside of the imaged region.
We apply the imaging techniques to pion, kaon, proton and Intermediate Mass 
Fragment (IMF) correlation functions.
Significant portions of the pion, proton and IMF 
sources extend to large distances ($r > 20$~fm).  The results of the imaging
show the inadequacy of common Gaussian parameterizations of the source.}
   \dedicate{\parbox{4in}{To Aleida, for her love, \\
	To my family, for their support, \\
	To my father, whom I will always remember.}}
   \dedication
   \acknowledgments{
	Mom and Dad, I owe you so much that I could not possibly list it all 
here.  Thank you for everything. 

	Aleida, thank you for being there by my side.   Your love 
and patience keeps me going.

	Doug, Jen and Steve, you remind me that there is a world outside
of science.

	All of my professors and teachers at EHS, Clarkson, UDel, the APS 
and MSU, thank you for teaching me so much and for making me realize 
there is so much more to learn. 

	Pawe{\l}, thank you for asking questions that aren't so easy to answer.

	Scott, Vladimir and Wolfgang, thank you for helping me try to 
answer Pawe{\l}'s questions.

	Mike, Eva and Tibor, making beer {\em is} more fun than work.

	Joelle, Frank, Declan and Alexander, working along side you 
is a pleasure.  I look forward to working with each of you again.}

   \tables

\end{frontmatter}

\chapter{INTRODUCTION}

\section{Heavy-Ion Collisions}

How does a colliding heavy-ion system evolve in space-time?  
This is an interesting question which we will discuss
in this introduction.  In the rest of this thesis, we will present work 
demonstrating how to access this space-time development  
using both theoretical and experimental techniques.
This work falls into two categories:  understanding transport-like models, 
with special emphasis on understanding their application to events at 
the Relativistic Heavy Ion Collider (RHIC)~\cite{tran:bro98}, and 
understanding the use of Hanbury-Brown/Twiss (HBT) intensity interferometry 
as a way of working backwards from the data to the end of 
a heavy-ion collision~\cite{HBT:bro97,HBT:bro98}.
So, why should we be interested in heavy-ion collisions and 
why should we care how such a system evolves in space-time?

In a heavy-ion collision, the two colliding nuclei create an excited, dense
and possibly thermalized, zone of nuclear matter in their wake.  We see a 
much larger version of this in the effectively infinite thermalized nuclear 
matter of neutron stars, accretion disks and supernovas.   We also expect 
that such matter existed moments after the Big Bang~\cite{QGP:mue85,QGP:har96}.
In all of these cases, a reasonable description can be built up using single 
nucleon-nucleon collisions.  Indeed, the systems created in heavy ion 
reactions are intermediate in size between single nucleon collisions and 
infinite thermalized nuclear matter and have features of both.
Both single nucleon-nucleon collisions and heavy-ion collisions are
easily accessible with current technology.  Creating and manipulating 
infinite thermalized nuclear matter in a controlled way is far beyond 
anything possible today -- we can not smash neutron stars together at will. 
In the absence of black hole/neutron star collider experiments, we must rely 
on extrapolation from finite nuclear systems.

\begin{figure}
	\centering
	\includegraphics{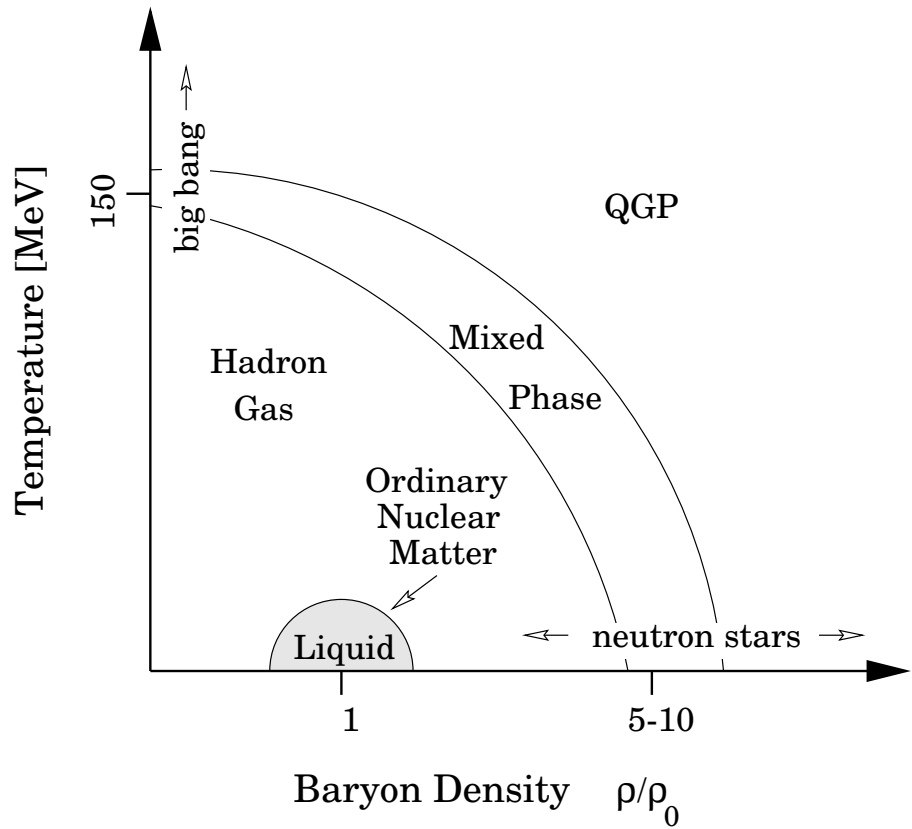}
	\caption[Phase diagram of nuclear matter.  In this figure, $\rho$ is 
	the baryon density and $\rho_0$ is normal nuclear matter density.  The 
	ground state of normal nuclear matter is the ``liquid'' phase.]
	{Phase diagram of nuclear matter.  In this figure, $\rho$ is the 
	baryon density and $\rho_0$ is normal nuclear matter density.  The 
	ground state of normal nuclear matter is the ``liquid'' phase.}
	\label{fig:PhaseDiag}
\end{figure}

Because the systems created in heavy-ion collisions are on the border of few 
particle systems and infinite thermalized nuclear matter, we expect that many 
of the features of both will 
show up in heavy-ion reactions.  In particular, the thermal features of 
infinite matter should reveal themselves in some form in heavy-ion 
collisions.  As an example, consider the liquid-gas phase-transition of 
nuclear matter -- it is predicted to reveal itself through the 
processes of fragmentation and multifragmentation \cite{lyn98}.  
Another phase-transition, the Quark-Gluon 
Plasma (QGP) phase-transition is predicted to reveal itself 
by ``fragmenting'' into disordered Chiral Condensates or quark 
droplets \cite{QGP:raj95,QGP:har96,QGP:mcl86,QGP:mue85}.
The QGP phase transition is predicted by lattice QCD 
\cite{QGP:det95,QGP:mcl86,QGP:mue85} 
and is implied by the hadronic model of Hagedorn \cite{hag65,hag68,hag71}
and by Chiral Perturbation Theory \cite{QGP:raj95}.  
This phase transition is currently generating 
great interest as it may already be happening at CERN-SPS 
energies \cite{QGP:rom98}
and should happen at both RHIC and the Large Hadron Collider (LHC) at 
CERN \cite{QGP:har96,QGP:mcl86,QGP:mue85}.
A phase diagram of nuclear matter  is shown in 
Figure~\ref{fig:PhaseDiag}.  In this diagram, we see the two important phase 
transitions -- the liquid gas phase transition and the transition to the QGP.  

\begin{figure}
	\centering
	\includegraphics{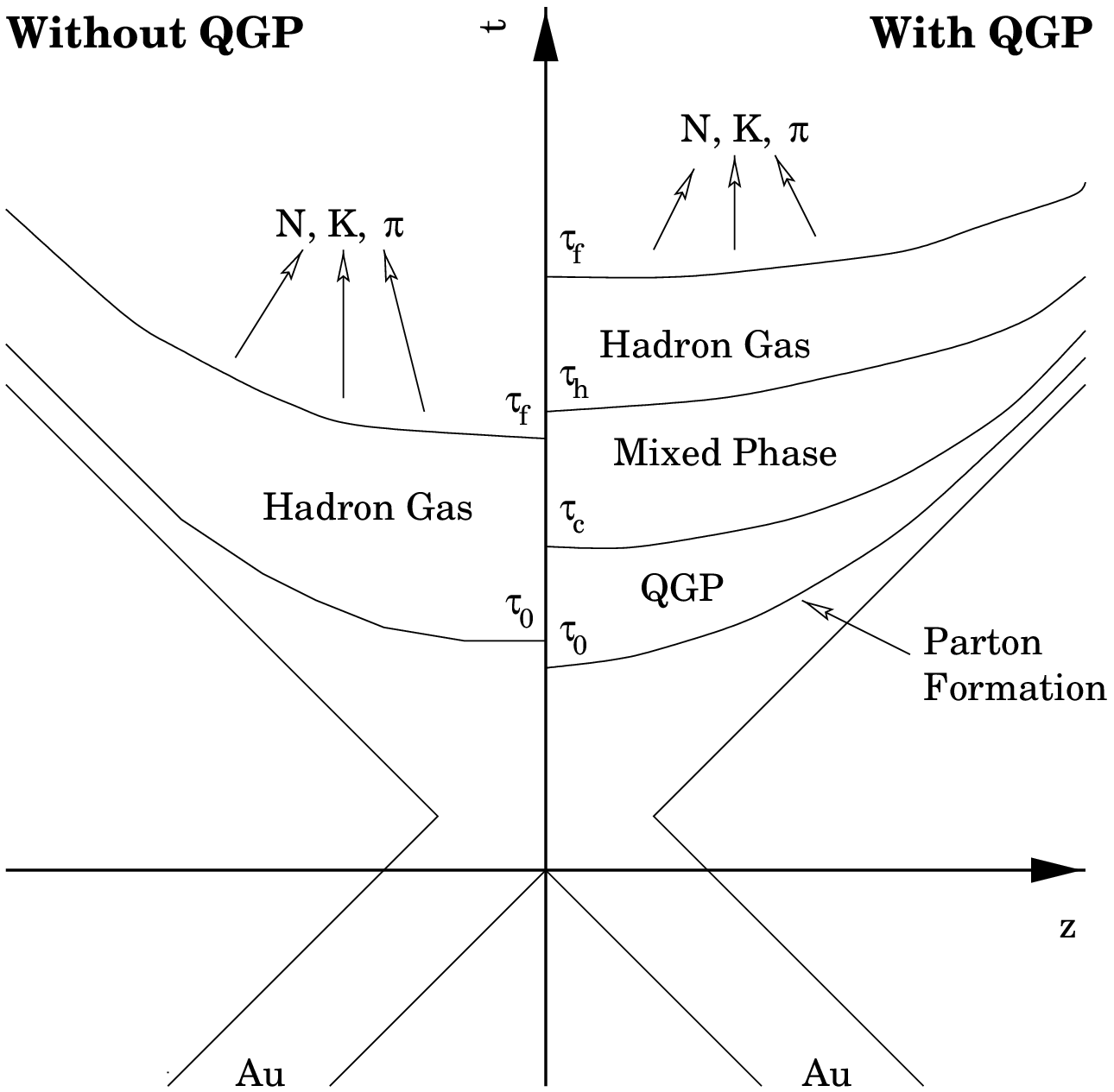}
	\caption[Space-time evolution of colliding heavy-ion system both 
	with and without the QGP.  In this 
	figure, the important hypersurfaces are labeled with the proper time
	in the system CM.   The time of full overlap of the 
	two nuclei occurs at proper time $\tau_0$.  The coexistence phase 
	begins at proper time $\tau_c$ and hadronization begins at $\tau_h$.
	The collision ends at the freeze-out proper time $\tau_f$ when
	the final state nucleons and mesons stream out to the detector.] 
	{Space-time evolution of colliding heavy-ion system both with and 
	without the QGP.  In this 
	figure, the important hypersurfaces are labeled with the proper time
	in the system CM.   The time of full overlap of the 
	two nuclei occurs at proper time $\tau_0$.  The coexistence phase 
	begins at proper time $\tau_c$ and hadronization begins at $\tau_h$.
	The collision ends at the freeze-out proper time $\tau_f$ when
	the final state nucleons and mesons stream out to the detector.}
	\label{fig:SpaceTimeQGP}
\end{figure}

Given that colliding heavy-ion systems are interesting, why study the
space-time evolution of a heavy-ion collision?  
Well, the existence of a phase transition dramatically affects the space-time 
development of the system which in turn modifies the final state particle 
characteristics.  Compare the two scenarios for a typical RHIC collision shown 
in Figure~\ref{fig:SpaceTimeQGP}:  on the left, the collision proceeds through
a purely hadronic phase and, on the right, the system undergoes a 
phase-transition to the Quark-Gluon Plasma.
Now, the existence of the phase-transition would lead to a 
drop in the pressure of the system at the phase-transition, softening the 
equation of state and leading to a disappearance of flow at 
the ``softest point'' 
\cite{QGP:ris95,QGP:ris96a,QGP:ris96b,QGP:ris96c,QGP:ris96d,QGP:ris97}.  
Additionally, a phase-transition
may lead to a long-lived system which would, in turn, lead to a larger
relative emission point distribution for identical particle pairs 
\cite{HBT:pra90}.  This could be detected using HBT interferometry and nuclear 
imaging.  The existence of a temporarily deconfined region would lead to 
other observable effects such as the color screening of the quarks in 
a $J/\psi$ particle.  This then allows them to disassociate, leading to  
so-called $J/\psi$ suppression~\cite{QGP:mat86}.
So, now it is clear that different physics leads to different space-time 
evolution of a collision.  What we need now are ways to get at this 
space-time evolution.

\begin{figure}
	\centering
	\includegraphics[width=\textwidth]{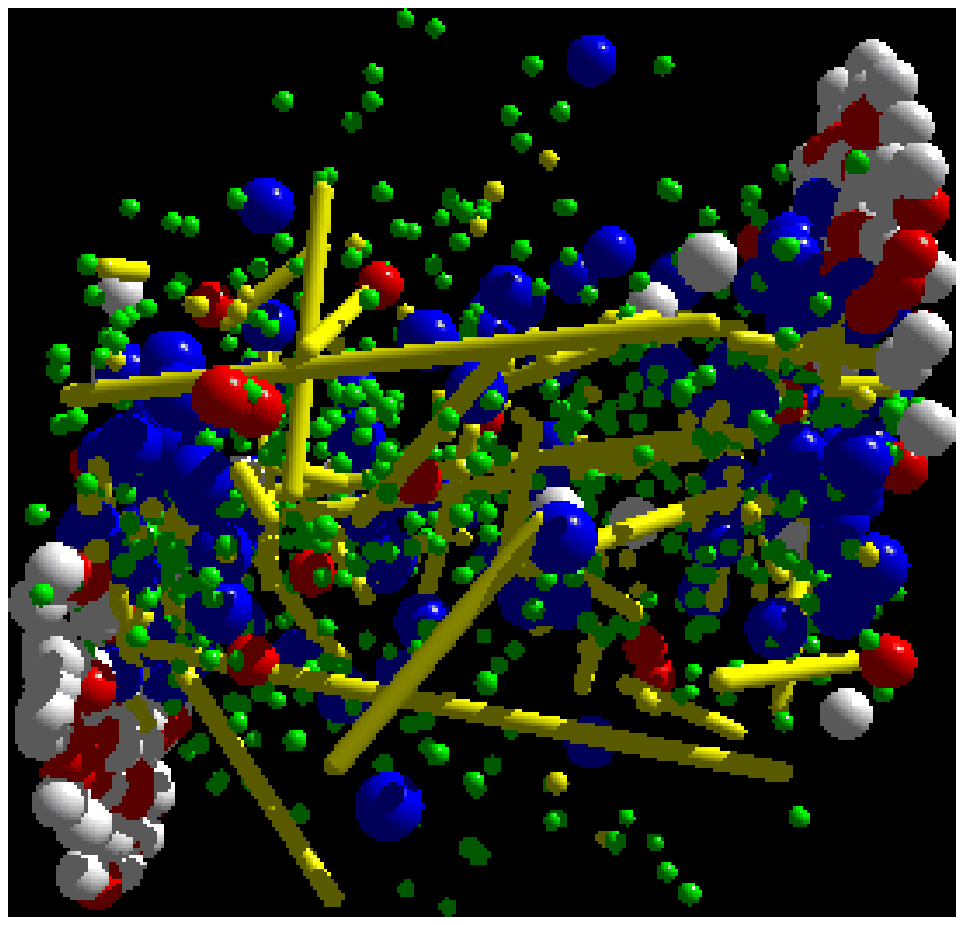}
	\caption[Sample UrQMD event: Au+Au at $E/A=200$~GeV and $b=5$~fm.
	In this picture, the small green spheres represent pions, the larger
	white and red spheres represent neutrons and protons, the larger blue
	spheres represent nucleon resonances and the yellow lines represent
	string excitations of the hadrons.  In this picture, one can clearly
	see the spectator matter leaving the central collision zone.  The 
	large number of strings in the central region shows that there is 
	a high energy density.] 
	{Sample UrQMD event: Au+Au at $E/A=200$~GeV and $b=5$~fm.
	In this picture, the small green spheres represent pions, the larger
	white and red spheres represent neutrons and protons, the larger blue
	spheres represent nucleon resonances and the yellow lines represent
	string excitations of the hadrons.  In this picture, one can clearly
	see the spectator matter leaving the central collision zone.  The 
	large number of strings in the central region shows that there is 
	a high energy density.}
	\label{fig:UrQMD_pict}
\end{figure}

We can access the space-time evolution of a heavy-ion collision in several
different ways.  Two ways in particular are quite fruitful and are the 
subject of this thesis:  1) modeling the reaction using transport-like models 
and 2) accessing the final state directly using nuclear imaging.
A large class of event generators are designed in order 
to study the phase-space\footnote{That is in both
coordinate and momentum representations simultaneously.} densities of the
particles as they evolve in time.  Models that provide the time evolution
of the phase-space particle densities are generally called transport
models.  In part, these models are useful because 
they provide a direct visualization of the collision (see for example the
output from an UrQMD Au+Au event at $b=5$~fm and $E/A=200$~GeV in 
Figure~\ref{fig:UrQMD_pict} from~\cite{web98}).  More importantly however, 
such models can easily incorporate the data from single 
nucleon-nucleon collisions in vacuum, allowing us to build up a transport 
theory consistent with the underlying, elementary physics.  
A different and complementary approach is to directly image the reaction.
By imaging the reaction, we directly get at the 
configuration space distribution of the system at freeze-out.  
Of course there are problems resulting from such an inversion of the 
experimental data resulting from the mixing of temporal evolution into the 
source in a nontrivial manner and the inherently difficulty of imaging 
resulting in rather crude images.  Both the subjects of transport theory and 
nuclear imaging are discussed at length in this thesis.

%
%
\section{Working Forwards -- Theory}

Modeling heavy-ion collisions has a long history, going back into the 1940's.
It has progressed to its current state by a series of improvements in our
transport models and through theoretical insights into transport theory as a 
whole.  However, the basic idea behind many of the models has remained more or 
less the same over the years:  nucleons travel along their classical paths 
in configuration space and collide when particles are within 
$\sqrt{\sigma_{\mbox{\sc tot}}/\pi}$ of one another.\footnote{We refer to the 
use of the cross section in this manner as the closest approach criteria.}
The total collision cross section, $\sigma_{\rm TOT}$, is often tuned to 
reproduce single particle spectra.
In other words, the nucleons are treated as an ensemble of ``billiard balls'' 
with radius $\sqrt{\sigma_{\rm TOT}/\pi}$ as they evolve in phase-space.
This picture works well for nucleons at intermediate to high energies
because they can be localized and interact on time scales much smaller than
their mean free path.  Should we expect this picture to work for the highest 
energy collisions?  In these collisions, the criteria for closest approach 
breaks down \cite{tran:kor95} and the dynamics are dominated by massless (or
nearly massless) particles which are both difficult to localize and may 
interact over large length and time scales.  
In other words, can the series of approximations used
to arrive at this picture be justified for a RHIC collision where the majority 
of the interacting particles are quarks and gluons (i.e. {\em partons})?
Indeed, can we even define the initial 
conditions for a RHIC collision, a job tantamount to 
rewriting the parton model in phase-space?

The first models of heavy-ion collisions were based on the Internucleon 
Cascade (INC) concept developed by Serber in 1947~\cite{tran:ser47}.  His idea
is simply to represent nucleons as ``billiard balls'' that travel 
along classical (relativistic) trajectories through configuration space.  The 
nucleons interact strictly through binary collisions between 
nucleons at the point of closest approach in configuration space.  This 
point of closest approach is defined through the total cross section to be
$\sqrt{\sigma_{\rm TOT}/\pi}$; this model is the origin of the closest 
approach criteria.  
Collisions are the only way, in this model, to modify the momentum portion 
of the phase-space distribution.
The INC concept has been extended into its modern-day incarnation by including
resonances \cite{tran:cug81,tran:cug82}.  Because the particles do 
not interact through mean-fields, or any other higher order mechanisms, 
this model can only reproduce single particle observables such as spectra.
Thus, it works best at energies where mean-field effects are small 
(i.e. $\gtrsim 1$~GeV).  The concepts 
first laid out in this model form a blueprint for all the successful 
heavy-ion transport models to follow.

There are several models that build upon the basic ideas laid out in the INC
and we will only discuss two classes of them:  Quantum Molecular Dynamics
and Boltzmann equation simulations.  There are many other types of models,
such as Time Dependent Hartree-Fock models, hydrodynamical models and thermal
models, just to name a few.  These models have varying degrees of success
however none are as successful at reproducing single and many particle 
observables at intermediate to high energies 
($\gtrsim 100$~MeV) as the classes of models that we discuss below.

Quantum Molecular Dynamics (QMD) models follow along much like the INC.  In 
QMD, as in the INC, each nucleon is treated as a ``billiard ball.''  The
particles are on-shell and all follow their classical equations of motion 
through configuration space.  However, instead of using the cross sections to 
determine how a collision proceeds, QMD uses nucleon-nucleon potentials.   
QMD's higher-energy incarnations, RQMD and UrQMD, supplement the 
nucleon-nucleon potentials with hard scattering through the cross sections 
using the closest approach criteria as in the INC~\cite{tran:sor91,tran:UrQMD} 
and with strings and resonances.  RQMD and UrQMD also sport two 
other features, not present in INC: a nucleon mean field and Pauli blocking.  
With the inclusion of the mean field, both RQMD and UrQMD can reproduce 
flow observables, which are sensitive to mean field effects.  The Pauli
blocking only really helps at the lower energies, and the lack of Pauli 
blocking in the INC partially explains why the INC can not work well 
below 150--200~MeV/A.  Both RQMD and UrQMD are significant 
advances over INC and are quite successful at reproducing single and multiple
particle observables.

In Boltzmann-Uehling-Uhlenbeck (BUU) based approaches, such as in the MSU-BUU
or the BEM models, one sets out to solve the BUU equation:
\begin{equation}
\begin{array}{rl}
\displaystyle\frac{\partial f}{\partial t}+\vec{v}\cdot\grad_r f-\grad_r 
U\cdot\grad_p f=&\displaystyle-\frac{1}{(2\pi)^6}\int\dn{3}{p_2}\dn{3}{p_{2'}}
d\Omega\frac{d\sigma}{d\Omega}v_{12}\\
&\times\displaystyle\left\{\left[f f_2(1-f_{1'})(1-f_{2'})-
f_{1'} f_{2'}(1-f_{1})(1-f_{2})\right]\right.\\
&\times\displaystyle\left.
\twopideltaftn{3}{\vec{p}+\vec{p}_2-\vec{p}_{1'}-\vec{p}_{2'}}\right\}.
\end{array}
\label{eqn:BUUeqn}
\end{equation}
In this equation, $f$ is the phase-space density.  The particles are all 
taken to be on-shell and primes 
denote quantities to be taken after a collision between particles $1$ and $2$.
This equation incorporates all of the important innovations
included in the QMD and INC models, namely Pauli blocking, a mean field and the
ability to fit results to known particle spectra.  The 
factor $d\sigma/d\Omega$ is the experimentally determined nucleon-nucleon 
cross sections, although in practice, it may be altered to account for 
in-medium effects such as screening.  The $(1-f)$ terms account for 
Pauli blocking in the final state;  if a cell in phase-space is occupied by
a Fermion, then in that region $f=1$ so $1-f=0$ making that collision 
term $0$.  Finally we have the mean-field, $U$, which produces a driving 
force through the term $\grad_r U \cdot\grad_p f$.  Now, the actual 
solving of the BUU equation varies with the model in question, 
but most models use the test-particle method.  In this method, one replaces 
the phase-space distribution of particles with an ensemble of test-particles
(essentially ``billiard balls'').  The test-particles follow classical 
trajectories that are modified by the mean-field driving force and interact 
using known cross sections in the same manner as the INC.  Most BUU-type 
models are capable of reproducing both single particle spectra 
and flow observables.

Clearly, there are several features common to all of these models:
classical relativistic kinetics, use of cross sections to constrain 
inter-particle interactions and full phase-space evolution of on-shell
particle densities.  How can we justify these features?  For the INC and
QMD based approaches, the justification is purely phenomenological.  However,
in BUU based approaches many features can be justified using known procedures 
and time-ordered non-equilibrium field theory 
\cite{neq:kad62,tran:ram86,neq:dan84a,neq:bez72,tran:kle97,tran:mro90,tran:mro94}.  
So, we can study these works and 
understand how to justify the various features of transport models.
In the standard derivations of the BUU equation the particles follow their 
classical trajectories only after applying the gradient approximation,
an approximation also know as the Quasi-Classical Approximation (QCA).  In
this approximation, one throws away short scale structure in favor of 
large-scale structure in the densities and collision integrals.  This washes 
out the quantum wanderings of particles from their classical paths.  To justify
this approximation, one needs the collision length scale to be much smaller
than the length scale of variation of particle densities.
Now the Lorentz dilation effects in an ultra-relativistic nuclear collision
can ruin this scale separation by simultaneously shrinking the mean 
free path (the nucleon density increases by a factor of $\gamma$)
and increasing the interaction time.
A complementary approximation that one typically makes is the so-called
Quasi-Particle Approximation (QPA).  In this approximation, one replaces the
full phase-space densities (i.e. in $E, \vec{p}, t,$ and $\vec{x}$) with 
distributions of on-mass shell particles 
(so $E\equiv \sqrt{{\vec{p} }^2+m^2}$).
This is equivalent to making all particles free particles with infinite 
lifetimes.  Most theorists recognize this as a problem because even in
intermediate energy collisions many off-shell and unstable particles exist
(the resonances in particular).  The persistence of off-shell and unstable
particles means that subsequent interactions are not independent, making the 
interactions effectively many-body interactions.  
An example of this is the Landau-Pomeranchuk effect in a QED plasma.
Suppose an electron is knocked off-mass-shell in a collision.  In the vacuum,
it would radiate a bremsstrahlung photon after some ``formation-time.''  In a
dense plasma, it is possible for that electron to be struck again, before the
``formation-time'' has elapsed, both resetting ``formation-time'' clock and
ensuring that that photon is not radiated.  
Thus, the subsequent electron interactions depend on the previous history of 
that electron.  In both QMD and BUU type 
models, off shell evolution of unstable particles is included in some form by 
introducing a life-time parameter -- when particles live too long, they are
decayed.  In the vacuum, particles are all on-shell
and in momentum eigenstates so are spread over all configuration space.
So, the QPA and QCA together act to scatter the particles as though they
are in the vacuum, at least on the length scale of the interactions.  

There is one feature that can not be justified on the basis of non-equilibrium
field theory under any circumstances:  the closest approach criteria.
In practice, colliding particles when they are within the closest approach 
radius (i.e. $\sqrt{\sigma_{\rm TOT}/\pi}$) of
one another is a purely phenomenological and conceptually 
simple way to implement the collision integrals in the BUU equations.  At 
lower energies, use of this criteria causes no problems.  However, at higher 
energies the closest approach radius acquires a frame dependence 
leading to the causality violations noted by Kortemeyer, et al. 
\cite{tran:kor95}.  These
violations grow more severe as the closest approach radius approaches the 
mean-free path of the particles in the simulation.
Kortemeyer, et al. suggest several ways to avoid the causality violations
but their solutions require a brute force suppression of the collisions
that result in the violations rather than addressing the validity of the 
closest approach criteria.  In the end, it is not clear whether the 
use of the closest approach criteria is a valid way to determine if two 
particles can collide.

Can these approximations be applied at RHIC energies to make a RHIC
BUU model?  Well, primary hadronic collisions in a typical nuclear reaction 
at RHIC will occur at $\sqrt{s}\sim 200A$ GeV.  Such collisions are so violent 
that the {\em partons}, i.e. the quarks and gluons, comprising the hadrons 
will become deconfined.  If the energy density is too low, the partons will 
immediately hadronize and the collision will presumably proceed as lower 
energy collisions do.  However, if the energy density is high enough,  the 
partons will remain deconfined and should form a quark-gluon plasma (QGP)
\cite{QGP:esk93,QGP:har96,gFT:nik86}.  In either case, {\em we will need a 
transport model that can handle the partons}.  There have been 
several attempts to build a pure parton transport theory
\cite{tran:gei96,tran:hen95,tran:hen96,tran:bla94}, but each has their 
problems.  Chief among these problems is that one either treats the soft 
long-range phenomena\footnote{In the case of \cite{tran:bla94}, the long-range 
modes are collective thermal modes.} or one treats the hard short-distance 
phenomena,\footnote{In the case of \cite{tran:gei96} the short-range modes 
are large-$Q^2$ partons.} but never both in the same framework.  Perhaps, by 
relaxing the need for a scale separation (and hence the QCA), we may be able 
to treat {\em both} hard and soft modes on equal footing.  
Additionally, we must relax the Quasi-Particle Approximation to allow for 
many-particle effects, such as the Landau-Pomeranchuk effect, which depend on 
particles being off-shell.
Although the parton model is usually formulated with on-shell partons (i.e. in
the Quasi-Particle Approximation), it has been known for some time that 
a proper covariant treatment of partons requires that the partons be allowed to 
be off-mass-shell \cite{QCD:lan77,QCD:saw93}.  In fact, by allowing the partons
to be off-shell, one can account for the apparent violation of the Gottfried 
sum rule and the relative depletion of Drell-Yan pairs at high $x_F$ in nuclear 
targets \cite{QCD:saw93}.

We do not have a phase-space treatment for QCD, but we have made several 
steps toward developing one for QED.  In particular, we discuss how to 
create a transport 
theory for massless particles in Chapter~\ref{chap:transport} and we discuss
issues related to constructing the parton model in phase-space in 
Chapter~\ref{chap:pips}.  The main 
result of both chapters is that the phase-space densities are convolutions of a
phase-space source and a phase-space propagator.  This 
``source-propagator'' picture should lead to an improved transport theory for
the massless partons at RHIC as it can handle both soft and hard modes 
simultaneously.
The ``source-propagator'' picture is covariant, so does not suffer from 
the causality violations of standard transport approaches.
So, in the end we may not have all the answers for what a transport model at 
RHIC would be, but we have made several steps in the right direction.  

%
%
\section{Working Backward -- Experiment}

Being able to watch a system evolve on the computer definitely helps to
visualize the events during a collision, but it pales in comparison 
to directly {\em imaging} the reaction.  The technique of intensity
interferometry allows us to take a large step toward this goal.
Astronomers have long recognized the value of intensity 
interferometry.\footnote{and of interferometry in general} 
In fact, it was a pair of astronomers who
developed the technique:  Robert Hanbury-Brown and Richard Twiss
\cite{HBT:hbj52,HBT:hbt54,HBT:hbt56a,HBT:hbt56b,HBT:hbt56c,HBT:hbt57a,HBT:hbt57b}.  
The application of intensity interferometry to nuclear 
collisions followed a few years after Hanbury-Brown and Twiss's initial 
discovery \cite{HBT:gol59,HBT:gol60}.  Until 
recently the goal of nuclear interferometry did not differ greatly from
Hanbury-Brown and Twiss's original goal;  they measured the radius of Sirius
while we typically measure the radius of the relative emission profile of 
particle pairs (the {\em source function}) in heavy-ion
reactions.  We have recently demonstrated how to move beyond simply extracting
source radii to extracting the entire source function from the 
experimental data \cite{HBT:bro97,HBT:bro98}. 

Intensity interferometry is based on a fairly simple observation.  
If we have two possible events, say detection of one pion in detector 1 
and detection of a second pion in detector 2, the probabilities of each 
occurring are uncorrelated if
the probability of both happening is the product of the probability 
of each happening individually:
\[P_{12}=P_1 P_2.\]
On the other hand, if they are correlated then this is not true.  
In this case, we can define the correlation function as the ratio
\[C_{12}=\frac{P_{12}}{P_1 P_2}.\]
Then $C_{12}=1$ if the two events are totally uncorrelated.

\begin{figure}
	\centering
	\includegraphics[width=\textwidth]{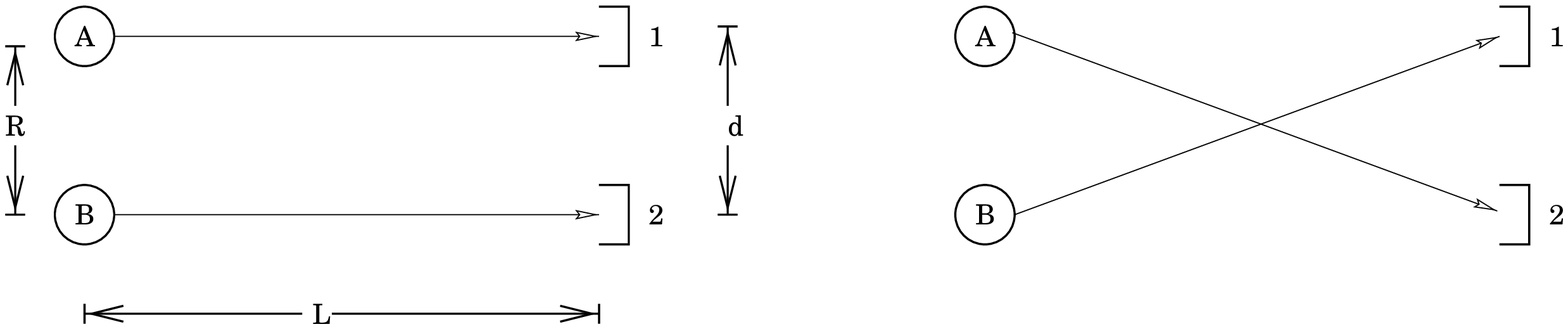}
	\caption[A simple example illustrating the HBT effect.] 
	{A simple example illustrating the HBT effect.}
	\label{fig:HBTeffect}
\end{figure}

It is useful to illustrate this point with an example.  This example is 
given in \cite{HBT:bay98} and pictured in Figure~\ref{fig:HBTeffect}.
In this figure, we have two sources of photons labeled A and B and two 
photon detectors labeled 1 and 2.  Now, suppose A and B emit the photons 
in spherical waves and bunched in time.\footnote{We bunch the photons in order 
to ensure that there is a short time scale on which the photons are 
correlated.} 
The amplitude for receiving a photon from A in detector 1 is
\[C_{A} \frac{\exp [{i r_{A1} k + i\phi_{A}}]}{r_{A1}}\]
so the total amplitude at detector 1 is (assuming $L\gg d,R$)
\[ A_1=\frac{1}{L}\left(C_A \exp[i r_{A1} k + i\phi_{A}]
+ C_B \exp [i r_{B1} k + i\phi_{B}]\right)\]
and similarly for detector 2.  Here, the phase $\phi$ is random and
changes with each bunch.
So, detector 1 receives photon hits with a probability (or intensity) of
\[\begin{split}
  I_1=|A_1|^2=&\frac{1}{L^2}\left( |C_A|^2+|C_B|^2
  + C_A^* C_B \exp [i((r_{B1}-r_{A1}) k + \phi_{B}-\phi_{A})]\right.\\
  &+ \left. C_A C_B^* \exp [-i((r_{B1}-r_{A1}) k + \phi_{B}-\phi_{A})]
  \right)
\end{split}\]
and similarly for detector 2.  Now averaging over many sets of bunches, each 
with different phases, we find the average intensity in any one detector to be
\[
  \expectval{I_1}=\expectval{I_2}=\frac{1}{L^2}
  \left(\expectval{|C_A|^2}+\expectval{|C_B|^2}\right).
\]
Now, if instead of averaging the intensities, we were to 
average the quantity $I_1 I_2$, we would
find that $\expectval{I_1 I_2}\neq \expectval{I_1}\expectval{I_2}$ as we would
expect for uncorrelated sources.  In fact, we would find 
\[
  \frac{\expectval{I_1 I_2}}{\expectval{I_1}\expectval{I_2}}=
  1+2\frac{\expectval{|C_A|^2}\expectval{|C_B|^2}}
  {\left(\expectval{|C_A|^2}+\expectval{|C_B|^2}\right)^2}
  \cos [k (r_{A1}-r_{A2}-r_{B1}+r_{B2})].
\]
This second term is the interesting one as it is purely quantum in origin 
and because it is sensitive to the separation of the two sources. 
In practice, the ratio $\expectval{I_1 I_2}/\expectval{I_1}\expectval{I_2}$
is defined as the correlation function $C_{12}$.

This example illustrates a simplified form of what Hanbury-Brown and Twiss 
\linebreak observed in their series of pioneering experiments into intensity 
interferometry \linebreak 
\cite{HBT:hbj52,HBT:hbt54,HBT:hbt56a,HBT:hbt56b,HBT:hbt56c,HBT:hbt57a,HBT:hbt57b}.  
The culmination of 
their experiments was the determination of the angular diameter of the star 
Sirius using a pair of World War II surplus searchlight mirrors and some
electronics \cite{HBT:hbt56a}.  The radius they found to be 
$0.0068''\pm 0.0005''$ or $3.1\times 10^{-8}$ radians.  At a distance 
of $2.7$ pc $= 8.5$ lyrs, Sirius is a mere $3.7$ times the radius of our sun.
As a result of their work, the technique of intensity interferometry became 
known as the HBT effect.

The HBT effect was first noted in subatomic physics in a study of the
Bose-Einstein symmetrization effects on pion emission in proton-antiproton
annihilation \cite{HBT:gol59}.  To explain the data Goldhaber, Goldhaber, Lee 
and Pais \cite{HBT:gol60} used a static source model for pion emission and
determined the effect on the angular distribution of pions due to 
symmetrization.  For a time, the effect was known as the GGLP effect.  While
Kopylev and Podgoretski first described how interferometry is sensitive
to the size of the emission region of a heavy-ion collision 
\cite{HBT:kop71,HBT:gri71a,HBT:kop72a,HBT:kop72b,HBT:kop73,HBT:kop74}, the GGLP
effect first was explained in terms of intensity interferometry by Shuryak 
\cite{HBT:shu73a,HBT:shu73b} and Cocconi \cite{HBT:coc74}.  In the late 70's 
and early 80's the ranks of particles where the HBT effect was observed grew 
to include protons, intermediate mass fragments (IMF's) and neutrons.  Today
interferometry is carried out with a wide array of particles, including (but 
not limited to) kaons, leptons, photons, and even pairs of unlike particles.

While the experimental community advanced dramatically in its ability to 
\sloppy\linebreak measure correlation functions, the amount of information 
gleaned from an individual correlation function has not advanced much at all;  
until recently, people still used the correlation function only to get the 
radius (or mean life-time) and $r=0$ intercepts of the source function. This 
is of course the
obvious thing to do because most correlation functions (at least for pions)
look like Gaussians after a Coulomb correction.  One can  
get a lot of information from a radius, especially when the data is
cut on the right kinematic variables.  Nevertheless, in many cases, one 
is replacing 50 points of a well defined function with one radius --
a great waste of information.  There are of course exceptions to this rule:
Pratt \cite{HBT:pra90} developed a series of codes to convert output
from transport models to correlation functions which can then be directly 
compared to data.  Beyond this, it is only lately that people have tried to 
go beyond simply fitting a radius:  Nickerson, Cs\"{o}rg\H{o} and Kiang
have attempted fitting correlation functions to a Gaussian plus an 
exponential halo \cite{HBT:nic97} and Wiedermann and Heinz
have proposed performing a moment expansion of the correlation functions
\cite{HBT:wie96}.  While both of these are valuable exercises, 
neither make full use of the data.  This is why my advisor, 
Pawe{\l} Danielewicz, and I proposed doing {\em nuclear imaging}.

Nuclear imaging amounts to reconstructing the relative emission 
distribution\footnote{a.k.a. the source function}
for the pairs used to construct the correlation function.
Imaging relies on the observation that the source function and the
correlation function are related through a simple integral equation which
can be inverted.  We discuss the methods for doing this inversion and the 
results we get from the inversion in Chapter~\ref{chap:HBT}.

%
%
\section{Overview of Thesis}

This thesis is about the tools we use to investigate the space-time 
developments of heavy-ion reactions.  Specifically, we discuss the 
application of transport theory to parton dominated collisions at RHIC and the
LHC and we discuss the application of imaging techniques to HBT intensity
interferometry.  Both lines of research are aimed at deducing the
features of the phase-space particle densities and how they evolve as a
heavy-ion collision proceeds.  


In the second chapter, we describe some of the things needed to derive a QCD 
transport theory.  We begin by defining the phase-space particle densities as 
understanding their time-evolution is the ultimate goal when building 
transport models.  We then describe the contour Green's functions and the 
other Green's functions that we need to perform actual calculations of the 
densities.  We will then spend the rest of the chapter deriving a QED 
transport theory valid for massless photons and electrons but without applying 
either the Quasi-Particle or the Quasi-Classical (or gradient) approximations. 
In doing so, we derive the Generalized Fluctuation-Dissipation theorem -- 
proving that a phase-space density is the convolution of a phase-space
source density with a phase-space propagator.   This is a general result, 
applicable to QCD as well as to particles with mass.  
We use this theorem to create QED phase space evolution equations and to 
illustrate the perturbative solution for the photon and electron densities 
that we discuss in Chapter~\ref{chap:pips}.
Finally, we describe how one might make a transport model based on these 
results.

We will need phase-space parton densities for input in a parton transport 
model so in the third chapter we begin the process of recasting the parton 
model in phase-space.  The parton model has two key components: 
factorization of the cross sections and evolution of the parton densities.  
The first component has
a QED analogy in the Weizs\"acker-Williams approximation: in the 
Weizs\"acker-Williams approximation, the cross section for a photon mediated 
process is the convolution of a photon density and the cross section for the 
photon induced sub-process.  We rederive the Weizs\"acker-Williams 
approximation in phase-space, applying it to both the photon cloud and 
electron clouds of a point charge.  In the process, we not only illustrate how 
to rewrite things in phase-space, but we investigate the roles of phase-space 
sources and propagators.  The second component, evolution of the parton 
densities, can also be investigated in QED.  The renormalization group 
evolution of the parton densities is equivalent to the summation of a class
of ladder diagrams in the Leading Logarithm Approximation.  We can study a 
simplified version of the ladder diagrams in QED by calculating the photon
and electron phase-space densities around a point charge.  Indeed, the 
electron and photons surrounding a point charge are point-like 
``constituents'' of the point charge, so they earn the right to be called 
QED partons.    
Finally, with all of this QED experience, we discuss 
the main features of the phase-space parton distribution of a nucleon in the 
Leading Logarithm Approximation.  

In the fourth chapter, we describe how nuclear imaging can be used to extract
the source function from correlation data.  We list the methods we have used
to perform this inversion, describing what works and how and why they work.
In particular, we will discuss the role of constraints in the stabilization 
of the inversion and the method Pawe{\l} Danielewicz developed which works 
well even without constraints.  Next, we discuss what
other quantities can be extracted from the images. 
Finally, we apply our inversion techniques to
various data sets and discuss what the images mean.  

We will conclude the thesis with a brief summary and a description of what work
needs to be done to follow up the lines of investigation opened here. 

There are also several appendices exploring various side and technical 
issues.  
Among these are discussion and derivations of the phase-space propagators, 
the measurables in a heavy-ion collision, the 
cross section in terms of phase-space densities, the Coulomb field
of a static point charge in phase-space, the gauge dependence of the 
photon distributions from Chapter \ref{chap:pips} and both wavepackets 
in phase-space and current densities in phase-space.

Throughout this thesis, we use natural units ($\hbar=c=1$) when
convenient, but we insert factors of $\hbar c$ when we need an energy or 
length in conventional units.  The signature of the metric tensor is 
$(+,-,-,-)$.

%
%


\chapter{EVOLUTION OF PHASE-SPACE PARTICLE DENSITIES}
\label{chap:transport}

How can we go beyond standard transport methods to produce a transport-like 
theory for massless (or nearly massless) particles in ultra-relativistic 
heavy-ion collisions?  However we do it, this theory must allow for 
off-shell evolution of the particle densities so we must relax the 
Quasi-Particle Approximation.  This theory must also deal with all hard 
and soft modes on equal footing.  Since the modes act on different length 
and time scales, the theory must not require scale separation or 
use the Quasi-Classical
Approximation.  The way we produce our transport-like theory is to go 
back to the original work on transport, identify where the QCA and QPA 
approximations are made, and replace them with more suitable approximations.  
We use QED for this study because it is not as complicated as QCD but 
still contains many of the relevant features of QCD.  By not making the 
standard approximations in our study of QED transport, we will find that the 
densities take a ``source-propagator'' form -- meaning a phase-space density 
is the convolution of a phase-space source density and a phase-space 
propagator.  As the reader will see, the phase-space source gives the 
quasi-probability\footnote{Strictly 
speaking, neither the phase-space sources nor propagators are true 
probabilities as they can be negative.  As with any other Wigner-transformed 
\sloppy\linebreak quantities, they must be smoothed over small phase-space 
volumes to render them positive definite.} density for creating a particle with 
a particular momentum.  The propagator then sends this particle from its 
creation point across a space-time displacement to the observation point.
This ``source-propagator'' form is a general result, not specific to QED, 
and it is encoded in the key result 
of this chapter: the phase-space Generalized Fluctuation-Dissipation Theorem.

To begin, we define the initial state through the density matrix and the 
particle phase-space densities as expectation values with the density matrix.
Because of the general nature of the density matrix we can simultaneously 
investigate anything from single states to ensembles of states.  The densities 
themselves are Wigner transforms of two-point functions such as 
$G^{\gtrless}(x_1,x_2)$.  We will discuss these densities, how they relate
to other possible definitions of the particle densities, and how one normally 
implements the Quasi-Particle Approximation.

To perform practical calculations of the 
$G^\gtrless$, we introduce the contour Green's functions.  These 
Green's functions are defined on a contour in the complex time 
plane.  By restricting the arguments of the contour Green's function to 
various branches of the contour, we can define other auxiliary Green's
functions such as Feynman's Green's functions.  
At this point, we also introduce the retarded Green's functions as they
will play a dominant role in later discussions.  The introduction of the 
complex time contour also leads to a simple scheme for perturbatively 
calculating the contour Green's functions.  This in turn leads to the 
Dyson-Schwinger equations which are the starting point for the derivation 
of the semi-classical transport equations.

After these preliminaries, we begin examining QED transport theory 
for massless particles.  In Section 
\ref{chap:transport}, we follow essentially the standard 
semi-classical transport equation derivation: we derive the
Kadanoff-Baym equations and formally solve them to get the Generalized
Fluctuation-Dissipation Theorem.  Unlike conventional derivations of transport 
theory, at this point we do not make the Quasi-Classical Approximation.
This approximation amounts to ignoring small-scale structure of the particle 
phase-space densities, resulting in much simpler collision integrals 
\cite{neq:dan84a,tran:mro94}.  By not making this approximation, 
we arrive at the Generalized
Fluctuation-Dissipation Theorem which codifies the ``source-propagator''
picture of the particle densities.  Crucial inputs to the theorem are the 
phase-space sources; we will discuss how to calculate them.  

With the sources and the Generalized
Fluctuation-Dissipation Theorem, we derive a set of phase-space QED
evolution equations.   These evolution equations describe the
evolution of the system in phase-space from the distant past to the present,
including all splittings, recombinations and scatterings. 
Furthermore, we can expand these evolution equations to get the
lowest order contributions to the particle densities or we can differentiate 
the evolution equations to get transport equations.  All of these results
are manifestly Lorentz covariant so do not suffer from the causality violation
of a more traditional approach.  However, our investigation is not as 
mature as conventional transport approaches and we are not at the stage where 
we can make quantitative predictions.

For those familiar with the common steps in deriving semi-classical transport 
equations from the Kadanoff-Baym equations, we suggest skipping past 
Section \ref{sec:standardstuff} to Section \ref{sec:newstuff}.

\section{The Density Matrix}

The systems that we study range from the very simple, i.e. binary 
collisions, to the very complex, i.e. interacting heavy-ion systems.  We
could deal with the potential complexity right up front by specifying the
incoming states, or we could cleverly
lump the complexity into a density matrix.  We choose the latter because it 
is both more general and simpler to do.

First we write the density matrix $\hat{\rho}$ as
\begin{equation}
   \hat{\rho}=\sum_{{\rm all\; states}}\ket{m}\bra{n}\rho_{mn}.
\end{equation}
Here, the states $\ket{m}$ and $\ket{n}$ can correspond to single particle
or many particle states, depending on the system of interest.  

Because the density matrix is so general, we can treat many different 
situations at the same time, all within one general framework.
For example, we can easily incorporate
a thermal population of states for work with infinite thermalized nuclear 
matter.  As a second example, we can account for correlated initial and 
final states as well as bound states using an appropriate choice of 
density matrix.  As a final example, 
we can choose suitable density matrices to give us
localized wavepackets of single particles in the initial state.
It is this last reason that we will take advantage of in this thesis.
In every subsequent chapter, the particles we consider will be localized in 
space and or momentum (or both!).  
In fact, we demonstrate how to make a wavepacket localized in phase-space in 
Appendix \ref{app:wavepacket}.  In this chapter and the following chapter 
we create particle phase-space densities that are 
localized both in momentum and coordinates and in Chapter \ref{chap:HBT} we 
will measure the localized sources of
particles created in heavy-ion reactions.

For now, we leave the density matrix in this general form.  The only condition
that we place on it is that we can perform a Wick decomposition on general 
expectation values.  This requirement is important for creating a 
perturbation expansion of the expectation values \cite{neq:dan84a}.  
The discussion of what density matrices allow a Wick decomposition is carried 
out elsewhere \cite{neq:dan84a,neq:cho85,tran:sch93}.

Now, in terms of this density matrix, we can define an arbitrary expectation 
value of an Heisenberg picture operator:
\begin{equation}
   \expectval{\Oper_H}=\frac{{\rm Tr}\left(\hat{\rho}\Oper_H\right)}
	{{\rm Tr}(\hat{\rho})}.
\end{equation}
As a simple example of both a density matrix and an expectation value, consider
the density matrix containing only the vacuum state, $\ket{0}$:
$\hat{\rho}=\ket{0}\bra{0}$.  The trace over this density matrix gives
the vacuum expectation value of the operator
\[
   \expectval{\Oper_H}=\frac{{\rm Tr}\left(\hat{\rho}\Oper_H\right)}
	{{\rm Tr}(\hat{\rho})}=\frac{\ME{0}{\Oper_H}{0}}{\OverLap{0}{0}}.
\]

\section{Particle Phase-Space Densities}

Since our ultimate goal is to follow the phase-space densities, we would like 
to define them precisely.  We will define them
as the Wigner transforms of certain two-point functions.  These two-point
functions, also known as the $>$ and $<$ Green's functions are for scalar 
bosons: 
\begin{align}
\daveseqn
   i\Ggreat{x}{y}  = & \left< \hat{\phi}(x)\hat{\phi}^{*}(y) \right>
	\label{eqn:gldensitya}\davetag{a}\\
   i\Gless{x}{y}  = & \left< \hat{\phi}^{*}(y)\hat{\phi}(x) \right>,
	\label{eqn:gldensityb}\davetag{b}\\
\intertext{for vector bosons (such as photons):}
   i\Dgreat{\mu}{\nu}{}{}{x}{y}  = & \left< \hat{A}_{\mu}(x) \hat{A}_{\nu}(y)
	\right> - \left< \hat{A}_{\mu}(x) \right> \left< \hat{A}_{\nu}(y) 
	\right>\label{eqn:gldensityc}\davetag{c}\\
   i\Dless{\mu}{\nu}{}{}{x}{y}  = & \left< \hat{A}_{\nu}(y)
	\hat{A}_{\mu}(x) \right> - \left< \hat{A}_{\mu}(x) \right> 
	\left< \hat{A}_{\nu}(y) \right>,\label{eqn:gldensityd}\davetag{d}\\
\intertext{and for fermions (such as electrons or nucleons):}
   i\Sgreat{\alpha}{\beta}{}{}{x}{y}  = & \left< \hat{\psi}_{\alpha}(x)
	\hat{\bar{\psi}}_{\beta}(y) \right>\label{eqn:gldensitye}\davetag{e}\\
   i\Sless{\alpha}{\beta}{}{}{x}{y}  = & - \left< \hat{\bar{\psi}}_{\beta}(y)
	\hat{\psi}_{\alpha}(x) \right>.\label{eqn:gldensityf}\davetag{f}
\end{align}
The field operators in these expressions are taken in the Heisenberg picture.
Note that, because of the equal time com\-mu\-ta\-tion re\-la\-tions for the 
interaction picture operators, if we write the above in the interaction 
picture, we find $\Ggreat{x}{y}= \Gless{y}{x}$ for both fermions and bosons.

These Green's functions are hermitian and they contain the complete 
single-particle information of the system.
For example, setting $x=y$ gives us the single particle density matrix.
Furthermore, Wigner transforming in the relative coordinate, we find the
off-mass shell generalization of the Wigner function for the particles -- in 
other words, the phase-space density.  Let us demonstrate for scalar fields:
\begin{equation}
\begin{split}
	f(x,p)&=i\Gless{x}{p}\\ &=\int\dn{4}{(x-y)}e^{i(x-y)\cdot
	p}i\Gless{x}{y}\\ &=\int\dn{4}{(x-y)}e^{i(x-y)\cdot
	p}\left< \hat{\phi}^{*}(y)\hat{\phi}(x)\right>.
\end{split}
\label{eqn:DefofDensity}
\end{equation}

We identify $f(x,p)$ with the number density of particle 
(or antiparticles) per unit volume in phase-space per unit invariant mass 
squared at time $x_0$:
\[
   f(x,p)=\frac{dn(x,p)}{d^3x \:d^3p \:dp^2}
\]
In particular, for $p_0>0$, $i\Gless{x}{p}$ is associated with the particle 
densities and $i\Ggreat{x}{p}$ is associated with the hole 
density.  For $p_0<0$, $i\Gless{x}{p}$ is the anti-hole density and 
$i\Ggreat{x}{p}$ is the anti-particle density.  The photon and electron 
densities are defined in the same way, however because of their more 
complicated spin structure, their Wigner functions carry indices.

Other, gauge invariant, definitions of the particle densities exist in the 
literature \cite{tran:zhu96,tran:vas87,tran:elz86a,tran:elz86b}.  However, 
while these distributions are gauge 
invariant, they do not obey simple Dyson-Schwinger equations and so it 
is difficult to derive transport theory from them.  Since 
all of the observables in which we are interested are gauge invariant and 
all of the equations involving the densities
that we derive are gauge covariant, we do not need to resort
to exotic definitions of the densities.

The off-shell Wigner function is related to the 
conventional Wigner function, $f_0(x,\vec{p})$, through the invariant mass
integration:
\begin{equation}
   f_0 (x,\vec{p})=\frac{dn(x_0,\vec{x},\vec{p})}{d^3x \:d^3p}=
   \int^\infty_{-\infty} dp^2 f(x,p).
\label{eqn:NormalWignerFtn}
\end{equation}
In the Quasi-Particle Approximation we assume that 
\begin{equation}
f(x,p)\approx \begin{cases}
	f_0 (x,\vec{p})\delta(p^2-m^2_*) & \text{for $p_0>0$}\\
	-(1-f_0 (x,\vec{p}))\delta(p^2-m^2_*) & \text{for $p_0<0$.}
\end{cases}
\end{equation}
In this approximation (which is quite a common approximation in 
transport-like models), $f(x,p)$ and $f_0(x,\vec{p})$ are interchangeable.
Here $m_*$ is the effective mass of the particle and it may be either the 
mass in free space or it may contain in-medium modifications.

Now, finding the particle densities are the ultimate goal of 
our work.  We will describe several ways to calculate them in the following 
chapters.  To this end, we will need several of the Green's functions in the 
next subsection.  Also, given that we measure the densities in any experiment, 
we discuss particle spectra (basically the momentum space density of 
particles) in great detail in the Appendix \ref{append:spectra}.

\section{Other Green's Functions}

In order to calculate the densities, we will need to introduce several other 
Green's functions.  The first of these, the contour Green's functions, are 
the most exotic as they are defined on a contour on the complex time plane.  
To see why such a contour is useful, we will first discuss the expectation
value of an arbitrary Heisenberg operator $\Oper_H(t)$.  With an understanding
of why this contour is used, we will define the contour Green's 
functions and all of the auxiliary Green's functions that the contour Green's 
functions encapsulate.

\subsection{Operator Expectation Values}

Consider the expectation value of an Heisenberg picture operator with one 
time argument:
\begin{equation}
   \expectval{\Oper_H(t)}=\frac{{\rm Tr}\left(\hat{\rho}\Oper_H(t)\right)}
	{{\rm Tr}(\hat{\rho})}.
\end{equation}
This operator could be anything from the energy density of the system to 
the number operator of a specific field provided that is a function of one 
time variable only.

\begin{figure}
	\centering
	\includegraphics[width=7.5cm]{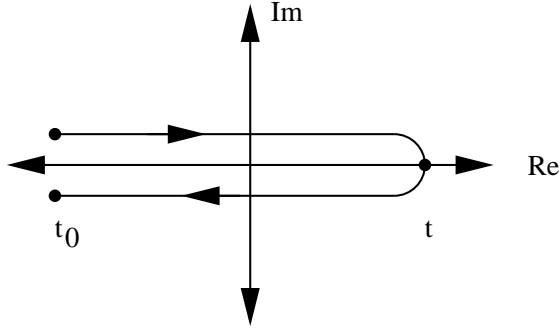}
	\caption[Contour in the complex time plane for evaluating operator 
	expectation values.  The upper branch corresponds to causal ordering 
	and the lower branch to anti-causal ordering. The arrows denote the 
	contour ordering enforced by the $\rm T$ operator.]
	{Contour in the complex time plane for evaluating operator 
	expectation values.  The upper branch corresponds to causal ordering 
	and the lower branch to anti-causal ordering. The arrows denote the 
	contour ordering enforced by the $\rm T$ operator.}
	\label{fig:contour}
\end{figure}

The simplest way to evaluate this operator is to rewrite the operator in 
the interaction representation.  Once in the interaction picture, we can 
perform a perturbative expansion of the time evolution operator and develop 
successive approximations to the expectation value.  The relation between an 
operator in the Heisenberg and \sloppy\linebreak interaction pictures is 
\begin{equation}
   \Oper_H(t)=\hat{U}(t_0,t)\Oper_I(t)\hat{U}(t,t_0)
\end{equation}
where $\hat{U}(t,t_0)$ is the interaction picture time evolution operator 
and $t_0$ is the time at which the two pictures coincide.  For $t>t_0$, the 
evolution operator is given in terms of the interaction part of the 
Hamiltonian in the interaction representation by
\begin{equation}
   \hat{U}(t,t_0)={\rm T}^c\left[\exp\left(-i\int^t_{t_0} dt' 
	\hat{H}_I^I(t')\right)\right].
\end{equation}
The operator ${\rm T}^{c(a)}$ simply orders the operators in the 
expectation value in a chronological (or anti-chronological) fashion.  In 
other words:
\begin{align}
\daveseqn
{\rm T}^c\left(\hat{A}(t_1) \hat{B}(t_2)\right)&=\theta(t_1-t_2)\hat{A}(t_1) 
	\hat{B}(t_2)+\theta(t_2-t_1)\hat{B}(t_2) \hat{A}(t_1)
	\davetag{a}\\
{\rm T}^a\left(\hat{A}(t_1) \hat{B}(t_2)\right)&=\theta(t_1-t_2)\hat{B}(t_2) 
	\hat{A}(t_1)+\theta(t_2-t_1)\hat{A}(t_1) 
	\hat{B}(t_2)\davetag{b}
\end{align}
So, we can write the expectation value of $\Oper_H(t)$ in the interaction 
picture as follows:
\begin{equation}\begin{array}{rl}
   \displaystyle\expectval{\Oper_H(t)}
	&=\displaystyle\expectval{\hat{U}(t_0,t)\Oper_I(t)\hat{U}(t,t_0)}\\
	&=\displaystyle\expectval{{\rm T}^a\left[
	\exp\left(-i\int^{t_0}_t dt' \hat{H}_I^I(t')
	\right)\right] \Oper_I(t) {\rm T}^c\left[
	\exp\left(-i\int^t_{t_0} dt' \hat{H}_I^I(t')
	\right)\right]}
\end{array}\label{eqn:longExpVal}\end{equation} 
Notice that the time ordering goes as follows from right to left:  
the rightmost time evolution operator takes things from $t_0$ forward 
in time to $t$ where the operator is evaluated then the second time 
evolution operator takes things from $t$, backwards in
time again to $t_0$.  We can simplify notation by introducing a contour 
in the complex time plane which runs from $t_0$ up to $t$ and back again 
to $t_0$ as shown in Figure~\ref{fig:contour}.

We can define ordering along this contour using the contour time-ordering 
operator, T, defined via
\begin{equation}
	{\rm T}\left(\hat{A}(t_1) \hat{B}(t_2)\right)=
	\theta(t_1,t_2)\hat{A}(t_1) \hat{B}(t_2)+
	\theta(t_2,t_1)\hat{B}(t_2) \hat{A}(t_1)
\end{equation}
where the contour theta function is given by
\[\theta(x_0,y_0) = \left\{
	\begin{array}{ll}
	1 & \mbox{if $x_0$ is later on the contour than $y_0$}\\
	0 & \mbox{otherwise}
	\end{array}
\right. \]
Using this notation, Equation~\eqref{eqn:longExpVal} simplifies to
\begin{equation}
	\expectval{\Oper_H(t)}=\expectval{{\rm T}\left[\exp\left(
	-i\int_{\cal C} dt' \hat{H}_I^I(t')
	\right)\Oper_I(t)\right]}
\label{eqn:OpExpVal_intpict}
\end{equation}
This idea of a contour that zig zags back and forth along the real time axis, 
encapsulating the various time orderings needed in an expectation value was 
first noticed by Schwinger \cite{neq:sch61}.
The idea was generalized by Danielewicz~\cite{neq:dan90} to 
account for operators with multiple time arguments.

Now we will not actually demonstrate how to solve for 
$\expectval{\Oper_H(t)}$, although it is discussed in several places, notably 
\cite{neq:sch61} and \cite{neq:fet71}.  However, our discussion is a 
useful motivation for the introduction of the contour ordering that 
we use to define the contour Green's functions in the next section.

\subsection{Contour Green's Functions}
\label{sec:greens}

Introducing the time ordering along a contour in the complex time 
plane is a clever way to express expectation values and
what makes the ordering so clever is the way the two branches
encode causal or anti-causal time orderings.  Let us take advantage of this 
feature and define the contour Green's functions as Green's functions that 
are ordered along the time contour:
\begin{equation}
\daveseqn
	i\Gprop{x}{y}=\left< T \hat{\phi}(x) \hat{\phi}^{*}(y) \right>
\label{eqn:contgreens1}\davetag{a}
\end{equation}
for scalar particles,
\begin{equation}
	i\Dprop{\mu}{\nu}{}{}{x}{y}=\left< T \hat{A}_{\mu}(x) 
	\hat{A}_{\nu}(y) \right>-\left< \hat{A}_{\mu}(x) \right>
	\left< \hat{A}_{\nu}(x) \right>
\label{eqn:contgreens2}\davetag{b}
\end{equation}
for photons and
\begin{equation}
	i\Sprop{\alpha}{\beta}{}{}{x}{y}=\left< T \hat{\psi}_{\alpha}(x) 
		\hat{\bar{\psi}}_{\beta}(y) \right>
\label{eqn:contgreens3}\davetag{c}
\end{equation}
for fermions.  For practical purposes, we must take the lower limit of the 
contour as $t_0\rightarrow -\infty$, where we specify the initial conditions 
in the density matrix.  Furthermore, we must take the upper limit of the 
contour as $t\rightarrow \infty$ to ensure that all of the time arguments 
of the contour Green's functions are between the limits $t_0$ and $t$ and thus
are on the contour.

All of the above Green's functions can be written in the interaction
picture in a manner analogous to Equation \eqref{eqn:OpExpVal_intpict}.  From 
this, Danielewicz \cite{neq:dan84a} has derived the set of Feynman rules for 
evaluating the contour Green's functions.  These rules differ slightly from
the Feynman rules for the S-matrices found in most field theory books, so we
tabulate the QED rules in the next section.  

The contour Green's function can be written in terms of the $\gtrless$ 
Green's functions as
\begin{equation}
   \Gprop{x}{y}=\theta(x_0,y_0)\Ggreat{x}{y}+\theta(y_0,x_0)\Gless{x}{y}
\label{eqn:greenrelation1}
\end{equation}
for both fermions and bosons.  By virtue of this, we have the relation
$\Gprop{x}{y}=\Gprop{y}{x}$.

\subsection{Contour Feynman Rules for Quantum Electrodynamics}
\label{sec:lagrangian}

In order to evaluate the contour Green's functions in the interaction picture,
we need a set of Feynman rules for these Green's functions.  These rules have 
been derived previously \cite{neq:dan84a} so we may just state them here.
The Feynman rules we state are the QED Feynman rules.  A similar set may be
written down for QCD or any other field theory.  
We use the field normalization conventions of \cite{gFT:akh65}.

The Feynman rules for the evaluation of the QED contour Green's functions in 
the interaction picture are: 
\begin{enumerate}
\item The vertex Feynman rules are summarized in Table \ref{table:vertex}.
\item The contour propagators are summarized in Table \ref{table:props}. 
\item Every closed fermion loop yields a factor of $(-1)$.
\item Every single particle line that forms a closed loop or is linked by the
        same interaction line yields a factor of $iG^{<}$.
\end{enumerate}
Notice that the second scalar coupling is second order in the coupling constant
while the rest of the couplings are of first order.   
\begin{table}
\caption[The vertex Feynman rules for scalar and spinor QED.]
{The vertex Feynman rules for scalar and spinor QED.}
\vspace*{4.5mm} 
\begin{center}
\begin{tabular}{|c|c|c|}
   \hline
      \parbox[b]{1.5in}{\hspace*{.1in}3~point \\ \hspace*{.1in} photon-scalar 
        \\ \hspace*{.1in} vertex \vspace*{.3in}} &
      \includegraphics{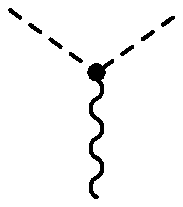} &
      \parbox[b]{1.7in}{\hspace*{.1in}
        $eZ\bothpartial_{\mu}=eZ(\leftpartial_{\mu}-
         \rightpartial_{\mu})$\vspace*{.3in}}\\
   \hline
      \parbox[b]{1.5in}{\hspace*{.1in}4~point \\ \hspace*{.1in}photon-scalar 
        \\ \hspace*{.1in} vertex \vspace*{.3in}} &
      \includegraphics{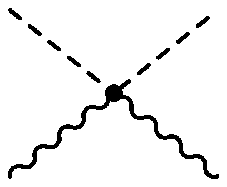} &
      \parbox[b]{1.7in}{\hspace*{.1in}$2ie^2Z^2g_{\mu\nu}$ \vspace*{.3in}}\\
   \hline
      \parbox[b]{1.5in}{\vspace*{.25in}\hspace*{.1in}fermion-photon \\ 
        \hspace*{.1in}vertex \vspace*{.3in}} &
      \includegraphics{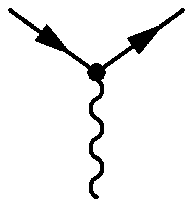} & 
      \parbox[b]{1.7in}{\hspace*{.1in}$-ie\gamma_{\mu}$ \vspace*{.3in}}\\
   \hline
\end{tabular}
\end{center}
\label{table:vertex}
\end{table}
\begin{table}
\caption[The contour scalar, photon, and electron propagators.]
	{The contour scalar, photon, and electron propagators.}
\vspace*{4.5mm} 
\begin{center}
\begin{tabular}[b]{|c|c|l|}
   \hline
      	scalar line &
      	\includegraphics{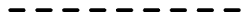} & 
      	$\displaystyle G(x_1,x_2)$\\
   \hline
      	photon line &
      	\includegraphics{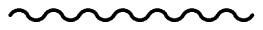} & 
	$\displaystyle D_{\mu\nu}(x_1,x_2)=4\pi g_{\mu\nu}G(x_1,x_2)$\\ 
   \hline
      	fermion line &
      	\includegraphics{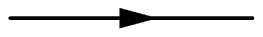} & 
      	$\displaystyle S_{\alpha\beta}(x_1,x_2)
		=-(-i\dirslash{\partial}+m)_{\alpha\beta}G(x_1,x_2)$\\ 
   \hline
\end{tabular}
\end{center}
\label{table:props}
\end{table}

\subsection{Auxiliary Green's Functions}

We define several auxiliary Green's functions in terms of the $>$ and $<$ 
Green's functions:  the retarded and advanced Green's functions and the 
Feynman and anti-Feynman propagators, and the spectral function.   
For the scalar particle retarded and advanced propagators, we have
\begin{equation}
   \Gmisc{x}{y}{\pm}=\pm\theta(\pm(x_0-y_0))(\Ggreat{x}{y}-\Gless{x}{y}).
\label{eqn:greenrelation2}
\end{equation}

For the Feynman and anti-Feynman propagators, we have: 
\begin{align}
   \daveseqn
   \Gcaus{x}{y}=&\theta(x_0-y_0)\Ggreat{x}{y}+\theta(y_0-x_0)\Gless{x}{y},
      \label{eqn:greenrelation3}\davetag{a}\\
   \Gacaus{x}{y}=&\theta(y_0-x_0)\Ggreat{x}{y}+\theta(x_0-y_0)\Gless{x}{y}.
      \label{eqn:greenrelation4}\davetag{b}
\end{align}
One can also obtain these Feynman and anti-Feynman propagators by restricting
the arguments of the contour propagators to be on one side of the contour in 
Figure~\ref{fig:contour}.  Finally, the Feynman and anti-Feynman propagators 
can also be written down as time ordered (or anti-time ordered) expectation 
values of the fields.  We state only those for the Feynman propagators: 
\begin{align}
\daveseqn
i\Gcaus{x}{y}  = & \left< T^{c} \phi(x)\phi^{*}(y) \right>\davetag{a}\\
i\Dcaus{\mu}{\nu}{}{}{x}{y}  = & \left< T^{c} A_{\mu}(x) A_{\nu}(y)
   \right> - \left< A_{\mu}(x) \right> \left< A_{\nu}(y) \right>\davetag{b}\\
i\Scaus{\alpha}{\beta}{}{}{x}{y}  = & \left< T^{c} \psi_{\alpha}(x)
   \bar{\psi}_{\beta}(y) \right>\davetag{c}
\end{align}

Finally, we have the spectral function:
\begin{equation}
\begin{split}
   \Gmisc{x}{y}{s}&=i\left(\Ggreat{x}{y}-\Gless{x}{y}\right)\\
	   &=\deltaftn{3}{\vec{x}-\vec{y}} 
	     \quad\text{ if $x_0=y_0$}
\end{split}
\end{equation}
Notice that the retarded and advanced Green's functions are actually the 
spectral function multiplied by a theta function:
\begin{equation}
\Gmisc{x}{y}{\pm}=\pm\theta(\pm(x_0-y_0))(\Gmisc{x}{y}{s}).
\label{eqn:ThatEqn}
\end{equation}
The spectral function can also be written as the expectation value of the 
(anti-)commutators of the (fermion)boson fields.

The Wigner transform of the spectral function plays an interesting 
role in transport.  By virtue of Equation~\eqref{eqn:ThatEqn}, the spectral function
determines how particles propagate.  Furthermore, given the interpretation of 
$G^\gtrless (x,p)$ in terms of particle and hole densities, the spectral 
density is the hole density minus the particle density.  In the Quasi-Particle
Approximation the spectral function also determines how far off shell particles
can get.  For example, in the most common implementation of the Quasi-Particle 
Approximation, 
\begin{equation}
   \Gmisc{x}{p}{s}=\sgn{p_0}\delta (p^2-m^2_*).
\end{equation}
Given that particles are rarely truly on shell in a nuclear reaction
(except in the final state), various schemes have been developed 
to accommodate the broadening of  
the spectral function.  Rather than go into them, we will keep away from the
Quasi-Particle Approximation when possible.

All of these auxiliary Green's functions will get used one way or another in 
the following chapters.  The Feynman propagators will get used 
in the perturbative expansion of the S-matrix in Chapter~\ref{chap:pips}
and are discussed in Appendix \ref{append:prop}.  On the other hand, the 
retarded functions are
used extensively in this chapter as they are most convenient
for deriving the Generalized Fluctuation-Dissipation Theorem and transport 
theory.  The spectral function, however, is rarely used in this work since it 
is used most often in the justification of the Quasi-particle Approximation.  

\section{Conventional Transport Theory}
\label{sec:standardstuff}

In this subsection, we follow the standard derivation of the transport
equations up to the point where we find the Generalized
Fluctuation-Dissipation Theorem.  The procedure is as follows: 1) find the
Dyson-Schwinger equations for the contour Green's functions, 2) apply the free
field equations of motion to get the Kadanoff-Baym equations and 3) solve the
Kadanoff-Baym equations to get the Generalized Fluctuation-Dissipation 
Theorem.

\subsection{Dyson-Schwinger Equations}

\begin{figure}
   \begin{center}
   \includegraphics{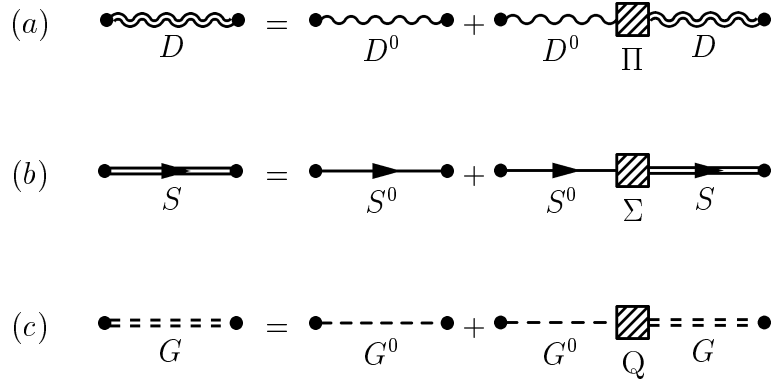}
   \end{center}
   \caption[The Dyson-Schwinger equations for the propagators.	
	Double lines represent the dressed Green's functions
        and single lines represent the non-interacting Green's functions.  
	The particle self-energies are the large square vertices.]
	{The Dyson-Schwinger equations for the propagators.  
	Double lines represent the dressed Green's functions
        and single lines represent the non-interacting Green's functions.  
	The particle self-energies are the large square vertices.}
        \label{fig:DSE}
\end{figure}

The Dyson-Schwinger equations encapsulate all of the nonperturbative effects 
in the field theory that that can be described at the level of two-point 
functions.\footnote{In other words, the nonperturbative effects that that can 
be described without resorting to three-point functions or higher order 
correlations.}  Using the Feynman rules for the QED contour Green's functions 
in Section \ref{sec:lagrangian}, we can write 
the Dyson-Schwinger equations for the photon, electron and scalar contour 
Green's functions:
\begin{align}
\daveseqn
\Gprop{1}{1'} & =  \Gnon{1}{1'} + 
      \intc d2\: d3\: \Gnon{1}{2} Q(2,3) \Gprop{3}{1'} 
      \label{eqn:DSE1} \davetag{a}\\
\Dprop{\mu}{\nu}{}{}{1}{1'} & =  \Dnon{\mu}{\nu}{}{}{1}{1'} + 
      \intc d2\: d3\: \Dnon{\mu}{\mu'}{}{}{1}{2} \Pi^{\mu'\nu'}(2,3) 
      \Dprop{\nu'}{\nu}{}{}{3}{1'} \label{eqn:DSE2} \davetag{b}\\
\Sprop{\alpha}{\beta}{}{}{1}{1'} & =  \Snon{\alpha}{\beta}{}{}{1}{1'} + 
      \intc d2\: d3\: \Snon{\alpha}{\alpha'}{}{}{1}{2} 
      \Sigma^{\alpha'\beta'}(2,3) \Sprop{\beta'}{\beta}{}{}{3}{1'} 
      \label{eqn:DSE3}\davetag{c}
\end{align} 
In these equations, we represent the coordinates by their index, 
i.e. $x_1\rightarrow 1$ and the time integrals are taken along the contour in
Figure \ref{fig:contour}.  We present the corresponding diagrams in 
Figures~\ref{fig:DSE}(a-c).  In Equations \eqref{eqn:DSE1}-\eqref{eqn:DSE3}, 
the non-interacting contour Green's functions have a $0$ superscript.

\begin{figure}
   \begin{center}
   \includegraphics[width=\textwidth]{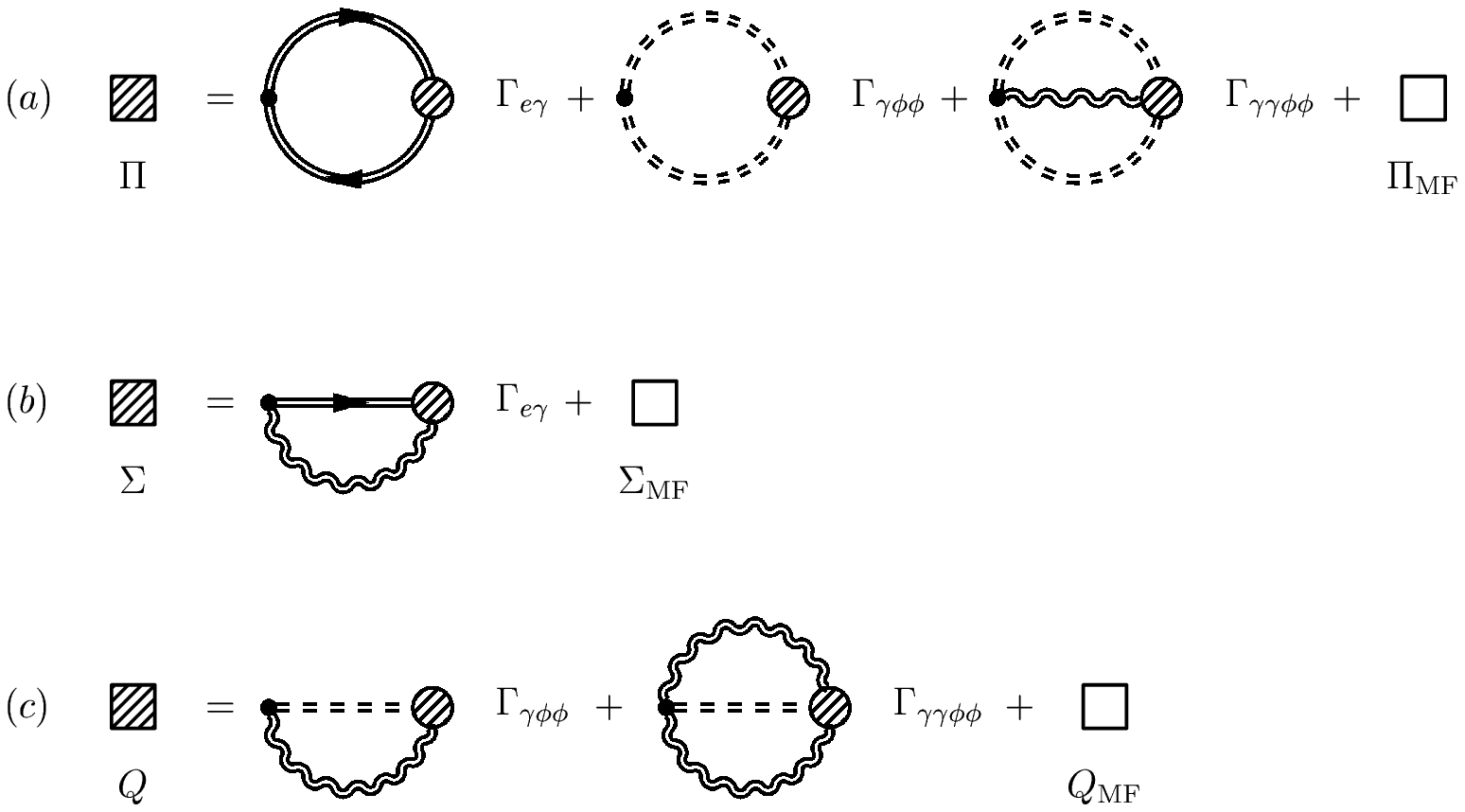}
   \end{center}
   \caption[The scalar and electron self energies and the photon polarization 
	tensor.  Bare vertices are represented by dots and 
	dressed vertices by blobs.  The self-energies  and the polarization 
	tensor are all represented by large square vertices.]
	{The scalar and electron self energies and the photon 
	polarization tensor.  Bare vertices are represented by dots and 
	dressed vertices by blobs.  The self-energies  and the polarization 
	tensor are all represented by large square vertices.}
   \label{fig:selfenergy}
\end{figure}

The self-energies describe all of the branchings and recombinations possible
for the photons, electrons and scalars.  The self-energies are:
\begin{align}
\daveseqn
\begin{split}
   Q(1,1')  = & i (eZ\bothpartial^\mu) \intc d2\: d3\:
   	\Gprop{1}{3} \Gamma_{\gamma \phi\phi}^{\nu}(2,3,1')
   	\Dprop{\mu}{\nu}{}{}{1}{3} \\
   &+i (2i \alpha_{em} Z^2 g^{\mu\nu}) \intc d2\: d3\: d4\: \Gprop{1}{2}
   	\Gamma_{\gamma\gamma \phi\phi}^{\mu'\nu'}(2,3,4,1')
   	\Dprop{}{}{\mu}{\mu'}{1}{3}\Dprop{}{}{\nu}{\nu'}{1}{4}\\
   &+Q_{\rm MF} (1)\deltaftn{4}{1,1'} 
\end{split}\label{eqn:selfen1}\davetag{a}\displaybreak[0]\\
\begin{split}
   \Pi_{\mu\nu}(1,1')  = & -i (-i e (\gamma_{\mu})_{\alpha\beta})
   	\intc d2\: d3\: \Sprop{}{}{\alpha}{\alpha'}{1}{2}
   	\Gamma_{\gamma e, \nu}^{\alpha' \beta'}(2,3,1')
   	\Sprop{}{}{\beta'}{\beta}{3}{1} \\
   & +i (eZ \bothpartial_\mu) \intc d2\: d3\: \Gprop{1}{2}
   	\Gamma_{\gamma \phi\phi, \nu}(2,3,1')\Gprop{3}{1}\\ 
   & +i (2 i \alpha_{em} Z^2 g_{\mu\mu'}) \intc d2\: d3\: d4\:
   	\Gprop{1}{2}\Gprop{3}{1}
   	\Gamma_{\gamma\gamma \phi\phi,\nu}^{\nu'}(2,3,4,1')
   	\Dprop{\mu'}{\nu'}{}{}{1}{4} \\
   &+\Pi_{\rm MF}(1) g_{\mu\nu} \deltaftn{4}{1,1'}
\end{split}\label{eqn:selfen2}\davetag{b}\displaybreak[0]\\
\begin{split}
   \Sigma_{\alpha \beta}(1,1')  = & i (-i e 
   	(\gamma^{\mu})_{\alpha \alpha'}) \intc d2\: d3\: 
   	\Sprop{\alpha'}{\beta'}{}{}{1}{2}
   	\Gamma_{\gamma e}^{\beta' \beta,\nu}(2,3,1')
   	\Dprop{\mu}{\nu}{}{}{1}{3}\\
   & + \Sigma_{\rm MF}(1)
   	\delta_{\alpha \beta}\deltaftn{4}{1,1'}
\end{split}\label{eqn:selfen3}\davetag{c}
\end{align}
In Figures~\ref{fig:selfenergy}(a-c), we show the diagrams corresponding 
to the non-mean-field terms in Equations 
\eqref{eqn:selfen1}-\eqref{eqn:selfen3}.  We define the contour delta function 
$\deltaftn{4}{x,y}$ by
\[
\deltaftn{4}{x,y}=\left\{ 
   \begin{array}{ll}
      \deltaftn{4}{x-y} & \mbox{for $x_0$, $y_0$ on the upper branch}\\
      0 & \mbox{for $x_0$, $y_0$ on different branches}\\
      -\deltaftn{4}{x-y} & \mbox{for $x_0$, $y_0$ on the lower branch}\\
   \end{array}\right. 
\]
Finally, there is another set of Dyson-Schwinger equations for the vertex 
functions.  Since we will truncate the vertices at tree level, we do not 
state the Dyson-Schwinger equations here.

\subsection{Kadanoff-Baym Equations}

The free-field contour Green's functions satisfy the equations of motion:
\begin{align}
\daveseqn
(\partial_{x}^2+M^2)\Gnon{x}{y}&=\deltaftn{4}{x,y}
	\label{eqn:greeneqmota}\davetag{a}\\
\partial_{x}^2\Dnon{}{}{\mu}{\nu}{x}{y}&=4\pi g_{\mu\nu}\deltaftn{4}{x,y}
	\label{eqn:greeneqmotb}\davetag{b}\\
(i\dirslash{\partial}_{x}-m_e)\Snon{}{}{\alpha}{\beta}{x}{y}&=
         \delta_{\alpha\beta}\deltaftn{4}{x,y}
	\label{eqn:greeneqmotc}\davetag{c}
\end{align}
Combining these with the Dyson-Schwinger equations, we have
\begin{align}
\daveseqn
(\partial_{1}^2+M^2)\Gprop{1}{1'} & =  \deltaftn{4}{1,1'} + 
      \intc d2\: Q(1,2) \Gprop{2}{1'} \label{eqn:fullgreeneqmota}\davetag{a}\\
\partial_{1}^2\Dprop{\mu}{\nu}{}{}{1}{1'} & = 
      4\pi g_{\mu\nu}\deltaftn{4}{1,1'}+
      4\pi\intc d2\: \Pi_{\mu}^{\;\;\nu'}(1,2) 
      \Dprop{\nu'}{\nu}{}{}{2}{1'} \label{eqn:fullgreeneqmotb}\davetag{b}\\
(i\dirslash{\partial}_{1}-m_e)\Sprop{\alpha}{\beta}{}{}{1}{1'} & = 
      \delta_{\alpha\beta}\deltaftn{4}{1,1'}+ 
      \intc d2\: \Sigma_{\alpha\beta'}(1,2) \Sprop{\beta'}{\beta}{}{}{2}{1'}. 
	\label{eqn:fullgreeneqmotc}\davetag{c}
\end{align}
There is a conjugate set of equations for 
\eqref{eqn:greeneqmota}--\eqref{eqn:greeneqmotc} and
\eqref{eqn:fullgreeneqmota}--\eqref{eqn:fullgreeneqmotc} with the 
differential operators acting on $1'$.

Restricting $t_1$ and $t_{1'}$ to lie on different sides of the
time contour in Figure \ref{fig:contour}, we arrive at the
Kadanoff-Baym equations.  
\begin{align}
\daveseqn
\begin{split}
(\partial_{1}^2+M^2)\Ggl{1}{1'}=&\displaystyle\int \dn{3}{x_2} Q_{\rm MF}
   (\vec{x}_1,\vec{x}_2,t_1) \Ggl{\vec{x}_2,t_1}{1'}\\
 &+\displaystyle\int^{t_1}_{t_0} d2\: \left(Q^{>}(1,2)-Q^{<}(1,2)\right) 
   \Ggl{2}{1'} \\
 &+\displaystyle\int^{t_1'}_{t_0} d2\: Q^{\gtrless}(1,2)\left(
   \Ggreat{2}{1'}-\Gless{2}{1'}\right) 
\end{split}\label{eqn:KBEalla}\davetag{a}\displaybreak[0]\\
\begin{split}
\frac{1}{4\pi}\partial_{1}^2\Dgl{\mu}{\nu}{1}{1'}=&\displaystyle
    \int \dn{3}{x_2}
    \Pi_{\rm MF}(\vec{x}_1,\vec{x}_2,t_1)
    \Dgl{\mu}{\nu}{\vec{x}_2,t_1}{1'}\\
 & +\displaystyle\int^{t_1}_{t_0} d2\: 
    \left(\Pi^{>\:\:\:\nu'}_{\:\:\:\mu}(1,2)
    -\Pi^{<\:\:\:\nu'}_{\:\:\:\mu}(1,2)\right) \Dgl{\nu'}{\nu}{2}{1'} \\
 & +\displaystyle\int^{t_1'}_{t_0} d2\: 
    \Pi^{\gtrless\:\:\:\nu'}_{\:\:\:\mu}(1,2)
    \left(\Dgreat{}{}{\nu'}{\nu}{2}{1'}-\Dless{}{}{\nu'}{\nu}{2}{1'}\right) 
\end{split}\label{eqn:KBEallb}\davetag{b}\displaybreak[0]\\
\begin{split}
(i\dirslash{\partial}_{1}-m_e)\Sgl{\alpha}{\beta}{1}{1'}=&
    \displaystyle\int \dn{3}{x_2} \Sigma_{\rm MF}(\vec{x}_1,\vec{x}_2,t_1)
    \Sgl{\alpha}{\beta}{\vec{x}_2,t_1}{1'}\\
 & +\displaystyle\int^{t_1}_{t_0} d2\: \left(\Sigma^{>}_{\alpha\beta'}(1,2)
    -\Sigma^{<}_{\alpha\beta'}(1,2)\right) \Sgl{\beta'}{\beta}{2}{1'} \\
 & +\displaystyle\int^{t_1'}_{t_0} d2\: \Sigma^{\gtrless}_{\alpha\beta'}(1,2)
    \left(\Sgreat{}{}{\beta'}{\beta}{2}{1'}-
    \Sless{}{}{\beta'}{\beta}{2}{1'}\right) 
\end{split}\label{eqn:KBEallc}\davetag{c}
\end{align}
Here the $>$ and $<$ self-energies have the same relation to
the contour self-energy that the $>$ and $<$ Green's functions have
to the contour Green's functions.  Again, there is a set
of conjugate equations with the differential operators acting on $1'$.

\subsection{Generalized Fluctuation-Dissipation Theorem}

Now we define the retarded and advanced self-energies for scalars:
\begin{equation} 
	Q^{\pm}(1,2)=Q_{\rm MF}\deltaftn{}{t_1,t_2}
	\pm\theta\left( \pm(t_1-t_2)\right) \left( Q^{>}(1,2)-Q^{<}(1,2) 
	\right)\label{eqn:pmselfen}
\end{equation}
The photon polarization tensor and electron self-energy are defined
in a similar manner.

Using these, the Kadanoff-Baym equations simplify:
\begin{align}
\begin{split}
   (\partial_{1}^2+M^2)\Ggl{1}{1'} =&
   	\displaystyle\int^{\infty}_{t_0} d2\: Q^{+}(1,2) \Ggl{2}{1'}\\
   &+\displaystyle\int^{\infty}_{t_0} d2\: Q^{\gtrless}(1,2)
   	\Gmin{2}{1'}
\end{split}\label{eqn:KBEall2a}\davetag{a}\displaybreak[0]\\
\begin{split}
   \frac{1}{4\pi}\partial_{1}^2\Dgl{\mu}{\nu}{1}{1'}=&
   	\displaystyle\int^{\infty}_{t_0} d2\: 
	\Pi^{+\:\:\:\nu'}_{\:\:\:\mu}(1,2)
   	\Dgl{\nu'}{\nu}{2}{1'} \\
   &+\displaystyle\int^{\infty}_{t_0} d2\: 
   	\Pi^{\gtrless\:\:\:\nu'}_{\:\:\:\mu}(1,2)
   	\Dmin{}{}{\nu'}{\nu}{2}{1'} 
\end{split}\label{eqn:KBEall2b}\davetag{b}\displaybreak[0]\\
\begin{split}
   (i\dirslash{\partial}_{1}-m_e)\Sgl{\alpha}{\beta}{1}{1'}=&
    	\displaystyle\int^{\infty}_{t_0} d2\: \Sigma^{+}_{\alpha\beta'}(1,2)
    	\Sgl{\beta'}{\beta}{2}{1'} \\
   &+\displaystyle\int^{\infty}_{t_0} d2\: 
	\Sigma^{\gtrless}_{\alpha\beta'}(1,2)
    	\Smin{}{}{\beta'}{\beta}{2}{1'}
\end{split}\label{eqn:KBEall2c}\davetag{c}
\end{align}
If we subtract the $>$ equations from the $<$ equations and
multiply the resulting equations by $\pm\theta(\pm(t_1-t_{1'}))$,
we get a second set of differential equations:
\begin{align}
\daveseqn
(\partial_{1}^2+M^2)\Gpm{1}{1'} &= \deltaftn{4}{1-1'}
   +\int^{\infty}_{t_0} d2\: Q^{\pm}(1,2)\Gpm{2}{1'}
   \label{eqn:KBEall3a}\davetag{a}\\
\frac{1}{4\pi}\partial_{1}^2\Dpm{\mu}{\nu}{1}{1'} &= \deltaftn{4}{1-1'}
   +\int^{\infty}_{t_0} d2\: \Pi^{\pm\:\:\:\nu'}_{\:\:\:\mu}(1,2)
   \Dpm{\nu'}{\nu}{2}{1'}
   \label{eqn:KBEall3b}\davetag{b}\\ 
(i\dirslash{\partial}_{1}-m_e)\Spm{\alpha}{\beta}{1}{1'}&=
    \deltaftn{4}{1-1'}
    +\int^{\infty}_{t_0} d2\: \Sigma^{\pm}_{\alpha\beta'}(1,2)
    \Spm{\beta'}{\beta}{2}{1'}
    \label{eqn:KBEall3c}\davetag{c}
\end{align}

Solving the initial value problem posed by Equations
\eqref{eqn:KBEall2a}--\eqref{eqn:KBEall2c} using Equations 
\eqref{eqn:KBEall3a}--\eqref{eqn:KBEall3c},  we find:
\begin{align}
\daveseqn
\begin{split}
   \Ggl{1}{1'} =& \int^{\infty}_{t_0} d2\: \int^{\infty}_{t_0} d3\:
   	\Gplus{1}{2}Q^{\gtrless}(2,3)\Gmin{3}{1'}\\
   & + \int \dn{3}{x_2}\dn{3}{x_3} \Gplus{1}{\vec{x}_2,t_0}
   	\Ggl{\vec{x}_2,t_0}{\vec{x}_3,t_0}\Gmin{\vec{x}_3,t_0}{1'}
\end{split}\label{eqn:FDTa}\davetag{a}\displaybreak[0]\\
\begin{split}
   \Dgl{\mu}{\nu}{1}{1'} =& \int^{\infty}_{t_0} d2\:
   	\int^{\infty}_{t_0} d3\:\Dplus{}{}{\mu}{\mu'}{1}{2}
   	\Pi^{\gtrless\:\:\:\mu'\nu'}(2,3)\Dmin{}{}{\nu'}{\nu}{3}{1'}\\ 
   & + \int \dn{3}{x_2}\dn{3}{x_3}
   	\Dplus{}{}{\mu}{\mu'}{1}{\vec{x}_2,t_0}
   	\Dglup{\mu'}{\nu'}{\vec{x}_2,t_0}{\vec{x}_3,t_0}
   	\Dmin{}{}{\nu'}{\nu}{\vec{x}_3,t_0}{1'}
\end{split}\label{eqn:FDTb}\davetag{b}\displaybreak[0]\\
\begin{split}
   \Sgl{\alpha}{\beta}{1}{1'}=&\int^{\infty}_{t_0} d2\:
   	\int^{\infty}_{t_0} d3\: \Splus{}{}{\alpha}{\alpha'}{1}{2}
   	\Sigma^{\gtrless}_{\alpha'\beta'}(2,3)
   	\Smin{}{}{\beta'}{\beta}{3}{1'}\\
   & + \int \dn{3}{x_2}\dn{3}{x_3}
   	\Splus{}{}{\alpha}{\alpha'}{1}{\vec{x}_2,t_0}
   	\Sgl{\alpha'}{\beta'}{\vec{x}_2,t_0}{\vec{x}_3,t_0}
   	\Smin{}{}{\beta'}{\beta}{\vec{x}_3,t_0}{1'}
\end{split}\label{eqn:FDTc}\davetag{c}
\end{align}
These equations are the Generalized Fluctuation-Dissipation Theorem.
They describe the evolution of a density fluctuation (given by the $>$
and $<$ Green's functions) from $t_0$ to $t_1$.  

\section{Phase-Space Generalized Fluctuation-Dissipation Theorem}
\label{sec:newstuff}

We now translate the Fluctuation-Dissipation Equations 
\eqref{eqn:FDTa}--\eqref{eqn:FDTc} into phase-space.  We will 
only illustrate this for the scalar equation because the photon and electron 
equations follow similarly.  First we extend the
integration region to cover all time:
\[
\begin{array}{ccl}
\Ggl{x_1}{x_{1'}} &=& \displaystyle\int \dn{4}{x_2}\: \dn{4}{x_3}\:
   \Gplus{x_1}{x_2}Q^{\gtrless}(x_2,x_3)\Gmin{x_3}{x_{1'}}\\
& & + \displaystyle\lim_{t_0\rightarrow -\infty} \int \dn{4}{x_2}\dn{4}{x_3}
   \deltaftn{}{t_0-x_{2 0}}\deltaftn{}{t_0-x_{3 0}}\\
& & \times\Gplus{x_1}{x_2}\Ggl{x_2}{x_3}\Gmin{x_3}{x_{1'}}.
\end{array}
\]

Next, we Wigner transform in the relative variable $x_1-x_{1'}$:
\begin{equation}
\begin{array}{ccl}
   \Ggl{x}{p}&=&{\displaystyle\int \dn{4}{x'}\dnpi{4}{p'}
      \tilde{G}^{+}(x,p;x',p')Q^{\gtrless}(x',p')}\\
   & &+ {\displaystyle\lim_{x'_0\rightarrow-\infty}\int\dn{3}{x'}\dnpi{4}{p'} 
      \tilde{G}^{+}(x,p;x',p')\Ggl{x'}{\vec{p}'}}
\end{array}
\label{eqn:inteqn}
\end{equation}
We recognize the Wigner transforms of the self-energy 
and initial particle density:
\begin{equation}
   Q^{\gtrless}(x,p)=\int\dn{4}{\tilde{x}} e^{ip\cdot\tilde{x}}
      Q^{\gtrless}(x+\tilde{x}/2,x-\tilde{x}/2)
\end{equation}
and
\begin{equation}\begin{array}{ccl}
   \lefteqn{\displaystyle\deltaftn{}{t_0-x_0}\Ggl{x}{\vec{p}}=}\\
   & & \displaystyle\int\dn{4}{\tilde{x}}
      e^{ip\cdot\tilde{x}}\deltaftn{}{t_0-(x_0+\tilde{x}_0/2)}
      \deltaftn{}{t_0-(x_0-\tilde{x}_0/2)}\Ggl{x+\tilde{x}/2}{x-\tilde{x}/2}.
\end{array}\end{equation}
The delta functions render the initial density independent of $p_0$.  
We have also defined the retarded propagator in phase-space:
\begin{equation}
   \tilde{G}^{+}(x,p;y,q)=\int\dn{4}{x'}\dn{4}{y'}e^{i(p\cdot x'-q\cdot y')}
   \Gplus{x+x'/2}{y+y'/2}\Gmin{x-x'/2}{y-y'/2}
\end{equation}
At this point, one usually applies the Quasi-Classical Approximation to 
Equation \eqref{eqn:inteqn} by expanding the propagators in gradients and throwing
away higher terms in the expansion and  eliminating the $\dn{4}{x'}$ integral. 
We do not do this.  

Next, we assume the translational invariance of the advanced and retarded
propagators.  This is reasonable at lowest order in the coupling
since the free field \sloppy\linebreak advanced and retarded propagators are 
translationally 
invariant.  However, this approximation neglects interference effects and 
it is likely that these terms are needed to accurately describe many-particle 
effects such as the Landau-Pomeranchuk effect \cite{neq:kno96,neq:kno98}.  
Making this approximation, the retarded propagator in phase-space becomes
\begin{equation}\begin{array}{ccl}
   \displaystyle
   \tilde{G}^{+}(x,p;y,q)&=&{\displaystyle\twopideltaftn{4}{p-q}\int
   \dn{4}{z}e^{ip\cdot z}G^{+}(x-y+z/2)\left(G^{+}(x-y-z/2)\right)^{*}}\\
   &\equiv & \twopideltaftn{4}{p-q}\Gplus{x-y}{p}.
\end{array}
\label{eqn:WigPropinGFDThm}
\end{equation}
We use $\Gplus{x-y}{p}$ in all subsequent calculations and in practice 
we only use the lowest order contribution to $\Gplus{x-y}{p}$.
This means that we dresses the $\gtrless$ propagators but not the $\pm$ 
propagators when we iterate Equation \eqref{eqn:FDTa}--\eqref{eqn:FDTc}.  
Thus, particles propagate as though they are in the vacuum.  In Appendix
\ref{append:prop} we calculate the lowest order contribution to 
$\Gplus{x-y}{p}$.  

Repeating this for the pho\-tons and elec\-trons, we ar\-rive at the 
phase-space Gen\-er\-al\-ized Fluc\-tua\-tion-Dissi\-pa\-tion Theorem:
\begin{align}
\daveseqn
\begin{split}
   \Ggl{x}{p}=&\int\dn{4}{y}\Gplus{x-y}{p} Q^{\gtrless}(y,p)\\
   &+ \lim_{y_0\rightarrow -\infty}\int\dn{3}{y}\Gplus{x-y}{p}\Ggl{y}{\vec{p}}
\end{split}\label{eqn:GFDTa}\davetag{a}\displaybreak[0]\\
\begin{split}
   \Dgl{\mu}{\nu}{x}{p}=&\int\dn{4}{y}\Dplus{\mu}{\nu}{\mu'}{\nu'}{x-y}{p}
        \Pi^{\gtrless \mu'\nu'}(y,p)\\
   &+\lim_{y_0\rightarrow -\infty}\int\dn{3}{y}
        \Dplus{\mu}{\nu}{\mu'}{\nu'}{x-y}{p}\Dglup{\mu'}{\nu'}{y}{\vec{p}}
\end{split}\label{eqn:GFDTb}\davetag{b}\displaybreak[0]\\
\begin{split}
   \Sgl{\alpha}{\beta}{x}{p}=&\int\dn{4}{y}
        \Splus{\alpha}{\beta}{\alpha'}{\beta'}{x-y}{p}
        \Sigma^{\gtrless}_{\alpha'\beta'}(y,p)\\
   &+\lim_{y_0\rightarrow -\infty}\int\dn{3}{y}
        \Splus{\alpha}{\beta}{\alpha'}{\beta'}{x-y}{p}
        \Sgl{\alpha'}{\beta'}{y}{\vec{p}}.
\end{split}\label{eqn:GFDTc}\davetag{c}
\end{align}
These equations have a clear meaning.  In the source terms, 
particles are created at point $y$ with momentum $p$ and they propagate 
with momentum $p$ out to point $x$.  In the terms with the initial conditions,
particles are initialized at point $\vec{y}$ at time $y_0\rightarrow -\infty$
with on-shell momentum $p$ and they propagate with momentum $p$ out to point 
$x$.  Thus, these equations describe the evolution of the particle phase-space
densities from $y_0\rightarrow -\infty$ to the time $x_0$, including
particle creation and absorption through the particle self-energies.
They also have ``source-propagator'' form, namely each term is a convolution 
of a phase-space source (or the initial conditions) and a phase-space 
propagator.  The derivation of these equations does not rely on the form of 
the self-energies and so these results should be immediately applicable to 
any system.  

\begin{figure}
   \begin{center}
   \includegraphics[totalheight=0.81\bagfigheight]{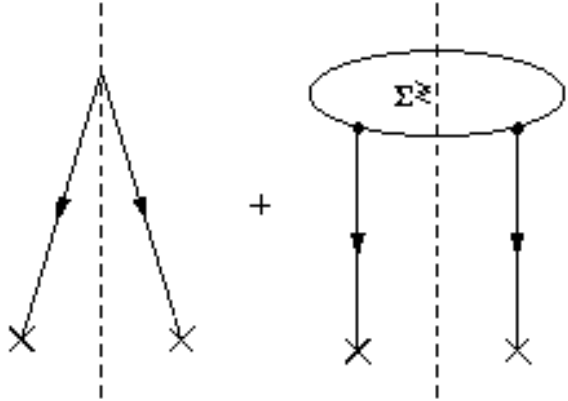}
   \end{center}
   \caption[Cut diagram for the particle densities in the Generalized
	Fluc\-tua\-tion-Dissi\-pa\-tion Theorem.]
	{Cut diagram for the particle densities in the Generalized 
	Fluc\-tua\-tion-Dissi\-pa\-tion Theorem.}
        \label{fig:GFDThm}
\end{figure}

The general form of the Generalized Fluctuation-Dissipation Theorem 
is shown in the cut diagrams in Figure \ref{fig:GFDThm}.  Since these diagrams
are slightly different from the contour diagrams and from traditional cut 
diagrams,\footnote{Meaning, the cut diagrams used to calculate exclusive 
reaction probabilities in Feynman's formulation of perturbation theory} 
we will describe what they mean.  Here, the cut line is the dashed line 
down the center.  The subdiagram on the left
encodes the initial conditions via the cutting of the propagators at the top
of the subdiagram.  The two propagators going down the the x'ed vertices
are the two retarded propagators that we Wigner transformed together in Equation
\eqref{eqn:WigPropinGFDThm}.  The x'ed vertices represent the two space-time 
arguments of the two-point function for the density.  The space-time
coordinates are Wigner transformed together.  On the right, the diagram
has much the same meaning except that now the initial conditions are replaced
with the cut self-energy.  In both subdiagrams, time flows downward toward
the future.  

\subsection{Sources}

The first step to\-ward getting the phase-space evo\-lu\-tion equa\-tions
from the Gen\-er\-al\-ized Fluc\-tua\-tion-Dissi\-pa\-tion Theorem is 
calculating the self-energies (i.e. the sources).  To do this, we insert 
Equations \eqref{eqn:pmselfen} and \eqref{eqn:greenrelation1} into the 
self-energy equations and keep only the lowest order approximation to the 
vertex functions.  Thus, we assume that the vertices are not dressed and 
are point-like.  So, we arrive at the creation and absorption rates:
\begin{align}
\daveseqn
   Q^{\gtrless}(1,1')=&i \alpha_{em} Z^2 \bothpartial_{1\mu} \Ggl{1}{1'}
   	\bothpartial_{1'\nu} \Dglup{\mu}{\nu}{1}{1'}
   	+Q_{\rm MF}^{\gtrless}(1) \deltaftn{4}{1-1'}
	\label{eqn:selfen4a}\davetag{a}\displaybreak[0]\\
\begin{split}
   \Pi^{\gtrless}_{\mu\nu}(1,1')=& i \alpha_{em} {\rm Tr}\left\{
   	\gamma_\mu\Sgl{}{}{1}{1'}\gamma_\nu\Slg{}{}{1}{1'}\right\}\\
   &+i \alpha_{em} Z^2 \bothpartial_{1\mu} \Ggl{1}{1'}
   	\bothpartial_{1'\nu} \Glg{1}{1'}
   	+\Pi_{\rm MF}^{\gtrless}(1)g_{\mu\nu}\deltaftn{4}{1-1'}
\end{split}\label{eqn:selfen4b}\davetag{b}\displaybreak[0]\\
   \Sigma^{\gtrless}_{\alpha\beta}(1,1')=&-i \alpha_{em}
        (\gamma_\mu)_{\alpha\alpha'}
   	\Sgl{\alpha'}{\beta'}{1}{1'}(\gamma_\nu)_{\beta'\beta}
   	\Dglup{\mu}{\nu}{1}{1'}
   	+\Sigma_{\rm MF}^{\gtrless}(1)\delta_{\alpha\beta}\deltaftn{4}{1-1'}
	\label{eqn:selfen4c}\davetag{c}
\end{align}
Here we neglect the second scalar term in the polarization tensor
and the second photon term in the scalar self-energy
because they enter with a factor $\alpha_{em}^2$ which is higher order than 
the other terms.  

The self-energies in \eqref{eqn:selfen4a}--\eqref{eqn:selfen4c} can be 
Wigner transformed.  Taking care to integrate the derivative scalar couplings 
by parts, we arrive at
\begin{align}
\daveseqn
\begin{split}
   Q^{\gtrless}(x,p)=&i \alpha_{em} Z^2\int\dnpi{4}{q_1}\dnpi{4}{q_2}
   	(q_1+q_2-i\bothpartial/2)_{\mu} \Ggl{x}{q_1}\\
   &\times(q_1+q_2-i\bothpartial/2)_{\nu} \Dglup{\mu}{\nu}{x}{q_2}
   	\twopideltaftn{4}{p-(q_1+q_2)}+Q_{\rm MF}^{\gtrless}(x) 
\end{split}\label{eqn:selfen5a}\davetag{a}\displaybreak[0]\\
\begin{split}
   \Pi^{\gtrless}_{\mu\nu}(x,p)=&i \alpha_{em} \int\dnpi{4}{q_1}\dnpi{4}{q_2}
   	{\rm Tr}\left\{\gamma_\mu\Sgl{}{}{x}{q_1}\gamma_\nu\Sgl{}{}{x}{q_2}
   	\right\}\twopideltaftn{4}{p-(q_1+q_2)}\\
   &+i \alpha_{em} Z^2\int\dnpi{4}{q_1}\dnpi{4}{q_2} 
   	(q_1+q_2+i\bothpartial/2)_{\mu}\Ggl{x}{q_1}\\
   &\times(q_1+q_2+i\bothpartial/2)_{\nu}\Ggl{x}{q_2}
   	\twopideltaftn{4}{p-(q_1+q_2)}+\Pi_{\rm MF}^{\gtrless}(x)g_{\mu\nu}
\end{split}\label{eqn:selfen5b}\davetag{b}\displaybreak[0]\\
\begin{split}
   \Sigma^{\gtrless}_{\alpha\beta}(x,p)=&-i \alpha_{em} 
        (\gamma_\mu)_{\alpha\alpha'}\int\dnpi{4}{q_1}\dnpi{4}{q_2}
   	\Sgl{\alpha'}{\beta'}{x}{q_1}(\gamma_\nu)_{\beta'\beta}
   	\Dglup{\mu}{\nu}{x}{q_2}\\
   &\times\twopideltaftn{4}{p-(q_1+q_2)}
   	+\Sigma_{\rm MF}^{\gtrless}(x)\delta_{\alpha\beta}
\end{split}\label{eqn:selfen5c}\davetag{c}
\end{align}

\section{QED Evolution Equations}
\label{sec:evolve}

We now insert the phase-space self-energies into the phase-space 
Gen\-er\-al\-ized Fluc\-tu\-a\-tion-Dissi\-pa\-tion theorem and rewrite these 
equations directly in terms of the particle and antiparticle densities.
\begin{align}
\daveseqn
\begin{split}
   \ggtrless{x}{p}=&\int\dn{4}{y}\dnpi{4}{q_1}\dnpi{4}{q_2}\Gplus{x-y}{p} 
   	\twopideltaftn{4}{p-(q_1+q_2)}\\
   &\times \alpha_{em} Z^2 (q_1+q_2-i\bothpartial/2)^{\mu}
	\ggtrless{y}{q_1}(q_1+q_2-i\bothpartial/2)^{\nu}
	\dgtrless{\mu}{\nu}{y}{q_2}\\
   &+ \int\dn{4}{y}\Gplus{x-y}{p} iQ_{\rm MF}^{\gtrless}(y)\\
   &+ \lim_{y_0\rightarrow -\infty}\int\dn{3}{y}\Gplus{x-y}{p}
   	\ggtrless{y}{\vec{p}}
\end{split}\label{eqn:evolve2a}\davetag{a}\displaybreak[0]\\
\begin{split}
   \dgtrless{\mu}{\nu}{x}{p}=&\int\dn{4}{y}\dnpi{4}{q_1}\dnpi{4}{q_2}
   	\Dplus{\mu}{\nu}{\mu'}{\nu'}{x-y}{p}\twopideltaftn{4}{p-(q_1+q_2)}\\
   &\times \left\{ \alpha_{em} {\rm Tr}\left[\gamma^{\mu'} 
   	\sgtrless{}{}{y}{q_1}\gamma^{\nu'}\sgtrless{}{}{y}{q_2}\right]\right.\\
   &+\left. \alpha_{em} Z^2 (q_1+q_2+i\bothpartial/2)^{\mu'}\ggtrless{y}{q_1}
   	(q_1+q_2+i\bothpartial/2)^{\nu'}\ggtrless{y}{q_2}\right\}\\
   &+\int\dn{4}{y}\Dplus{\mu}{\nu}{\mu'}{\nu'}{x-y}{p}
   	i\Pi_{\rm MF}^{\gtrless}(y)g^{\mu'\nu'}\\
   &+\lim_{y_0\rightarrow -\infty}\int\dn{3}{y}
   	\Dplus{\mu}{\nu}{\mu'}{\nu'}{x-y}{p}\dgtrlessup{\mu'}{\nu'}{y}{\vec{p}}
\end{split}\label{eqn:evolve2b}\davetag{b}\displaybreak[0]\\
\begin{split}
   \sgtrless{\alpha}{\beta}{x}{p}=&\int\dn{4}{y}\dnpi{4}{q_1}\dnpi{4}{q_2}
   	\Splus{\alpha}{\beta}{\alpha'}{\beta'}{x-y}{p}
   	\twopideltaftn{4}{p-(q_1+q_2)}\\
   &\times \alpha_{em} (\gamma^\mu)_{\alpha'\alpha''} 
   	\sgtrless{\alpha''}{\beta''}{y}{q_1}(\gamma^\nu)_{\beta''\beta'} 
   	\dgtrless{\mu}{\nu}{y}{q_2}\\
   &+\int\dn{4}{y}\Splus{\alpha}{\beta}{\alpha'}{\beta'}{x-y}{p}
   	\left(\pm i\Sigma_{\rm MF}^{\gtrless}(y)\right)\delta_{\alpha'\beta'}\\
   &+\lim_{y_0\rightarrow -\infty}\int\dn{3}{y}
   	\Splus{\alpha}{\beta}{\alpha'}{\beta'}{x-y}{p}
   	\sgtrless{\alpha'}{\beta'}{y}{\vec{p}}.
\end{split}\label{eqn:evolve2c}\davetag{c}
\end{align}
These equations 
simultaneously describe all ``partonic'' splittings, recombinations and
scatterings from the distant past to the present.  Note that an implementation 
of these equations would be very different from the conventional 
transport approach.  First, these splittings and recombinations occur in all 
cells of coordinate space.  This is a very different from the conventional 
approach where particles interact only when they are within 
$\sqrt{\sigma_{\rm TOT}/\pi}$ of each other \cite{tran:kor95,tran:ko87,tran:gei92a,tran:gei92b,tran:gei94,tran:gei95,tran:gei96,tran:kle97}.  Because the 
approach in this thesis
is both non-local and Lorentz covariant, implementing it would avoid the 
causality violating problems implicit in conventional approaches.
Second, the particles in our approach do not 
follow straight-like trajectories.  Instead, they have a ``probability'' 
distribution for propagating to a certain point.  This idea is elaborated on
somewhat in the next chapter and discussed in detail in 
Appendix~\ref{append:prop}.

Equations \eqref{eqn:evolve2a}--\eqref{eqn:evolve2c} are the
phase-space QED analog of Mahklin's evolution equations 
\cite{neq:mak95a,neq:mak95b,neq:mak98}.  A
QCD version of the phase-space evolution equations should reduce to Makhlin's
equations when integrating out the coordinate dependence.  
Geiger \cite{tran:gei96} has derived a set of QCD transport equations based on
Makhlin's work.  While his derivation is very similar to our derivation of the
\sloppy\linebreak phase-space evolution equation,  he uses a variant of the 
Quasi-Classical Approximation tailored toward the DGLAP partons in order to
simplify his collision integrals.  The QCD version of the transport equations
we derive in Section \ref{sec:transport} would reduce to his semi-classical
equations if one applies this approximation.

There are several ways to solve Equation 
\eqref{eqn:evolve2a}--\eqref{eqn:evolve2c}
but we propose only two methods in the following 
subsections.  The first method is a perturbative scheme which
we use to derive the time-ordered version of the results of Sections 
\ref{sec:pdist}--\ref{sec:edist} in the next chapter.  The
second method is to derive transport equations from Equations 
\eqref{eqn:evolve2a}--\eqref{eqn:evolve2c}.  

\section{Perturbative Solutions}
\label{sec:nonequilibedist}

We can perform a coupling constant expansion on Equations 
\eqref{eqn:evolve2a}--\eqref{eqn:evolve2c} and get the leading contributions 
to the particle densities.  We show this for
the photons and electrons surrounding a classical point charge. 
The discussion here is mainly technical and is designed to show how to perform
a perturbative calculation in phase-space.  There is 
an expanded discussion of these densities in the next chapter;
there we describe the sources and propagators for 
the photon and electron densities around a point charge.

We begin by stating the initial densities\footnote{Unlike Feynman's 
perturbation theory for exclusive amplitudes, we can only specify the 
initial particle densities here.} and listing our assumptions.
In the initial state, we assume there is one massive scalar particle serving
as the photon source.  If we view only photons with a wavelength much larger 
than the spread of the scalar wavepacket then the scalar particle density is
\[\gless{y_0=-\infty,\vec{y}}{\vec{p}}={\cal N} 
\theta(p_0)\deltaftn{3}{\vec{p}-\vec{p_i}}\delta(p^2-M^2) 
\deltaftn{3}{x_0\vec{p}/p_0-\vec{x}}\]
This form is only needed to make the correspondence between the results here 
and the results in Chapter~\ref{chap:pips} and the form is discussed in 
Appendix~\ref{append:classdens}.  
The initial electron and photon particle densities are all zero:
\[\sless{\alpha}{\beta}{y_0=-\infty,\vec{y}}{\vec{p}}=
\dless{\mu}{\nu}{y_0=-\infty,\vec{y}}{\vec{p}}=0\]
Finally, the other assumptions that we make are that we neglect all mean 
fields and drop the gradients in the scalar-photon coupling.  

\subsection{A Photon Distribution}

\begin{figure}
   \begin{center}
   \includegraphics[scale=.85]{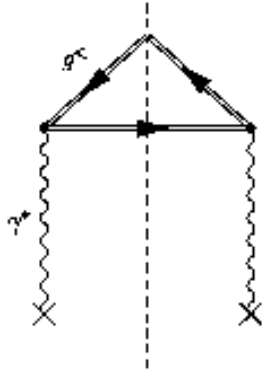}
   \end{center}
   \caption[Cut diagram for the time-ordered non-equilibrium photon density.]
	{Cut diagram for the time-ordered non-equilibrium photon density.}  
   \label{fig:noneqEPD}
\end{figure}

Since the scalar field only couples to the photons, the lowest order
contribution to the photon density comes from the photons directly coupling to
the initial scalar density.  The cut diagram for this process has the form of 
the left subdiagram in the Generalized Fluctuation-Dissipation theorem 
of Figure \ref{fig:GFDThm} and is shown in Figure \ref{fig:noneqEPD}. 
In Figure \ref{fig:noneqEPD}, the photon self-energy is the
triangular source current loop.  

For positive energy photons, we can write down the density directly from 
Equation \eqref{eqn:evolve2b}:
\[\begin{array}{ccl}
\dless{\mu}{\nu}{x}{p}&=&\displaystyle\int\dn{4}{y}\dnpi{4}{q_1}\dnpi{4}{q_2}
   \Dplus{\mu}{\nu}{\mu'}{\nu'}{x-y}{p}\twopideltaftn{4}{p-(q_1+q_2)}\\
   & & \times \alpha_{em} Z^2 (q_1+q_2)^{\mu'} \gless{y}{q_1}
   (q_1+q_2)^{\nu'}\gless{y}{q_2}.
\end{array}\]
Now, $\Ggreat{x}{p} = \Gless{x}{-p}$ because 
$\gtrless$ propagators obey the relation $\Ggreat{x}{y}
= \Gless{y}{x}$ at lowest order in the coupling.  Thus, we can 
switch one of the $\gless{y}{q}$ to $\ggtr{y}{-q}$, changing it from an 
initial state antiscalar to a final state scalar (or an initial state hole).  
Doing so, we have 
\begin{equation}
\label{eqn:noneqEPD1}
\begin{array}{ccl}
\dless{\mu}{\nu}{x}{p}&=&\displaystyle\int\dn{4}{y}\dnpi{4}{q_1}\dnpi{4}{q_2}
   \Dplus{\mu}{\nu}{\mu'}{\nu'}{x-y}{p}\twopideltaftn{4}{p-(q_1-q_2)}\\
   & & \times \alpha_{em} Z^2 (q_1-q_2)^{\mu'} \gless{y}{q_1}
   (q_1-q_2)^{\nu'}\ggtr{y}{q_2}.
\end{array}
\end{equation}

Comparing Equation \eqref{eqn:noneqEPD1} and Figure \ref{fig:noneqEPD} we can 
further understand the correspondence between the cut diagrams and the 
perturbative solution.  The factor of $\gless{y}{q_1}$ for the initial scalar 
density is the cut upper double line in Figure \ref{fig:noneqEPD}.  The
other factor of $\ggtr{y}{q_2}$ then is the final scalar density and is 
represented by the lower cut double line.  

The entire photon source can be associated with the Wigner transform of the 
scalar current density after a sum over the final scalar momenta.  This is
discussed in Appendix \ref{append:current}.  Because of this correspondence, 
Equation \eqref{eqn:noneqEPD1} is the non-equilibrium, time-ordered, analog 
of the Wigner transform of the photon vector potential in 
Section \ref{sec:pdist}.


\subsection{An Electron Distribution}

\begin{figure}
    \begin{center}
    \includegraphics[scale=.85]{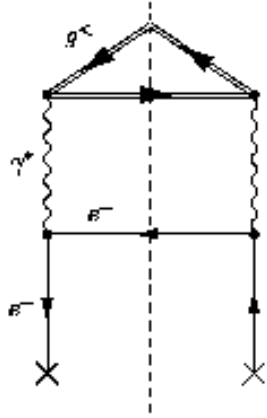}
    \end{center}
    \caption[Cut diagram for the time-ordered non-equilibrium electron 
	density.]{Cut diagram for the time-ordered non-equilibrium electron 
	density.}
    \label{fig:noneqEED}
\end{figure}

Since the electrons only couple to the photons, the lowest order contribution
to the electron density comes from a photon splitting into an electron-positron
pair.  The cut diagram for this is shown in Figure \ref{fig:noneqEED}.  As one
can see, the electron self-energy is everything above the two electron 
propagators.  From 
Equations \eqref{eqn:evolve2a}--\eqref{eqn:evolve2c} we have:
\[\begin{array}{ccl}
\sless{\alpha}{\beta}{x}{p}&=&\displaystyle\int\dn{4}{y}\dnpi{4}{q_1}
   \dnpi{4}{q_2}\Splus{\alpha}{\beta}{\alpha'}{\beta'}{x-y}{p}
   \twopideltaftn{4}{p-(q_1+q_2)}\\
   & &\displaystyle\times \alpha_{em} (\gamma^\mu)_{\alpha'\alpha''} 
   \sless{\alpha''}{\beta''}{y}{q_1}(\gamma^\nu)_{\beta''\beta'} 
   \dless{\mu}{\nu}{y}{q_2}.\\
\end{array}\]
Using $\sless{\alpha}{\beta}{x}{q}=\sgtr{\beta}{\alpha}{x}{-q}$, we find
\begin{equation}\label{eqn:noneqEED1}\begin{array}{ccl}
\sless{\alpha}{\beta}{x}{p}&=&\displaystyle\int\dn{4}{y}\dnpi{4}{q_1}
   \dnpi{4}{q_2}\Splus{\alpha}{\beta}{\alpha'}{\beta'}{x-y}{p}
   \twopideltaftn{4}{p-(-q_1+q_2)}\\
   & &\displaystyle\times \alpha_{em} (\gamma^\mu)_{\alpha'\alpha''} 
   \sgtr{\beta''}{\alpha''}{y}{q_1}(\gamma^\nu)_{\beta''\beta'} 
   \dless{\mu}{\nu}{y}{q_2}.\\
\end{array}\end{equation}

As with the photon density, this equation maps directly to its analog in
Chapter~\ref{chap:pips}, specifically \eqref{eqn:elecdist}.  Together the
electron density here and the photon density in \eqref{eqn:noneqEPD1} show 
that a we can solve the phase-space evolution equations perturbatively.

\section{The QED Semi-classical Transport Equations}
\label{sec:transport}

While we can solve the evolution equations perturbatively, this does not 
lend itself towards the more complex calculations
needed to model a nuclear collision.  
In this section, we find a set of transport equations from the integral 
equations in \eqref{eqn:evolve2a}--\eqref{eqn:evolve2c}.  The QCD version
of this section might be what is needed to construct a parton transport model.
We will find the transport equations by writing two equations of 
motion for the phase-space retarded propagator.  Applying these equations to
the phase-space evolution equations, we derive two sets of coupled 
integro-differential equations.
The first set of equations are the transport equations and the second set 
are the ``constraint'' equations of Mr\'{o}wczy\'{n}ski and 
Heinz~\cite{tran:mro94,tran:zhu96}.  The transport equations are what is 
normally solved in a transport approach.  The second ``constraint'' equations, 
supplement the first by describing the mass shift of the particles in medium.  

The equation of motion for the non-interacting 
retarded massless scalar propagator is
\[\partial^2G^{+}(x)=\deltaftn{4}{x}.\]
The conjugate equation is 
\[\partial^2(G^{+}(x))^{*}=\deltaftn{4}{x}.\]
Multiplying both sides of the first equation by $(G^{+}(y))^{*}$,  
both sides of the 
second equation $G^{+}(y)$ and Wigner transforming in the relative 
space-time coordinate, we find two equations:
\begin{align}
\daveseqn
        (k+ i\partial/2)^2\Gplus{x}{k}=&\int\dn{4}{x'}e^{ix'\cdot k}
                \left(G^{+}(x- x'/2)\right)^{*} \deltaftn{4}{x+ x'/2}
		\davetag{a}\label{eqn:rettrana}\\
        (k- i\partial/2)^2\Gplus{x}{k}=&\int\dn{4}{x'}e^{ix'\cdot k}
                \left(G^{+}(x+ x'/2)\right) \deltaftn{4}{x- x'/2}
		\davetag{b}\label{eqn:rettranb}
\end{align}
Inserting the retarded propagator in the energy-momentum representation 
(with $m_e=0$) and adding and subtracting Equations \eqref{eqn:rettrana} 
and \eqref{eqn:rettranb}, we find the equations of motion for the retarded 
propagator:
\begin{align}
\daveseqn
   k\cdot\partial\Gplus{x}{k}=&\frac{2}{\pi}\theta (x_0)\delta (x^2)
 	\sin{(2 x\cdot k)}\label{eqn:retprop1}\davetag{a}\\
   (\partial^2/4-k^2)\Gplus{x}{k}=&\frac{2}{\pi}\theta (x_0)\delta (x^2)
	\cos{(2 x\cdot k)}.\label{eqn:retprop2}\davetag{b}
\end{align}
Taylor series expanding the sine or cosine and keeping only the lowest order 
is equivalent to performing the gradient expansion in the Quasi-Classical 
Approximation.

Now, we apply the $k\cdot\partial$ and 
$(\partial^2/4-k^2)$ operators to the particle densities in Equation 
\eqref{eqn:evolve2a}--\eqref{eqn:evolve2c}.  On the right hand side, 
these differential operators
act on the retarded propagators, so we can use their equations of motion to 
simplify the results.  For scalars we find
\begin{align}
\daveseqn
\begin{split}
\lefteqn{p\cdot\partial\ggtrless{x}{p}=\displaystyle
   \int\dn{4}{y}\dnpi{4}{q_1}\dnpi{4}{q_2}
   \frac{2}{\pi}\theta(x_0-y_0)\delta((x-y)^2)}\hspace*{.25in}\\
   &\displaystyle\times\sin{(2(x-y)\cdot p)} 
   \twopideltaftn{4}{p-(q_1+q_2)}\\
   &\displaystyle\times \alpha_{em} Z^2 
   (q_1+q_2-i\bothpartial/2)^{\mu}\ggtrless{y}{q_1} 
   (q_1+q_2-i\bothpartial/2)^{\nu}\dgtrless{\mu}{\nu}{y}{q_2}\\
   &+ \displaystyle\int\dn{4}{y} \frac{2}{\pi}\theta(x_0-y_0)\delta((x-y)^2)
   \sin{(2(x-y)\cdot p)} iQ_{\rm MF}^{\gtrless}(y)\\
   &+ \displaystyle\lim_{y_0\rightarrow -\infty}\int\dn{3}{y}
   \frac{2}{\pi}\theta(x_0-y_0)\delta((x-y)^2)
   \sin{(2(x-y)\cdot p)}\ggtrless{y}{\vec{p}}
\end{split}\label{eqn:transport1a}\davetag{a}\displaybreak[0]\\
\begin{split}
\lefteqn{(\partial^2/4-k^2)\ggtrless{x}{p}=\displaystyle\int\dn{4}{y}
   \dnpi{4}{q_1}
   \dnpi{4}{q_2}\frac{2}{\pi}\theta(x_0-y_0)\delta((x-y)^2)}\hspace*{.25in}\\
   &\times \displaystyle\cos{(2(x-y)\cdot p)} 
   \twopideltaftn{4}{p-(q_1+q_2)}\\
   &\times \displaystyle\alpha_{em} Z^2 (q_1+q_2-i\bothpartial/2)^{\mu}
   \ggtrless{y}{q_1}
   (q_1+q_2-i\bothpartial/2)^{\nu}\dgtrless{\mu}{\nu}{y}{q_2}\\
   &+ \displaystyle\int\dn{4}{y} \frac{2}{\pi}\theta(x_0-y_0)\delta((x-y)^2)
   \cos{(2(x-y)\cdot p)} iQ_{\rm MF}^{\gtrless}(y)\\
   &+ \displaystyle\lim_{y_0\rightarrow -\infty}\int\dn{3}{y}
   \frac{2}{\pi}\theta(x_0-y_0)\delta((x-y)^2)
   \cos{(2(x-y)\cdot p)}\ggtrless{y}{\vec{p}}
\end{split}\label{eqn:transport1b}\davetag{b}
\end{align}
Now because of the delta functions, the boundary conditions at 
$y_0\rightarrow -\infty$ only contribute when
$|\vec{x}-\vec{y}|$ goes to $\infty$, implying that we need 
$\ggtrless{x}{p}$  as $\vec{x}\rightarrow \infty$.  The 
densities are zero there, so they drop out from these equations.

The transport equations for the photons and electrons are 
\begin{align}
\begin{split}
\lefteqn{p\cdot\partial\dgtrless{\mu}{\nu}{x}{p} = 
   \displaystyle\frac{2}{\pi}\int\dn{4}{y}
   \theta(x_0-y_0)\delta((x-y)^2)\sin{(2(x-y)\cdot p)}}\hspace*{.25in} \\
&\times\left\{ \displaystyle\int\dnpi{4}{q_1}\dnpi{4}{q_2}
   \twopideltaftn{4}{p-(q_1+q_2)}\left\{ \alpha_{em} 
   {\rm Tr}\left[\gamma^{\mu}\sgtrless{}{}{y}{q_1}\gamma^{\nu}
   \sgtrless{}{}{y}{q_2}\right]\right.\right.\\
&+\left.\left.\alpha_{em} Z^2 (q_1+q_2+i\bothpartial/2)^{\mu} 
   \ggtrless{y}{q_1}(q_1+q_2+i\bothpartial/2)^{\nu}\ggtrless{y}{q_2}
   \right\}\right.\\
&\left. +  g_{\mu\nu}i\Pi_{\rm MF}^{\gtrless}(y)\right\}
\end{split}\label{eqn:transport2a}\davetag{c}\displaybreak[0]\\
\begin{split}
\lefteqn{p\cdot\partial\sgtrless{\alpha}{\beta}{x}{p} = 
   \displaystyle\frac{2}{\pi}(\dirslash{p}+
   i\dirslash{\partial})_{\alpha\alpha'}
   (\dirslash{p}-i\dirslash{\partial})_{\beta\beta'}}\hspace*{.25in}\\
&\times\displaystyle\int\dn{4}{y}\theta(x_0-y_0)\delta((x-y)^2)
        \sin{(2(x-y)\cdot p)}\\
&\times\left\{ \displaystyle\int\dnpi{4}{q_1}\dnpi{4}{q_2}
   \twopideltaftn{4}{p-(q_1+q_2)}\right.\\
&\times\displaystyle\left.\alpha_{em} (\gamma^\mu)_{\alpha'\alpha''} 
   \sgtrless{\alpha''}{\beta''}{y}{q_1}(\gamma^\nu)_{\beta''\beta'} 
   \dgtrless{\mu}{\nu}{y}{q_2}\right.\\
&+\left. \delta_{\alpha'\beta'}\left(\pm i\Sigma_{\rm MF}^{\gtrless}(y)
   \right)\right\}.
\end{split}\label{eqn:transport2b}\davetag{d}
\end{align}
These equations almost have the form of the Boltzmann equation:  the left side
clearly is the Boltzmann transport operator and the right side is almost the
collision integrals.  If we were to expand the sines in the collision 
integrals and keep only the lowest term, we would recover the collision 
integrals.  Furthermore, if we were to do this same approximation to the QCD 
version of \eqref{eqn:transport2a} we would arrive at Geiger's semi-classical 
QCD transport equations \cite{tran:gei96}.

We also state the constraint equations:
\begin{align}
\begin{split}
\lefteqn{(\partial^2/4-k^2)\dgtrless{\mu}{\nu}{x}{p} = 
   \displaystyle \frac{2}{\pi}\int\dn{4}{y}
   \theta(x_0-y_0)\delta((x-y)^2)\cos{(2(x-y)\cdot p)}}
   \hspace*{.25in}\\
&\times\left\{ \displaystyle\int\dnpi{4}{q_1}\dnpi{4}{q_2}
   \twopideltaftn{4}{p-(q_1+q_2)}\left\{ \alpha_{em} {\rm Tr}
   \left[\gamma^{\mu} 
   \sgtrless{}{}{y}{q_1}\gamma^{\nu}\sgtrless{}{}{y}{q_2}
   \right]\right.\right.\\
&+\left.\left.\alpha_{em} Z^2 (q_1+q_2+i\bothpartial/2)^{\mu} 
   \ggtrless{y}{q_1}(q_1+q_2+i\bothpartial/2)^{\nu}\ggtrless{y}{q_2}
   \right\}\right.\\
&\left. +  g_{\mu\nu}i\Pi_{\rm MF}^{\gtrless}(y)\right\}
\end{split}\label{eqn:transport2c}\davetag{e}\displaybreak[0]\\
\begin{split}
\lefteqn{(\partial^2/4-k^2)\sgtrless{\alpha}{\beta}{x}{p} = 
   \displaystyle\frac{2}{\pi}(\dirslash{p}
   +i\dirslash{\partial})_{\alpha\alpha'}
   (\dirslash{p}-i\dirslash{\partial})_{\beta\beta'}}\hspace*{.25in}\\
&\times\displaystyle\int\dn{4}{y}\theta(x_0-y_0)\delta((x-y)^2)
   \cos{(2(x-y)\cdot p)}\\
&\times\left\{ \displaystyle\int\dnpi{4}{q_1}\dnpi{4}{q_2}
   \twopideltaftn{4}{p-(q_1+q_2)}\right.\\
&\times\displaystyle\left.\alpha_{em} (\gamma^\mu)_{\alpha'\alpha''} 
   \sgtrless{\alpha''}{\beta''}{y}{q_1}(\gamma^\nu)_{\beta''\beta'} 
   \dgtrless{\mu}{\nu}{y}{q_2}\right.\\
&+ \left. \delta_{\alpha'\beta'}\left(\pm i\Sigma_{\rm MF}^{\gtrless}(y)
   \right)\right\}.
\end{split}\label{eqn:transport2d}\davetag{f}
\end{align}
If we were we to derive the constraint equation 
for massive particles, we would find that $(\partial^2/4-k^2)\rightarrow
(\partial^2/4-k^2+m^2)$.  Therefore, 
the constraint equations give rise to the in-medium mass shift for the 
photons and electrons and thus the RHS of the constraint equations for massless
particles can be interpreted as an ``in-medium'' mass.  Note that despite the
presence of this ``in-medium'' mass, particles still propagate on the 
light-cone.   
Finally, we have not written the various constants in terms of their
renormalized values.  Dressing the particle densities by solving the evolution
equations (which are nonperturbative) should, to some extent, be equivalent 
to using renormalized couplings.


\section{Summary and Implications for QCD Parton Transport Theory}

The ``source-propagator'' picture must apply to QCD partons since the
derivation of the phase-space Generalized Fluctuation-Dissipation Theorem 
does not depend on the form of the self-energies but
rather on the form of the Dyson-Schwinger equations for the contour 
propagators in~\eqref{eqn:DSE1}--\eqref{eqn:DSE3}.
It would then seem that if we find the QCD self-energies and define the 
parton distributions appropriately, we may construct QCD phase-space parton 
evolution equations.  However, before we could do this we must assess whether 
we need to dress the phase-space propagators and vertices and we must 
implement renormalization.  

In the present work, we would dress the particle densities by iterating the 
phase-space evolution equations but we would not dress the phase-space 
propagators or vertices.  Hopefully, dressing the particle densities is 
sufficient to incorporate any needed higher order or many particle effects.  
One simple form of dressing mentioned above is the in-medium mass shift.
Given this, it may prove necessary to give particles an effective mass and 
in this event, we would need the phase-space propagator for non-zero mass.
However,  we do not know the analytic form of the retarded phase-space 
propagators for particles with non-zero mass.  We are currently 
investigating propagation in this case and a summary of what we have so far is
in Appendix \ref{append:prop}.

The issue of implementing renormalization will require some work as there is 
not a well-developed understanding of renormalization in non-equilibrium 
quantum mechanics.  In momentum-space perturbation theory, renormalization 
is used to correct some parameters (e.g. a particle's mass) to make them 
correspond to their observed values.  Some of these 
corrections can be ascribed to many-particle effects that are effectively 
dealt with by dressing the densities, propagators and/or vertices.  
Nevertheless, there may be divergencies that need to be removed in our
formulation of non-equilibrium perturbation theory but, at the present, 
we have not yet encountered any. The issue of renormalization
brings up one other question.  Usually momentum-space renormalization is 
interpreted as removing physics at one momentum scale in favor of another 
scale.  It is not clear what this means in phase-space.  When renormalizing 
in phase-space, are we removing physics at a certain length scale, a certain 
momentum scale, both, or neither?  Is renormalization a form of smoothing in 
phase-space, akin to the gradient approximation? 


In any event, these two issues are intricately intertwined and their 
investigation is beyond the scope of the present work.  Nevertheless, in the 
absence of a phase-space evolution equation, we can still use the
Generalized Fluctuation-Dissipation Theorem as insight to build models.

\chapter{PARTONS IN PHASE-SPACE}
\label{chap:pips}

How can we rewrite the QCD Parton Model in phase-space?  This is a 
necessary step if one is to connect the quark and gluon phase-space densities
in a transport approach to the Parton Distribution Functions (PDF's) measured 
experimentally.  Two of the key components of the parton model are 
factorization of QCD cross sections and evolution of the parton densities.  
Both of these components can be studied within the Weizs\"acker-Williams 
approximation, the QED analog of the parton model.

Factorization in the QCD Parton Model is the idea that the cross section for 
a reaction involving a hadron can be written as the convolution of an 
elementary parton/target cross section and the Parton Distribution Function 
of the partons in the hadron \cite{QCD:alt77,QCD:qui83}.   The QED 
Weizs\"acker-Williams approximation follows exactly along this track:
a cross section in the Weizs\"acker-Williams approximation is the convolution
of the Effective Photon Distribution with the elementary photon/target cross
section \cite{gFT:wei34,gFT:wil34,jackson,ber88}.  We will extend the 
Weizs\"acker-Williams approximation to include electrons.
The analogy between the factorization in the parton model and the 
Weizs\"acker-Williams approximation makes even 
more sense when one realizes that both
photons and electrons are the point-like constituents of a dressed QED
point charge;  in this sense, photons and electrons are the {\em QED partons}
of the point charge.  Thus, by learning how to write the 
Weizs\"acker-Williams Approximation in phase-space, we will be 
showing factorization in parton model cross sections in phase-space.
Now, factorization does fail when there are interference terms in the S-matrix
squared and in Appendix~\ref{sec:probeedist} we discuss an example the failure
of factorization in phase-space.
My advisor, Pawe{\l} Danielewicz, and I were not the first to consider
writing cross sections in phase-space; Remler \cite{tran:rem90} rewrites 
transition probabilities in phase-space in 
his discussion of simulating many-particle 
systems in phase-space.  Remler's work is not immediately
applicable to partons because his work only applies to particles with a large 
mass.

In the parton model, evolution describes how the the parton densities change 
via parton splitting and radiation.  Evolution is modeled 
by using evolution equations, which are a set of coupled 
integro-differential equations for the quark and gluon densities, or by 
summing over a class of ladder diagrams in the Leading Logarithm 
Approximation.  One such ladder is shown in Figure \ref{fig:noneqNthgen}.  
We can study these ladder diagrams by building up a simplified QED parton 
ladder.  
The photon ``parton distribution'' is the boosted
Coulomb field of the point charge and constitutes the first leg in the QED
ladder.  The electron ``parton distribution'' is the virtual electron 
distribution from photons virtually splitting into an electron and a 
final state positron.  The electron is the second leg of the ladder and 
the positron is the first rung.  Since the QED coupling constant, 
$\alpha_{em}$, is small, only one rung is needed to describe the electron 
densities and no rungs are needed for the photons.  The QCD coupling, 
$\alpha_{s}$, is much larger implying that a large number of rungs will be 
needed to reasonably approximate the parton distributions.  Thus we can
only expect the QED ladder to give some of the qualitative features of the
full QCD problem.

Let us outline this chapter.  The first two sections, Sections \ref{sec:pdist}
and \ref{sec:edist}, outline the calculation
of the QED phase-space ``parton distributions'' of a point charge.  
We begin Section~\ref{sec:pdist}
by writing the Weizs\"acker-Williams Approximation in phase-space.  We do this 
in several steps.  First, we write the
reaction rate density for the ``partonic subprocess,''  namely the reaction
rate for absorbing a free photon.  By writing this rate in phase-space, we
illustrate how we convert a  momentum-space reaction rate to one in 
phase-space.  Next, we write the reaction probability for photon
exchange in phase-space.  Comparing the full reaction probability to the
reaction rate density for absorbing a photon, we can identify the 
phase-space Effective Photon Distribution.  This photon distribution is the 
effective photon number density in phase-space and it has the form of a 
phase-space source folded with a phase-space propagator.
Following this, we calculate the photon number
density surrounding a classical point charge and explain how the photon's 
phase-space propagator and phase-space source work.  Finally, we comment
on the implications of this section for the QCD parton model.  We will find 
that we understand how partons propagate, but since our photon source is 
point-like we do not learn anything about QCD parton sources. 
In Section \ref{sec:edist}, we continue the study of the QED parton 
distributions of a point charge by studying the first link in a parton ladder:
a virtual photon splitting into a virtual electron and on-shell positron. 
We start our analysis by generalizing the
phase-space Weizs\"acker-Williams Approximation to include electrons and
writing down the ``Effective Electron Distribution.''  This Effective Electron 
Distribution takes the ``source-propagator'' form.  While this  ``partonic''
splitting leads to a complicated form of the electron source, the shape of the
source is mostly determined by the underlying photon (the ``parent parton'')
distribution.
We calculate the electron distribution explicitly for a classical point charge 
and discuss how the electron propagates from the source to the observation 
point.

As a practical application of this study, in Section \ref{sec:QCD}
we examine the configuration
space structure of the parton cloud of a nucleon.  In principle, one should
Wigner transform the quark or gluon wavefunctions of a nucleon.
Since we do not know the quark or gluon
wavefunctions of a nucleon, such a specification is not possible and we must
result to model building.  One might envision constructing a model 
phase-space parton density of a nucleon
by multiplying the momentum space density (the Parton Distribution Function) 
and the coordinate space density of the partons 
\cite{tran:gei92a,tran:gei92b,tran:gei94,tran:gei93,tran:gei95}.  
This approximation neglects correlations between the momentum and position in 
the parton density which are present in the phase-space density 
\cite{Wig:tat83,Wig:lee95,tran:car83}.  One might insert these correlations 
using uncertainty principle based arguments 
\cite{QGP:mue89,tran:gei92a,tran:gei92b,tran:gei94,tran:gei93,tran:gei95}.  
This has intuitive appeal, but such a prescription is ad-hoc at best. 
We can approach this problem in a more systematic manner using 
some physical insight from the momentum-space renormalization-group 
improved parton model and our understanding of how particles propagate in 
phase-space.  In the renormalization-group improved parton model, one 
calculates the parton densities by evolving the densities in momentum 
scale, $Q^2$ (which we take to be 
the parton virtuality), and in
longitudinal momentum fraction, $x_F$.  This evolution is equivalent to 
evaluating a certain class of ladder diagrams and these diagrams  
can be re-cast in the form of the phase-space Generalized 
Fluctuation-Dissipation Theorem.  
Thus, we can discuss the shape of the parton phase-space densities of 
an hadron in the large-$Q^2$ limit or in the small-$x_F$ limit using 
a simple model for the nucleon and the phase-space propagators.
We argue that neither large-$Q^2$ partons nor small-$x_F$ partons extend
beyond the nucleon bag in the transverse direction.
We also argue that the large-$Q^2$ partons extend out an 
additional\footnote{The nucleon has 4-momentum $P_\mu=(P_0,P_L,\vec{0}_T)$.} 
$\hbar c/x_F P_L$ from the bag surface in the longitudinal direction.  This is 
in line with what others have estimated 
\cite{QGP:mue89,tran:gei92a,tran:gei92b,tran:gei94,tran:gei93,tran:gei95}.  
Furthermore, we estimate that the small-$x_F$ partons extend at least an
additional $\hbar c\sqrt{-q^2}$ from the bag 
so the small-$x$ parton cloud is substantially larger
than the large-$Q^2$ cloud.  

The reaction probabilities that we calculate in this section are for exclusive
reactions.  The interaction picture Feynman rules for the S-matrix needed for 
such calculations are found in many field 
theory books \cite{gFT:akh65,gFT:bog79,gFT:itz80,gFT:ste93}.  The densities 
we find are directly related to the densities we calculated in the previous 
chapter by the summation over all final states.  This is elaborated on 
somewhat in Appendix~\ref{append:spectra} where we discuss measurables of a 
heavy-ion reaction.  Furthermore, we can map our results directly to cross 
sections in the way outlined in Appendix \ref{sec:crosssection}.

\section{Photons as QED Partons}
\label{sec:pdist}

\begin{figure}
   \begin{center}
   \includegraphics[totalheight=2.6in]{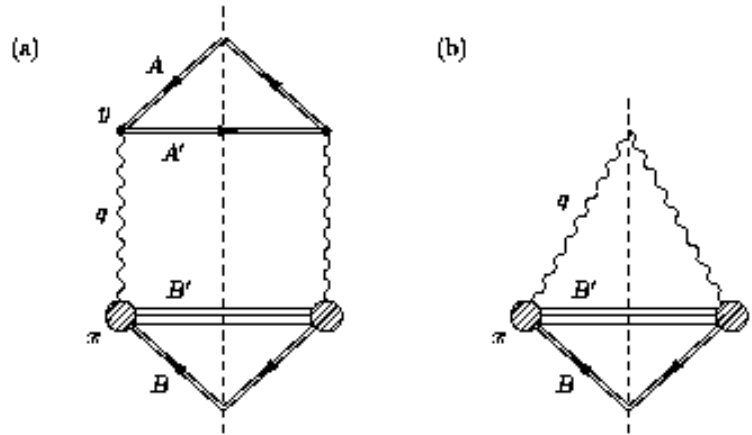}
   \end{center}
   \caption[Cut diagrams for photon exchange and free photon absorption.
	(a) Cut diagram for current A to exchange a photon with current B.  
	(b) Cut diagram for current B to absorb a free photon.  In both 
	figures, the photon/current B interaction is unspecified
      	and is represented with a blob.]
        {Cut diagrams for photon exchange and free photon absorption.  (a) Cut 
	diagram for current A to exchange a photon with current B.  (b) Cut 
	diagram for current B to absorb a free photon.  In both figures, the 
	photon/current B interaction is unspecified and is represented with 
	a blob.}
   \label{fig:dis}
\end{figure}

If we are to interpret photons as QED partons, we must write the photon 
exchange process reaction rate in a factorized, parton model-like, form.  
In other words, we want to write the cross section of the process of 
photon exchange (pictured in Figure \ref{fig:dis}(b)) as a convolution of the 
cross section for free photon absorption (pictured in Figure \ref{fig:dis}(a))
with an Effective Photon Distribution, and in phase-space.  We can then go 
on to study the properties of the QED version of a phase-space parton 
distribution with the example of the photon distribution of a point charge.  
Not only will we
rewrite the Weizs\"acker-Williams Approximation in phase-space in this work, 
but we will also show that the phase-space photon density has the form of a 
phase-space source convoluted with a phase-space propagator.  This 
formal structure of the phase-space densities is a general property as
we saw in Chapter \ref{chap:transport}, so it is not a real surprise to find 
it here.

\subsection{Photon Absorption}
\label{sec:pabsrate}

We begin this section by finding the photon/current B reaction rate, 
$\ReactRateonearg{\gamma B\rightarrow B'}$;   
this reaction rate is our ``partonic'' subprocess reaction rate.
Our derivation demonstrates how to rewrite
the reaction probability in terms of phase-space quantities.  
The high point in this calculation occurs in
Equation \eqref{eqn:itsaWT} when we identify the Wigner transforms of B's 
current and of the photon field.  This type of identification lets us rewrite
the reaction probabilities in phase-space.   

To find $\ReactRateonearg{\gamma B\rightarrow B'}$ we write the S-matrix for 
the process in Figure \ref{fig:dis}(b):
\begin{eqnarray*}
S_{\gamma B\rightarrow B'}
&= & \int\dn{4}{x}\ME{0}{A^{\mu}(x)}{\vec{q},\lambda}
       \ME{B'}{j_{\mu}(x)}{B}\\
&= & \int\dn{4}{x}\dnpi{4}{k}e^{-i k\cdot x}\ME{0}{A^{\mu}(x)}{\vec{q},\lambda}
       \ME{B'}{j_{\mu}(k)}{B}. 
\end{eqnarray*}
Here $\ME{0}{A^{\mu} (x)}{\vec{q},\lambda}=\sqrt{\frac{4\pi}{2|q_0|V}}
\pol{\mu}{\lambda}e^{iq\cdot x}$ is the free photon wave function 
(with $q^2=0$) and $j_{\mu}$ is the current operator for the probe particle B. 
We leave both the initial and final states of $B$ unspecified so 
the final state may be a single particle or 
several particles (as pictured in Figure \ref{fig:dis}(b)). 

If we now square the S-matrix and average over photon polarizations, we find:
\begin{eqnarray*}
\Smatrix{\gamma B\rightarrow B'}
& = & \int\dn{4}{x}\dn{4}{x'}
      \dnpi{4}{k}\dnpi{4}{k'}e^{-i(k\cdot x-k'\cdot x')}\\
&   & \times\displaystyle\frac{1}{2}\sum_{\lambda=\pm} 
      \ME{0}{A^{\mu}(x)}{\vec{q},\lambda} 
      \ME{\vec{q},\lambda}{A^{*\nu}(x')}{0} 
      \ME{B'}{j_{\mu}(k)}{B}\ME{B}{j^{\dagger}_{\nu}(k')}{B'}.
\end{eqnarray*}
On writing the coordinates and momenta in terms of the relative and average
quantities (i.e. $\tilde{k} = k-k'$ and $K=\frac{1}{2}(k+k')$),
and taking advantage of the momentum conserving delta
functions in the current matrix elements, $\Smatrix{\gamma B
\rightarrow B'}$  becomes      
\begin{equation}
\begin{split}
\Smatrix{\gamma B\rightarrow B'}=
& \int\dn{4}{X}\dn{4}{\tilde{x}}
    \dnpi{4}{K}\dnpi{4}{\tilde{k}}
    e^{-i(K\cdot\tilde{x}+\tilde{k}\cdot X)}
    \frac{1}{2}\sum_{\lambda=\pm}
    \ME{0}{A^{\mu}(X+\tilde{x}/2)}{\vec{q},\lambda} \\
& \times \ME{\vec{q},\lambda}{A^{*\nu}(X-\tilde{x}/2)}{0}
    \ME{B'}{j_{\mu}(K+\tilde{k}/2)}{B}
    \ME{B}{j^{\dagger}_{\nu}(K-\tilde{k}/2)}{B'}.
\end{split}
\label{eqn:itsaWT}
\end{equation}
There are two Wigner transforms in this equation:  the Wigner transform of the
photon field (the $\tilde{x}$ integral) and the Wigner transform of B's current
(the $\tilde{k}$ integral). 

Now we rewrite the S-matrix in terms of the phase-space quantities and define
the reaction rate density: 
\begin{equation}
\begin{array}{ccl}
\displaystyle
\Smatrix{\gamma B\rightarrow B'}
& = & \displaystyle\int\dn{4}{x}\dnpi{4}{k}\frac{\pi}{V|q_0|}\sum_{\lambda=\pm}
    \poldn{\mu}{\lambda}\polstardn{\nu}{\lambda}
    \twopideltaftn{4}{q-k}\Jcurrent{\mu}{\nu}{x}{k}{B}\\
& \equiv & \displaystyle\int\dn{4}{x}\ReactRate{\gamma B\rightarrow B'}{x}{q}.
\end{array}
\label{eqn:Ssquareabs}
\end{equation}
Here the Wigner transform  of the current is
\begin{equation}
   \Jcurrent{\mu}{\nu}{x}{q}{B}\equiv\int\dnpi{4}{\tilde{q}}
      e^{-i\tilde{q}\cdot x}\ME{B'}{j^{\mu}(q+\tilde{q}/2)}{B}
      \ME{B}{j^{\dagger \nu}(q-\tilde{q}/2)}{B'}.
\label{eqn:current}
\end{equation}
Since B's Wigner current is proportional to the reaction rate, it seems 
natural to give them both the same physical interpretation: as a 
``probability'' density,\footnote{Because the Wigner current is the Wigner 
transform of a quantum object, it may not be positive definite 
\cite{Wig:tat83,Wig:lee95,tran:car83} 
so it can not be strictly interpreted as a probability.} for
absorbing a free photon with momentum $q$ at space-time point $x$.  Now, it
may not be clear where the spatial structure of the reaction rate comes from,
especially since the incident photon is completely delocalized in space (it is
in a momentum eigenstate).  To give the reaction rate spatial structure, we
must localize either the initial or final states of $B$ with a 
wavepacket.\footnote{In other words, choose the appropriate density matrix.}  

\subsection{Photon Exchange}

We have the probability density,$\ReactRate{\gamma B\rightarrow B'}{x}{q}$, 
for free photon absorption in phase-space.  We now need the reaction rate 
for one-photon exchange (see Figure \ref{fig:dis}(a)) in phase-space in order
to extract the Effective Photon Density.  We do it two different ways:
in terms of the  Wigner transforms of the currents A and B and the photon 
propagator and in terms of the Wigner transform of the photon vector 
potential. The first form of the reaction rate has a clear physical 
interpretation in terms of photon emission, propagation, and absorption.  
However, it is the second form which can be brought into a factorized form.
 
The S-matrix for Figure \ref{fig:dis}(a) is 
\begin{equation} 
   S_{\rm AB\rightarrow A'B'}=\int d^4x d^4y \ME{A'}{j^{A \mu}(x)}{A}
      \Dcaus{\mu}{\nu}{}{}{x}{y}\ME{B'}{j^{B \nu}(y)}{B}.
\label{eqn:ABgoes2AB}
\end{equation}
Taking the absolute square of this S-matrix and rewriting it in terms of 
Wigner transformed currents and propagators, we find 
\begin{equation}
\Smatrix{AB\rightarrow A'B'}=\int\dn{4}{y}\dn{4}{x}\dnpi{4}{q}
        \Jcurrent{\mu}{\nu}{y}{q}{A} \Dcaus{\mu}{\nu}{\mu'}{\nu'}{y-x}{q}
        \Jcurrent{\mu'}{\nu'}{x}{q}{B}.
\label{eqn:photexch}
\end{equation}
Here, the Wigner transform of the photon propagator is
\begin{equation}\begin{array}{ccl}
   \Dcaus{\mu}{\nu}{\mu'}{\nu'}{x}{q}
   &=& \displaystyle\int\dn{4}{\tilde{x}}e^{i \tilde{x}\cdot q}
      D^{c}_{\mu\nu}(x+\tilde{x}/2)
      D^{c*}_{\mu'\nu'}(x-\tilde{x}/2) \\
   &=& (4\pi)^2g_{\mu\nu}g_{\mu'\nu'}\Gcaus{x}{q}
\end{array}\label{eqn:PhotProp}\end{equation}
and $\Gcaus{x}{q}$ is the Wigner transform of the scalar Feynman propagator. 
We outline the derivation and properties of $G^c(x,q)$ in Appendix 
\ref{append:prop}. 

Equation \eqref{eqn:photexch} has an obvious physical meaning:  
1)~current A makes a photon with momentum~$q$ at space-time point~$y$, 
2)~the photon propagates from~$y$ to~$x$ with momentum~$q$ and 
3)~current B absorbs the photon at space-time point~$x$.  
The spatial structure of the integrand of \eqref{eqn:photexch} can come from
localizing either $A$ or $B$.

Now we take a detour and calculate the Wigner transform of the vector
potential of the current $A$.  This detour will lead us to a form of the 
reaction probability amenable to a parton model-like interpretation.  
In terms of the current density and propagator,
the vector potential with causal boundary conditions 
is\footnote{Jackson uses the retarded boundary conditions for the vector 
potential because he discusses classical fields.} \cite{jackson}:
\begin{equation} 
   A^\mu(x)=\int\dn{4}{y}D^c_{\mu\nu}(x-y)J_A^\nu (y).
\label{eqn:vectpot}
\end{equation}
The Wigner transform of this is:
\begin{equation}
\begin{array}{ccl}
\Afield{\mu}{\nu}{x}{q} 
   &=&\displaystyle\int\dn{4}{\tilde{x}} e^{i\tilde{x}\cdot q} 
      A_\mu(x+\tilde{x}/2)A_\nu^{*}(x-\tilde{x}/2)\\ 
   &=&\displaystyle\int\dn{4}{y}\Jcurrent{\mu'}{\nu'}{y}{q}{A}
      \Dcaus{\mu'}{\nu'}{\mu}{\nu}{x-y}{q}.
\label{eqn:Afield}
\end{array}
\end{equation}
The Wigner transform of the vector potential has a ``source-propagator'' form.
Current $A$ (the photon source) creates the photon 
with momentum $q$ at
space-time point $y$ and the propagator takes the photon from $y$~to~$x$.  
Let us put this in Equation \eqref{eqn:photexch},
\begin{equation}
        |S_{AB\rightarrow A'B'}|^2=\int d^4 x \frac{d^4 q}{(2\pi)^4}
        A_{\mu\nu} (x,q) J^{\mu\nu}_B(x,q).
\label{eqn:halfway}
\end{equation}
Stated this way, the spatial structure of the integrand of this equation
comes from either localizing $B$ or from the spatial structure in the Wigner
transform of the photon vector potential.

Equation \eqref{eqn:halfway} is nearly factorized because current $B$ is 
proportional to the photon/current B reaction rate and, as we see in the 
next section, the vector potential is proportional to the phase-space 
Effective Photon Distribution.

\subsection{The Weizs\"acker-Williams Approximation}

We have the reaction rate for photon exchange, 
${\cal W}_{AB\rightarrow A'B'}(x,q)$, and the reaction rate for
absorption of a free photon, ${\cal W}_{\gamma B\rightarrow B'}(x,q)$. 
Let us compare them and extract the phase-space Effective
Photon Distribution.  
First, we decompose B's Wigner current into photon polarization vectors, 
allowing us to rewrite $\ReactRate{\gamma B\rightarrow B'}{x}{q}$ in terms of 
$\Jcurrent{\mu}{\nu}{x}{q}{B}$.  Knowing this, we can identify 
the Effective Photon Distribution.  

\subsubsection{Current Decomposition} 
 
If the photon probing $\Jcurrent{\mu}{\nu}{x}{q}{B}$ is sufficiently 
delocalized in space (in other words, $\partial_{\sigma} A_{\mu\nu}(x,q)
\ll q_{\sigma}A_{\mu\nu}(x,q)$), the momentum-space cutting 
rules tell us that we can expand $\Jcurrent{\mu}{\nu}{x}{q}{B}$ in terms of the
photon polarization vectors \cite{gFT:bud75}:\footnote{This particular 
decomposition of the current is specific to scalar current densities.  
Budnev et al. \cite{gFT:bud75} write down the polarization vectors for other 
cases.}
\begin{equation}
\begin{array}{ccl}
   \displaystyle\Jcurrent{\mu}{\nu}{x}{q}{B} 
   & = & \displaystyle\sum_{\lambda=\pm}\pol{\mu}{\lambda}
      \polstar{\nu}{\lambda}J_{trans}(x,q)\\
   & + & \displaystyle\pol{\mu}{0}\polstar{\nu}{0}J_{scalar}(x,q)\\
   & + & \displaystyle\frac{q^{\mu}q^{\nu}}{q^2}J_{long}(x,q).
\end{array}
\label{eqn:hadrontensor}
\end{equation}
Here, $\epsilon_\mu(0)$ is the scalar (i.e. time-like) polarization vector:
$\epsilon_\mu(0)=p_{B\mu}-q_\mu q\cdot p_B/q^2$, where $p_B$ is the momentum 
of B.  The transverse polarization vectors,
$\epsilon_\mu(\pm)$, span the hyperplane perpendicular to $\epsilon_\mu(0)$
and $q_\mu$.  Now, if $A_{\mu\nu}(x,q)$ is not delocalized, then
Equation~(\ref{eqn:hadrontensor}) should be modified to include
gradients\footnote{These gradients come from Wigner transforming terms
proportional to the relative photon momentum.} in $x$.  However, if we were to
include those gradients here, we may not be able to map 
$\Jcurrent{\mu}{\nu}{x}{q}{B}$ to ${\cal W}_{\gamma B\rightarrow B'}$. 

Since $\pol{\mu}{\lambda}\polstardn{\mu}{\lambda'}=\delta_{\lambda\lambda'}$,
it is simple to find the separate currents in (\ref{eqn:hadrontensor}) 
in terms of $\Jcurrent{\mu}{\nu}{x}{q}{B}$:
\[
   J_{scalar}(x,q)=\poldn{\mu}{0}\polstardn{\nu}{0}
   \Jcurrent{\mu}{\nu}{x}{q}{B}
\]
and 
\[
   J_{trans}(x,q)=\frac{1}{2}\sum_{\lambda=\pm}\poldn{\mu}{\lambda}
      \polstardn{\nu}{\lambda}\Jcurrent{\mu}{\nu}{x}{q}{B}
\]
The longitudinal piece, $J_{long}(x,q)$, vanishes due to current conservation. 

\subsubsection{The Effective Photon Distribution}

If we insert (\ref{eqn:hadrontensor}) 
into Equation (\ref{eqn:photexch}), the 
reaction probability is a sum of two terms:
\begin{equation}
\begin{array}{ccl}
   \Smatrix{AB\rightarrow A'B'} 
   & = & \displaystyle\int\dn{4}{x}\dnpi{4}{q}\Afield{\mu}{\nu}{x}{q}
      \sum_{\lambda=\pm}\pol{\mu}{\lambda}\polstar{\nu}{\lambda}
      J_{trans}(x,q)\\
   & + & \displaystyle\int\dn{4}{x}\dnpi{4}{q}\Afield{\mu}{\nu}{x}{q}
      \pol{\mu}{0}\polstar{\nu}{0}J_{scalar}(x,q).
\end{array}
\label{eqn:factorized}
\end{equation} 
The two terms in (\ref{eqn:factorized}) describe transverse 
and scalar photon exchange between currents A and B, respectively.  

Noting that if 
$J_{trans}(x,q)$ has  a weak $q^2$ dependence, then
\[J_{trans}(x,q)\propto\ReactRate{\gamma B\rightarrow B'}{x}{q}.\]  
In other words, $J_{trans}(x,q)$ is proportional to the reaction rate for the
``partonic'' subprocess.
Therefore, the transverse term of (\ref{eqn:factorized}) can be written as
\begin{equation}
   \Smatrix{AB\rightarrow A'B'}=\frac{1}{4\pi}\int \dn{4}{x}
     \frac{Vd^3 q}{(2\pi)^3} \frac{d q^2}{2\pi} 
      \EPdist\ReactRate{\gamma B\rightarrow B'}{x}{q},
\label{eqn:wwapprox}
\end{equation}
provided we identify the transverse Effective Photon Distribution 
as\footnote{We can make a similar identification for the scalar term.}
\begin{equation}
   \EPdist=\sum_{\lambda=\pm}\pol{\mu}{\lambda}\polstar{\nu}
   {\lambda}\Afield{\mu}{\nu}{x}{q}.
\label{eqn:wwapprox2}
\end{equation}
So, Equation \eqref{eqn:wwapprox} 
generalizes the Weizs{\"a}cker-Williams method to phase-space.  The Effective 
Photon Distribution in \eqref{eqn:wwapprox2} is the spin summed photon Wigner 
function.  In other words, it is the phase-space number density of 
photons at time $x_0$ per unit $q^2$.   Now,
when we assume that $J_{trans}(x,q)$ has  a weak $q^2$ dependence we are in
essence assuming that the photons are good quasi-particles.  We say this for 
the following reason:  since $J_{trans}(x,q)$ is nearly independent of 
$q^2$, we can perform the $q^2$ integral in Equation~\eqref{eqn:wwapprox} by 
pulling it past the reaction rate
to act solely on the Effective Photon Distribution.  In that case, we have
\[
\frac{dn_\gamma (x,\vec{q})}{\dn{3}{x}\dn{3}{q}}=\int \frac{dq^2}{2\pi}\EPdist
\]
which is the photon quasi-particle density (see 
Equation~\eqref{eqn:NormalWignerFtn}).  This should not be a surprise 
since we wanted a photon distribution that we could fold with a reaction
rate for on-shell particles.  We comment that conventional formulations of 
the parton model factorization give a 
similar result for the full parton model cross section:
the full parton model cross section is a folding of the
Parton Distribution Function (essentially the momentum-space quasi-particle 
density of partons) with a hard parton cross section (with on-shell parton).
While making partons quasi-particles is a questionable goal for transport
applications (as partons must be allowed to evolve in $q^2$ to properly 
include many-particle effects), a definition in terms of quasi-particles 
gives us the connection to the experimentally determined Parton Distribution 
Functions.  Finally, we state that the factorization we have achieved
here is possible because their are no interference terms in our expression for 
the S-matrix squared.  In Appendix \ref{sec:probeedist}, we explore the case 
of lepton pair production in the strong electromagnetic field of two passing 
ions.  In that case, there is interference between the photon fields of the 
two ions and factorization is not possible.

Now, while classical derivations of the Weizs{\"a}cker-Williams method begin 
with finding the photon power spectra from the Poynting flux
\cite{jackson,gFT:wei34,gFT:wil34}, a quantum mechanical
derivation follows along the lines of what we do here
\cite{gFT:bud75,ber88}.  Were we
to perform the spatial integrals in \eqref{eqn:factorized}, we would find that
the exponentials in the Wigner transforms conspire to make several
delta functions.  The resulting delta function integrations are trivial and we
would quickly recover the momentum-space result.

Finally, we comment that multiplying the photon phase-space density by the 
projector $\sum_{\lambda=\pm}\pol{\mu}{\lambda}\polstar{\nu}{\lambda}$ in
\eqref{eqn:wwapprox2} does {\em not} render the photon distribution gauge 
invariant.   
Were we in momentum space, the projection would render the distribution gauge 
invariant \cite{gFT:bud75}.  However because we are in phase-space, 
when we gauge transform $\Afield{\mu}{\nu}{x}{q}$ and apply the projector,
terms proportional to $q_{\mu}$ are removed but terms 
proportional to $\partial/\partial x^{\mu}$ are not. 
If the distribution is sufficiently delocalized than the gradients in $x$ become
negligible and we might find an ``approximately'' gauge invariant 
distribution.  Of course we could always find a pathological gauge that still
makes the result gauge dependent.  A truly gauge invariant virtual photon 
distribution is introduced in Appendix \ref{sec:gauge}.  This gauge invariant
distribution reduces to \eqref{eqn:wwapprox2} when $A_{\mu\nu}(x,q)$ is
sufficiently delocalized.

\subsection{The Photon Cloud of a Point Charge}
\label{sec:classEPD}

We know we can calculate the QED analog of a phase-space Parton Distribution
Function -- the phase-space Effective Photon Distribution.  
The phase-space Effective Photon Distribution is a quasi-particle distribution
and the important part of this distribution
is the Wigner transform of the vector potential, 
$\Afield{\mu}{\nu}{x}{q}$.  In this section, we calculate $A_{\mu\nu}(x,q)$ for 
the  simple case of a classical point charge radiating photons.  
If we localize the source's wavepacket and view it on a length scale larger 
than its localization scale, we can treat its density as a delta function.   
In this case, the shape of the photon distribution is determined the photon 
propagation and we can use this calculation to illustrate how
partons propagate in phase-space.  Now field theory
books tell us we should use the Feynman propagator to propagate the 
photons, however we will describe photon propagation via the Wigner
transform of the retarded (time-ordered) propagator.  We will argue that this
is a valid procedure and we will 
summarize the behavior of the propagator.  A complete discussion and derivation
of the analytic form of the propagator is contained in 
Appendix \ref{append:prop}.
As a result of the discussion in the appendix, the photon propagates a 
distance of roughly $R_\|\sim 1/|q_0|$ in the direction parallel to its 
3-momentum and $R_\perp\sim 1/\sqrt{|q^2|}$ in 
the direction perpendicular to its 3-momentum.
We demonstrate this behavior by plotting the coordinate-space distribution
of photons with $q^2\ll q_0^2$ (making the photons collinear with the source) 
and with $q^2 \sim q_0^2$.  In Appendix~\ref{append:static} we examine the
additional case of a static point charge (i.e. $\vec{v}=0$).  This case is not
relevant for QCD partons since a QCD parton source must be taken in the limit
$|\vec{v}|\rightarrow c$.

\subsubsection{The Photon Source}

For our source
current, we assume the source particle's wavepacket is localized on the
length scales that the photons can resolve so we can replace the source 
current with the current of a point particle.\footnote{We discuss when this 
replacement is valid in Appendix \ref{append:current}.}
The source particle follows a classical trajectory $x_{\mu}=x_{0}v_{\mu}$ 
with four-velocity $v_{\mu}=(1,v_L,\vec{0}_T)$, $v_L \approx c$ 
and $\gamma=1/(1-v_L^2)\gg 1$.  Ignoring the recoil caused by 
photon emission, the current of the point charge 
is \cite{jackson} 
\[j_{\mu}(x)=e v_{\mu} \: \deltaftn{3}{\vec{x}-x_0\vec{v}}.\]
The Wigner transform of this is the classical Wigner current:
\begin{equation}
\begin{array}{rl}
   \Jcurrent{\mu}{\nu}{x}{q}{classical} 
   & = \displaystyle \int \dn{4}{\tilde{x}} e^{iq\cdot\tilde{x}}
      j_\mu(x+\tilde{x}/2) j_\nu^\dagger(x-\tilde{x}/2)\\
   & = 2\pi \alpha_{em} \: v_{\mu} v_{\nu}\: \delta (q\cdot v)
      \deltaftn{3}{\vec{x}-x_0\vec{v}}.
\end{array}
\label{eqn:classcurrent}
\end{equation}
Here $e^2=\alpha_{em}$ is the QED coupling constant.

The current has several easy to interpret
features.  The first delta function sets $q\cdot v=0$.  This ensures that the 
emitted photons are space-like and that current is conserved.  It also 
insures that, when $q^2\rightarrow 0$, the
photons become collinear with the emitting particle 
($q_0 = v_L q_L \approx q_L$ making $q^2\approx {\vec{q_T}}^2\approx 0$).  
This delta function arises because
we neglect the recoil of the point charge as it emits a
photon.  The second delta function insures
that the source is point-like and follows its classical trajectory.  

This source has one other feature of note:  it allows for emission of both
positive and negative energy photons. In the following work, we consider 
creating only positive energy photons so we insert a factor
of $2\theta(q_0)$ in (\ref{eqn:classcurrent}).  This amounts to constraining 
the source's initial energy to be greater its final energy.  

\subsubsection{The Photon Propagator}
\label{sec:simpleprop}

For the photon propagator, we take the result in 
Equation~\eqref{eqn:Afield} and 
replace the scalar Feynman propagator with the retarded propagator.  We have 
several very good reasons why we can do this.  The first is that, in the 
momentum representation, the retarded propagator and the Feynman propagator are
identical for particles with positive energy 
(normally they differ in their $+i\varepsilon$
prescription).  The second is that the Feynman rules for the S-matrix 
can be formulated equivalently in terms of either propagator \cite{gFT:leh59}.
The final reason why we may make this replacement is that one can view 
Equation~\eqref{eqn:Afield} as 
an approximation of Equation~\eqref{eqn:noneqEPD1} where only one exit channel 
dominates.  In Equation~\eqref{eqn:noneqEPD1}, the propagator must be retarded 
by the time-ordering requirements implicit in our discussion in 
Chapter~\ref{chap:transport}.

Given that the switch is allowed, the scalar part\footnote{The vector part 
($g_{\mu\nu}g_{\mu'\nu'}$ in the Lorentz gauge) of 
the propagator has been removed for clarity.} of the retarded phase-space 
propagator is given in Appendix~\ref{append:prop}:
\[   
\Gplus{x}{p}=\frac{1}{\pi}\theta (x_0)\theta (x^2) \theta (\lambda^2) 
      \frac{\sin{(2\sqrt{\lambda^2})}}{\sqrt{\lambda^2}} 
\]
Here $\Gplus{x}{p}$ is the scalar part of the photon propagator, $x$ is the 
space-time displacement the photon with momentum $q$ traverses and the 
Lorentz invariant $\lambda^2$ is $\lambda^2=(x\cdot q)^2-x^2 q^2.$
Looking at the equation for $\Gplus{x}{p}$, there are some obvious features:  
the photons must propagate
inside the light-cone and forward in time.  However the rest of the features
of the propagator are tied up in the dependence on $\lambda^2$.  In  
Appendix \ref{append:prop} (especially Section \ref{sec:retWorks}), we go into 
great detail understanding what this $\lambda^2$ dependence means and we will 
summarize the results here.

\begin{figure}
   \begin{center}
   \includegraphics[totalheight=\bagfigheight]{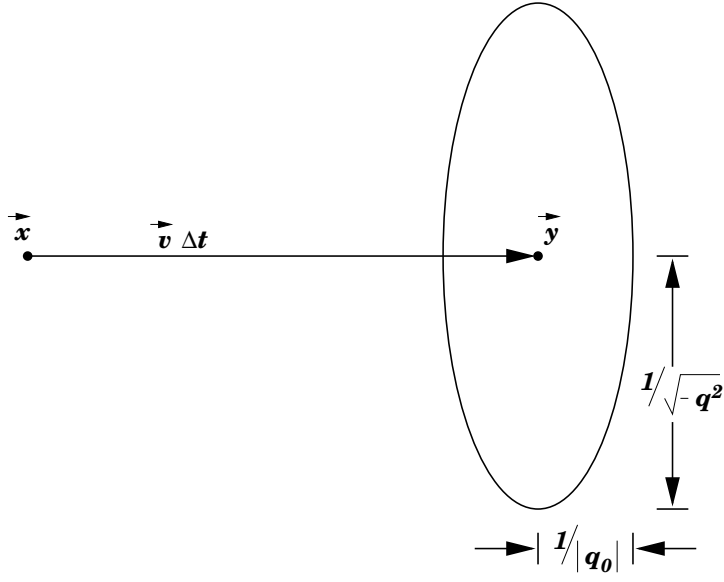}
   \end{center}
   \caption[A schematic of phase-space propagation for off-shell particles.]
	{A schematic of phase-space propagation for off-shell particles.}
   \label{fig:howprop}
\end{figure}

Given that our source can only make photons with $q^2\leq 0$, we only 
describe space-like and on-shell propagation.  
The case of time-like propagation is discussed in the appendix.  
Now, what we found was that massless particles in phase-space do not 
follow their classical trajectory.  In fact, if we were to characterize their
trajectory by the start and end points of the trajectory, then what we found
is not a fixed end point at $\vec{x}+\vec{v}\Delta t$, but rather an entire
end {\em zone}.  The size of the zone depends on the virtuality and energy
of the photon.  This is sketched for photons with space-like momentum in
Figure~\ref{fig:howprop}.  As one can see, the photon can end up anywhere in the
end zone; the transverse size of the zone is set by the off-shellness of the 
photon and the longitudinal size is set by the energy of the photon.  For 
on-shell photons the situation is similar except both the transverse and 
longitudinal size is set by the energy of the photon.  In actual fact,
the propagator is a bit more complicated then what we lay out here, but these
estimates of the propagation distance will serve us well in understanding the
photon density to follow.

\subsubsection{The Photon Density}

Now we put the photon source and propagator together into the photon density.  
We concentrate 
our efforts on $A_{\mu\nu}(x,q)$ because all of the spatial dependence of the
photon distribution is tied up in the Wigner transform of the vector potential.
Inserting the classical
current and the retarded photon propagator in the 
Lorentz gauge into (\ref{eqn:Afield}), we find
\begin{equation}
   \Afield{\mu}{\nu}{x}{q} = 4\pi \alpha_{em} \: v_{\mu} v_{\nu}
      \theta (q_0) \delta (q\cdot v)\int\dn{4}{y}\Gplus{x-y}{q} \:
      \deltaftn{3}{\vec{y}-y_0\vec{v}}.
\label{eqn:retintoA}
\end{equation}
The delta function integrals 
in Equation~\eqref{eqn:retintoA} are trivial, 
however the remaining proper time integral can not be done analytically.
We find
\begin{equation}\begin{split}
   \Afield{\mu}{\nu}{x}{q}=&\frac{(8\pi)^2 \alpha_{em} \gamma 
      \theta(q_0)\delta(q\cdot v)}
      {\sqrt{-q^2}}v_{\mu}v_{\nu}\\&\times\twoargftn{A}{2|x\cdot q|}
      {2\sqrt{-q^2\gamma^2\left((x\cdot v)^2-x^2 v^2\right)-(x\cdot q)^2}},
\end{split}
\label{eqn:Amunu}
\end{equation}  
where the dimensionless function $\twoargftn{A}{a}{b}$ is given by
\begin{equation}
   \twoargftn{A}{a}{b}=\int_{a}^{\infty}d\tau \frac{\sin{\tau}}
      {\sqrt{b^2+\tau^2}}.
\label{eqn:dimlessA}
\end{equation}  

There are two interesting cases that are easy to explore: that of 
photons nearly collinear to the source particle 
(i.e. $q_0\approx q_L\gg |\vec{q}_T|$), and that of photons with a large 
transverse momentum (i.e. $|\vec{q}_T|\sim q_0,q_L$).  Since there is a 
$1/\sqrt{-q^2}$ singularity in the photon density and nearly on-shell
photons (i.e. $q^2\rightarrow 0$) are collinear, there will be many more
collinear photons than any other kind.  

\begin{figure}
   \begin{center}
   \includegraphics[width=\textwidth]{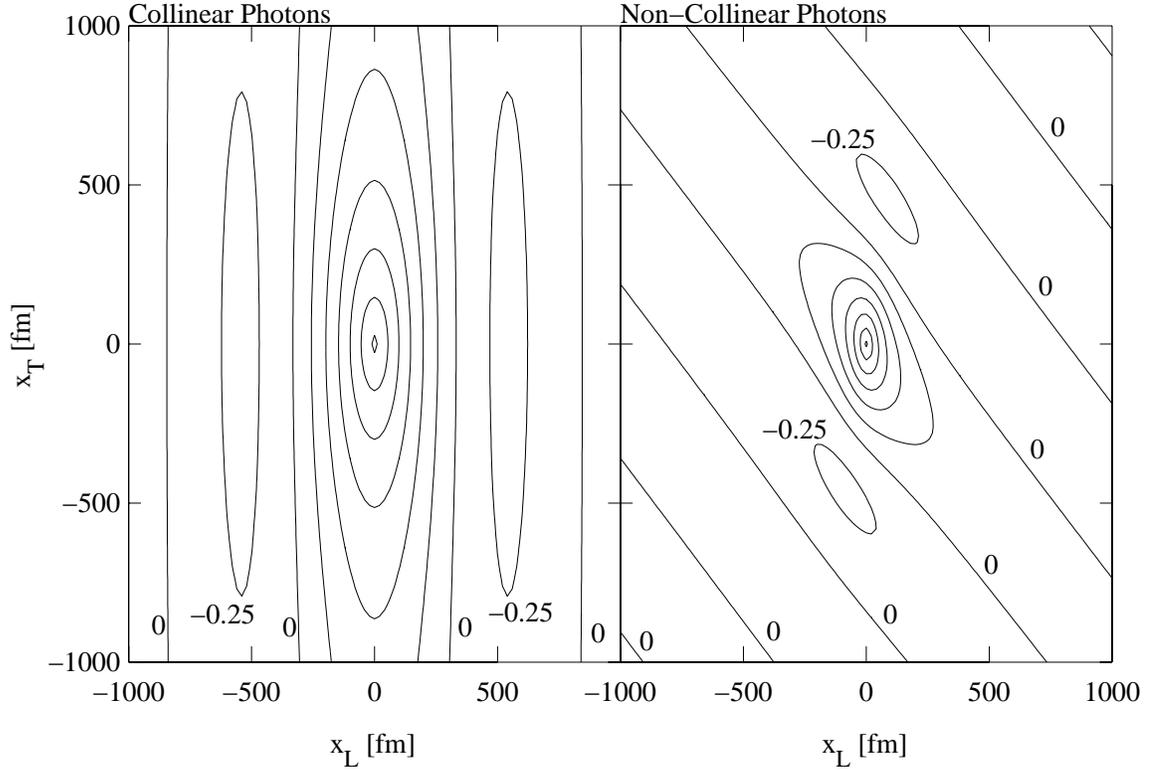}
   \end{center}
   \caption[Plots of the photon phase-space density.  Both figures are plots 
	of the dimensionless function ${\cal A}$ corresponding to the 
	photon number density of a point charge with 3-velocity 
	$\vec{v}=(v_L,\vec{0}_T)$ where $v_L=0.9c$. 
        These slices of the phase-space density have
        $q_{\mu}=(m_{e},m_e/v_L,\vec{0}_T)$ (left) and 
        $q_{\mu}=(m_{e},m_e/v_L,0.56\mbox{ MeV/c},0) $ (right).
        In both plots, only the negative and zero contours are labeled.  The 
        positive contours increase in increments of $0.25$.]
	{Plots of the photon phase-space density.  Both figures are plots of 
	the dimensionless function ${\cal A}$ corresponding to the 
	photon number density of a point charge with 3-velocity 
	$\vec{v}=(v_L,\vec{0}_T)$ where $v_L=0.9c$. 
        These slices of the phase-space density have
        $q_{\mu}=(m_{e},m_e/v_L,\vec{0}_T)$ (left) and 
        $q_{\mu}=(m_{e},m_e/v_L,0.56\mbox{ MeV/c},0) $ (right).
        In both plots, only the negative and zero contours are labeled.  The 
        positive contours increase in increments of $0.25$.}
   \label{fig:photondist}
\end{figure}

Two plots, representative
of collinear photons and high $\vec{q}_T$ photons, are shown in
Figure \ref{fig:photondist}.  The left is a plot of the
dimensionless function $\twoargftn{A}{a}{b}$ for collinear photons with
$q_{\mu}=(m_{e},m_e/v_L,\vec{0}_T)$.  On the right is a plot of 
$\twoargftn{A}{a}{b}$ for photons with
transverse momentum comparable to their transverse momentum and energy, 
$q_{\mu}=(m_{e},m_e/v_L,0.56 {\rm MeV/c},0)$.    
The characteristic energy scale of QED is $m_e$, so we choose this scale for 
the momenta to plot.  In both plots, we chose $v_L=0.9c$ to illustrate the 
Lorentz contraction of the distribution.
The oscillations exhibited by both photon distributions are expected for
a Wigner transformed density \cite{Wig:tat83,Wig:lee95,tran:car83}.  To 
obtain an equivalent classical distribution, one should smear this 
distribution over a unit volume of phase-space.  

Both cuts through the photon distribution show Lorentz
contraction.  For the collinear photons, this contraction occurs in
the longitudinal direction.  We can account for the contraction with the 
behavior of the retarded propagator.  We expect that the  
width will be $\sim R_\| = \hbar c/|q_0|$ parallel to  $\vec{q}$
and $\sim R_\perp = \hbar c/\sqrt{|q^2|}$ perpendicular to $\vec{q}$.  
For the collinear photons, $\vec{q}$ is in the longitudinal direction and
$q_0=\gamma\sqrt{|q^2|}$ so the longitudinal width is 
$\sim R_L= \hbar c/\gamma\sqrt{|q^2|}$.  
In other words, the collinear photon distribution is a
``Lorentz contracted onion'' centered on the moving point source.  
The inner layers of this ``onion'' correspond to higher $|q^2|$ photons.  
However, we must emphasize that the contraction is
{\em not} due to the movement of the source, but rather due to kinematics of 
the photon's creation and the propagation of the photon. 
To illustrate this point, one only needs to look at the high transverse
momentum photons: their distribution is tilted.  In the case plotted on the
right in Figure \ref{fig:photondist}, the photon momentum points 
$45^{\circ}$ to the longitudinal direction, coinciding with the tilt
of the distribution.
Furthermore, the width of the distribution is $\sim R_\| = \hbar c/|q_0|$ 
along this tilted axis and $\sim R_\perp = \hbar c/\sqrt{|q^2|}$ 
perpendicular to this tilted axis.

\subsection{What the Photons Tell Us about QCD Partons}

QCD parton model cross sections can be written in phase-space as a folding of 
the phase-space Parton Distribution Function with the reaction rate for the
partonic sub-process.  
A phase-space Parton Distribution Functions is the
number of quasi-particle partons per unit volume in phase-space.  The 
phase-space Parton Distribution Functions is simply related to the 
underlying quark and gluon phase-space densities through integration over the
off-shellness of the particles.  The phase-space Parton Distribution Functions 
have a ``source-propagator'' form and 
possibly may be defined in a gauge invariant manner as discussed in 
Appendix~\ref{sec:gauge}.  It may not be necessary
to resort to gauge invariant parton distributions however, provided the 
parton evolution equations are gauge covariant.
If the phase-space parton source produces only positive
energy partons or if we use time-ordered field theory then the partons
propagate from their source using the Wigner transform of the retarded
propagator.  This retarded propagator propagates off-shell partons up to
roughly $R_\| = \hbar c /{\rm min}(|q_0|,|\vec{q}|)$
parallel to the parton three-momentum and 
$R_\perp = \hbar c/\sqrt{|q^2|}$ perpendicular to the parton \sloppy\linebreak 
3-momentum.    Both of these estimates are valid only in frames
with $q_0, \vec{q} \neq 0$.  When either $q_0=0$ or $\vec{q}=0$, propagation 
is cut off at $\sim R_{\|,\perp} = \hbar c/\sqrt{|q^2|}$.  On-shell (i.e.
$q^2=0$) partons tend to follow their classical trajectory, with deviations
from that trajectory of order $1/|q_0|$.

Despite what we have learned, we know next to nothing about 
QCD parton sources in phase-space.  We use a point source here while,
because of the finite size of the valence quark bag, a nucleon has spatial 
structure on the length scales of interest.   Furthermore, QCD partons 
radiate other partons making Figure \ref{fig:dis}a an entire ladder diagram
and this alters the source.  We gain more insight into the phase-space sources 
in the next few sections. 

\section{Electrons as QED Partons}
\label{sec:edist}

\begin{figure}
   \begin{center}
   \includegraphics[totalheight=2.6in]{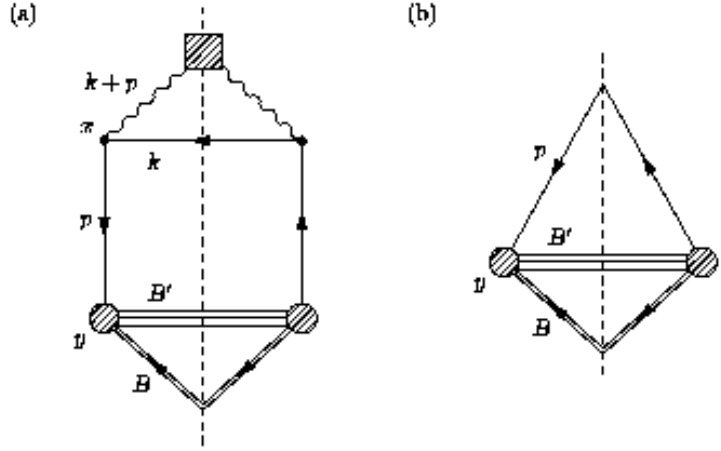}
   \end{center}
   \caption[Cut diagrams for a photon splitting into a positron-virtual electron
	pair and for a free electron interacting with a probe.  (a) Cut
	diagram for creating an electron-positron pair by photon splitting.  
	The virtual electron interacts with the  
      	probe particle, $B$.  The square vertex represents the photon source.  
      	(b) Cut diagram for a free electron interacting with the probe 
	particle.]
	{Cut diagrams for a photon splitting into a positron-virtual electron
	pair and for a free electron interacting with a probe. (a) Cut 
	diagram for creating an electron-positron pair by photon splitting.  
	The virtual electron interacts with the  
      	probe particle, $B$.  The square vertex represents the photon source.  
      	(b) Cut diagram for a free electron interacting with the probe 
	particle.}
   \label{fig:splitting}
\end{figure}

In the QCD parton model, the Parton Distribution Functions can be found by 
summing a class of ladder diagrams and the simplest of these has only one rung,
corresponding to a single partonic splitting.
The QED analog of the first rung of such a ladder is shown in 
Figure~\ref{fig:splitting}(a) where we see a virtual photon splitting into 
an electron-positron pair.
Probing the electron distribution spawned by this process occurs in three steps:
1)~a virtual photon splits into an electron-positron pair with the positron
on-shell, 2)~the virtual electron propagates from the splitting point toward 
the probe particle and 3)~the electron interacts with the probe.  
In step 1, we assume that the leading contribution to the virtual electron 
distribution comes from photon splitting.  We put the positron on-shell in 
order to sum over its states -- thus encapsulating all possible emission 
contributions to the electon density.  We can give the rate for 
Figure \ref{fig:splitting}(a) a parton-model like form by 
associating steps 1 and 2 with the ``Effective Electron Distribution''
and step 3 with the electron/probe reaction rate (see Figure
\ref{fig:splitting}(b)).  

Let us outline this section.  In Subsection~\ref{sec:3a}, we demonstrate 
that the reaction rate for the virtual electron exchange process in 
Figure~\ref{fig:splitting}(b) factorizes in phase-space.  In the process, 
the electron phase-space density acquires the ``source-propagator'' form.
In Subsection~\ref{sec:3b}, we calculate the electron distribution of a point
charge.  The electron source shape is determined mostly by the shape of the
parent photon distribution.  As with the photon distribution, we use the 
retarded propagator here instead of the Feynman propagator.  We assume the
electrons are massless throughout this section because quarks in QCD are 
effectively massless ($\Lambda_{QCD} \sim m_q
\ll p_0,|\vec{p}|$).  Finally, in Subsection~\ref{sec:3d} we discuss 
the implications of this section.  

\subsection{Factorization}
\label{sec:3a}

First we want to show that the process in Figure~\ref{fig:splitting} can be 
factorized in phase-space, as needed for a parton model-like form.
The S-matrix for the process in Figure~\ref{fig:splitting}(a) is:
\begin{equation}
        S_{\rm \gamma B\rightarrow \bar{e} B'} = 
                \int \dn{4}{x}\dn{4}{y}
                A_{\mu}(x)\psi_{\bar{e}}(x,s) e\gamma^{\mu} S^c(x-y)
                {\cal V}_{\rm B e\rightarrow B'} (y).
\label{eqn:eSmatrix}
\end{equation}
The spatial structure of the electron source comes from localizing the
photon vector potential, $A_\mu(x)$.  The spatial structure of the ``partonic''
subprocess comes from ${\cal V}_{\rm B e\rightarrow B'} (y)$, the 
electron/probe interaction in Figure~\ref{fig:splitting}(b).
In Equation~(\ref{eqn:eSmatrix}),
 $\psi_{\bar{e}}(x,s)=\int\dnpi{4}{k}v(k,s) e^{i k\cdot x}
\frac{f^{*}(k)}{\sqrt{2 k_0 V}}$ is the final state positron
wavepacket and $S^{c}(x-y)=\int\dnpi{4}{p} e^{-i p
\cdot (x-y)}\frac{\not{p}+m_e}{p^2+m_e^2+i\epsilon}$ is the electron's
Feynman propagator.  

We square $S_{\rm \gamma B\rightarrow \bar{e} B'}$ and write it in
terms of phase-space quantities:
\begin{equation}
\begin{split}
       \Smatrix{\gamma B\rightarrow \bar{e} B'}=&\alpha_{em}
               \displaystyle\int\dn{4}{x}\dn{4}{y}
               \dnpi{4}{p}\dnpi{4}{q}\dnpi{4}{k}\Afield{\mu}{\nu}{x}{q}
               f(x,k)\\
       &\times \twopideltaftn{4}{k+p-q}{\rm Tr} \left\{ \frac{1}{2} 
               (\dirslash{k}-m)\gamma^{\mu}
               S^{c}(y-x,p){\cal V}_{\rm B e\rightarrow B'}(y,p)
               \gamma^{\nu}\right\}
\end{split}
\label{eqn:wignerSelec}
\end{equation}
Here, $S^{c}(y-x,p)$ is the Wigner transform of the two electron Feynman 
propagators and it can be written in terms of the Wigner transform of the 
scalar Feynman propagator, $\Gcaus{x}{q}$:
\[\begin{array}{rl}
   \Scaus{\alpha}{\alpha'}{\beta}{\beta'}{x}{p}
   &= \displaystyle \int \dnpi{4}{\tilde{p}} e^{-i\tilde{p}\cdot x}
      S^c_{\alpha\beta}(p+\tilde{p}/2)\bar{S}^c_{\alpha'\beta'}(p-\tilde{p}/2)\\
   &= (\dirslash{p}+i\dirslash{\partial}+m_e)_{\alpha\beta}
      (\dirslash{p}-i\dirslash{\partial}+m_e)_{\alpha'\beta'}\Gcaus{x}{p}.
\end{array}\]
Also in Equation~(\ref{eqn:wignerSelec}), ${\cal V}_{\rm B e\rightarrow B'}(y,p)$ is
the Wigner transform of the electron/probe interaction and $f(x,k)$ is the
phase-space density of final state positrons.

Now we want to sum over the full set of positron final states.  The positron
is on-shell and in a momentum eigenstate\footnote{This makes the positron 
momentum weight-function $f^{*}(k)\propto \delta^4 (k-k_f)$, 
with $k_f^2=m_e^2$, and the positron phase-space density 
$f(x,k)=(2V|k_{f0}|)^{-1}\twopideltaftn{4}{k-k_f}$.} so we sum over the 
positron momentum and spin.
Furthermore, we can separate off the spinor structure of the electron 
propagator and shift the derivatives to act on 
${\cal V}_{\rm B e\rightarrow B'}$.  In the end we find
\begin{equation}
\begin{split}
        \Smatrix{\gamma B\rightarrow \bar{e} B'} =& \displaystyle\alpha_{em}
                \int\dn{4}{x}\dn{4}{y}\dnpi{4}{p}\dnpi{4}{q}\dnpiinvar{3}{k_f}\\
        &\displaystyle\times\Afield{\mu}{\nu}{x}{q}
                \Gcaus{y-x}{p}\twopideltaftn{4}{k_f+p-q}\\
        &\displaystyle\times {\rm Tr} \left\{ \frac{1}{2} 
                (\dirslash{k}_f-m)\gamma^{\mu}
                (\dirslash{p}+i\dirslash{\partial}/2+m_e)
                {\cal V}_{\rm B e\rightarrow B'}(y,p)
                (\dirslash{p}-i\dirslash{\partial}/2+m_e)\gamma^{\nu}\right\}.
\end{split}
\label{eqn:elecSmatrix}
\end{equation}

Since ${\cal V}_{\rm B e\rightarrow B'}(y,p)$ is separated from the electron
propagator, the reaction probability is factorized.  We could explicitly 
calculate the rate for the partonic subprocess $eB\rightarrow B'$, but 
we are only interested in the shape of the distribution 
as a function of electron momentum.  We can guess the form of the electron
density just by looking at Equation (\ref{eqn:elecSmatrix}),
without performing the explicit rate density calculation. 
The electron density is
\begin{equation}\begin{array}{ccl}
   \displaystyle\EEdist &\propto &\displaystyle\alpha_{em} \int \dn{4}{x} 
        \Gcaus{y-x}{p} 
      \int \dnpiinvar{3}{k} \Afield{\mu}{\nu}{x}{k+p}\\
      &\equiv &\displaystyle\int \dn{4}{x} \Gcaus{y-x}{p} \Sigma(x,p).
\end{array}\label{eqn:elecdist}\end{equation}
Here, $\Gcaus{x}{p}$ is the Wigner transform of the scalar propagator.
Equation (\ref{eqn:elecdist}) has the ``source-propagator''
form: $\Sigma$, the integral of $\Afield{\mu}{\nu}{x}{k+p}$ over the 
positron momentum, plays the role of the partonic source.  
Because the emitted positron's wavepacket has no configuration space structure 
(it is in a momentum eigenstate), the spatial structure of the source comes 
solely from the parent photon's phase-space distribution.

At this stage, we see several important features of the source.
First, we note the $d^3k/|k_0|$ in the positron momentum integral.
This factor weights positron emission toward small $k_0$ and is the origin
of the small--$x_F$ singularity used in BFKL evolution
\cite{QCD:fie89,QCD:dok91,QCD:CTEQ,QCD:tun94,QCD:lae94}.  
Second, we note that the entire spatial
dependence of the electron source comes from the parent photon distribution. 
These two points are especially important for the partons in 
Section~\ref{sec:QCD} so they are elaborated on in the next subsection.

\subsection{The Electron Cloud of a Point Charge}
\label{sec:3b}

Our main interest is with how the ``parton ladder'' (in our case, the source 
is only one rung of the ladder) shapes the electron distribution.  First, we
discuss the electron's source and how both the parent photon distribution and 
the cut positron rung effect it.  Second, we discuss the interplay of 
the electron creation and propagation.
Because the electron has positive energy, we
can use either the retarded or Feynman phase-space propagator.  
We choose to use the retarded propagator.
We describe the Feynman propagator in Appendix \ref{append:prop}.  

\subsubsection{The Electron Source}
\label{sec:esource}

In our electron source, we use the photon distribution of
Equation~(\ref{eqn:Amunu}).  This is not a QCD parton-like distribution as the
photon source is point-like.  Nevertheless, because the electron
source is a short parton ladder, it contains many of the
general features that one expects from a QCD parton source.  
In particular it contains the 
$1/|k_0|$ singularity from the integration over final state positrons 
(in other words, the cut positron rung).  We
discuss the effects of this integration and we detail both the shape of the 
source and how this shape depends on the photon distribution.  

Up to irrelevant factors, the electron source is 
\begin{equation}
   \Sigma(x,p) \propto \alpha_{em}\int \dn{4}{k} \theta(-k_0)
      \theta(p_0+k_0)\deltaftn{}{k^2-m_e^2} \deltaftn{}{q\cdot v} 
      \frac{\twoargftn{A}{a}{b}}{\sqrt{-(k+p)^2}}
\label{eqn:edist_w1}
\end{equation} 
where $a=2|x\cdot (k+p)|$ and $b=2\sqrt{-(k+p)^2\gamma^2
((x\cdot v)^2-x^2v^2)-(x\cdot (k+p))^2}$.  The longitudinal and temporal
positron momentum integrals can be done with the aid of the delta functions, 
leaving the transverse momentum integrals:
\begin{equation}
\begin{array}{rl}
   \Sigma(x,p) \propto & \displaystyle\alpha_{em}\theta(p\cdot v) 
      \int_{|\vec{k}_T|<k_{T {\rm max}}}\frac{\dn{2}{k_T}}
      {\sqrt{k_{T {\rm max}}^2-k_T^2}} \\
      & \times\displaystyle\left\{\frac{\theta(p_0+k_{0+})
         \twoargftn{A}{a_{+}}{b_{+}}}{\sqrt{-(k_{+}+p)^2}} 
      +\frac{\theta(p_0+k_{0-})\twoargftn{A}{a_{-}}{b_{-}}}
        {\sqrt{-(k_{-}+p)^2}} \right\}.
\end{array}
\label{eqn:edist_w3}
\end{equation}
Here, $k_{T {\rm max}}^2=\gamma^2 (p\cdot v)^2-m_e^2$ and
\begin{equation}
\begin{array}{rcl}
   k_{0 \pm} & = & -\gamma(\gamma p\cdot v\mp v_L\sqrt{k_{T{\rm max}}^2-
        \vec{k}_T^2})\\
   k_{L \pm} & = & -\gamma(\gamma v_L p\cdot v\mp \sqrt{k_{T{\rm max}}^2-
        \vec{k}_T^2}).
\end{array}
\end{equation}

Now, in a parton ladder we expect to find a factor of $d^3k/|k_0|$ for each
cut rung.  Here is no exception, one can see that 
$d^4k \theta (-k_0)\delta(k^2-m^2_e)$ gives us this factor.  However, 
because we neglect the recoil of the source, we have an additional 
$\delta(q\cdot v)$, turning this factor into 
$d^2k/\sqrt{k_{T{\rm max}}^2-k_T^2}$.  Because of this factor, the positron's
tend to have $|\vec{k}_T|\ll |k_L|$, making the positron momentum collinear 
(or anti-collinear) to the point charge's velocity.  The small photon 
momentum forces the electron to be emitted with momentum opposite the 
positrons.  Incidentally, one finds this same behavior of the momenta of the 
leg and rung partons of a real QCD parton ladder.  There, the collinearity of 
the momenta of the partons with the hadron momentum gives rise to the so-called
collinear or infra-red singularities.

Now, the shape of $\Sigma$ comes from the parent photon
distribution.  In our simple case, we can actually estimate the 
$\left<q_\mu\right>$ that gives the dominant contribution to $\Sigma$.  
For $v_L\approx 1$, we estimate that the average 
$\left< k_{\mp\mu} \right>$ is given by
\[
	\left< k_{\mp\mu} \right>=\gamma^2 (p\cdot v) \left(-1\pm \frac{1}{2}, 
	-1\pm \frac{1}{2}, \cos(\theta_T)/\gamma,\sin(\theta_T)/\gamma\right)
\]   
For the purposes of illustration, we choose to emit the positron in the 
direction $\hat{k}_T\cdot\hat{x}_T =\cos(\theta_T)=\frac{1}{\sqrt{2}}$.  
By momentum conservation, the dominant photon momentum is
$\left<q_{\pm\mu}\right>=p_\mu+\left<k_{\pm\mu}\right>$.

\begin{figure}
    \begin{center}
    \includegraphics[width=\textwidth]{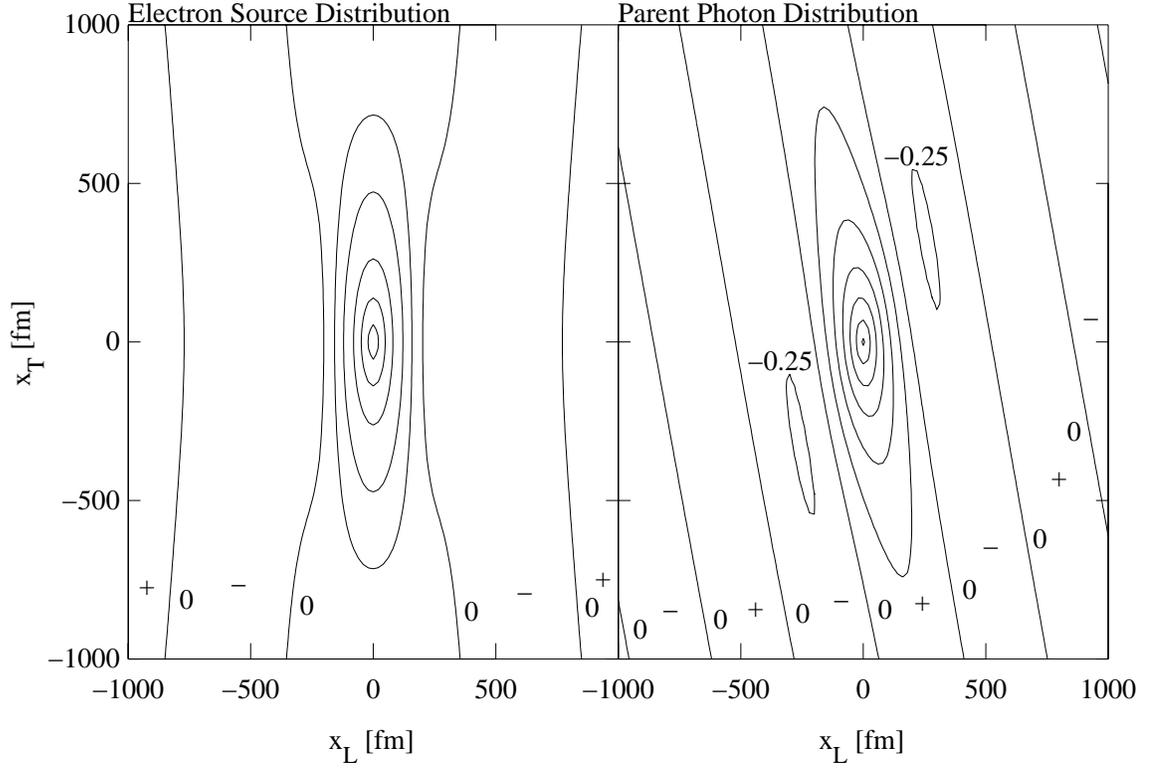}
    \end{center}
    \caption[Plots of the electron source and underlying photon distribution.
	On the left:the electron source for electrons with 
	$p_{\mu}=(2.0, 2.05, \vec{0}_T)$ MeV/c. 
	In this figure, only the zero contours are labeled.  
	The positive contours are (in arbitrary units) 1.0, 2.5, 5.0, 7.5 and 
	10.0.  On the right: the virtual photon distributions corresponding to 
	one of dominant contributions to electron source.  
	These photons have a momentum of $\left<q_{+\mu}\right>=
	(0.956, 1.063, 0.045, 0.045)$ MeV/c.  
	The other root has similar momentum and a similar distribution.
	In this figure, only the negative and zero contours are labeled.
	The positive contours increase in increments of 0.25 (in arbitrary 
	units).]
	{Plots of the electron source and underlying photon distribution.
	On the left:the electron source for electrons with 
	$p_{\mu}=(2.0, 2.05, \vec{0}_T)$ MeV/c. 
	In this figure, only the zero contours are labeled.  
	The positive contours are (in arbitrary units) 1.0, 2.5, 5.0, 7.5 and 
	10.0.  On the right: the virtual photon distributions corresponding to 
	one of dominant contributions to electron source.  
	These photons have a momentum of $\left<q_{+\mu}\right>=
	(0.956, 1.063, 0.045, 0.045)$ MeV/c.  
	The other root has similar momentum and a similar distribution.
	In this figure, only the negative and zero contours are labeled.
	The positive contours increase in increments of 0.25 (in arbitrary 
	units).}
    \label{fig:source}
\end{figure}

On the left in Figure~\ref{fig:source}, we plot the electron source for 
$p_{\mu}=(2.0, 2.05, \vec{0}_T)$ MeV/c electrons from a point 
charge moving to the right with $v_L=0.9c$.  We choose this $p_\mu$ because 
it is both collinear with the point charge and because it is space-like 
($p^2<0$).  Our source can emit both $p^2>0$ and $p^2\le 0$ electrons, however 
the typical QCD parton in a parton ladder
is either space-like or on-shell.\footnote{depending on how one formulates the 
parton model}  On the right in Figure~\ref{fig:source}, we
also plot the photon distribution corresponding to the dominant 
$\left<q_\mu\right>$.  Note
that both the source and the photon
distribution have approximately the same width in both the longitudinal and
transverse directions.  The tilt in the photon distribution gets averaged away
in the $\vec{k}_T$ integrals in Equation \eqref{eqn:edist_w3}.

\subsubsection{The Electron Density}

Now we put elements of the electron distribution together.
According to Equation~\eqref{eqn:edist_w3}, we need the Wigner transform of the
Feynman propagator.  However, since the electrons have positive energy we can 
replace the Feynman propagator with the retarded propagator as we did for the 
photons.  We discuss both the retarded and Feynman phase-space propagators 
in Appendix \ref{append:prop}.

\begin{figure}
    \begin{center}
    \includegraphics[width=4in]{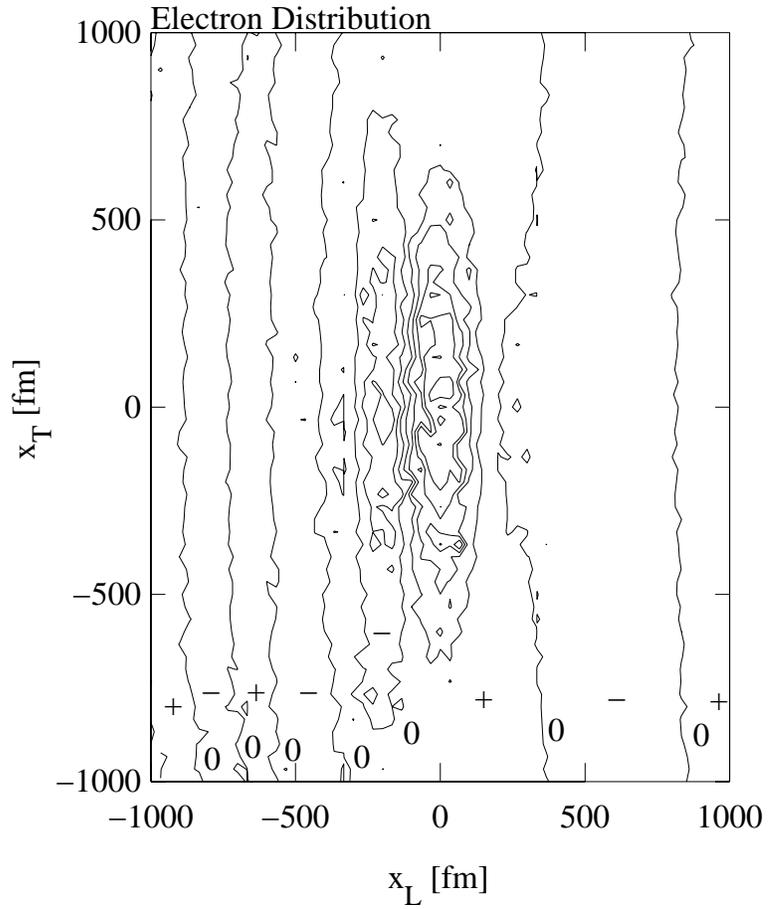}
    \end{center}
    \caption[Coordinate space dependence of the electron
 	phase-space density at four momentum 
	$p_{\mu}=(2.0, 2.05, \vec{0})$~MeV/c.  Only 
 	the negative and zero contours are labeled.  The positive contours are 
 	in increments of 1.0 (in arbitrary units).  The sign of the contours 
	in each region are denoted by $\pm$ signs.]
	{Coordinate space dependence of the electron
 	phase-space density at four momentum 
	$p_{\mu}=(2.0, 2.05, \vec{0})$~MeV/c.  Only 
 	the negative and zero contours are labeled.  The positive contours are 
 	in increments of 1.0 (in arbitrary units).  The sign of the contours 
	in each region are denoted by $\pm$ signs.}
    \label{fig:edist4}
\end{figure}

We are interested in electrons that have momenta that are both
space-like and collinear with the source (for comparison with QCD partons), 
so we plot the coordinate space distribution of electrons with 
$p_\mu=(2.0, 2.05, \vec{0}_T)$~MeV/c in Figure~\ref{fig:edist4}.
The point source
is moving to the right with velocity $0.9c$.  Both the source and the
underlying photon distribution for these electrons is shown in
Figure~\ref{fig:source}.  To perform the four-dimensional spatial integral in 
Equation~\eqref{eqn:edist_w3}, we use a Monte-Carlo integration scheme 
\cite{pre92}.  This integration scheme, being probabilistic by nature,
returns both the integral at a point and the error on the integral at
that point.  The nonzero data
points never had a relative error greater than 20\%, but due to this error, the
location of the zero contours is uncertain by $\sim 30$~fm. 

Comparing the electron distribution with the source, we see that
the electron distribution is elliptical with longitudinal and transverse 
widths comparable to what one
expects by adding the source width in Figure~\ref{fig:source}a 
to the estimates for the propagation distance
in Equations~(\ref{eqn:retlimits}).
Unlike the electron source distribution, the electron distribution
is not symmetric about $x_L=0$. 
This is caused by the positron recoil because, were there no positron recoil, 
we would have a delta function to insure $p_0=p_L v_L$ (as we found for the
photons).  Because of the positron recoil, the delta function is
widened and the additional spread in energy causes electrons with the chosen 
momentum to propagate forward preferentially.   

\subsection{What the Electrons Tell Us About QCD Partons}
\label{sec:3d}

In this section, we learned several things about the phase-space densities 
for massless QCD partons.  Owing to the fact that the simplest parton
ladder contains one rung representing a single partonic splitting, we
learned how both the parent parton and cut rung affect the parton
distribution.  The shape of the parent parton distribution determines
the spatial structure of the parton source.  
The integral over the final states of the parton represented by the cut 
rung of the parton ladder has a $1/k_0$ weight -- giving the parton a low
$k_0$ (or small longitudinal momentum) and giving rise to the expected 
collinear singularity.

\section{The Parton Cloud of a Nucleon}
\label{sec:QCD}

We cannot calculate the phase-space Parton Distribution Functions
without a set of QCD phase-space evolution equations.  
Although both the work in Chapter~\ref{chap:transport} and the recent work by
Makhlin and Surdutovich \cite{neq:mak98} are steps toward this goal, neither
are sufficient.  In the renormalization group improved parton model, one 
specifies the Parton Distribution Functions along some curve in the $x_F$-$Q^2$
plane and then evolves in $x_F$, $Q^2$ or both, mapping out the entire PDF for 
all $x_F$ and $Q^2$.  This evolution is equivalent to summing over a class of 
ladder diagrams in the Leading Logarithm Approximation.  Because the 
Leading Logarithm Approximation can be translated into phase-space, 
many of the insights from the Leading Logarithm Approximation in momentum-space
can be reused in phase-space.  More precisely, using the momentum ordering 
in the Leading Logarithm Approximation and 
a simple model of the nucleon we can estimate the size of the sea parton
distribution as a function of parton momentum.

\subsection{QCD Parton Model and Leading Logarithm Approximation}

\begin{figure}
   \begin{center}
   \includegraphics[totalheight=3in]{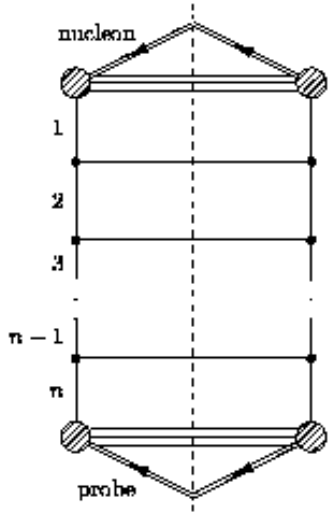}
   \end{center}
   \caption[Cut diagram for probing the $n^{th}$ generation of partons in a 
	typical cascade.]
	{Cut diagram for probing the $n^{th}$ generation of 
	partons in a typical cascade.}
   \label{fig:noneqNthgen}
\end{figure}

Typically the Parton Distribution Functions are calculated using 
either  Dok\-shit\-zer-Gribov-Lipatov-Altarelli-Parisi (DGLAP), 
Balitsky-Fadin-Kurayev-Lip\-a\-tov (BFKL), or Gribov-Levin-Ryskin (GLR) 
evolution equations, all of which are equivalent to applying 
a Leading Logarithm Approximation (LLA) 
\cite{QCD:gri83,QCD:lae94,QCD:dok91}.  
In the LLA, we assume the parton is
produced in a cascade represented by the ladder diagram in Figure 
\ref{fig:noneqNthgen}.  The probability of emitting the $n^{th}$ parton with 
longitudinal momentum fraction $x_{Fn}$ and transverse momentum $q_{nT}^2$ from
this cascade goes like \cite{QCD:lae94}
\begin{equation}
dP_n=\frac{N_c \alpha_s}{\pi}\frac{d x_{Fn}}{x_{Fn}}\frac{d q_{nT}^2}{q_{nT}^2}.
\label{eqn:emitprob}
\end{equation} 
Thus, by ordering the momentum properly as we go down the ladder, we can pick 
up the largest logarithmic contributions to the $n^{th}$ parton's density.

Most hadron colliders probe regions where the data are well described with
Parton Distribution Functions calculated within the
Dokshitzer-Gribov-Lipatov-Altarel\-li-Parisi (DGLAP) evolution scheme.
DGLAP evolution is equivalent the Leading Logarithm Approximation in $1/q^2$
(LLA($Q^2$)).  New experiments at HERA are beginning to see evidence that
Balitsky-Fadin-Kurayev-Lipatov (BFKL) type evolution is necessary to describe
the Parton Distribution Functions at small-$x_F$ \cite{QCD:aid95}.   BFKL-type
physics is believed to be responsible for the rise in the number of partons as
$x_F\rightarrow 0$, however this rise can also be partially described by
DGLAP-type physics \cite{QCD:lae94,gFT:nik86,QCD:aid95}.  BFKL evolution is 
equivalent the Leading Logarithm Approximation in $1/x_F$ (LLA($x_F$)).  
Unlike DGLAP and BFKL evolution, Gribov-Levin-Ryskin (GLR) type evolution 
does not have a simple momentum ordering because one sums terms with
varying powers of $1/x_F$ and $1/q^2$ \cite{QCD:gri83,QCD:lae94}.  Because of 
the simplicity of the ladder 
structure and the momentum ordering needed to pick up the largest
contributions, we will discuss both DGLAP and BFKL
type partons in phase-space.

We can apply the QCD parton model and LLA in phase-space if both are modified 
appropriately.  Assume that we are working in a regime where $\alpha_s \ll 1$, 
so we can apply perturbation theory, and assume that all elementary particles 
are massless.  Assume also that the probe in Figure~\ref{fig:noneqNthgen}
is localized on
the length scale of the parton cloud.  This assumption is equivalent to saying 
the parton lifetime is large compared to the interaction time.

\begin{figure}
   \begin{center}
   \includegraphics{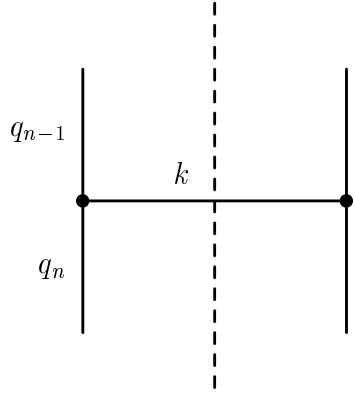}
   \end{center}
   \caption[Typical rung of the LLA ladder.]{Typical rung of the LLA ladder.}
   \label{fig:rung}
\end{figure}

Now, if we find the same singularities in both phase-space and
momentum-space, then we know that the LLA will give the
dominant contribution to the particle densities in phase-space.
From what we have seen from the photon and electron density calculations and 
from the Generalized Fluctuation Dissipation Theorem, the parton densities
have the form
\begin{equation}
	f (x,p)=\int\dn{4}{y}G^{+}(x-y,p)\Sigma (y,p).
\label{eqn:neGeneric}
\end{equation}
The self-energy, $\Sigma$, is given by the parton ladder 
in Figure~\ref{fig:noneqNthgen} and 
the $n^{th}$ segment of $\Sigma$ is shown in Figure \ref{fig:rung}.
In momentum-space, the cut rung gives a $d^3k/|k_0|$ 
which leads to the $dx_F/x_F$ in Equation (\ref{eqn:emitprob}).
To see how the factor of $d^3k/|k_0|$ arises in phase-space, one needs only
look at the  electron source in Section \ref{sec:esource}.  The electron 
source has exactly
the form of the segment in Figure \ref{fig:rung} and in that calculation we
found exactly this factor of $d^3k/|k_0|$.
The fact that we find the same factor of $d^3k/|k_0|$ in both the
energy-momentum representation and in phase-space simply reflects the fact
that the cut parton density is proportional to $\thetaftn{k_0}\delta(k^2)$ in 
both cases and we sum over final parton states.
The factor of $d q^2/q^2$
in Equation \eqref{eqn:emitprob} comes from the integration over the 
leg's propagator, $1/q^2$. In phase-space, the $1/q^2$ poles are tied up in 
the Wigner transform of the retarded propagator, but they are still there:
\[
   \Gplus{x}{q_1-q_2}=\int\dnpi{4}{(q_1-q_2)} e^{-i x \cdot (q_1-q_2)}
   \frac{1}{q_1^2+i\epsilon q_{10}}
   \frac{1}{q_2^2-i\epsilon q_{20}}.
\] 
Thus, this segment of the parton ladder produces the same divergencies in 
phase-space and  mo\-men\-tum-space.  Whatever orderings are needed to 
produce the leading contributions in momentum space will produce the same 
leading contributions in phase-space.

\begin{figure}
   \begin{center}
   \includegraphics[totalheight=0.87\bagfigheight]{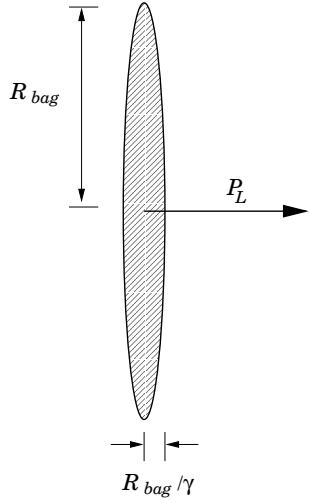}
   \end{center}
   \caption[Schematic of the relativistic nucleon's valence quark distribution]
	{Schematic of the relativistic nucleon's valence quark distribution.}
   \label{fig:Nbag}
\end{figure}

Our self-energy has the same ladder structure as the electron source in
Section \ref{sec:edist}, so we know the spatial structure of the $n^{th}$
parton's source is given by the $n-1^{th}$ parton's distribution.  Iterating
back to the $0^{th}$ parton (a valence quark), we see that the shape of the
valence distribution sets the shape of the sea parton source.  So, we take the 
valence quark wavefunction to be uniformly spread throughout a bag 
with radius $R_{bag}$ as illustrated in Figure~\ref{fig:Nbag}.  Since we are 
interested in high-energy collisions, we take the nucleon to be moving to 
the  right with 4-momentum $p_\mu=(P_0,P_L,\vec{0}_T)$ with 
$P_0\approx P_L \gg M_N$.  Thus, this nucleon has 4-velocity  
$v_\mu=(1,v_L,\vec{0}_T)$ and the bag is contracted in the 
longitudinal direction by a factor of $\gamma=1/\sqrt{1-v_L^2} \gg 1$.
We assume the partons lose memory of the original valence quark
momentum as one goes down the ladder.   
Thus, any momentum/coordinate correlations in the 
source function should be washed out by the spatial 
integrations in Equation (\ref{eqn:neGeneric}).  One might expect that 
the sea partons forget the shape of the nucleon bag as well, but we show
that the partons cannot propagate far enough from the original source for this
to happen.

\subsection{Large-$Q^2$ (DGLAP) Partons}
 
In the large-$Q^2$ regime,  the parton density is low but 
$\alpha_s (Q^2) \ln (Q^2/\Lambda_{QCD}^2) \gtrsim 1$.  Here the largest 
contribution to the leading log ladder comes from large $Q^2$ 
logarithms.\footnote{$Q^2$ can be taken as the typical momentum 
scale of the process.  In the case of a DIS probe, this is the momentum 
transferred by the probe.}  To get the largest contributions 
from these logs, we order the momenta as we move down the ladder:
\[
   -q_{n}^2 \gg -q_{n-1}^2 \gg \ldots \gg -q_{1}^2 \gg 1/R^2_{bag}
   \approx \Lambda_{QCD}^2.
\]  
Here $q_{i}^2$ is the virtuality of the $i^{th}$ leg.  The kinematics at each 
leg-rung vertex ensure that the momentum fraction carried by each leg is 
also ordered:
\[
   1 \ge x_{F1} \ge \ldots \ge x_{F{n-1}} \ge x_{Fn}.
\]
Whether a rung or leg is a quark or gluon is irrelevant, provided $k^2=0$ and 
the $q^2$ ordering holds.  Now, given that the proton has longitudinal 
momentum $P_L$ and the rungs and legs are massless, each generation of partons 
must have energy $q_{n0}\approx x_{Fn} P_L$ and transverse momentum of 
$q_T^2\approx -q^2 \ll x_F^2 P_L^2$. 

Let us figure out the general features of the parton cloud.  
The retarded propagator lets the $n^{th}$ parton propagate out to 
$R_{n\perp}\sim \hbar c/\sqrt{-q_n^2}$ 
transverse to the parton momentum and to 
$R_{n\|}\sim \hbar c/q_{n0}= \hbar c/x_{Fn} P_L$ 
parallel to the the parton momentum.  The parton momentum is approximately 
parallel to the nucleon momentum, since $x_FP_L \gg p_T$.  The partons can 
never get far from the bag in the transverse direction because 
$R_{n\perp}\ll R_{bag}$, so the
transverse spread of the partons will be dominated by 
the bag size: $\Delta R_{T} \sim R_{bag}$.  
On the other hand, the longitudinal spread of the partons is roughly given by 
$\Delta R_{L} \sim R_{bag}/\gamma +\hbar c/x_F P_L$, so can be dominated 
by the longitudinal propagation distance 
$R_\|$ if $x_F\ll M_N R_{bag}/\hbar c$. 
In fact, for very small $x_F$ (i.e. $x_F\sim M_N R_{bag}/\gamma\hbar c$) the 
spread of the partons can meet or exceed the nucleon bag 
radius.  Furthermore, the actual distribution may be 
somewhat broader due to the propagation of the virtual partons between the
subsequent emissions along the ladder.  

\begin{figure}
   \begin{center}
   \includegraphics[totalheight=\bagfigheight]{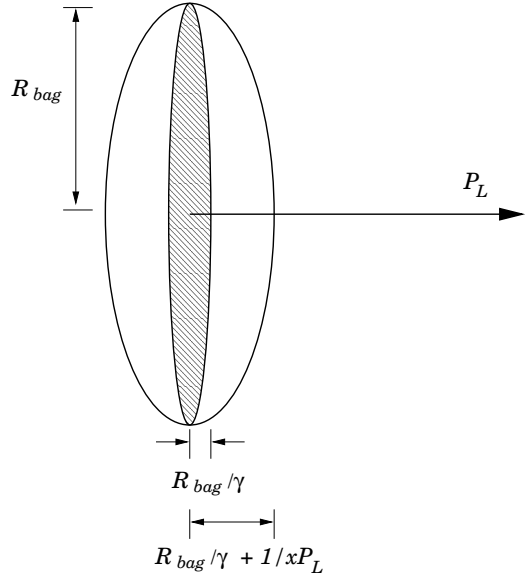}
   \end{center}
   \caption[Schematic of the cloud of large $Q^2$ partons]
	{Schematic of the cloud of large $Q^2$ partons.}
   \label{fig:DGLAPbag}
\end{figure}

So in our picture, which is summarized in Figure~\ref{fig:DGLAPbag}, the sea 
large-$Q^2$ parton distributions have the same transverse size as 
the parent nucleon's transverse size, but the longitudinal size can be 
significantly bigger than the parent's longitudinal size and even approaching 
the parent's transverse size.  Furthermore, the drop off in the parton
density in the longitudinal direction occurs at the characteristic radius of
$\sim \hbar c/x_FP_L$.  This picture of the nucleon is consistent with the
uncertainty principle based arguments of A.~H.~Mueller \cite{QGP:mue89}, 
later user by Geiger to initialize the parton distributions in
his Parton Cascade Model 
\cite{tran:gei92a,tran:gei92b,tran:gei94,tran:gei93,tran:gei95}.

\subsection{Small-$x_F$ (BFKL) Partons}

In the small-$x_F$ regime, the parton density 
is high and $\alpha_s (Q^2) \ln (1/x_F) \gtrsim 1$.  The small-$x_F$ partons 
are mostly gluons.  In this regime, the leading
logs come from the $1/x_F$-type singularities, i.e. from the cut rungs. 
Since leading logs come from the $1/x_F$ singularities, the largest 
contributions come about by strongly ordering the longitudinal momentum 
fraction as one moves down the ladder
\cite{QCD:bal78,QCD:kur76,QCD:kur77,QCD:lip76}:
\[
  1 \gg x_{F1} \gg \ldots \gg x_{F{n-1}} \gg x_{Fn}.
\] 
BFKL evolution has only a weak dependence on the virtuality of the 
partons as we move down the ladder, so we assume $q^2$ to be fixed:
$q_{n-1}^2\approx q_{n}^2\gg 1/R_{bag}^2$. This does not significantly effect 
the results of the analysis \cite{QCD:bar93}.

\begin{figure}
   \begin{center}
   \includegraphics[totalheight=1.1\bagfigheight]{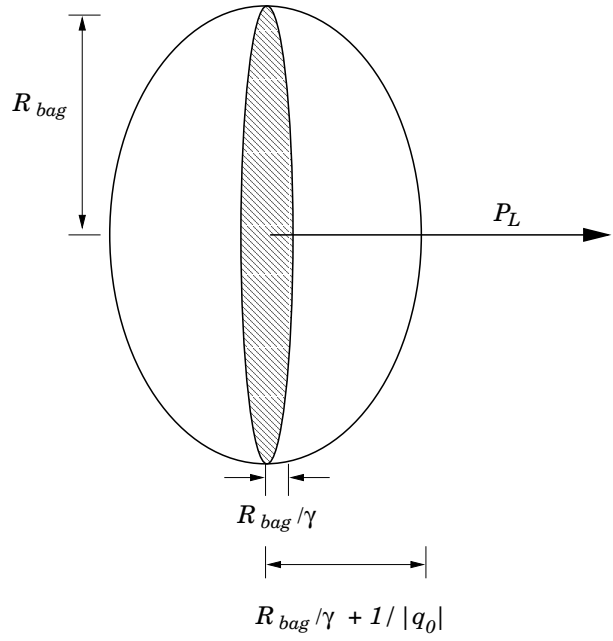}
   \end{center}
   \caption[Schematic of the cloud of small $x_F$ partons]
	{Schematic of the cloud of small $x_F$ partons.}
   \label{fig:BFKLbag}
\end{figure}

Now we must understand how the transverse momentum and energy of each parton 
leg changes as we go down the ladder.  A well known effect of iterating the 
BFKL kernel (equivalent to moving down the ladder) is that the transverse 
momentum undergoes a random walk in $\ln(q^2_T)$ 
\cite{QCD:lae94,QCD:bal78,QCD:kur76,QCD:kur77,QCD:lip76}.  In fact, 
after iterating through a sufficiently large number of rungs, the spread in 
the $q_T$ distribution is given by:
\[
\left< \left( \ln (\frac{q_{nT}^2}{q_{1T}^2}) \right)^2 \right> = 
   C \ln (\frac{1}{x_F})
\]
where $C=\frac{N_c\alpha_s}{\pi} 28 \zeta (3) = 32.14\alpha_s$.  Thus, 
$q_{nT}^2$ can be orders of magnitude larger or smaller than $q_{1T}^2$.
This is more clearly seen by rewriting $q_{nT}^2$ as
\begin{equation}
        q_{nT}^2 \sim q_{1T}^2 e^{\pm 5.7 \sqrt{\alpha_s \ln (1/x_F)}}.
\label{eqn:BFKLqT}
\end{equation}
We will consider the extreme cases of the transverse momentum and comment on 
the typical case, $q^2_{nT}\sim q^2_{1T}$.

If the random walk results in a large transverse momentum, we will have
$q^2_{nT}\gg q^2_{1T}\sim -q^2\sim (x_{Fn} P_L)^2$.  Thus, the $n^{th}$ parton
will have 3-momentum in the transverse direction.  We know that the parton can
only propagate to a distance of roughly $R_{\|}\sim \hbar c/|q_0|=\hbar
c/\sqrt{q^2+q^2_T+(x_F P_L)^2}$ in the direction parallel to $\vec{q}$.  Since
$\sqrt{q^2+q^2_T+(x_F P_L)^2} \approx |q_T|$ and $\hbar c/|q_T|\ll R_{bag}$, the
parton cannot travel far from the original source in the transverse direction.
On the other hand, the parton's longitudinal spread can be larger than the
longitudinal bag size.  The parton can propagate to a distance of 
$R_{\perp} \sim \hbar c/\sqrt{-q^2}$ in the
direction perpendicular to $\vec{q}$, so we can expect a longitudinal spread of
the parton distribution of $\Delta R_L \sim R_{bag}/\gamma + \hbar
c/\sqrt{-q^2}$.  Since $R_{bag}\gg\hbar c/\sqrt{-q^2}$, this additional spread
can not match the spread of the DGLAP partons.

If the random walk results in a small transverse momentum, we will have
$q^2_{nT}\ll q^2_{1T}\sim -q^2\sim (x_{Fn} P_L)^2$.  In this case, the $n^{th}$
parton will have 3-momentum in the longitudinal direction.  As in the case of
the DGLAP partons the additional transverse spread is 
$\Delta R_T\sim \hbar c/\sqrt{-q^2} \ll R_{bag}$ and so is negligible.  
The additional longitudinal spread is
$\Delta R_L\sim\hbar c/|q_0|\approx\hbar c/\sqrt{q^2+(x_F P_L)^2}$.  This may 
be significantly larger than the spread of the DGLAP partons because the partons
have space-like momenta making $q^2+(x_FP_L)^2<(x_FP_L)^2$.

Summarizing both possibilities, the BFKL parton distribution has a 
transverse spread of $\Delta R_T \sim R_{bag}$, but a widely varying 
longitudinal spread, ranging from 
$\Delta R_L \sim R_{bag}/\gamma +\hbar
c/\sqrt{-q^2} \ll R_{bag}$ to $\Delta R_L \sim R_{bag}/\gamma +\hbar
c/\sqrt{q^2+(x_FP_L)^2}\gg R_{bag}$ for partons with space-like momentum. 
Presumably the typical case, when $q^2_{nT}\approx q^2_{1T}$, lies between 
these two extremes so the BFKL parton
distribution ranges from much smaller than the DGLAP distribution to a lot
larger than the DGLAP distribution.  In all cases, since $|q_0|, \sqrt{-q^2},
(x_FP_L)\gg\Lambda_{\rm QCD}$, the DGLAP and BFKL distribution longitudinal 
widths must not be as large as the transverse width.
A picture of BFKL partons is illustrated in Figure~\ref{fig:BFKLbag}.

The fact that the longitudinal extent of the BFKL cloud can be so much larger
than the longitudinal width of the nucleon bag has implications for the 
small-$x_F$ parton distribution of a nucleus.  
Because the longitudinal width of the small-$x_F$ distribution is so large,
the small-$x_F$ partons (which are mostly gluons) can see the color charge of 
any other nucleon in a longitudinal tube centered on the parent nucleon.  
This suggests that we should treat the 
nucleus as a whole as a source of color charge for the small-$x_F$ partons in 
the spirit of McLerran-Venugopalan model \cite{ven95}.  
Specifically, the Lorentz contracted nucleus is replaced with a semi-infinite 
sheet of fluctuating color charge.  In this limit, the gluon distribution 
function per unit area is the semi-classical Weizs\"acker-Williams 
distribution for gluons scaled by the density of charge squared fluctuation 
per unit area.  In practice in a nuclear collision, this approach may only be 
useful when the nuclei are far apart.
As the nuclei approach, the BFKL partons from each nucleus begin interacting 
and exciting shorter wavelength modes.  These shorter wavelength modes will 
not see the nuclei as sheets of color charge, but rather objects extended in 
the longitudinal direction.  Eventually then, the McLerran-Venugopalan approach
must break down.

The large longitudinal extant of the small-$x_F$ cloud may have another 
consequence:  in a zero impact parameter nucleon-nucleon collision, we would 
find that the soft (BFKL) partons interact much earlier than the 
harder (DGLAP) partons because of their greater longitudinal spread.  
This, coupled with the large density of small-$x_F$ partons, leads to earlier 
entropy production and stopping of the soft partons.  In fact, this is likely
to be part of the cause of the high stopping and early entropy production 
in the Geiger's PCM model 
\cite{tran:gei92a,tran:gei92b,tran:gei94,tran:gei93,tran:gei95}.  However, 
it is known that small-$x_F$ partons couple weakly to themselves and to 
the rest of the system \cite{QCD:mcl97,QCD:lae94}, so this may end up having
no observable consequences on the rest of the system.

%

\section{Summary}

The parton model rests firmly on the concepts of factorization and on
evolution of the parton densities.
Using conventional S-matrix perturbation theory on simple QED processes, 
we showed the reaction rates (and hence the cross sections) can be 
factorized into a parton model-like form.  In other 
words, they take the form of a reaction rate density convoluted with a 
phase-space Parton Distribution Function.  This phase-space PDF is 
the quasi-particle parton number density and has the form of a phase-space 
source folded with a phase-space propagator.  Our work with the 
Weizs\"acker-Williams Approximation demonstrates that the Parton 
Distribution Functions can be defined in phase-space.
Since parton evolution is equivalent to summing over a class
of ladder diagrams, we examined the first segment of a QED parton ladder.
This ladder is simple as it includes only one 
``partonic'' splitting: a virtual photon splitting into an electron-positron 
pair.  Not only does this simple ladder exhibit the $1/x_F$ singularity that
we would expect from cutting the rung of a parton ladder, but the study of 
this ladder shows that the shape of a parton's distribution is controlled to a
large extent by the shape of its parent's distribution. 

As a side benefit of this study, we were able to discuss 
how the phase-space propagators work.
We found that the retarded propagator propagates particles to 
distances of $\sim R_\| =\hbar c/\mbox{min}(|q_0|,|\vec{q}|)$ parallel to the 
particle's momentum and to 
distances of $\sim R_\perp =\hbar c/\sqrt{|q^2|}$
perpendicular to the particle's momentum when $q^2\neq 0$.  When $q^2=0$, the
particles tend to follow their classical paths with deviations from this path
being of order $1/|q_0|$.
Furthermore, the retarded propagator can only send
particles forward in time and inside the light-cone.

In the end, we have made progress toward specifying the initial phase-space 
parton distributions of a relativistic nuclear collision.  
Regardless of the kinematical regime, the transverse spread of a parton 
distribution is dominated by the bag radius $\sim 1$ fm.  
The longitudinal spread of a parton distribution varies from 
roughly $\sim R_{bag}/\gamma+\hbar c/x_FP_L$ for moderate to large $x_F$ 
(i.e. for DGLAP partons) and from  
$\Delta R_L\sim R_{bag}/\gamma + \hbar c/\sqrt{-q^2}$ to 
$\Delta R_T\sim R_{bag}/\gamma + \hbar c/\sqrt{q^2+(x_FP_L)^2}$
for small $x$ (i.e. BFKL partons).  
Since the small $x_F$ partons have a large longitudinal spread
and a high density, we expect the small $x_F$ partons to interact much earlier
than the large $x_F$ partons in a typical nuclear collision.  This may cause
earlier entropy production and higher stopping than one expects in models 
that include only DGLAP parton distributions such as HIJING 
\cite{QGP:wan91,QGP:wan92,QGP:wan95} and others.

\chapter{NUCLEAR IMAGING}
\label{chap:HBT}

Imaging techniques have been applied to a wide variety of problems, from
extracting license plate numbers from blurred photos of speeding cars to 
imaging the interior of the earth.  The typical linear imaging problem is
to extract an image from experimental data where the data is the 
convolution of the sought-after image with some kernel.  In the study of
nuclear reactions we have one such linear imaging problem -- inverting
the Pratt-Koonin equation:
\begin{equation}
   C_{\vec{P}}(\vec{q})-1=\int\dn{3}{r} K(\vec{q},\vec{r})S_{\vec{P}}(\vec{r}).
\label{eqn:PrattKoonin}
\end{equation}
Here, the data is the two-particle correlation function, $C$, and the 
image we seek is the source function, $S$.
The source function is the relative distribution of emission points for 
the pair of particles: 
\begin{equation}
S_{\vec{P}}(\vec{r})=\int\dn{3}{R}d t_1 dt_2 
	D(\vec{P},\vec{R}+\vec{r}/2,t_1)D(\vec{P},\vec{R}-\vec{r}/2,t_2).
\label{eqn:sourcedef}
\end{equation}
Here $D$ is the normalized single particle source and $D$ can be identified
with the distribution of last collision points in space, time and momentum 
of the particles in a nuclear collision.   The source function and single 
particle sources are discussed briefly in Section \ref{sec:source} and in detail
in Appendix \ref{append:spectra}.
In \eqref{eqn:PrattKoonin}, the kernel in the convolution is 
$K(\vec{q},\vec{r})$.  For identical pairs, the kernel can be written 
\begin{equation}
   K(\vec{q},\vec{r})=\left|\phi^{(-)}_{\vec{q}}(\vec{r})\right|^2-1.
\label{KPhi}
\end{equation}
where, as we discuss in Appendix \ref{append:spectra}, 
$\phi^{(-)}_{\vec{q}} (\vec{r})$ is the two particle relative wavefunction.  
In \eqref{eqn:PrattKoonin}, $\vec{q}$ is the relative momentum between the 
particle pair, $\vec{r}$ is the separation of emission points in the pair 
rest frame and $\vec{P}$ is the total pair momentum.

At first glance, imaging appears easy: we could discretize $C_{\vec{P}}$
and $S_{\vec{P}}$ and invert the resulting matrix equation.  However, in 
practice this does not work as small variations in data, even 
within statistical or systematic errors, can generate huge changes
in the reconstructed source.  This stability problem is well known in 
other fields and vast literature exists on its resolution 
\cite{img:boe85,img:smi85,pre92,img:bal80,img:bac67}.  In fact, many imaging
problems \cite{img:ber80,pre92} that are now routinely solved would be 
considered ill-posed in the sense of Hadamard who discussed stability in
inversion as early as 1923 \cite{img:had23}.  
One would think then that we could take a well known imaging method, such as 
the Maximum Entropy Method (MEM) which is often used in astronomical imaging
\cite{img:ski85,img:bev85,pre92}, and apply it directly to our case.
MEM assumes
the most likely image on large length scales, corresponding to small 
amplitude noise, and uses that information to stabilize the image on the 
shorter length scales, corresponding to the point-like stars.  
MEM is very successfully applied to astronomical problems but it is not an
appropriate approach for nuclear imaging.  The reason is that the kernels 
often  encountered by astronomers mainly blur the stars, mixing them into  
uniform noisy background.  In short, the data is 
rather singular and the kernel is smooth.  In our case, the situation is 
quite the opposite;  the source in \eqref{eqn:PrattKoonin} is 
highly peaked at small to intermediate distances and drops to zero only at 
high distances and the kernel is more like a Fourier transform than a 
blurring function.  In other words, our kernel is singular while our data is 
smooth.  In fact, when the kernel does blur, we tend to lose a lot of 
information.

Given our unique situation, we have tried a variety of techniques
to invert \eqref{eqn:PrattKoonin}.  These techniques have been applied 
mainly to the angle-averaged Pratt-Koonin equation and the techniques are
discussed in the Section \ref{sec:inverting}.  These techniques vary from 
directly Fourier 
Transforming the correlation function to more sophisticated methods that seek 
to arrive at ``most likely'' source.  These more sophisticated methods include 
the Lucy-Richardson algorithm, where we seek the ``most likely'' source 
through an iterative scheme, and others that take the ``most likely'' source 
to be the source that minimizes the $\chi^2$ fit 
to the correlation function.  These last methods include brute force 
minimization of the $\chi^2$, solving an algebraic equation for the source
that produces the smallest $\chi^2$, and an improvement of this method we 
term Optimized Discretization.  
The most successful of these techniques take advantage of specific 
properties of either the kernel or the data.  In all cases, if we exploit 
certain properties of the images, i.e. use {\em constraints}, we can further 
stabilize the imaging.  Use of constraints implies discarding spurious 
solutions not meeting the criteria for a valid source.  The use of 
constraints in this role was first recognized by Tikhonov \cite{img:tik63}  
and we will illustrate the use of constraints in the 
Section \ref{sec:inverting}.

While the images we obtain are interesting in their own right, we will 
discuss how other pieces of information can be extracted from them in 
Sections \ref{sec:otherstuff} and \ref{sec:integralofsource}.  In 
particular, we can estimate both the average phase-space density at freeze-out 
and the entropy of the system at freeze-out from the images.  
In the past, the average pion phase-space distribution at freeze-out has been 
estimated using the correlation function \cite{HBT:ber94b,HBT:ber94c}.
However, because we use the source function directly we can make the 
estimate for any type of particle.  There is 
another piece of information one can extract from the images -- the integral
of the source over space out to a specific distance.  Given that the source 
function is normalized to 1, if we integrate the source out to a set distance
and do not get 1, then we know that a certain amount of the source must
lie at greater distances.  Since our images can not extend out indefinitely,
we can use this quantity to signal for large distance (or long time) emission 
of pairs we can not otherwise image.

Our imaging work has already yielded an array of new results as
we have inverted several data sets from the AGS experiment E877 and from the
Michigan State Cyclotron.  The results of these inversions are contained in
Sections \ref{sec:pions}, \ref{adata} and \ref{sec:IMFstuff}.
From the AGS, we analyzed both pion correlations at two different 
beam energies and kaons at the highest of those two beam energies.  Our 
analysis shows that the kaon source is much narrower than the two pion sources 
and that the pion sources show evidence for a tail, possibly from long-lived 
resonance decay.  While the pion and kaon sources we found are consistent 
with Gaussian sources, both the proton and IMF sources are definitely 
non-Gaussian.  We also see clear emission lifetime effects in the 
sources of both IMF and proton correlations at MSU energies.  While we expect 
that lifetime effects dominate the IMF sources, it was a surprise to see 
evidence for a tail in the proton sources.  In all cases, our imaging allow us 
to test the positivity of the sources, a condition necessary to interpret the
sources semi-classically.

\section{What Is the Source Function?}
\label{sec:source}

To understand what the images will tell us, we must understand what the 
source function is, what the source function is {\em not} and the
coordinates we use to express the source function.  

\begin{figure}
\begin{center}\includegraphics[width=\textwidth]{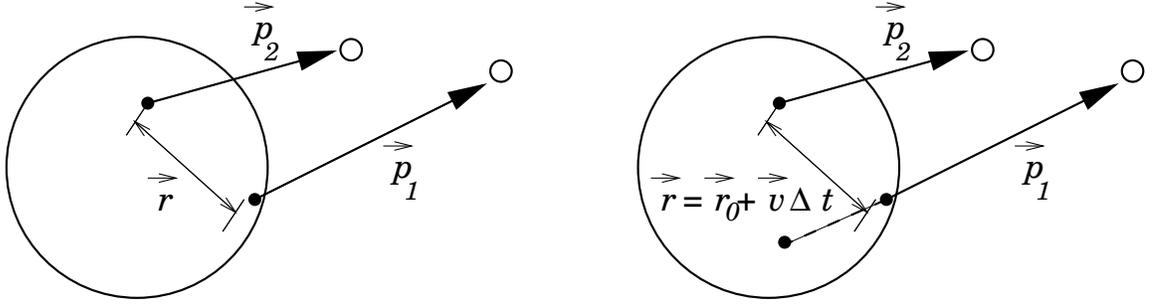}\end{center}
\caption[Two emission scenarios that give the same separation of emission 
	points.  On the left, the pair is emitted simultaneously with a 
	relative separation $\vec{r}$.  On the right, the first particle is 
	emitted with a relative velocity $\vec{v}$ and the second particle is 
	emitted $\Delta t$ later a distance of $\vec{r_0}$ from the first 
	particle.  The combined spatio-temporal separation of emission is 
	$\vec{r}= \vec{r_0}+\vec{v}\Delta t$, giving a separation identical to 
	that in the scenario on the left.]
	{Two emission scenarios that give the same separation of emission 
	points.  On the left, the pair is emitted simultaneously with a 
	relative separation $\vec{r}$.  On the right, the first particle is 
	emitted with a relative velocity $\vec{v}$ and the second particle is 
	emitted $\Delta t$ later a distance of $\vec{r_0}$ from the first 
	particle.  The combined spatio-temporal separation of emission is 
	$\vec{r}= \vec{r_0}+\vec{v}\Delta t$, giving a separation identical to 
	that in the scenario on the left.}
\label{fig:emission}
\end{figure}

The source function is the probability distribution for 
emitting a particle pair a distance $\vec{r}$ from one another, in the pairs 
center of mass frame.  First, because it is a probability distribution,
if we sum over all possible emission points, then we must obtain 1:
	\[\int \dn{3}{r} S_{\vec{P}}(\vec{r})=1.\]
This will prove useful in our discussion in Section \ref{sec:integralofsource}.
Second, the source says nothing directly about the time separation between
the particle emissions.  This is clearly seen in Equation~\eqref{eqn:sourcedef}
as the temporal information in the single particle sources is integrated over.
As a specific example of this problem, we can not distinguish between the two 
scenarios pictured in Figure \ref{fig:emission}: simultaneous emission with 
a pair separation $\vec{r}$ and sequential
emission with a time separation $\Delta t$ combined with a spatial separation 
$\vec{r_0}$ in a model independent way, if $\vec{r}= \vec{r_0}+\vec{v}\Delta t$.
The fact that the temporal information is entangled into the source function 
in such a nontrivial manner cast some doubt on efforts to simultaneously fit 
the radius of a source function and an effective emission time.  Finally, the 
source is {\em model independent}.  The only assumption that goes into the 
inversion process is an assumption of the validity of the Pratt-Koonin 
equation.  In other words, we only require a factorization of final 
state interactions of the pair from the evolution of the pair in the excited, 
colliding, nuclear system.

The source function {\em must not} be confused with the single particle 
sources, \linebreak $D(\vec{x},t,\vec{p})$.  However, the source function and 
single particle sources are related through Equation \eqref{eqn:sourcedef}.  
The single particle sources tell us where and with what momentum 
the particle are created relative to the system as a whole.  The source 
function can not tell us where either of the particles are created, just 
how far apart they were when the second one was 
emitted.  Nevertheless, because the source can be written in terms of the single
particle sources, model source functions can be constructed from any transport 
model that gives the distribution of the last collision points in space, time 
and momentum of the particles in the collision.

\begin{figure}
\begin{center}\includegraphics{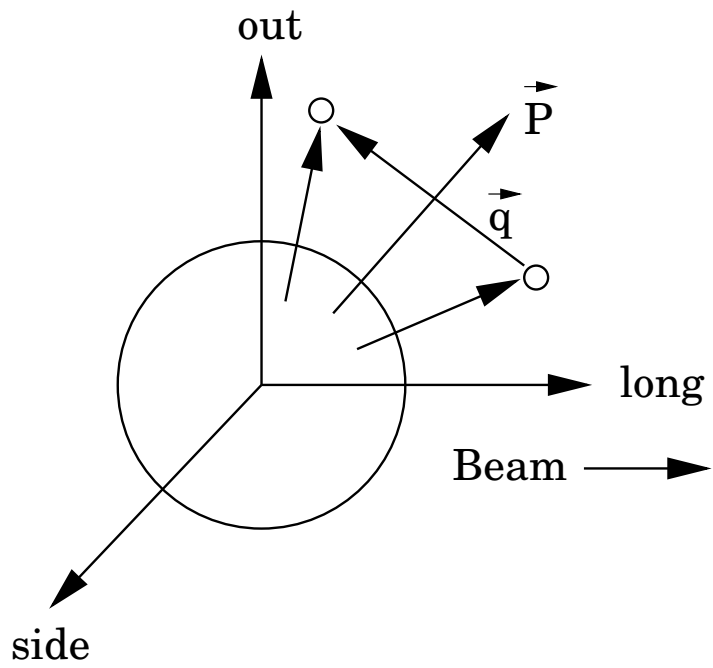}\end{center}
\caption[The directions we use in the analysis of $S$ 
	in the pair center of mass frame are
	outward along the transverse momentum of the
	pair, longitudinal along the beam, and the remaining
	direction we term trans\-verse.]
	{The directions we use in the analysis of $S$ 
	in the pair center of mass frame are
	outward along the transverse momentum of the
	pair, longitudinal along the beam, and the remaining
	direction we term trans\-verse.}
\label{fig:HBTcoords}
\end{figure}

The source function is given  in the pair center of mass frame in the 
coordinates sketched in Figure~\ref{fig:HBTcoords}.
We use the coordinates in Figure~\ref{fig:HBTcoords} as they are commonly used
in the analysis of correlation data.  
We use the CM frame of the pair 
to simplify the form of the Pratt-Koonin equation.
In an arbitrary frame, the source function obtains a dependence on the 
separation time of the pair emission.  This might appear to aid in unfolding 
the temporal structure of pair emission, but this time dependence is folded into
the spatial separation in the $\vec{P}$ direction in a non-trivial manner.  
In the end, the Pratt-Koonin equation says nothing more than it does in the CM 
frame, it just does so in a more complicated way.

\section{Restoring the Source Function}
\label{sec:inverting}

There are several ways that we can approach our imaging problem 
and in this section we will discuss each  of the ones we have considered.
These ways range from the simplest Fourier inversion, which is only applicable 
to particles with no final state interactions, to more sophisticated methods 
that seek the ``most likely'' source.  The first of these other methods is 
an iterative scheme called the Lucy-Richardson algorithm.  In this scheme
the ``most likely'' source is one that minimizes a quantity similar to the 
information entropy.  The rest of our methods are based on Bayesian arguments
that suggest the ``most likely'' source is one that minimizes 
the $\chi^2$ fit.  The least sophisticated (and least successful) of these
is the brute force minimization of the $\chi^2$.  A more fruitful approach
is to create an algebraic solution for the $\chi^2$ minimum.  Now, none
of these approaches use any features of the data or kernel to stabilize the 
inversion.  As we will see, our most promising method, the Optimized 
Discretization method, uses the error on the data and the behavior of the 
kernel to choose the resolution that produces the best source.  One can go 
further than just taking advantage of the kernel to taking advantage of known 
properties of the source itself.  We describe some of these properties
and then use them to further constrain the inversion.  We demonstrate the 
role these constraints have with an example.  However, before discussing the 
various methods, we will recast our inversion problem into the angle-averaged 
version of the problem.

\subsection{The Angle-Averaged Pratt-Koonin Equation}

To begin, we write the angle averaged version
of the Pratt-Koonin equation 
as this is the version we use in the data analysis at the end of the chapter. 
Although we stick to angle averaged work throughout this chapter, 
our results are applicable to the full three dimensional case.

If we introduce $C(\vec{q}) = \sqrt{4\pi} \sum_{\ell m} C^{\ell m} (q) 
\, {\rm Y}^{\ell m} (\Omega_{\vec{q}})$,
and an~analogous representation for~$S$ 
then the individual moments satisfy
\begin{equation}
   C_{\vec{P}}^{\ell m}(q) - \delta^{\ell 0} \, \delta^{m 0}
	= 4 \pi \int_0^\infty dr \, r^2 \, K_\ell (q, r) \,
	S_{\vec{P}}^{\ell m} (r) \, .
\label{CPl}
\end{equation}
Here the~spin-averaged kernel~$K$ depends only on the angle 
between~$\vec{q}$ and~$\vec{r}$, and not on the separate directions of these
vectors:
\begin{equation}
   K_\ell(q,r)=\frac{1}{2}\int^1_{-1} d(\cos\theta) K(\vec{q},\vec{r}) 
	P^\ell(\cos\theta).
\end{equation}

Due to the symmetry of~$S$ and~$C$, only even~$\ell$ appear in
the angular expansion of these functions.  Since both functions
are real, the moments satisfy
$(C^{\ell m})^* = (-1)^m \,
C^{\ell \, - m}$.
The~relation between the angular moments in~\eqref{CPl} may help in 
analyzing three-dimensional data.

As a specific case, relation \eqref{CPl} shows that
the angle-averaged correlation function $C^{00} (q) \equiv
C(q)$ is directly related to
the~angle-averaged source $S^{00} (r) \equiv S(r)$:
\begin{equation}
   C_{\vec{P}}( q ) -1 \equiv {\cal R}_{\vec{P}}( q ) =
   4 \pi \int_0^\infty d{r} \, r^2 \,
   K( q , r) \, S_{\vec{P}} ( r ) \, .
\label{Kav}
\end{equation}


\subsection{Fourier Transformation}

One way to invert \eqref{Kav} or \eqref{eqn:PrattKoonin}
is to take advantage of the fact that, when final state interactions can 
be neglected, the imaging problem becomes a Fourier inversion problem.  
This means that our intuition regarding Fourier transforms can be applied to 
the imaging problem.  

When we can neglect the final state interactions, the relative wavefunction 
simplifies dramatically.
In particular, for like-charged pions\footnote{We can safely ignore the 
nuclear forces between the pions even for small relative momentum.} emitted 
from small sources, their relative Coulomb wavefunction 
factorizes \cite{HBT:pra90}:
\begin{gather*}
   \left|\phi^{\rm Coul}_{\vec{q}}(\vec{r})\right|^2\approx
	G^2(q)\left|\phi^{\rm free}_{\vec{q}}(\vec{r})\right|^2 \\
   \intertext{where}
   G(q)=\frac{2\pi\eta}{e^{2\pi\eta}-1} \quad\text{   and   } \quad
	\eta=\frac{\alpha_{em}m_\pi}{2q}
\end{gather*}
and the free wavefunction in the pair rest frame is 
$\phi^{\rm free}_{\vec{q}}(\vec{r})
=(\sqrt{2})^{-1} (e^{i\vec{r}\cdot\vec{q}}+e^{-i\vec{r}\cdot\vec{q}})$.
Thus, the Coulomb effects can be ``corrected'' by dividing out
the Gamow factor $G^2(q)$ and the 
inversion becomes a Fourier cosine transform:
\begin{equation}
   S_{\vec{P}} (\vec{r}) = {1 \over \pi^3} \int \dn{3}{q} \,
   \cos{\left(2{\vec{q}\cdot\vec{r}}\right)} \left( C_{\vec{P}}(\vec{q}) -1 
   \right) \, .
\label{SPr=}
\end{equation}

The angle averaged case follows just  as simply as $K( q, r) = 
\sin{(2 q r)}/(2qr)$ and
the source, $S_{\vec{P}}$, is an inverse Fourier 
sine transform of~$C$~\cite{HBT:bro97}.
For the angle-averaged source one finds:
\begin{equation}
  r \, S_{\vec{P}}(r) = {2 \over \pi^2} \, \int_0^{\infty} dq \,
  q \, \sin{(2qr)} \, (C_{\vec{P}} (q) - 1) \, .
\label{rSr2}
\end{equation}
In fact, for any $\ell m$, we have
\begin{equation}
   S_{\vec{P}} ^{\ell m} (r) = {
   (-1)^{\ell/2} \, 4 \over \pi^2} \int_0^\infty dq \, q^2 \
   j_{\ell} (2 q r)
   \, (C_{\vec{P}}^{\ell m} (r) - \delta^{\ell 0} \, \delta^{m 0})
   \, .
\label{SPl}
\end{equation}

In practice, the~integration in \eqref{rSr2} must be cut off at
some suitably chosen value of $q_{\rm max}$.  This upper limit must be 
chosen so that the integral covers the region where the correlation 
in~\eqref{Kav} is dominated by the two-particle interference.
The magnitude of this cut-off determines the smallest feature that 
we can resolve in the source.  This happens because a feature of size 
$\Delta r$ will contribute maximally to Fourier modes with frequency 
$\sim 1/\Delta r$.  If the maximum mode one restores has frequency 
$q_{\rm max}$ then the smallest feature that can be resolved has size 
$\sim 1/2q_{\rm max}$.  Similarly, since the data is typically binned 
in relative momentum with size $\Delta q$, the source can only be imaged 
to a distance of $~1/\Delta q$.

We will demonstrate the application of Fourier inversion on a two 
pion correlation function in Section~\ref{sec:pions}.  Despite the simplicity 
of this approach, it does suffer from some problems.  First, it only 
works for cases where the final state interactions can be neglected.  This 
means that it can only really apply to pion and photon correlations.  In fact 
its application to pions is limited to the class of reactions where 
the source is small.  For larger sources, typically a ``finite size Coulomb 
correction'' is used, wherein instead of dividing by the Gamow factor, one 
divides the correlation function by the square of the Coulomb wavefunction 
smoothed over a finite 
size Gaussian.  It is doubtful that such a correction does anything more 
than confuse the analysis.  Second, Fourier inversion is very sensitive to 
noise in the correlation function.  Simply put, a bump around a specific
relative momentum $q_0$ appears as a standing wave of frequency $q_0$ in the
imaged source.  This can lead to nonsensical behavior in the tails of the 
source where the sources amplitude is comparable to the magnitude of the
standing wave.  Together these problems lead us to consider more sophisticated 
approaches to imaging the source function.

\subsection{General Case}

Typically in an~experiment, the correlation function~$C_{\vec{P}}$ is
determined at discrete values of the magnitude of relative
momentum $\lbrace q_i \rbrace_{i = 1, \ldots, M}$ for
directionally-averaged function, or on a~mesh
in the momentum space $\lbrace \vec{q}_i \rbrace_{i = 1,
\ldots, M}$  when no averaging is done.  With each
determined value~$C_i^{\rm exp}$ some
error $\Delta C_i$ is
associated.  It is this set of
values $\lbrace C_i^{\rm exp} \rbrace_{i=1, \ldots, M}$, that
we use in determining the source function. In the 3-dimensional
case we may introduce a~rectangular mesh in the space of
relative particle separation and assume that the source
function is approximately constant within different cells of
the mesh.  In the angular expansion \eqref{CPl}, 
we may also assume
that the spherical expansion coefficients vary slowly within the
$r$-intervals.
In~the angle-averaged case, discretization amounts simply
to the representation $S
\simeq \sum_{j=1}^N S_j \, g_j(r)$, where~$N$ is the number of
intervals in~$r$, $g_j(r) = 1$ for $r_{j-1} < r <
r_j$, and $g_j(r) = 0$ otherwise,  with $r_j = j \,
\Delta r$.  On~inserting
a~discretized form of~$S$ into Equation~\eqref{eqn:PrattKoonin} or~\eqref{Kav},
we find a~set of equations for
the correlation functions $\lbrace C_i^{\rm th}
\rbrace_{i=1, \ldots, M}$, in~terms of $\lbrace S_j
\rbrace_{j=1, \ldots, N}$, 
\begin{equation}
   C_i^{\rm th} - 1 \equiv {\cal R}_i^{\rm th} = \sum_{j=1}^N
   K_{ij} \, S_j \, ,
\label{eqn:avedbinnedPKeqn}
\end{equation}
where, in the angle-averaged case,
\begin{equation}
   K_{ij} = 4 \pi
   \int_{r_{j-1}}^{r_j} dr \, r^2 \, K(q_i, r) \, .
\label{Kij}
\end{equation}

Now, supposing that the correlation function was measured with high accuracy
(i.e. that the experimental uncertainties are negligible compared to the data) 
and that the data do not contain a noise component, the 
Equation~\eqref{eqn:avedbinnedPKeqn} can be inverted using standard Maximum 
Likelihood methods such as the Lucy-Richardson algorithm.  
In this method, the source is found through the 
iteration of the equation \cite{img:var93}
\begin{equation}
   S^n_j=\frac{S^{n-1}_j}{\sum_i K_{ij}} 
   \sum_i K_{ij} \frac{C_i^0}{C_i^{n}}
\end{equation}
where 
\begin{equation}
   C_i^{n}=\sum_j K_{ij} S^{n-1}_j
\end{equation}
and $S^0_j$ is an initial guess for the source and $C^0_i$ is the correlation 
data.  One can see that when the $n^{\rm th}$ iteration generates a source 
close to the correct one, then $C_i^n\approx C_i^0$ making 
$S_j^n\approx S_j^{n-1}$.  So in this method, the ``most likely'' source is 
the end result of this iteration and should be close to the true source
in the sense of the Kullbeck-Leibler information divergence \cite{img:var93}:
\[
\sum_j S^{\rm true}_j \log{(S^{\rm true}_j/S^{n}_j)}
\]
This method is general but the convergence is very slow.  Furthermore
it relies on the notion of highly accurate data, which is often not the case 
in practice.   Finally, it is difficult to implement constraints or to 
estimate the error in the imaged sources.  Luckily images of
comparable or better quality can be achieved in much shorter times using
the Optimized Discretization method discussed below.

We now adopt a more Bayesian outlook and take the ``most likely'' source to be
the one that has the highest probability given the correlation data.  For
Gaussian distributed errors, the probability density for $S$ to be the ``most
likely'' source of $C$ is \cite{pre92,dag95}
\begin{equation}
	f(S|C)\propto \exp{(-\frac{1}{2}\chi^2)}
\label{eqn:BayesProb}
\end{equation}
where the $\chi^2$ is given by
\begin{equation}
   \chi^2=\sum_{i=1}^M \frac{(C_i^{\rm th} - C_i^{\rm exp})^2}{\Delta^2 C_i}
    = {\rm min} \, .
\label{chi}
\end{equation}
Clearly the probability density is largest when the source gives the minimum
$\chi^2$.  We can adjust the probability density in \eqref{eqn:BayesProb} 
by adding factors representing a priori knowledge of the source.  For example,
if we knew that the source is positive, we could add a factor of 
$\prod_{j}\theta {(S_j)}$
to the probability density.  The addition of factors such as this, which 
encode prior knowledge of the source, lead to {\em constraints} and the 
discussion in Subsection~\ref{sec:constraints}.

At this point, the obvious thing to do is to search the $\chi^2$ surface
for a minimum by varying the values of the source function in 
Equation~\eqref{eqn:avedbinnedPKeqn}.
This is in fact the very first thing we tried however there are problems with 
this method.  In a typical inversion, the source function has roughly 10-15 
points so the search for a minimum $\chi^2$ takes place over a 10--15 
dimensional parameter space.  This makes the method slow.  Furthermore,
in a typical search, one is not always sure the minimum one finds
is a true minimum or just a local minimum.  This means there is a chance that
any solution found is spurious.

Instead, we look for an algebraic way to find the source that minimizes the 
$\chi^2$.  If we do not constrain the space within which we search
for~$\lbrace S_j \rbrace_{j=1, \ldots, N}$, then we can get a~set of
linear equations for the values by functionally differentiating
\eqref{chi} with respect to~$\lbrace S_j\rbrace_{j=1, \ldots, N}$,
\begin{equation}
   \sum_{ij} \frac{1}{\Delta^2 C_i} \left(K_{ij} \, S_j -
   {\cal R}_i^{\rm exp} \right) K_{ik} = 0 \, ,
\end{equation}
or in a matrix form
\begin{equation}
   K^\top \, B \, (K \, S - {\cal R}^{\rm exp}) = 0 \, ,
\end{equation}
where $B_{il} = \delta_{il}/\Delta^2C_i$.
This matrix equation can be solved for $S$:
\begin{equation}
   S = \left( K^\top \, B \, K \right)^{-1} \, K^\top \, B \,
   {\cal R}^{\rm exp} \, .
\label{S=}
\end{equation}
So, in Equation~\eqref{S=} we have an algebraic equation for the source 
function and a new method for determining the source.
In the following subsections, we will examine this method and its enhancement,
the Optimized Discretization method.

We can determine the error on the source function 
by applying standard methods of error propagation to \eqref{S=},
\begin{equation}
   \Delta^2 S_j = \left( K^\top \, B \, K \right)^{-1}_{jj} \, .
\label{DS}
\end{equation}
The $N \times N$ matrix $K^\top \, B \, K$ in (\ref{S=}) is
symmetric, positive definite and may be diagonalized,
\begin{equation}
   \left( K^\top \, B \, K \right)_{kj} \equiv \sum_{i=1}^M
   \frac{1}{\Delta^2 C_i} \, K_{ik} \, K_{ij}
   = \sum_{\alpha = 1}^N \lambda_\alpha \, u_i^\alpha \,
   u_j^\alpha \, ,
\label{diag}
\end{equation}
where $\lbrace u^\alpha \rbrace_{\alpha = 1, \ldots, N}$ are
orthonormal and
$\lambda_\alpha \ge 0$.  With (\ref{DS}) and~(\ref{diag}),
the~square errors for individual values of~$S$ are
\begin{equation}
   \Delta^2 S_j = \sum_\alpha {(u_j^\alpha)^2  \over
   \lambda_\alpha} \, .
\label{DSl}
\end{equation}

We see in \eqref{DSl} that the errors for the source
diverge (or the~inversion problem becomes unstable) if~one or
more of the eigenvalues~$\lambda$ approaches zero.  In particular,
this happens when~$K$ maps an~investigated spatial region to
zero. A~specific case is when one of of the particles is neutral 
so $|\Phi|^2\approx 1$, cf.\ Equation~\eqref{KPhi}.  Moreover, instability 
can arise when one demands too high a resolution for a given set of 
measurements.  In such a~case, what might happen is that~$K$ smoothes out
variations in $S$, so we lose this information in the correlation function.
If we then try to restore $S$, we find that
we cannot restore~$S$ uniquely at high resolution. However at lower 
resolution, we might still be able to restore it.
Unlike typical numerical methods,
we need a singular rather than a smooth
kernel~\cite{pre92,img:hoo80} for our inversion problem to be tractable.
Finally, a~$\lambda$ close to zero can
be reached by accident for an~unfortunate choice of
$\lbrace r_k \rbrace_{k = 1, \ldots, M}$ in a  given  measurement.

\subsection{Optimized Discretization}

To make progress from here, we take advantage of the behavior
of our kernel.  In particular, we ask whether the fixed size binning in r is 
optimal in the algebraic approach.  For example, in the pp case,
the correlation function is dominated by the Coulomb interaction
at low-relative momenta and by the strong interaction and
antisymmetrization at intermediate momenta.  The~different
momentum regions should give access to large and short distances
within the source, respectively, with the resolution decreasing
at the large distances, rather than being fixed.
This suggests that, by varying the size of the discretization
interval of $S$, we can optimize the kernel to best restore a given source.
Since we do not know the source ahead of time, we must specify an
``model source'' in order to choose the best kernel.  This kernel
is then used in the actual inversion process in Subsections \ref{adata} 
and \ref{sec:IMFstuff}.

The~first stage of analysis involves the values of relative momenta
$\lbrace q_i \rbrace_{i =1 , \ldots , M}$, \sloppy\linebreak 
where correlation function was measured, and errors on these 
measurements \sloppy\linebreak
$\lbrace \Delta C_i \rbrace_{i =1 , \ldots , M}$, but not the
values themselves.
Specifically, we~vary the edges of the intervals for source discretization,
$\lbrace r_j \rbrace_{i =1 ,
\ldots , N}$,
demanding that the sum of errors relative to some ``model
source'' is minimized at fixed~$N$ and $r_0 = 0$,
\begin{equation}
\sum_{j=1}^N \left| {\Delta S_j \over S^{\rm mod}_j } \right|
= {\rm min} \, ,
\label{Dmin}
\end{equation}
where the $\lbrace \Delta S_j \rbrace_{j =1 , \ldots , N}$ stem
from Equation~\eqref{DSl}.   The specific choice of $S^{\rm mod}$ 
does not bias the inversion as it only ensures that there is sufficient
resolution where we know it is needed.  In fact, we~find a~rather weak
sensitivity of the results to fine details of $S^{\rm mod}$
in~(\ref{Dmin}), so we just use a~simple exponential form
$S^{\rm mod} \propto
\exp{(-r/R_0')}$, $S^{\rm mod}_j = S^{\rm mod}((r_{j-1}
+ r_j)/2)$, with~$R_0'$ of the order of few~fm.
The~exponential form is consistent with a~possible tail in the 
source due to prolonged decays. 

Features of the squared wavefunction in~\eqref{KPhi}
and the binning in q appear to have the greatest effect
on determining the best set \rj.  Nevertheless, it is
important to use {\em relative} errors, with some
sensible~$S^{\rm mod}$ in~\eqref{Dmin}.
If~absolute errors are taken, then the $r^2$ weight
from angle-averaging in~\eqref{Kij} favors large $r$'s.
The net result is that we learn that the source is close to
zero at large $r$ to a very high accuracy; we do not need
imaging to tell us this.
Our practical observation is that the~sum of
relative errors in~\eqref{Dmin},
rather than the sum of squares, is preferred for minimization;
the~sum of squares pushes \rj  inwards, leaving little
resolution at high~$r$.

\begin{figure}
\begin{center}\includegraphics[width=\HBTfigwidth]{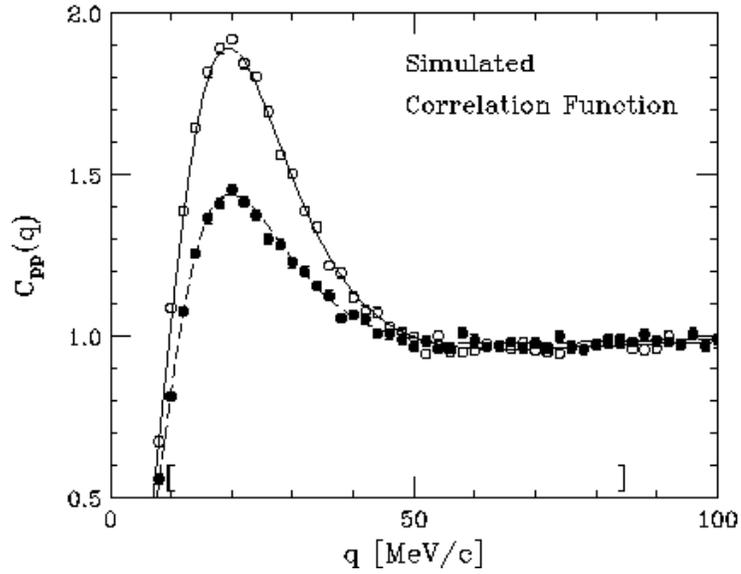}\end{center}
\caption[Comparison of original correlation function and restored correlation
	function.  The solid line is the correlation function from the source 
	in Equation \eqref{SG} and the dashed line is from the source in 
	Equation \eqref{SGo}.  We obtained the wavefunctions in the kernel 
	in \eqref{Kav} by solving the Schr\"odinger equation with the REID93
	potential \cite{sto94}.  The symbols represent the correlation 
	functions with added random noise;  the noise has a rms magnitude 
	of~0.015.  The~square brackets above the horizontal axis indicate 
	the range of~$q$ we used to restore source.]
	{Comparison of original correlation function and restored correlation
	function.  The solid line is the correlation function from the source 
	in Equation \eqref{SG} and the dashed line is from the source in 
	Equation \eqref{SGo}.  We obtained the wavefunctions in the kernel 
	in \eqref{Kav} by solving the Schr\"odinger equation with the REID93
	potential \cite{sto94}.  The symbols represent the correlation 
	functions with added random noise;  the noise has a rms magnitude 
	of~0.015.  The~square brackets above the horizontal axis indicate 
	the range of~$q$ we used to restore source.}
\label{simcor}
\end{figure}

To illustrate how well this imaging procedure works, we
take a relative pp source of a~Gaussian form
\begin{equation}
   S(r) = {1 \over (2 \pi R_0^2)^{3/2}} \exp{\left(- \,
   {r^2 \over 2 R_0^2} \right)} \, ,
\label{SG}
\end{equation}
and generate a~correlation function~$C$ at relative momenta $q$
separated by $\Delta q = 2$~MeV/c.  We use the folding~(\ref{Kav})
with the wavefunctions in the kernel calculated by solving the
Schr\"odinger equation with the regularized Reid soft-core potential 
REID93~\cite{sto94}.  This simulated correlation function is shown in
Figure~\ref{simcor}.  We~take $R_0 = 3.5$~fm
in the source and we add random Gaussian-distributed errors
to the correlation function from the folding.  The~rms
magnitude of the error is~0.015, which is representative of the
pp data of
Reference~\cite{HBT:gon90} analyzed in~\cite{HBT:bro97}.  We then attempt to
restore the source by discretizing it with~7 intervals of
fixed size, $\Delta r = 2$~fm, for $r
= (0-14)$~fm.  We use a $q$--interval similar to the one used in
Reference \cite{HBT:bro97}, i.e. $10$~MeV/c~$<q<86$~MeV/c.
Note that, were the inversion problem a Fourier-transform,
we could use more than~7 equally spaced $r$-intervals narrower than 2~fm.

\begin{figure}
\begin{center}\includegraphics[width=\HBTfigwidth]{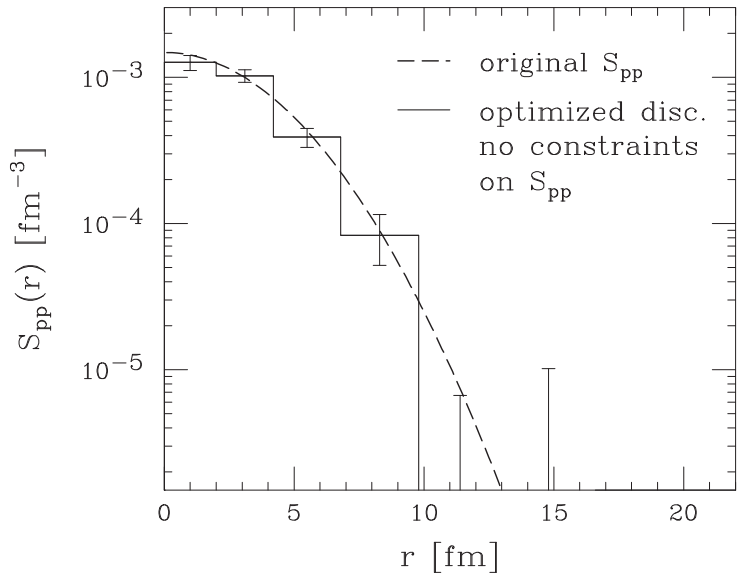}\end{center}
\caption[This plot is the same as Figures \ref{nocon2} and 
	\ref{con2}, except that the source is not constrained and
	is restored with the Optimized Discretization method.]
	{This plot is the same as Figures \ref{nocon2} and 
	\ref{con2}, except that the source is not constrained and
	is restored with the Optimized Discretization method.
	}
\label{nocono}
\end{figure}

\begin{figure}
\begin{center}\includegraphics[width=\HBTfigwidth]{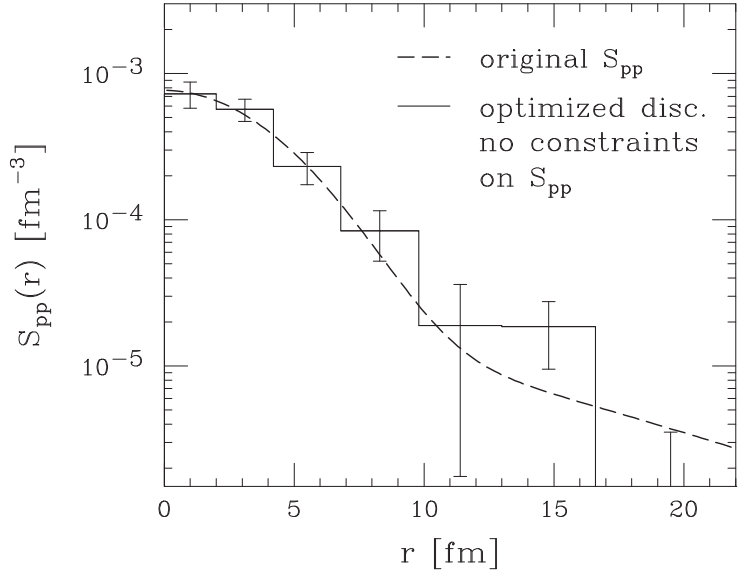}\end{center}
\caption[The solid histogram is the relative pp source
	restored using the Optimized Discretization method.  The correlation 
	function used is the solid data in Figure \ref{simcor}.  The original 
	source function is shown with the dashed line.]
	{The solid histogram is the relative pp source
	restored using the Optimized Discretization method.  The correlation 
	function used is the solid data in Figure \ref{simcor}.  The original 
	source function is shown with the dashed line. 
	}
\label{noconot}
\end{figure}

The results of applying  our procedure to the simulated pp correlation 
function of the preceding section, are shown in~Figure~\ref{nocono}, for $N=7$.
The~optimal intervals for discretization typically increase in size with~$r$.
For example, the first interval in Figure~\ref{nocono} is 2~fm wide and the
sixth is 3.6~fm wide.  The figure clearly shows that we can satisfactorily 
restore the source without imposing any constraints.  Figure~\ref{noconot} 
shows the results from a~similar restoration of the source with an
exponential tail:
\begin{equation}
	S(r) = {1 \over 2} \, {1 \over (2 \pi R_0^2)^{3/2}}
	\exp{\left(- \, {r^2 \over 2 R_0^2} \right)} + {1 \over 2} \,
	{15 \over 4 \pi^5 R_1^4} \, {r \over \exp{(r/R_1)} - 1}\, , 
\label{SGo}
\end{equation}
where~$R_0 =3.5$~fm and $R_1 = 6$~fm.  We show the~correlation function
corresponding to the restored source in Figure~\ref{simcor}, both with and 
without errors and random noise with an rms magnitude of~0.015.  Since~the
same~$N$ and the same $\lbrace q_i,\Delta C_i
\rbrace_{i=1,\ldots,M}$ are used in the inversion,
we find the~same optimal \rj used in Figure~\ref{nocono}.
The~restored source gives evidence for
the tail in the source, despite of the fact that the magnitude of the
tail is lower by~2 orders of magnitude compared to the
maximum at $r=0$.  Comparing Figures~\ref{nocono}
and~\ref{noconot},
we see that our method can discriminate between the two source shapes.
If we impose additional constraints to the optimized discretization
method, the agreement between the restored and original source
functions improves.


So, while imposing constraints on the source stabilizes the inversion 
(as we will show in the next subsection), we have developed an~imaging 
method that can yield very satisfactory
results even without any constraints.
Indeed, one may want to see directly whether the
data are consistent with positive definite sources.

\subsection{Constraints}
\label{sec:constraints}

In the Optimized Discretization method, we used the specific behavior of the 
kernel and the data to improve the inversion.  
We can also use the source itself, or at least known properties 
of the source, to stabilize the inversion further.  
We do this by adding constraints as first suggested by Tikhonov 
\cite{img:tik63}.  
In practice this amounts to Monte Carlo sampling the error on the 
experimental data to construct a test correlation function, inverting 
the test correlation, and testing whether the test source obeys a known 
set of constraints.  If the source is not acceptable, we discard it.   
We repeat this sampling until we have enough statistics to report the source
and its error.

\begin{figure}
\begin{center}\includegraphics[width=.9\HBTfigwidth]{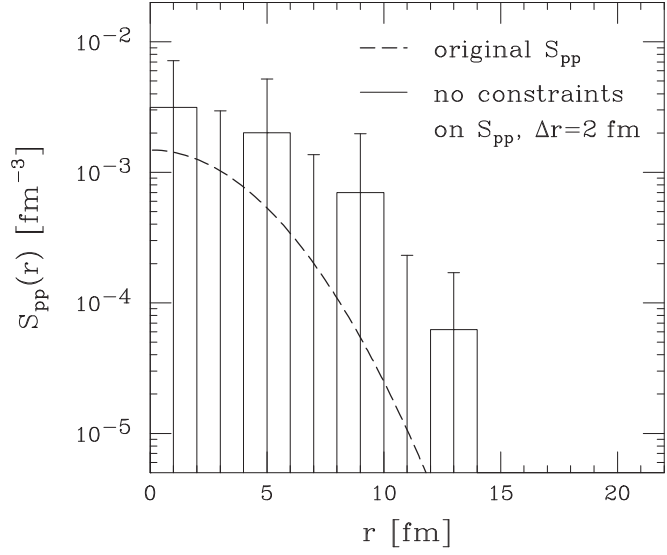}\end{center}
\caption[Comparison of the original source with the source restored without 
	constraints.  The solid histogram is the source 
	function~$S$ restored from the simulated correlation function 
	with the open symbols in Figure~\ref{simcor}.
	The dashed line is the original source function in \eqref{SG}
	used to generate the correlation function.
	We used fixed intervals of $\Delta r = 2$~fm 
	for discretizing the source function.]
	{Comparison of the original source with the source restored without 
	constraints.  The solid histogram is the source 
	function~$S$ restored from the simulated correlation function 
	with the open symbols in Figure~\ref{simcor}.
	The dashed line is the original source function in \eqref{SG}
	used to generate the correlation function.
	We used fixed intervals of $\Delta r = 2$~fm 
	for discretizing the source function.}
\label{nocon2}
\end{figure}

Without using either Optimized Discretization or constraints, it is 
difficult to obtain stable images.  To illustrate how serious an issue of
stability is, take the Gaussian source used in 
Figure \ref{nocono}.  Applying the straight algebraic approach yields the 
results in Figure \ref{nocon2}.  Clearly the
errors for restored source far exceed the original source function.
In~fact, every second value of the restored source is
negative.  Incidentally, this is one of our more fortunate 
simulations, since all of the errors actually fit on the plot.

\begin{figure}
\begin{center}\includegraphics[width=.9\HBTfigwidth]{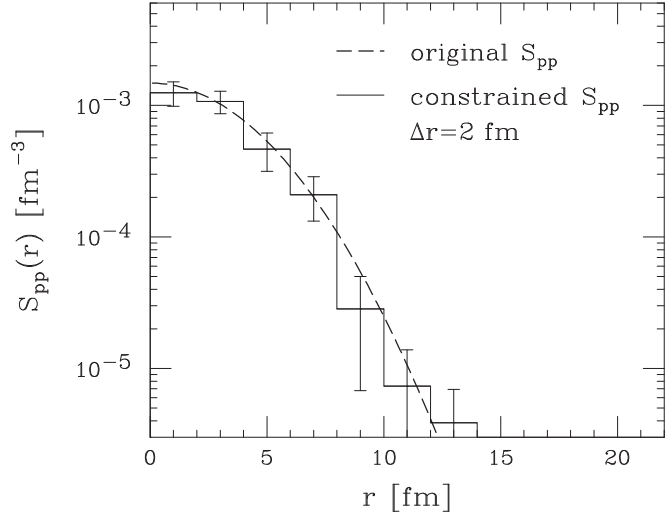}\end{center}
\caption[This plot is the same as Figure \ref{nocon2}, except that the restored
	source is constrained to be positive and is normalized to one.]
	{This plot is the same as Figure \ref{nocon2}, except that the restored
	source is constrained to be positive and is normalized to one.
	}
\label{con2}
\end{figure}

We next illustrate the dramatic stabilizing effect that
the constraints have on the imaging,
as it was first discussed by Tikhonov~\cite{img:tik63}.
We carry out the inversion using the same
correlation function and errors that we used
for Figure~\ref{nocono}.
We impose the constraints that the imaged source is positive
definite, i.e. $S_j \ge 0$, as  expected in the
semi-classical limit, and that the source is
normalized: within the restored region
$1 \ge 4 \pi \int_0^{r_N} dr \, r^2 \, S \approx 4 \pi
\sum_{j=1}^N S_j \int_{r_{j-1}}^{r_j} dr \, r^2$.
Following the general strategies~\cite{dag95} for
estimating values with errors under constraints,
we carry out our imaging by sampling the values of the
correlation
of function according to the errors $\lbrace \Delta C_i
\rbrace_{i = 1, \ldots, N}$ (equal to 0.015 in our case) and
by applying~(\ref{S=}).  This amounts to the
replacements in~(\ref{S=}): $C_i
\rightarrow C_i + \Delta C_i \, \xi_i$, $i = 1,
\ldots, N$, with $\xi$'s drawn from the standard normal
distribution.  We~accept only those samplings where the
constraints
are met.  With these samplings, we calculate the average
source values
and the average dispersions.  The~results are shown in
Figure~\ref{con2} together with the original source. They
now compare favorably to the original source.

Clearly usage of constraints helps to stabilize the images, so 
what constraints can be used during imaging?  We have already 
mentioned two: normalization and positivity.  
One other that we have used takes advantage of the fact that the source for 
like particle pairs is actually the convolution of the single particle 
sources in Equation \eqref{eqn:PK2}.  Fourier transforming Equation \eqref{eqn:PK2}
we find that the Fourier transform of the source must be positive definite:
\begin{equation}
  \int \dn{3}{r} e^{i \vec{r}\cdot\vec{q}} S(\vec{r})=|g(\vec{q})|^2\geq 0
\end{equation}
where the $g$'s are Fourier transforms of the time-integrated single particle 
sources (see Equation~\eqref{eqn:sourcedef}).  For the angle averaged source, 
this constraint is written in terms of a spherical Bessel function:
\begin{equation}
  \int_0^\infty \dn{}{r} r^2 S(r) j_0(qr)\geq 0.
\end{equation}
Another constraint
one might use~\cite{img:tik63,img:ber80,img:hoo80} is an assumption of
smoothness of the source, permitting inversion with
more points in the image then there are in the data.
Cutting off the integral in the Fourier-transform
method~\cite{HBT:bro97} at $q_{\rm max}$ implies the constraining
assumption
that~$S$ varies slowly on the scale of $1/(2q_{\rm max})
\approx 1.2$~fm in~\cite{HBT:bro97}, which is reasonable given the
range of strong interactions.
Finally we comment that, in imaging terms, the common
Gaussian parameterization of
sources in heavy-ion collisions is a very extreme constraint
for stabilizing the inversion.

\section{Generalized Chaoticity Parameter}
\label{sec:integralofsource}
Just as important as what is in the images, there is  
a quantity that characterizes what is not in the images.
This parameter is the generalized chaoticity parameter
and is defined as the integral of the source over a region where
the source is significant:
\begin{equation}
   \lambda (r_N) = \int_{r < r_N} \dn{3}{r} \, S(\vec{r}) \, .
\label{lambda}
\end{equation}
This symbol, $\lambda(r_N)$, is a generalization of the 
the chaoticity~$\lambda$ used to parameterize high-energy
$\pi \pi$ correlations.  The standard chaoticity parameter 
is defined by fitting the $\pi \pi$ correlation function to a Gaussian:
\begin{equation}
C(q) - 1 = {2 \pi \over q} \int_0^\infty dr \, r \,
\sin{\left(2qr \right)} S(r) \simeq \lambda \exp{\left(- \,
{4 q^2 R_0^2 } \right)} \, .
\label{Cql}
\end{equation}
When we uses this parameterization of the correlation function, we are assuming
this parameterization of the source:
\begin{equation}
S(r) \approx { \lambda \over (2 \pi R_0^2)^{3/2}} \exp{\left(- \,
{r^2 \over 4 R_0^2} \right)}.
\label{SGl}
\end{equation}

Our chaoticity parameter generalizes the one in (\ref{Cql}) and 
(\ref{SGl}) in two ways.
First, the integral defining the generalized chaoticity parameter
extends only up to some cut-off, $r_N$.  The conventional definition of
$\lambda$ can be recovered from ours by extending $r_N\rightarrow \infty$.
Second, because our chaoticity parameter is defined in terms of the imaged
source function, rather than a Gaussian fit to the correlation function, it can
apply to any particle pair, not just pion pairs.

The importance of our definition of the generalized chaoticity parameter lies
in the fact that some particles in the reaction,
such as pions or protons, can stem from long-lived resonances
and be emitted far from any other particles.
Thus, they contribute primarily to~$S$ at large~$r$, outside the imaged region.
For large~$r$ and moderate-to-high~$q$, the kernel $K$ 
averages the large-$r$ tails in~$S$ to zero so they would not contribute to
deviations of $C$ from~1 in Equation~(\ref{eqn:PrattKoonin}) or~(\ref{Cql}).  
In~practice, it should be
only possible to directly detect the tails in~$S$ by
investigating the low-$q$
correlation functions for charged particles.  For the analyzed
pp data~\cite{HBT:gon90}, this $q$-region is either not
available or is associated with large systematic errors.
Imaging can only extend up to $r_N \sim 20$~fm and
we expect~$\lambda (r_N) < 1$.   In many $\pi \pi$
measurements, e.g.~\cite{HBT:bar97}, the~resolution 
allows one to image regions of comparable sizes,
typically 10--20~fm \cite{HBT:bro97}.
In~contrast to the $pp$ and $\pi \pi$ data, the~data on
IMF, such as~\cite{HBT:ham96}, often extend to low values of
relative velocity.  This permits imaging up to relative separations
as large as~50~fm.  For a~discussion of this, see Section~\ref{sec:IMFstuff}.

\section{Freeze-Out Density, Average Phase-Space Occupancy and Entropy}
\label{sec:otherstuff}

The source function and generalized chaoticity parameter both indirectly
tell us quite a bit about the reaction dynamics.
For example, the source function tells us the relative emission profile, but 
in the pair frame.  It would be an improvement if we had this profile in the 
system frame.  We show how to get this in this section.
Of course, we really want is to see the time-evolution of the entire 
phase-space density.  This is not directly accessible from the source, but we 
can get at the average phase-space density.  From this, we can estimate 
the entropy.

To get at the source function in the system frame we must make some 
assumptions.  For a~rapid freeze-out, 
the single particle source is given by $D(\vec{p}, \vec{r}, t) \simeq
f(\vec{p}, \vec{r}) \, \delta( t - t_0)$ where $f$
is the Wigner function of the particles.  For
weak {\em directional} correlations between the
total and relative momentum of pairs
and between the spatial and momentum variables,
the~momentum average of~$S$ approximates
the~relative distribution of emission points for any two particles
from the reaction, and not just for the particles with close
momenta.  
Under these conditions, the~relative distribution for any 
two particles is
\begin{equation}
   {\cal S} (\vec{r}) = {\int \dn{3}{P} \dn{3}{p} \dn{3}{R} 
   f(\vec{P}/2 + \vec{p}, \vec{R} + \vec{r}/2) \,
   f(\vec{P}/2 - \vec{p}, \vec{R} - \vec{r}/2) \over
   \int \dn{3}{p_1} \dn{3}{r_1} f(\vec{p}_1 ,\vec{r}_1) \,
   \int \dn{3}{p_2} \dn{3}{r_2} f(\vec{p}_2 ,\vec{r}_2) } \, .
\label{sr}
\end{equation}
Rewriting and expanding the numerator in~(\ref{sr}):
\begin{align}
   \lefteqn{f(\vec{P}/2 + \vec{p}, \vec{R} + \vec{r}/2) \,
	f(\vec{P}/2 - \vec{p}, \vec{R} - \vec{r}/2)} \notag \hspace{4em}&\\
   = & \int \dn{3}{r_1'} \, f(\vec{p}_1 ,\vec{r}_1') 
 	\int \dn{3}{r_2'} \, f(\vec{p}_2 ,\vec{r}_2') 
	{f(\vec{P}/2 + \vec{p}, \vec{R} + \vec{r}/2) \,
	f(\vec{P}/2 - \vec{p}, \vec{R} - \vec{r}/2)\over
	\int \dn{3}{r_1'} f(\vec{P}/2 + \vec{p} ,\vec{r}_1') 
 	\int \dn{3}{r_2'} f(\vec{P}/2 - \vec{p} ,\vec{r}_2')} \notag\\
    \begin{split}
	= &\int \dn{3}{r_1'} f(\vec{p}_1 ,\vec{r}_1') 
 	   \int \dn{3}{r_2'} f(\vec{p}_2 ,\vec{r}_2') 
	   \left(1 + \vec{p} \, {\partial \over \partial \vec{p}'}
	   + \ldots \right) \\
        &\times \left.{f(\vec{P}/2 + \vec{p}', \vec{R} + \vec{r}/2)\,
	   f(\vec{P}/2 - \vec{p}', \vec{R} - \vec{r}/2)\over
	   \int \dn{3}{r_1'} f(\vec{P}/2 + \vec{p}' ,\vec{r}_1') \,
	   \int \dn{3}{r_2'} f(\vec{P}/2 - \vec{p}' ,\vec{r}_2')} \,
	   \right|_{\vec{p}' = 0} \, .
       \end{split}
\label{ff}
\end{align}
The gradient term must be proportional to a~combination of the
vectors $\vec{P}$, $\vec{r}_1$, and $\vec{r}_2$ and for the
weak directional correlations it would average to zero under
the integration in~\eqref{sr}.  Inserting~\eqref{ff}
into~(\ref{sr}) and keeping the leading term, we~obtain
\begin{equation}
   { \cal S} (\vec{r}) \simeq {1 \over N^2} \int \dn{3}{p_1} \,
	\dn{3}{p_2} {d N \over \dn{3}{p_1}} \,
	\, {d N \over \dn{3}{p_2}} \, \gamma_{P} \,
	S_{\vec{P}}(\vec{r} + \hat{n}_{\vec{P}} \, (\gamma_{P} - 1) (
	\hat{n}_{\vec{P}} \, \vec{r})) \, ,
\label{rs}
\end{equation}
where $N$ is particle multiplicity and $\hat{n}_{\vec{P}} =
\vec{P}/|\vec{P}|$. The~argument of~$S_{\vec{P}}$ has been written in
the CM frame of an emitted~pair and $\gamma_{P}$ is the Lorentz
factor for the transformation from the system frame to the pair
CM frame.  
In general, the relative distribution of emission
points for any two particles with~$r \rightarrow 0$, when
multiplied by~$N-1$, gives an~average freeze-out density.
Thus, if the assumptions
above are valid, this density may be obtained by multiplying
the~average~(\ref{rs}) of~$S_{\vec{P}}(r \rightarrow 0)$ by~$N-1$.

We can estimate the phase-space occupancy at freeze-out regardless of any 
correlations between momentum of coordinate variables or of the validity of
instantaneous freeze-out.  The~product of the $r \rightarrow 0$
source function and the momentum distribution yields the configuration-space
average of the phase-space occupancy at freeze-out,
\begin{equation}
   \langle f \rangle (\vec{p}) = {(2 \pi)^3 \over 2 s+1}\, {E_p \over m} 
   \, {d N \over \dn{3}{p}} \, S_{2 \vec{p}} (r \rightarrow 0)\, ,
\label{f}
\end{equation}
as is discussed in~\cite{HBT:ber94b,HBT:ber94c}.
Equation~\eqref{f} can then be used to determine the phase-space
average of the occupancy at freeze-out, 
\begin{equation}
   \langle f \rangle = \int \dn{3}{p} (\langle f\rangle (\vec{p}))^2 \, /
   \int \dn{3}{p} \, \langle f \rangle (\vec{p}), 
\label{fave}
\end{equation}
and to estimate the entropy per particle,
\begin{equation}
   {S \over A} \approx - \, {\int \dn{3}{p} \left(\langle f \rangle
   (\vec{p}) \, \log{\left(\langle f \rangle (\vec{p})\right)} -
   \left(1 - \langle f \rangle (\vec{p}) \right) \,
   \log{\left(1 - \langle f \rangle (\vec{p})\right)} \right)
   \over \int \dn{3}{p} \langle f \rangle (\vec{p})} \, .
\label{entropy}
\end{equation}
Now, the average phase-space occupancy at freeze-out and the entropy per pion
are often calculated for pions because this can be done directly from the 
correlation function \cite{HBT:ber94b,HBT:ber94c}.  What is new here is that,
because we use the source function directly, 
Equations \eqref{f}--\eqref{entropy} apply to any type of particle.
In particular, we will calculate the phase-space occupancy and entropy per 
proton in Section~\ref{adata}.

\section{Pions and Kaons}
\label{sec:pions}

As~a~specific example of the source extraction, we present 
the inversion of the angle-averaged $\pi^-$ source function determined
from the central 10.8~GeV/c Au~+~Au data of Reference \cite{HBT:mis96}
and the angle-averaged $\pi^-$ and $K^+$ source functions determined
from the preliminary central 11.4~GeV/c Au~+~Au data of 
Reference \cite{HBT:tho98b}.
The comparison of the data sets will show the lack of variation of the 
$\pi^-$ sources with energy and will show the dramatic difference between
the $\pi^-$ and $K^+$ sources.  Before presenting this discussion, we will
compare inversions of the \cite{HBT:mis96} data set using the Optimized 
Discretization method and direct Fourier transformation of Coulomb corrected 
data, demonstrating the consistency of the inversions.

\subsection{Comparison of Inversion Methods}

\begin{figure}
\begin{center}\includegraphics[angle=270,width=\textwidth]{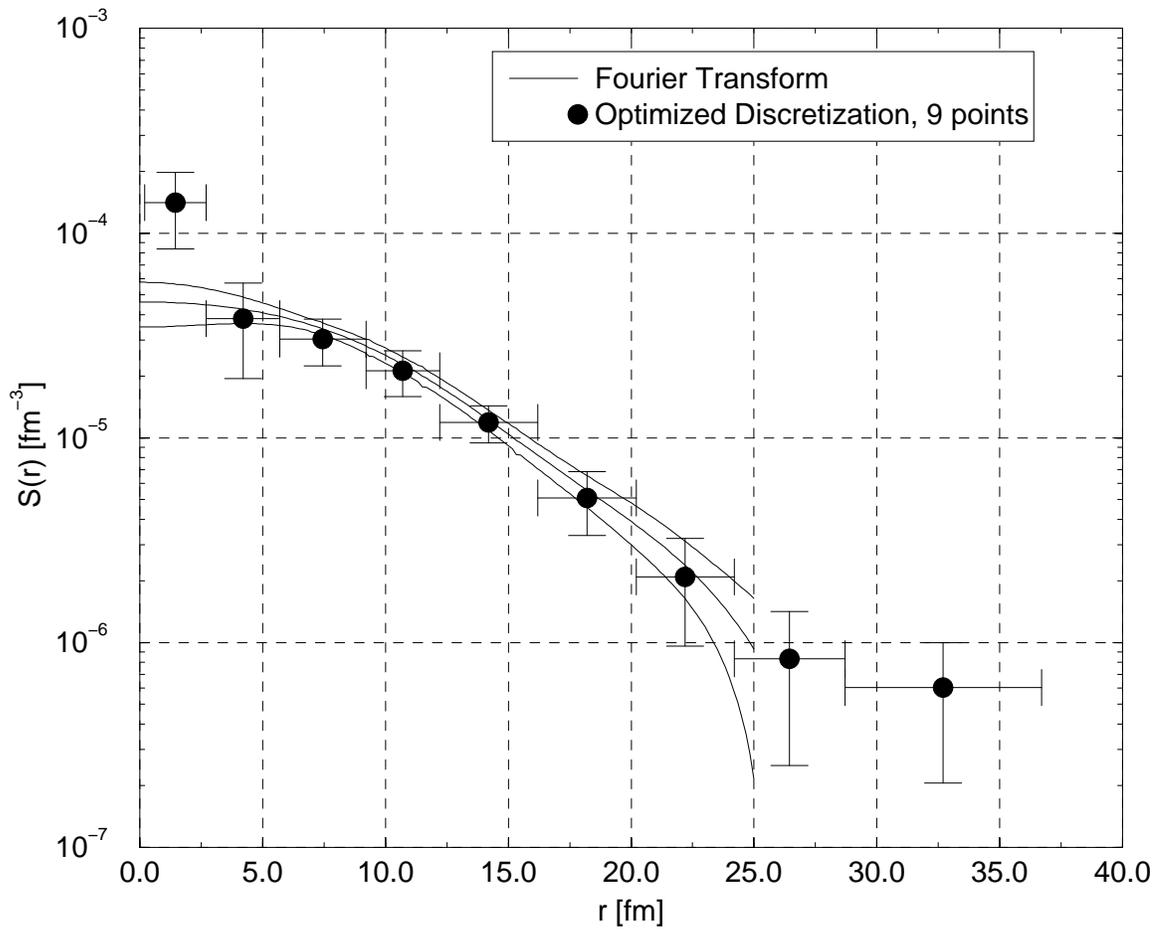}\end{center}
\caption[Source function for negatively charged pions from Reference 
	\cite{HBT:mis96}.]
	{Source function for negatively charged pions from Reference 
	\cite{HBT:mis96}.}
\label{pipi-}
\end{figure}

The data of \cite{HBT:mis96} has been Coulomb corrected for the pair
Coulomb force.  So in the Pratt-Koonin equation, we may use the kernel for a
non-interacting pair, $K(q,r)=\sin{(2 q r)}/(2 q r)$.  Thus, the inversion can
be done by performing a Fourier transform.  
In Figure~\ref{pipi-}, we show the results from this 
Fourier transform of the correlation function and from using the 
Optimized Discretization method.
Unlike in  \cite{HBT:bro97}, we do note correct the data for the Coulomb 
interaction between the pions and the source as in \cite{HBT:bay96}.  

In the Fourier transformed data, the upper and lower lines represent the error
band surrounding the average source.
We carried out the Fourier inversion in \eqref{rSr2} for~Figure~\ref{pipi-} 
up to~$q_{\rm max} \simeq 50$~MeV/c giving a~resolution in
the relative distance of $\Delta r \sim 1/2q_{\rm max} \sim 2.0$~fm.  
The~largest $r$ that we should be able to image 
follows from $1/2 \, \Delta q \approx 20$~fm, where $\Delta q$ is the momentum
resolution for the data ($\Delta q = 5$~MeV/c in~the case
of~\cite{HBT:mis96}).

For the Optimized Discretization results, we again use the non-interacting pion 
kernel.  Ideally we would have liked to do the
inversion using the full Coulomb wavefunctions for the kernel as this would
increase the overall accuracy of the plot, however the data was corrected
using the ``finite size Coulomb correction'' and a finite resolution 
correction which we can not unfold.

The results from both inversions are quite consistent except for the lowest r 
bin in the Optimized Discretization set.  This point is at $r=1.45$~fm
and at that range, the inter-pion force should receive significant 
contribution from the nuclear force and neither inversion 
results should reflect the actual source.  In the tail region, where past the
edge of the Fourier inverted data, the Optimized Discretization data tends to 
flatten a bit.  This is probably due to the large size of the bins rather than
an actual flattening of the source.  So, in the end it is comforting to see that
the source comes out the same with both methods.

\subsection{Comparison of $\pi^-$ and $K^+$ Sources}

\begin{figure}
\begin{center}\includegraphics[angle=270,width=\textwidth]{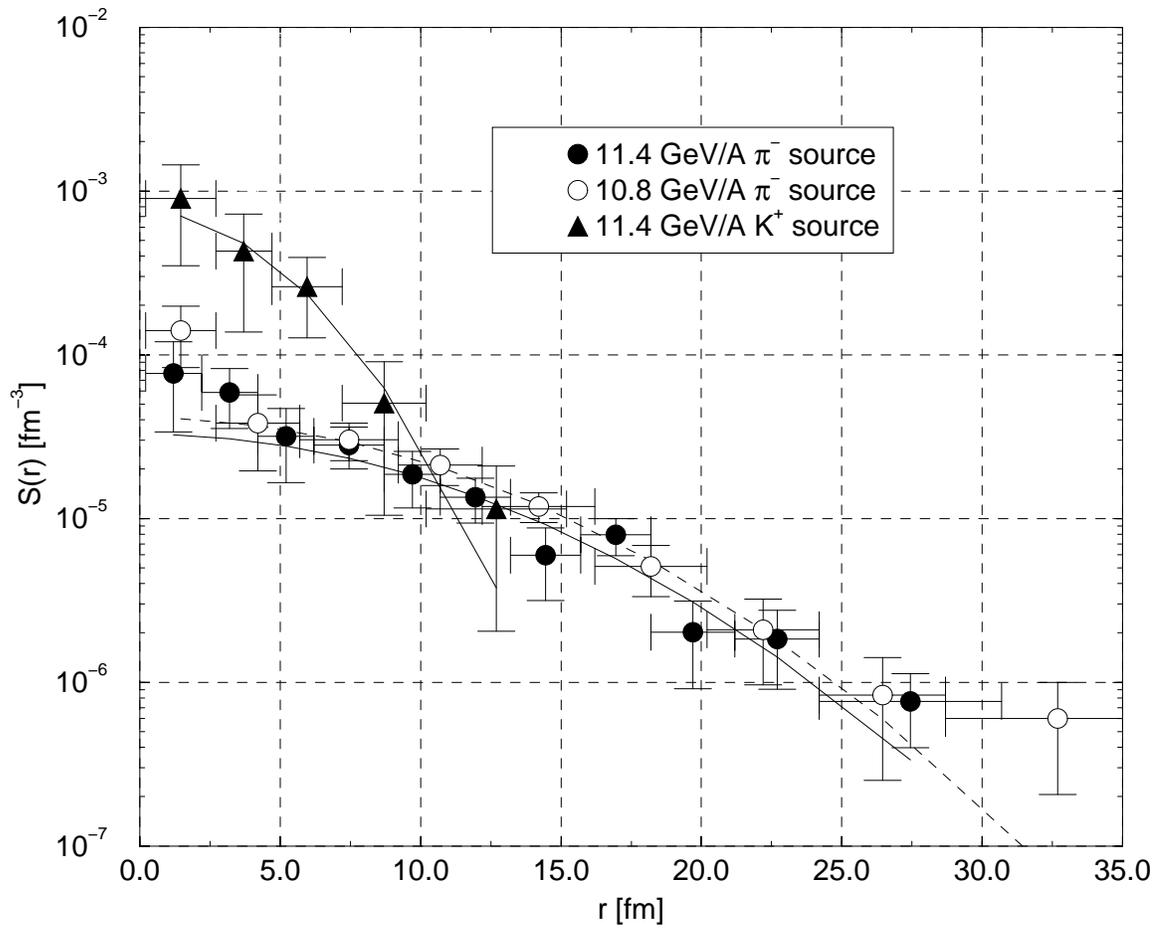}
\end{center}
\caption[Comparison of the negatively charged pion source functions from the 
Au~+~Au reaction at $11.4$ and $10.8$ GeV/A and positively charge kaon source 
functions from the same reaction at $11.4$ GeV/A.]
	{Comparison of the negatively charged pion source functions from the 
Au~+~Au reaction at $11.4$ and $10.8$ GeV/A and positively charge kaon source 
functions from the same reaction at $11.4$ GeV/A.}
\label{fig:allpi-andK+}
\end{figure}

We now discuss the Optimized Discretization results from the $10.8$ GeV/A pion 
data of \cite{HBT:mis96} and the preliminary $11.4$ GeV/A pion and kaon data 
sets from \cite{HBT:tho98b}.
The plots of the source functions for these data are shown in 
Figure \ref{fig:allpi-andK+}.  The $10.8$ GeV/A $\pi^-$ data set was Coulomb 
corrected and inverted in the manner described above.  The $11.4$ GeV/A pion 
data set was treated differently since it is not Coulomb corrected for the pair
Coulomb force.  We restored this source using the full Coulomb 
wavefunctions and the kernel
\begin{equation}
K_0(q,r)=\sum_{\ell\,\,{\rm even}} \frac{|g_\ell(r)|^2}{(2\ell+1)}-1.
\end{equation}
The kaon data set was also restored using this kernel and the full Coulomb 
wavefunctions.  In both the pion and kaon cases, the inter-meson nuclear forces
were neglected as they are only important at distances $\lesssim 1$~fm for the 
momenta of interest.

Examining Figure \ref{fig:allpi-andK+}, several things are apparent.  First, 
we note that both $\pi^-$ data sets are consistent despite 
the difference in beam energy and the difference in centrality cuts
($\sigma/\sigma_{\rm geom}< 10\%$ for the $10.8$ GeV/A data versus 
$\sigma/\sigma_{\rm geom}< 4\%$ for the $11.4$ GeV/A data).  Because the higher 
energy data set has roughly five times the statistics of the lower energy set, 
we can restore with higher resolution.  

The next thing we notice is that the kaon source is 
more compact than the pion sources.  
Because the temporal and spatial evolution of the single particle sources is
entangled in the source function in a non-trivial way 
(see Equation \eqref{eqn:sourcedef}), the difference in 
sources sizes can be attributed to either life-time effects, emission time 
effects, real differences in the source size or a combination of the three.
First, a sizeable fraction of the pions are the products of resonance decays 
such as the $\omega$, $\rho$, $\eta$ and $\eta'$ (the rest are produced 
directly) while the majority of the kaons are 
produced directly or via the decay of $K^*$'s.  Since both the $\eta$ and 
$\eta'$ decay at distances much greater than $35$ fm (the $\eta$ lifetime is 
$1.6\times 10^5$ fm/c and the $\eta'$ lifetime is $980$ fm/c), decays from 
these resonances do not contribute to the shape of the pion source.  
Comparing the lifetimes of the $\rho$ ($1.3$ fm/c) and the $\omega$ ($23$ fm/c),
it would seem that $\omega$ decay would have the greatest effect on the shape 
of the pion source, provided they are produced in sufficient numbers.  
Similarly, the decay of the $K^*$ would have a great effect on the kaon source
as it's lifetime ($3.9$ fm/c) is comparable to the size of its source.
Second, the difference in the pion and kaon sources may be attributed to 
emission time effects.  Given that the pion mass is roughly $1/3$ the kaon mass,
the pions will have a larger average velocity.
Comparing a kaon pair and a pion pair, both created with 
a similar time separation, the early pion can travel much farther the early kaon
could -- extending the pion source relative to the kaon source.  
Finally, the pions may be emitted for a larger source region than the 
kaons.  The cross sections for the pions to interact with the particles in the 
system are, on the average, larger than the kaon cross sections.  This means 
that the pions couple to the system more strongly and so are more effected by 
the system's evolution.  Since the colliding system expands as it evolves, 
the source size of the pions could be attributed to the pions being emitted 
at a later stage in the system's evolution than the kaons.

Unentangling which combination of effects is responsible for the difference 
between the $\pi^-$ and $K^+$ sources could be accomplished either through
modeling or further experimentation.  We have not yet performed model 
calculations for this reaction but calculations for similar reactions at CERN
energies have been performed by Sullivan and coworkers \cite{HBT:sul93} and 
many of their qualitative results are applicable at AGS energies.  First, they 
find that most of the kaons stem from the decay of $K^*$'s and string 
fragmentation.  There are few (if any) strings at AGS energies, so it is 
likely that our kaons stem solely from $K^*$ decays.  Second, they find that 
a majority of pions are produced in secondary collisions or via short-lived 
resonance decays (meaning those with lifetimes $\leq 2$ fm/c).  However, they 
also found that a large 
fraction of pions are made from long-lived resonance decays (with lifetime
$> 2$ fm/c).  The long-lived resonance contribution dominates at distances
of about 10 fm onwards with the longest-lived resonances (the $\eta$ and 
$\eta'$) producing 1 out of every 4 pions in the central rapidities at SPS 
energies.  So, it would seem that a combination of effects can explain
why the kaon source is narrower that the pion sources.  
We could also unravel which combination of scenarios gives the 
difference in source sizes by using the full three-dimensional correlation 
function.  With the full source function, we would look 
for elongation in the $\vec{P}$ direction.  A difference in emission 
time of the pair would show up as an elongation in the direction of the 
average relative velocity of the pair and this average velocity is parallel to 
the total momentum of the pair.

Let us now move beyond discussion of the size of the pion and kaon sources 
and examine
the integrals of the sources and what we can learn from Gaussian fits to the 
sources.  Now, all three data sets appear Gaussian so we can fit the sources
with the Gaussian parameterization 
\[ S(r) = \frac{\lambda_{\rm fit}}{(2\sqrt{\pi} R_0)^3} \exp{\left(-\left(
\frac{r}{2 R_0}\right)^2\right)}\]
Results from the fits are listed in Table \ref{table:piK}.
In this table, we have also tabulated the integrals of the sources over the 
entire imaged region.  In all three cases, the integrals of the source, a.k.a.
the generalized chaoticity parameter, are
consistent with the fit parameter $\lambda_{\rm fit}$.  The fit parameter
$\lambda_{\rm fit}$ is usually identified as the chaoticity parameter.  The 
fact that the integral of the kaon source is consistent with $1$ tells us that
the entire source is within the imaged region.  The fact that both pion sources
are {\em not} consistent with $1$ tells us that much of the pion sources must
lie {\em outside} of the imaged region.  In fact, we could estimate that roughly
$40\%$ of the pion pairs have one or both pions emitted farther than $35$ fm 
from the center of the reaction zone.  What could account for this long-distance
emission?  The most obvious answer is resonance production of the pions.  The
$\omega$ decays with a lifetime of $23$ fm/c so it might account for a tail
near the edge of the image -- higher resolution correlations (especially near
$q\approx 0$) might show evidence for this long-distance decay.  The $\eta$ and
$\eta'$ decays are much longer -- the lifetimes are $1.6\times 10^5$ fm/c and 
$980$ fm/c respectively.  Pion production from either of these resonances could
not be detected directly with our imaging method but could account for 
$\lambda(35 {\rm fm})$ being less than one.

\begin{table}
\caption[Listing of Gaussian fit parameters and the integrals of the source for
the pion and kaon sources.]
	{Listing of Gaussian fit parameters and the integrals of the source for
the pion and kaon sources.}
\vspace*{\baselineskip}
\begin{center}
\begin{tabular}{|cccr@{$\pm$}l|}\hline
\multicolumn{1}{|c}{} &
\multicolumn{1}{c}{$R_0$ [fm]} &
\multicolumn{1}{c}{$\lambda_{\rm fit}$} &
\multicolumn{2}{c|}{$\lambda(35 {\rm fm})$} \\ \hline
$K^+$ (11.4 GeV/A)   & 2.76 & 0.702 & 0.86 & 0.56 \\
$\pi^-$ (11.4 GeV/A) & 6.42 & 0.384 & 0.44 & 0.17 \\
$\pi^-$ (10.8 GeV/A) & 6.43 & 0.486 & 0.59 & 0.22 \\ \hline
\end{tabular}
\end{center}
\label{table:piK}
\end{table}

\section{Protons}
\label{adata}

Now we apply the Optimized Discretization method to
anal\-yze the pp cor\-re\-la\-tion data~\cite{HBT:gon90,HBT:gon91a}
from the $^{14}$N +$^{27}$Al reaction at 75~MeV/nucleon, that we imaged in
a naive fashion in~Reference~\cite{HBT:bro97}.
Since this method does not require a positive definite source to stabilize 
the image, we are able to
lift this constraint and verify whether the data favor positive
definite sources.  Further, we do not need to normalize the sources
to one within the imaged region.  We also compare the pp sources
from data to those from the transport model~\cite{tran:dan95}, over
a~large range of relative separations and magnitude of the
sources.  Past experiences in comparing semi-classical transport
models to single-particle and correlation data
have been mixed~\cite{HBT:gon90,HBT:gon91a,HBT:gon93,HBT:han95,HBT:gaf95}, 
for this particular reaction and others in this energy range.

In Reference~\cite{HBT:gon90,HBT:gon91a}, the~low relative-momentum 
pp-correlations were determined for pairs emitted around 
$\theta_{\rm lab} = 25^\circ$ from the $^{14}$N + $^{27}$Al reaction at 
75~MeV/nucleon, in three intervals of the total momentum:
270--390, 450--780, and 840--1230~MeV/c.  The~highest lab momenta
interval corresponds to the highest proton momenta
in the participant CM for this reaction.  These momenta are higher than the
average for participant protons and directed
rather forward. Transport calculations~\cite{HBT:bro97,tran:dan95}
show that the highest momenta bin is mostly populated by
pairs from the semi-central to peripheral collisions.
The~intermediate momenta interval
corresponds to the magnitude of typical momenta
of participant nucleons in the forward NN CM hemisphere.
The transport calculations show that these pairs stem mainly
from the semi-central collisions. The~lowest
lab momenta interval has both participant and target spectator
contributions.  In the latter case, the transport calculations
show that pairs are mostly from the semi-central to central
collisions.  According to the transport model, the~average
emission times for protons in the three momenta
intervals, from the first contact of the nuclei,
are $\sim 35$, $\sim 80$, and $\sim 110$~fm/c, respectively.

\begin{figure}
\begin{center}\includegraphics[width=5.5in]{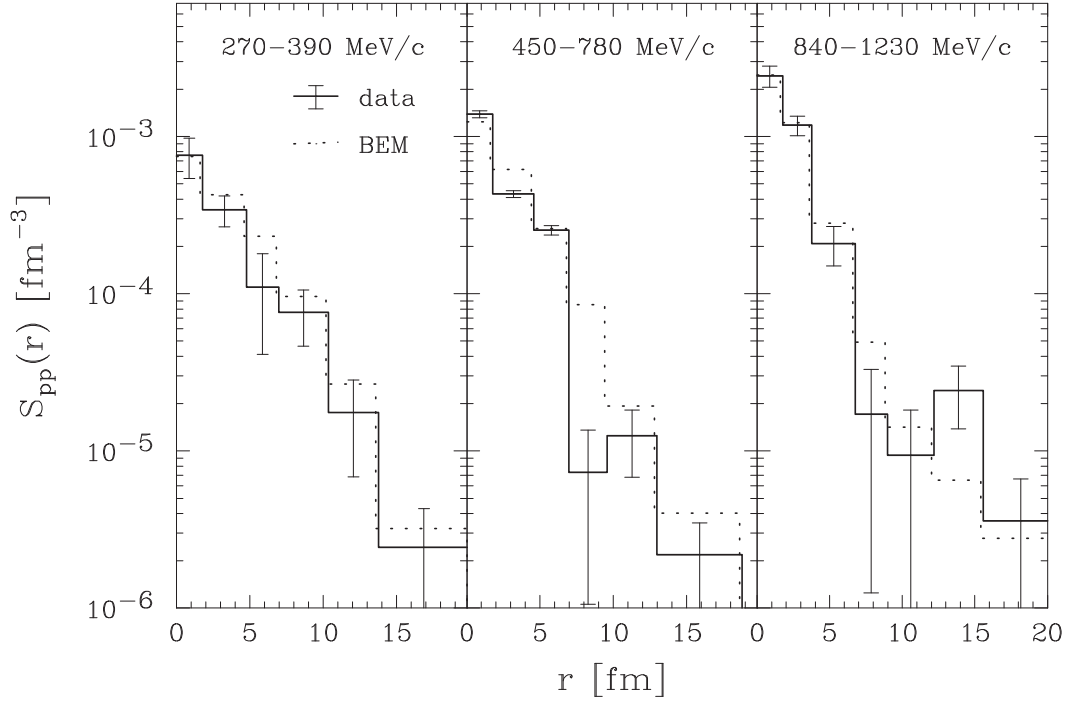}\end{center}
\caption[Relative proton source from the $^{14}$N + $^{27}$Al
	reaction at 75~MeV/nuc\-leon, in the vicinity of $\theta_{\rm
	lab} = 25^\circ$, in the three to\-tal mo\-men\-tum in\-ter\-vals
	of 270--390~MeV/c (left panel), 450--780~MeV/c
	(center panel), and 840--1230~MeV/c (right panel).  
	Solid lines are the source values extracted from the 
	data~\protect\cite{HBT:gon91a} and the dotted lines are the 
	source values obtained in the Boltzmann-equation calculation.]
	{Relative proton source from the $^{14}$N + $^{27}$Al
	reaction at 75~MeV/nucleon, in the vicinity of $\theta_{\rm
	lab} = 25^\circ$, in the three to\-tal mo\-men\-tum in\-ter\-vals
	of 270--390~MeV/c (left panel), 450--780~MeV/c
	(center panel), and 840--1230~MeV/c (right panel).  
	Solid lines are the source values extracted from the 
	data~\protect\cite{HBT:gon91a} and the dotted lines are the 
	source values obtained in the Boltzmann-equation calculation.}
\label{pps}
\end{figure}

The results of analyzing data using the Optimized
Discretization method are presented in Figure~\ref{pps}
The~angle and spin averaged kernel that we used to produce 
these images is~\cite{HBT:bro97}
\begin{equation}
   K_0 (q, r) = {1 \over 2} \sum_{j s \ell \ell'} (2 j +1) \,
   \left( g_{js}^{\ell \ell'} (r) \right)^2 -1 \, ,
\end{equation}
where $g_{js}^{\ell \ell'}$ is the radial wave function
with outgoing asymptotic angular momentum~$\ell$.
While we only show the source obtained using wavefunctions for the 
REID93 potential, the~values for the NIJM2~\cite{sto94} differ
only by a~fraction of $1/1000^{\rm th}$.  
Both sources imaged with and without constraints are similar; given their 
errors we only plot the constrained source in the figure.
The~results obtained without constraints are generally consistent with 
positive definite source functions.  

In~Figure~\ref{pps}, we compare the constrained results from the 
data to the distributions of relative separation of last collision 
points for protons with similar  momentum from the transport
model~\cite{HBT:bro97,tran:dan95}.  Clearly, the~semi-classical model
can only yield positive-definite source functions.  Again, we can see 
the focusing of the experimental distribution at low $r$ as the pair 
momentum increases.  The~large-$r$ tails in the distribution at different 
momenta cannot be accommodated with the Gaussian parameterizations used to
describe the low-$r$ behavior of the sources~\cite{HBT:gon90,HBT:gon91a}.

Generally (see Figure~\ref{pps}),  the~Boltzmann-equation model (BEM) yields 
relative emission point distributions that are similar to the imaged data, 
including the dependence on total pair momentum.  In~fact,  the~maximae 
around $q =20$~MeV/c (such as in Figure~\ref{simcor}) are nearly the 
same height as the data (see Figure~2 
in~\cite{HBT:gon93}).  Such findings are somewhat surprising
for the low and intermediate total momentum intervals.
While BEM adequately describes high-momentum wide-angle
single-particle spectra of protons, which correspond to
the highest total-momentum interval (see Figure~1 in~\cite{HBT:gon93}),
the~model overestimates the single-particle proton spectra by as much
as 1.5--5 in the two lower momentum intervals
\cite{HBT:gon93}.\footnote{For other comparisons of the transport theory to 
single-particle data from the same or similar reactions 
see~\cite{HBT:gon90,tran:dan91,HBT:han95}; overall proton multiplicities 
are typically overestimated by a~factor of~2, possibly due to excessive 
stopping within the semi-classical transport model in this energy range.}  
Looking closer at Figure~\ref{pps}, we find that the distributions
from data are somewhat sharper at low-$r$ in the two lower
total-momentum intervals than the the distributions from the model.
In the next section, we reveal a~serious
discrepancy when we go beyond a~point by point examination.

\begin{table}
\caption[Comparison of the integral of the relative pp source function, 
	$\lambda(r_N)$, for the restored and BEM sources 
	in three total momentum gates.]
	{Comparison of the integral of the relative pp source function, 
	$\lambda(r_N)$, for the restored and BEM sources 
	in three total momentum gates.
	The restored sources use the data of Reference \cite{HBT:gon90}.  
	The integrals are truncated at the distance $r_N$.}
\vspace*{\baselineskip}
\begin{center}
\begin{tabular}{|lr@{$\pm$}lr@{$\pm$}lcc|}
\hline
\multicolumn{1}{|c}{$P$-Range} & 
\multicolumn{5}{c}{$\lambda(r_N)$} &
\multicolumn{1}{c|}{$r_N$} \\ \cline{2-6}
\multicolumn{1}{|c}{[MeV/c]} &
\multicolumn{2}{c}{unconstrained} &
\multicolumn{2}{c}{constrained} &
\multicolumn{1}{c}{BEM} &
\multicolumn{1}{c|}{[fm]} \\ \hline
270-390  & 0.69  & 0.22  & 0.69  & 0.15  & 0.98 & 20.0 \\
450-780  & 0.560 & 0.065 & 0.574 & 0.053 & 0.91 & 18.8 \\
840-1230 & 0.65  & 0.37  & 0.87  & 0.14  & 0.88 & 20.8 \\
\hline
\end{tabular}
\end{center}
\label{lambdapp}
\end{table}

\subsection{Integral of Proton Source and Its Implications}

In Table \ref{lambdapp}, we have tabulated $\lambda(r_N)$ in each momentum
interval for the constrained and unconstrained sources as
well as sources from the BEM. Comparing the results of the BEM to the
constrained source, we only find agreement in the highest \sloppy\linebreak 
momentum interval.
It seems that, when compared to the model,  significant portions of the 
source are {\em missing} from the
imaged regions.  This  discrepancy is especially
pronounced in the intermediate-momentum region.
Nevertheless, it is comforting that the
transport model describes the features
of the relative source for the high momentum protons  
since it properly describes~\cite{HBT:gon93} the 
high-momentum single-particle spectra. 
Now, in~BEM no IMFs are
produced and the~IMFs may decay over an~extended time,
contributing to large separations in the relative emission function,
as they move away from the reaction region.
Of course these decays produce some final IMFs,
contributing to the relative IMF~sources at distances similar
to those for the pp~sources.  It~may be interesting to see 
whether a~significant portion of the relative
IMF sources in Section~\ref{sec:IMFstuff} 
can extend beyond $\sim 20$~fm, as is apparent for the pp sources.

The disagreements between the data and
calculations in both the values of $\lambda(r_N)$ and the
single-particle spectra~\cite{HBT:gon93}, 
for the lower momenta, reveal
unphysical features of low-momentum proton emission
in the transport model.  The coarse agreement between the
measured correlation function and the function
calculated using the model in Reference~\cite{HBT:gon93}
in the lower total-momentum intervals is coincidental.  
Since the images show $\lambda(r_N)<1$, some of the strength of $S$ is 
shifted out to large $r$ and a large source results in a correlation function 
with a sharper shape.   Thus, the BEM correlation function can match the 
height of the sharper
correlation peak, while not matching the shape of the peak.
For other systems in the
general energy range, disagreements were found even for the
height of~$C$~\cite{HBT:gon91a,HBT:han95,HBT:gaf95}.

\begin{figure}
\begin{center}\includegraphics[width=5.5in]{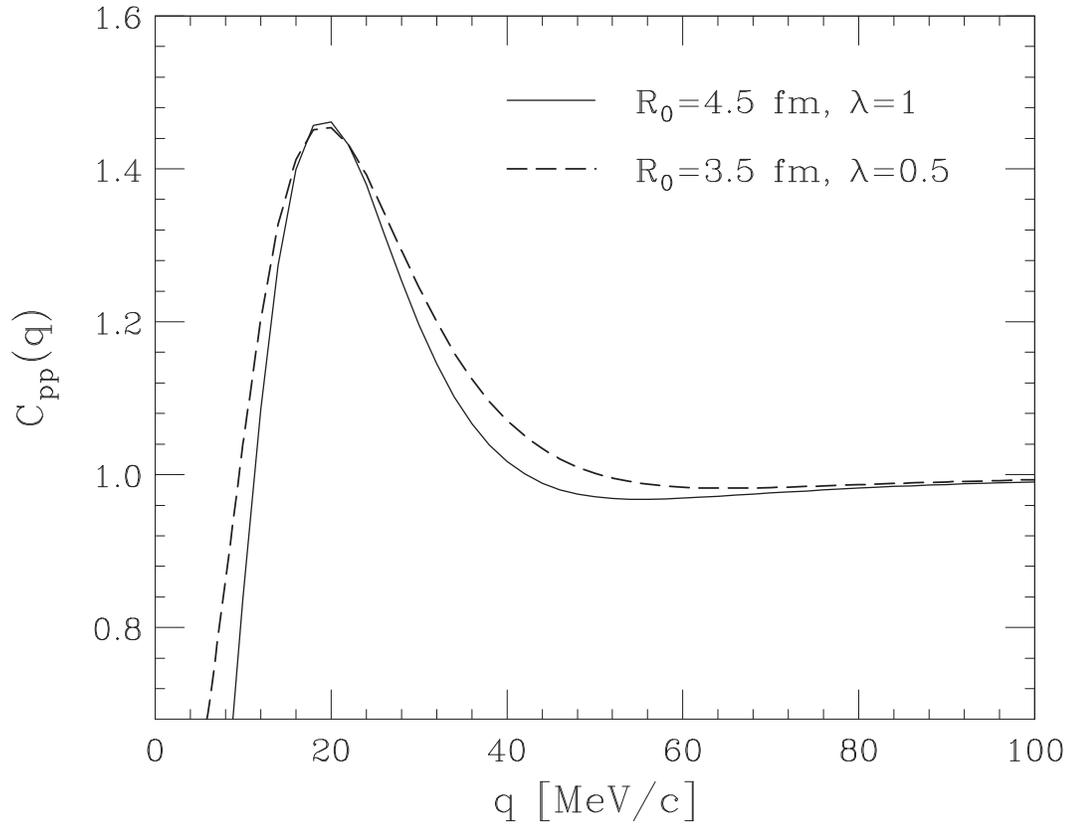}\end{center}
\caption[The two-proton correlation functions 
	for $R_0 = 4.5$~fm and $\lambda = 1$ and
	for $R_0 = 3.5$~fm and $\lambda = 0.5$.
	The source is the Gaussian in Equation~(\protect\ref{SGl}).]
	{The solid line is the two-proton correlation function 
	for $R_0 = 4.5$~fm and $\lambda = 1$ while the dashed line is 
	for $R_0 = 3.5$~fm and $\lambda = 0.5$.
	The source is the Gaussian in Equation~(\protect\ref{SGl}).}
\label{2cor}
\end{figure}

These conclusions make us question the sensibility of attempting to fit 
the \sloppy\linebreak magnitude of pp (or nn) correlation functions at the 
maximum \cite{HBT:boa90,HBT:gon90,HBT:gon91a} by adjusting
the radius $R_0$ while keeping $\lambda = 1$ in a
Gaussian parametrization of the
source function (such as in Equation~\eqref{SGl}).  
When one fits $\pi\pi$ correlation functions at intermediate relative momenta,
one varies {\em both} the strength and extent of the source function
at low~$r$: $\lambda$ is read off from the
magnitude of the correlation function at low~$q$ and $R_0$ is
read off from the width of the correlation function in~$q$.
However, because of the resonant nature of the low-momentum NN
interaction the~magnitude of NN
correlation functions are determined by the strength of the
source within the resonance peak of the wavefunction.
That amount is both effected by the strength of the source at low-$r$ 
{\em and} the low-$r$ source falloff.  This is
illustrated in Figure~\ref{2cor} which shows pp correlation
functions for the source in~\eqref{SGl}:  the~same maximum height
can be obtained using $R_0 = 4.5$~fm and $\lambda = 1$
as using $R_0 = 3.5$~fm and $\lambda=0.5$.

Given that the resonance peak in the $^1S_0$ wavefunction is quite narro 
(it has an outer radius of $\sim 2.5$~fm), and the source falloff cuts off
large-$r$ contributions to the integration in Equation~\eqref{eqn:PrattKoonin},
the low-$r$ limit of $S$ is proportional to the $C-1$ at the maximum
(to a $\pm 20 \%$ level) for virtually all low-$r$ falloffs that may be encountered in practice (namely, $R_0=2.5-6.0$~fm):
\begin{equation}
C_{\vec{P}} (20 \, {\rm MeV/c}) \approx  1 +  520\,{\rm fm^3} \,
S_{\vec{P}} (r \rightarrow 0).
\label{cps}
\end{equation}
For the~two sources in Figure~\ref{2cor}, we get 
about the same value of $S(r \rightarrow 0)$ and therefore 
about the same maximum height in~$C$.  
With the same maximum height, 
the~source falloff is reflected in the width of the
maximum in Figure~\ref{2cor}.

\subsection{Nucleon Freeze-Out Density, Proton Phase-Space Occupancy and 
the Entropy per Nucleon}

Now that we have discussed the pp data, let us see what we can learn
about the final, freeze-out, conditions of the reaction.  From the source,
we can estimate the nucleon freeze-out density, the average proton phase-space 
occupancy at freeze-out and entropy per nucleon.  First, we need to know 
which momentum gate best represents the average situation in nearly central 
collisions.  Transport calculations~\cite{tran:dan95} indicate that the 
measured~\cite{HBT:gon91a} coincidence cross sections for 
the $^{14}$N + $^{27}$Al reaction are dominated by nearly central
collisions with~$b \sim 2.8$~fm.  The chance of detecting two particles 
at a~wide angle simultaneously is large only for such collisions.
The~rms nucleon CM momentum in these collisions is $\sim
185$~MeV/c.  At~25$^\circ$ this corresponds to~$\sim
320$~MeV/c nucleon laboratory momentum, or $\sim 640$~MeV/c
total momentum for a~pair.  Thus, the~results for the
intermediate-momentum gate in~Figure~\ref{pps} best
represent the~average situation in central collisions.

We can estimate the freeze-out nucleon configuration-space density.
We start with Equation~\eqref{rs} and use the~assumptions of 
Section~\ref{sec:otherstuff}, together with the presumption that the 
relative spatial distributions of other particles to protons is similar 
to that between two protons.  We further assume the participants to have a 
total mass of 18 and to be in a fireball geometry with~$b \approx
2.8$~fm.  Given this, the proton relative spatial distribution gives 
the~average nuclear density in the vicinity of any emitted proton of 
$17 \times S(r \rightarrow 0) \simeq 17 \times 
0.0015$~fm$^{-3} = 0.025$~fm$^{-3} = 0.16\,
n_0$, where $n_0$ is the average nuclear density. The~directional 
space-momentum correlations due to 
collective motion, to shadowing, or to~emission that is most likely not 
instantaneous make this value actually an~upper limit on the freeze-out
configuration-space density.

We can also estimate the proton phase-space density at freeze-out and, from 
that, the entropy per nucleon at freeze-out.
References~\cite{HBT:gon91a,HBT:gon93} give the inclusive proton
cross-sections in the $^{14}$N + $^{27}$Al reaction, but only
at two angles and the cross sections include large contributions from
peripheral events.  Under these circumstances, we~use the~thermal
distribution $dN_{\rm th}/\dn{3}{p}\propto 1/(z^{-1}\,{\rm e}^{p^2/2mT}+1)$,
for the central events, in~formula~\eqref{f}.
Here~$z$ is set from the requirement of maximum entropy.  For $\sim 9$
participant protons at~$b = 2.8$~fm in the $^{14}$N + $^{27}$Al
reaction, that~requirement gives $z \sim 1.10$ and $T \approx 10.2$~MeV.
Use of the thermal momentum distribution for the $^{14}$N + $^{27}$Al
reaction in Equations~\eqref{fave} and~\eqref{entropy}
yields $\langle f \rangle \approx 0.23$ and $S/A
\approx 2.7$.  For a~distribution with non-equilibrium
features, these values should represent the~lower limit on the
average occupation and the upper limit on the entropy.  Indeed, 
when applied to the transport model, using a~thermal distribution
yields an~entropy about 0.5 per nucleon higher than
the entropy calculated directly within the model.

\section{Intermediate Mass Fragments}
\label{sec:IMFstuff}

We now turn to the~analysis of IMF sources.  We~choose the correlation 
data of Hamilton et al. \cite{HBT:ham96},  from central 
$^{84}$Kr + $^{197}$Au reactions at 35, 55, and 70~MeV/nucleon, because 
these data give us the opportunity to examine the variation of sources 
with beam energy.

Hamilton et al. collected pairs in the angular range of 
$25^\circ < \theta_{\rm lab} < 50^\circ$ in order to limit 
contributions from target-like residues.
They tabulate the correlation functions in terms of the reduced velocity
\begin{equation}
	v_{\rm red} = {v \over (Z_1 + Z_2)^{1/2}} \, ,
\end{equation}
under the assumptions that the pair Coulomb correlation dominates the 
fragment correlation and that the fragments were approximately symmetric,
$Z/A \approx 1/2$.   Under these assumptions and the additional 
assumption that three body effects can be neglected, the kernel is
\begin{equation}
   K_0 (q,r) = \theta(r - r_c) \,\left( 1 - r_c/r \right)^{1/2} - 1 \, ,
\label{KCou}
\end{equation}
The~distance of closest approach in~\eqref{KCou} for symmetric fragments is 
approximately
\begin{equation}
   r_c = {2 \, Z_1 \, Z_2 \, (A_1 + A_2) \, e^2 \over A_1 \, A_2
   \, m_N \, v^2} \approx {e^2 \over m_N \, v_{\rm red}^2} \, .
\label{rc}
\end{equation}

\begin{figure}
\begin{center}\includegraphics[width=1.1\HBTfigwidth]{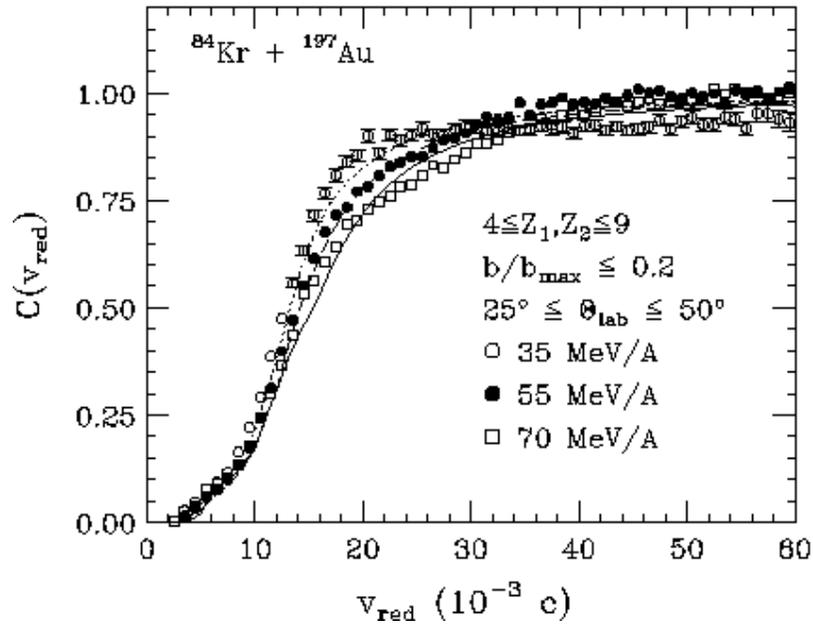}\end{center}
\caption[Fragment-fragment velocity correlation function in central
	$^{84}$Kr + $^{197}$Au reactions.  The symbols show the data of 
	Reference \cite{HBT:ham96} and the lines show the
	imaged source function.  The 35~MeV/nucleon data is represented by 
	the open circles and dotted line, the 55~MeV/nucleon data by
	solid circles and dashed line, and 70~MeV/nucleon by
	open squares and solid line.]
	{Fragment-fragment velocity correlation function in central
	$^{84}$Kr + $^{197}$Au reactions.  The symbols show the data of 
	Reference \cite{HBT:ham96} and the lines show the
	imaged source function.  The 35~MeV/nucleon data is represented by 
	the open circles and dotted line, the 55~MeV/nucleon data by
	solid circles and dashed line, and 70~MeV/nucleon by
	open squares and solid line.}
\label{corIMF}
\end{figure}

The correlation functions at the three beam energies are shown
in Figure~\ref{corIMF}.   We have reduced the normalization of the 
correlation function at 35~MeV/nucleon by~5\%, compared 
to~\cite{HBT:ham96} to better satisfy the condition that $C \rightarrow
1$ at large $v_{\rm red}$ (this also allows 
$C \approx 1$ for $v_{\rm red} = (0.05-0.08)c$, a requirement of
the authors in~\cite{HBT:ham96}).  When we examine 
Equations~\eqref{eqn:PrattKoonin},\eqref{KCou}, \eqref{rc} and Figure~\ref{corIMF}, 
an issue becomes apparent:  only the region of the source 
with $r > r_c$ contributes to the correlation function at a given 
$v_{\rm red}$.  As~$v_{\rm red}$ increases from~0, the~distance of
closest approach~$r_c$ decreases, with more and more inner
regions of the source~$S$ contributing to~$C$.  The~low-$v_{\rm red}$ 
correlation functions at the three beam energies in
Figure~\ref{corIMF} are quite similar, suggesting that the tails
of the source functions are similar.  Differences occur at
higher $v_{\rm red}$, indicating differences in the inner
regions of the source.

To image of the IMF sources, we optimize \rj as in the pp
case, but we add the constraint $r_1 \ge r_1^{\rm min}$.  We do this
because the Coulomb interaction in (\ref{eqn:PrattKoonin}) does not dominate
when the measured fragments are in close contact.
The Coulomb correlation alone cannot be relied upon to get information 
on the most inner portion of the source.  The~typical touching distance 
for the fragments measured in~\cite{HBT:ham96} is $r_t \sim 5$~fm; we 
chose a minimum imaging distance $r_1^{\rm min} = 7.0$~fm which ensures that 
there is more volume in the lowest bin outside $r_t$, than inside~$r_t$.

\begin{figure}
\begin{center}\includegraphics[width=\HBTfigwidth]{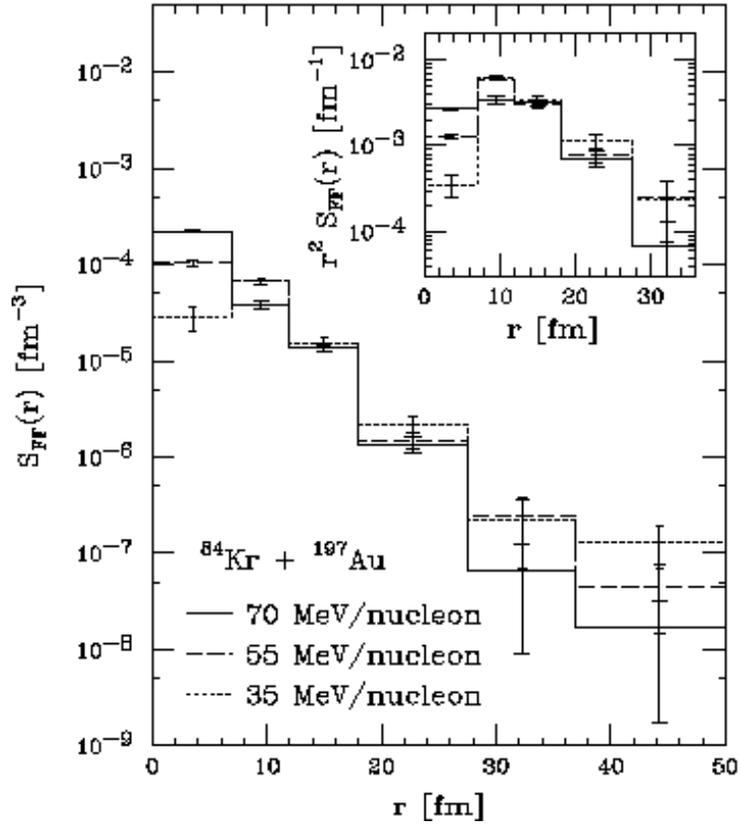}\end{center}
\caption[Relative source for IMFs emitted from central
	$^{84}$Kr + $^{197}$Au reactions from the data of
	Reference \cite{HBT:ham96} at 35 (dotted line), 55~(dashed line), and 
	70~MeV/nucleon (solid line).  The insert shows the source
	multiplied by~$r^2$.]
	{Relative source for IMFs emitted from central
	$^{84}$Kr + $^{197}$Au reactions from the data of
	Reference \cite{HBT:ham96} at 35 (dotted line), 55~(dashed line), and
	70~MeV/nucleon (solid line).  The insert shows the source
	multiplied by~$r^2$.}
\label{sorf}
\end{figure}

The results from our imaging are shown in Figure~\ref{sorf}.
Given the errors in the figure, the~tails
of the sources at the three energies are not very different.
However, we observe significant variation with energy at
short relative distances ($r < 12$~fm) with the source undergoing 
a~larger change between 55 and
75~MeV/nucleon, than between 35 and 55~MeV/nucleon.

In Table \ref{lambdaIMF} we tabulate the generalized chaoticity parameter for 
the IMF sources with $r_N=20$~fm and over the whole image.  The $\lambda_{\rm TOT}$
for all three energies are all consistent with one, within errors.  Since the 
$\lambda(20 {\rm fm})$ are all 20\%-30\% lower than $\lambda_{\rm TOT}$, we see 
that a large part of IMF emission occurs at distances that are not imaged with 
the protons.  Nevertheless, the $\lambda(20 {\rm fm})$ for the 
the~low- and high-momentum pp sources in section
\ref{adata} are roughly equal to  the IMF $\lambda(20 {\rm fm})$ but the 
the intermediate-momentum pp $\lambda(20 {\rm fm})$ is lower.  
It~should be mentioned that no complete quantitative
agreement should be expected, even if the data were from the
same reaction and pertained to the same particle-velocity
range.  This is because more protons than IMFs can stem from secondary
decays.  Besides, the~velocity gained by a~proton in a~typical decay is large
compared to the relevant relative velocities in pp correlations, but the
velocity gained by an~IMF can be quite small on the scale of
velocities relevant for the IMF Coulomb correlations.
Thus, the~IMF correlations may reflect the primary parent sources, which 
are concentrated around the origin, rather than the final sources.

The tails of the IMF sources extend so far that they
must be associated with the time extension of
emission.  Even so, it~is interesting to ask how far into the 
center of the source must we go to see where the effects due
to the spatial extent of the primary source.
In~\cite{HBT:ham96}, the combinations of single-particle source radii and 
lifetimes that gave acceptable descriptions of their data 
have radii varying between 5 and 12~fm.  In general, the~combination of 
spatial extent and lifetime effects should give rise to a~bone-like
shape of the relative source, with the source elongation being due to 
the emission lifetime.  With this, one
could try to separate the temporal and spatial effects using
a~three-dimensional source restoration.  In~the angle-averaged source, 
the~part dominated by lifetime effects should fall off as an~exponential 
divided by the square of the separation, $r^2$, as a function of relative 
separation $r$.  On the other hand, the~part of the relative
source dominated by spatial effects may fall off at a~slower pace or 
even be constant.  In~Figure~\ref{sorf}, we see
that the~sources change weakly with~$r$ at 35 and
55~MeV/nucleon and faster at 70~MeV/nucleon,   
within the range where sources vary with energy ($r \lesssim 12$~fm).
For reference, in the insert to Figure~\ref{sorf}
we show the IMF source multiplied
by~$r^2$.  We see an~edge at $r \sim 11$~fm at 35 and at
55~MeV/nucleon which disappears at 70~MeV/nucleon.  This is
consistent with an~emission from the spatial region of a~radius
$R \sim 11/\sqrt{2} \sim 8$~fm that becomes more diffuse, and
possibly spreads out, with the increase in energy.
The~disappearance of the sharply
pronounced spatial region at 70~MeV/nucleon agrees with the general
expectation on the IMF production in central symmetric
collisions or from central sources in asymmetric collisions
that the IMF yields maximize towards 100~MeV/nucleon~\cite{lyn98}.

\begin{table}
\caption[Comparison of the integrals of the IMF source function, 
	$\lambda(r_N)$, for different truncation points, $r_N$,
	in three total momentum gates.
	The restored sources use the data of Reference \cite{HBT:ham96}.]
	{Comparison of the integrals of the IMF source function, 
	$\lambda(r_N)$, for different truncation points, $r_N$,
	in three total momentum gates.
	The restored sources use the data of Reference \cite{HBT:ham96}.}
\vspace*{\baselineskip}
\begin{center}
\begin{tabular}{|lr@{$\pm$}lr@{$\pm$}l|}
\hline
\multicolumn{1}{|c}{beam energy} & 
\multicolumn{2}{c}{$\lambda_{\rm TOT}$} &
\multicolumn{2}{c|}{$\lambda(20 {\rm fm})$} \\ 
\multicolumn{1}{|c}{[MeV/nucleon]} &
\multicolumn{2}{c}{ } &
\multicolumn{2}{c|}{ }  \\ \hline
35  & 0.96 & 0.07 & 0.72 & 0.04  \\
55  & 0.97 & 0.06 & 0.78 & 0.03  \\
70  & 0.99 & 0.05 & 0.79 & 0.03  \\
\hline
\end{tabular}
\end{center}
\label{lambdaIMF}
\end{table}

\section{Summary}

Imaging the source functions from two particle correlation data is 
possible and we have introduced several methods for doing the inversion.
Our most successfull methods use specific features of the kernel 
or data to aid in the inversion.  For example, in  
the method of Optimized Discretization, we use the behavior of the kernel and
the error on the correlation function to 
adjust the resolution to minimize the relative errors of the source.  
The~fact that we can actually estimate errors in this method
gives our method a~significant advantage over the Maximum Entropy Method.  
We tested this method by restoring assumed {\em compact} pp sources and found
that the~quality of the restored source is comparable to the restored
source we obtain by imposing the constraints of positivity and normalization.  
Imposing these constraints in our new method further reduces 
the source errors, but is no longer required.  
This method allows one to study the {\em long-range} source structure 
by adjusting the overall size of the imaged region and
the resolution at large distances.

The robustness of our imaging method gave us the capability to search 
for features of the sources that can not be seen in any other analysis.
For example, we can look for the resonance contribution to 
pion sources in \cite{HBT:mis96} or check the positive 
definiteness of proton Wigner sources in the $^{14}$N + $^{27}$Al 
reaction at 75~MeV/nucleon \cite{HBT:gon90,HBT:gon91a}.  
In~principle, unraveling the quantal negative values of a~Wigner
function is not far fetched.  In fact, negative values of Wigner functions
have been observed in interfering atomic beams~\cite{Wig:kur97}.
Admittedly, if we had discovered such values in the heavy-ion reactions,
we would first have to check for a~possible breakdown of the assumptions
leading to~(\ref{eqn:PrattKoonin}) and on systematic errors in data,   
before concluding on a~success.  The~extensive averaging in
the reactions\footnote{The averaging includes averaging over the impact 
parameter, the~central position for the source and emission times 
($\vec{R}$, and $t_1$ and $t_2$, respectively, in Equation~(2) 
in~\cite{HBT:bro97}), and the total pair momentum~$\vec{P}$.} makes it
unlikely that a  genuinly quantal oscillation in the source
function would  survive except in the very tail of the function.

Our analysis of pion and kaon source in Au-Au collisions at $\sim 11$~GeV/A 
show that the kaons stem from a much smaller source region than the pions.  This
is likely to be due to a combination of emission time and source size effects.
Both the pions and the kaons are produced directly and from the decay of short
lived resonances, so their sources should be dominated by the shape of emitting
region.  However, because the pions are lighter and couple more strongly to 
the system, they are subject to various dynamic effects that can extend their
source function relative to the kaon source function.
When examining the generalized chaoticity parameters of the sources, we find
the kaons have a $\lambda$ consistant with one, indicating that we have imaged
the entire source.  We also find that the pion $\lambda$ is {\em not} consistant
with one, indicating that a large fraction of pions are emitted at large 
distances ($\gtrsim 35 {\rm fm}$).  This large distance emission is likely due 
to contributions from the decay from long-lived resonances such as the $\eta$
and $\eta'$.

We found that our imaged proton and IMF sources change significantly with 
the total pair momentum, becoming sharpest for the largest momenta in the CM.
Significant portions of the imaged proton source are missing from the
imaged region\footnote{The imaged region corresponds to relative 
distances with $r< 21$~fm.} at typical participant momenta in the CM, but 
not at the highest momenta.  The chaoticity parameter (the integral of the 
source) from the Boltzmann-equation model \cite{tran:dan91,tran:dan95} 
agrees with the data, at the highest momenta but the integral is close to 
one in the participant and target-emission momenta. Nevertheless, the model 
yields the correct height of the maximae of the correlation 
functions \cite{HBT:gon93}.  This is because the right combination of  
source normalization and sharpness in the~model can yield the right 
value of~$S$ at short separations, $S_{\vec{P}} (r \rightarrow 0)$,
and this primarily determines the height of~$C$.  Gaussian-source fits to 
the height of the pp correlation function
\cite{HBT:boa90,HBT:gon90,HBT:gon91a} are of a limited value because
considerable source strength may lie at large relative separations.

In our analysis of midrapidity IMF sources in central $^{84}$Kr
+ $^{197}$Au reactions at different beam energies, we found
a~significant variation of the sources with energy at short
distances, but not at large distances.  Considerable portions
of the IMF sources extend to large distances ($r > 20$~fm) just 
like the lower total-momentum pp sources.  It~would be very 
interesting to image both the IMF and pp sources in one reaction.  
There is a deficiency of our fragmentation analysis method, namely
our method lacks three-body Coulomb effects.  When weak, these 
effects could be included as a first-order perturbation.

While we have made significant progress in the inversion of angle-averaged 
correlation functions, we have much work to do in the area of inverting 
full three dimensional data.   Unravelling the emission time effects from the 
three-dimensional sources would be possible because the
three-dimensional sources would be distorded in the direction of the pair total
momentum.  In fact, in the case where lifetime effects dominate the 
source, we would expect the source to be nearly bone-like in shape 
\cite{HBT:bro97}.   In any event, the Optimized Discretization method 
should prove helpful in inverting three-dimensional data.
Finally, the strategy of letting the errors and the kernel choose what 
source they can image is novel not just for the problem of
inverting correlations but the inversion problem in general~\cite{pre92}.

\chapter{CONCLUSIONS}

Do we fully understand the space-time development of a heavy-ion collision?
Definitely not, but the two sets of techniques that we study in this thesis, 
transport models and nuclear imaging, can play an important role in 
deducing this space-time evolution at the next generation of
nuclear colliders at Brookhaven and CERN.
Transport models give us access to the space-time development
of collisions by giving the phase-space densities of particles as their output.
We studied the application of these models to the massless partons in
RHIC collisions.  Crucial input to a parton transport model are the initial 
parton densities.  Since it is desirable to be able to connect these 
phase-space densities to the experimentally determined Parton Distribution 
Functions, we studied some of the issues associated with constructing a 
phase-space Parton Model.
In contrast to this theoretical effort, nuclear imaging gives us direct 
access to the space-time development by allowing us to reconstruct the 
two-particle relative emission distribution from experimental data.
Both of these studies are still in their infancy but have yielded many 
important results.

\section{Transport Theory for Partons}

Transport models have aided immensely in understanding the space-time 
evolution of reactions at intermediate energies and we can only hope
that they will again aid us as we move into the realm of ultra-relativistic
nuclear collisions at RHIC and the LHC.  For this to happen, we must have 
a transport theory applicable to massless {\em partons}.  We have taken 
several steps to this goal by providing a scheme that does not rely on either 
the Quasi-Particle or Quasi-Classical approximations.  
Instead, the scheme rests upon the phase-space Generalized-Fluctuation 
Dissipation Theorem which states that the phase-space densities are 
convolutions of the phase-space source of the particle with the phase-space 
propagator.  We illustrated several ways to calculate the densities in this
new framework.  These ways include a coupling 
constant expansion and solving semi-classical transport equations.  In both
cases, because we have not made either the QCA or QPA we can make controlled
approximations based on the strengths of couplings or the size of
densities.

In the future, we can hope to advance this approach in several ways.  
The first job is to understand how renormalization will work in phase-space
in particular and non-equilibrium field theory in general.  Will we need
to dress the vertices and the phase-space propagators or will dressing the 
densities and letting the masses and couplings run be enough?  If we are
to renormalize the masses, then we need analytic results for the phase-space
propagators for particles with mass.  We will also need this propagator
if we are to discover if, where and how standard semi-classical transport 
theory breaks down for lower
energy collisions.  Also along this track, it would be very useful to 
understand the interplay and limits of the QCA and QPA in this phase-space
approach.  Finally, we would like to investigate what it would take to 
implement bound states in this phase-space approach.

\section{Parton Model in Phase-space}

A transport model for RHIC or LHC nuclear collisions will need input 
phase-space parton distributions.  We would like to be able to constrain 
these input distributions using the momentum-space Parton Distribution 
Functions.  Given this, it seems reasonable to wonder if we can rewrite the
entire Parton Model in phase-space.  As a first step toward this, we examine
whether two key tenets of Parton model, namely factorization and evolution,
will work in phase-space.  We demonstrated the concept of factorization
by deriving the QED analog of the parton model, the Weizs\"acker-Williams 
method, in phase-space.  We also demonstrated that parton ladder diagrams 
can be evaluated in phase-space.  This is useful because these ladders 
are the basis of the 
Leading Logarithm Approximation which is, in turn, equivalent to  
evolution in the renormalization group improved parton model.  The ladder 
diagram we considered was a simple QED ladder consisting of a point charge 
radiating a photon which subsequently splits into an on-shell final-state 
positron and virtual electron.  Despite its simplicity, 
it contains many of the features
that we expect from a full-blown QCD parton ladder such as both $1/x_F$ and
$1/Q^2$ type singularities.  Furthermore, investigation of this simple ladder
shows us how a realistic source distribution shapes the particle phase-space
densities.  Armed with this understanding, we investigated the size and 
shape of the parton cloud of a nucleon.  We found that the sea parton 
distributions are roughly the same size as the valence quark bag in the 
transverse direction and ranged from marginally larger to vastly larger than 
the valence quark bag in the longitudinal direction.


In the future, we will need quantitative calculations of the parton 
phase-space densities
for input into a transport model.  This means a couple of things:  this
S-matrix based approach must be connected with the time-ordered non-equilibrium
approach and we must implement renormalization in the time-ordered
non-equilibrium approach.  Some of this work has already been done by Makhlin
and Surdutovich \cite{neq:mak95a,neq:mak95b,neq:mak96,neq:mak98} 
however it must be reworked in phase-space.

\section{Nuclear Imaging}

Directly accessing the space-time development of nuclear reactions
through intensity interferometry and nuclear imaging is possible.  Results from
this approach go far beyond standard Gaussian fits by reconstructing the 
entire two-particle source functions.  
We found several methods to perform nuclear imaging and the best ways rely on 
specific features of the source, the kernel, or the data itself.  
In fact, the method of
Optimized Discretization uses the behavior of the kernel and the error on 
the correlation function data to choose the best resolution for the source.
This method is further improved by using the constraints that the source is
known to obey.  We applied the inversion methods to several data sets 
including $K^+K^+$, $\pi^-\pi^-$, pp and IMF correlations.
An important quantity derived from these sources is the generalized chaoticity
parameter which can be used to characterize the amount of the source that 
lies within, or outside of, the imaged region.  The source functions 
provides other information, namely the freeze-out density and the entropy per 
nucleon.  In total the images provide a great deal more information than 
previous Gaussian fits and related approaches.

We would like to extend our imaging to invert three-dimensional correlation
pion and proton data sets.  Such results would be immensely useful in 
separating the temporal and directional dependence of the sources.  Of course
there is still much to do for the angle-averaged inversions.  We are in the
process of analyzing other pion and proton data sets at different energies
and with higher resolution.  Furthermore, we are developing a set of 
computer programs that we hope will become generally available for nuclear 
imaging. 

\section{Final Remarks}

The nuclear physics community has made great strides in understanding the
space-time development of nuclear collisions.  Nevertheless, our
understanding will be tested when we begin to see results from 
RHIC and the LHC.  Is there a quark-gluon plasma?  How will it evolve?
Our studies of parton transport theory and nuclear imaging should help answer 
these and many more questions.

\coversheet{APPENDICES}
\appendix

\chapter{PHASE-SPACE PROPAGATORS}
\label{append:prop}

Quantum particles do not propagate between two space-time points 
by traveling along straight line classical trajectories.
Instead, a particle with fixed 4-momentum can propagate from a point to
anywhere in some space-time region defined by its 4-momentum.  
The propagation is controlled by the phase-space 
propagators and, in this appendix, we study the retarded and 
Feynman phase-space propagators in a detail not possible in the main text.
Here, we state our phase-space propagators for scalar particles,
discuss the symmetries of these propagators, detail both how the retarded and
Feynman propagators work, give the derivation of their analytic expressions
and, finally, discuss what is needed to treat particles with nonzero mass.
The Feynman propagator for particles with nonzero mass has already been 
discussed by Remler \cite{tran:rem90} so our discussion here is brief.  
The Dirac and vector propagators differ
from the scalar propagators by the addition of spin projectors 
so we do not need to discuss them separately.

We define a phase-space propagator as the Wigner transform of two 
translationally invariant propagators
(such as in Equation \eqref{eqn:WigPropinGFDThm} of Chapter 
\ref{chap:transport} or Equation \eqref{eqn:PhotProp} of Chapter 
\ref{chap:pips}):
\begin{align}
\daveseqn
   \Gprop{x}{p} 
   = & \displaystyle\int\dnpi{4}{p'} e^{-i x\cdot p'} G(p+p'/2)
      G^{\dagger}(p-p'/2) \label{eqn:propdefa}\davetag{a}\\ 
   = & \displaystyle\int\dn{4}{x'} e^{i x'\cdot p} G(x+x'/2)
      G^{\dagger}(x-x'/2) \label{eqn:propdefb}\davetag{b}
\end{align}
The vacuum propagators in momentum space are \cite{gFT:bog79}:
\begin{align}
\daveseqn
G^{\pm}(p)=&-\left( p^2-m^2\pm i\epsilon p_0\right)^{-1}
	\label{eqn:Gpma}\davetag{a}\\
G^{\stackrel{c}{a}}(p)=&-\left( p^2-m^2\pm i\epsilon \right)^{-1}
	\label{eqn:Gpmb}\davetag{b}
\end{align}
In configuration space, the propagators are \cite{gFT:bog79}:
\begin{align}
\daveseqn
G^{\pm}(p)=&\frac{1}{2\pi}\theta {(\pm x_0)} \deltaftn{}{x^2}
	\label{eqn:Gpmaa}\davetag{a}\\
G^{\stackrel{c}{a}}(p)=&\frac{1}{4\pi}\left(\deltaftn{}{x^2}
	\mp{\cal P}\frac{i}{\pi x^2} \right)
	\label{eqn:Gpmbb}\davetag{b}
\end{align}
As one can see from the configuration space listing, the time ordering is 
explicit in the retarded propagator while there is no time ordering in the 
Feynman propagators.  This feature is preserved in the phase-space propagators
even after the Wigner transforms in \eqref{eqn:propdefa}--\eqref{eqn:propdefb}.

\section{The Propagators for Zero Mass Scalar Particles}

We now present the Feynman and retarded propagators for particles with 
zero mass.  The advanced
and anti-Feynman propagators can be recovered using the symmetry
relations discussed in the next section.
The propagators are:
\begin{align}
\daveseqn
   \begin{split}
   \Gcaus{x}{p}=&
        \frac{1}{4\pi} \left[ \mbox{ sgn} (x^2) + 
        \mbox{ sgn} (p^2) + 2 \mbox{ sgn} (x\cdot p)\right]\\
   &\times\displaystyle\left\{ \theta (\lambda^2)
        \frac{\sin{(2\sqrt{\lambda^2})}}{\sqrt{\lambda^2}} -
        \theta (-\lambda^2) \frac{\exp{(-2\sqrt{-\lambda^2})}}
        {\sqrt{-\lambda^2}}\right\}
   \end{split}\label{eqn:Gcausdef}\davetag{a}\\
   \Gplus{x}{p}=&\frac{1}{\pi}\theta (x_0)\theta (x^2) \theta (\lambda^2) 
      \frac{\sin{(2\sqrt{\lambda^2})}}{\sqrt{\lambda^2}} 
      \label{eqn:Gretdef}\davetag{b}
\end{align}
Here the Lorentz invariant $\lambda^2$ is given by 
$\lambda^2=(x\cdot p)^2-x^2 p^2$.  
The retarded phase-space propagator is used throughout 
Chapters \ref{chap:transport} and \ref{chap:pips} while the Feynman phase-space
propagator is mentioned only in Chapter \ref{chap:pips}.
We discuss how the propagators 
work in the Sections \ref{sec:retWorks} and \ref{sec:feynWorks}
and derive the analytic form of each of the propagators 
in the Sections \ref{sec:retDerivation} and \ref{sec:feynDerivation}.

\section{Symmetries of the Propagators}

Here we list how the phase-space propagators behave under several 
coordinate transforms.  Some of these relations are quite useful because
they relate several of the propagators together through simple coordinate
reflections.

A time reversal transform in coordinate space is equivalent to a 
simultaneous reflection in time and energy in phase-space.  Under time
reversal the $+$ and $-$  propagators 
change into one another while the Feynman and anti-Feynman propagators
remain unchanged:
\begin{align}
\daveseqn
   \Gplus{x_{0},\vec{x}} {p_{0},\vec{p}}=&\Gmin {-x_{0},\vec{x}}
      {-p_{0},\vec{p}}\davetag{a}\\
   G^{\stackrel{a}{c}} (x_{0},\vec{x},p_{0},\vec{p}) =&
      G^{\stackrel{a}{c}} (-x_{0},\vec{x},-p_{0},\vec{p}).\davetag{b}
\end{align}

A parity transform in coordinate space is equivalent to a simultaneous
reflection in a space coordinate and the corresponding momentum coordinate.
Under a parity transformation, all of the propagators remain unchanged:
\begin{align}
\daveseqn
   G^{\pm} (x_{0},\vec{x},p_{0},\vec{p}) =&
      G^{\pm} (x_{0},-\vec{x},p_{0},-\vec{p})\\
   G^{\stackrel{a}{c}} (x_{0},\vec{x},p_{0},\vec{p}) =&
      G^{\stackrel{a}{c}} (x_{0},-\vec{x},p_{0},-\vec{p}).
\end{align}

The Feynman propagators have another (rather amusing) {\em argument
switching} symmetry.  Here all the space-time components are switched with the
the corresponding momentum-energy components:
\begin{equation}
   G^{\stackrel{a}{c}}(x,p)=G^{\stackrel{a}{c}}(p,x)
\end{equation}

Finally, the Feynman and anti-Feynman propagator are related through
a  complete reflection of all of the space or momentum coordinates:
\begin{align}
\daveseqn
   \Gcaus{x}{p}=&\Gacaus{-x}{p}\davetag{a}\\
   \Gcaus{x}{p}=&\Gacaus{x}{-p}.\davetag{b}
\end{align}

\section{Detail: the Retarded Propagator}
\label{sec:retprop}

The phase-space retarded propagator naturally arose in our discussion of 
the phase-space Generalized Fluctuation-Dissipation theorem in 
Chapter \ref{chap:transport} and in the discussion of time-ordered 
amplitudes in Chapter \ref{chap:pips}.  In this section we will describe 
how the scalar propagator works and how we derive the analytic expression
for it at lowest order.  A simplified version of the discussion of how 
the propagator functions is contained in Section \ref{sec:simpleprop}.

\subsection{How it Works}
\label{sec:retWorks}

The Wigner transform of the retarded propagator gives the weight for 
a particle with four-momentum $q_\mu$ to propagate across the space-time 
separation $\Delta x_\mu = x_\mu-y_\mu$.  Without loss of generality, let 
us give the particle a the momentum $q_\mu=(q_0,q_L,\vec{0})$ and estimate
how far the retarded propagator can send the particle.  
First, the retarded propagator has two theta functions, one that enforces 
causality and one that forces propagation inside the light cone.  
The rest of the interesting features of the propagator are tied up in the 
dependence on $\lambda^2$.  Since 
$\Gplus{\Delta x}{q}\propto \theta (\lambda^2) 
\sin{(\sqrt{\lambda^2})}/{\sqrt{\lambda^2}}$, 
the particle does not propagate much farther than the inequalities
$0\le \sqrt{\lambda^2} \lesssim 1$ allow.  To see what these constraints mean, 
we investigate the $q^2>0$, $q^2<0$, and $q^2=0$ cases separately.  

To study the $q^2>0$ case, we position ourselves in the frame where 
$q'_\mu=(q'_0,\vec{0})$.
In this frame, the $\lambda^2$ constraint translates into a restriction on the 
spatial distance a particle can propagate:
\[
   0\le {q'_0}^2 |\Delta \vec{x}'|^2 \lesssim 1.
\]
Combined with the light-cone constraint, $\Delta\vec{x}'$ is constrained to 
\begin{equation}
   |\Delta\vec{x}'| \lesssim \left\{ 
   \begin{array}{cl}
      \Delta x'_0 & \mbox{for small} \; \Delta x'_0 < 1/|q'_0| \\
      1/|q'_0| & \mbox{for large} \; \Delta x'_0 > 1/|q'_0|.
   \end{array}
   \right.
\end{equation} 
To find a cutoff for $\Delta x'_0$, we realize that, for a given $q'_\mu$, the
propagator gives the ``probability'' distribution for propagating across the
space-time displacement $\Delta x'_\mu$.  Thus, we can integrate 
$\Gplus{\Delta x'}{q'}$ over all space and over time up to some cutoff time
$\tau$, giving us the total ``probability'' for propagating to time $\tau$.  
We find that the propagation probability becomes unimportant for $\tau\gtrsim
1/|q'_0|$ and this sets a cutoff in $\Delta x'_0$: 
\begin{equation}
        \Delta x'_0 \lesssim 1/|q'_0|. 
\label{eqn:constraintsa}
\end{equation}
Together, these three constraints define the space-time region where the 
particle can propagate.
When we move back to the frame with 
$q_\mu=(q_0,q_L,\vec{0}_T)$, the region contracts in the temporal and
longitudinal directions.  From Equation~(\ref{eqn:constraintsa}), the 
limits of the propagation region are 
\begin{align}
\daveseqn
        |\Delta\vec{x}_T|\lesssim& R_\perp=\frac{1}{\sqrt{|q^2|}}\davetag{a}\\
        |\Delta x_L|\lesssim& R_\|=\displaystyle\frac{1}{|q_L|}\davetag{b}\\
        |\Delta x_0|\lesssim& R_0=\displaystyle\frac{1}{|q_0|}\davetag{c}
\end{align}

We study the $q^2<0$ case in a similar manner.  In the frame with
$q'_\mu=(0,q'_L,\vec{0}_T)$, the $\lambda^2$  constraint implies 
\[
        0\le {q'}_L^2 (\Delta {x'}_0^2-\Delta {x'}_T^2)\lesssim 1.
\]
Combining this with the light-cone constraint immediately gives us a limit on
$\Delta x'_L$:
\begin{equation}
\daveseqn
        |\Delta x'_L|\lesssim \frac{1}{|q'_L|}.\davetag{a}
\end{equation}
As with the $q^2>0$ case, we can integrate the propagator to find the total
``probability'' for propagating to the time $\tau$.  In this case, the
propagation probability is important only for  $\tau\lesssim 1/|q'_L|$, giving
us a limit in $\Delta x'_0$ of:
\begin{equation}
        |\Delta x'_0|\lesssim \frac{1}{|q'_L|}.\davetag{b}
\end{equation} 
The limit on $|\Delta \vec{x}'_T|$ then follows directly from the light-cone
constraint:
\begin{equation}
        |\Delta \vec{x}'_T|\lesssim \frac{1}{|q'_L|}.\davetag{c}
\end{equation} 
Again, these constraints define a the space-time region where the particle can
propagate.  
Boosting back to the frame with $q_\mu=(q_0,q_L,\vec{0}_T)$, again the
longitudinal and temporal spread gets Lorentz contracted:
\begin{align}
\daveseqn
        |\Delta \vec{x}_T | \lesssim& R_\perp =\displaystyle\frac{1}
        	{\sqrt{|q^2|}}\davetag{a}\\
        |\Delta x_L | \lesssim& R_\| =\displaystyle\frac{1}{|q_0|}
		\davetag{b}\\
        |\Delta x_0 | \lesssim& R_0 =\displaystyle\frac{1}{|q_L|}.\davetag{c}
\label{eqn:retlimits}
\end{align}

Now we study the $q^2=0$ case.  With $q^2=0$, $\lambda^2$ becomes
\begin{equation}
        \lambda^2=|\Delta x\cdot q|
        =|q_0||\Delta\vec{x}\cdot\hat{q}-\Delta x_0|\lesssim 1.
\label{eqn:followsclasspath}
\end{equation}
On other words, high energy particles tend to follow their classical path 
while low energy particles can deviate from their classical path.  
Expression (\ref{eqn:followsclasspath})
then gives a measure of the deviation from the classical path.

\subsection{Derivation}
\label{sec:retDerivation}

The Wigner transform of $G^{+}$ is easiest to do in coordinate space.  In
coordinate space, $G^{+}(x)=\frac{1}{2\pi}\theta(x_0)\delta(x^2)$, so the 
Wigner transform integral in Equation~\eqref{eqn:propdefb} is a series of delta 
function integrals.  Performing the first delta function integral and 
simplifying the theta functions, we find
\[
   \Gplus{x}{p}=\frac{\theta(x_0)}{(2\pi)^2}\int^{2x_0}_{-2x_0} d x'_0
      \sqrt{4x^2+{x'_0}^{2}}\int_{4\pi}d\Omega_{\vec{x}'} 
      e^{i x'\cdot p}\delta(x'\cdot x).
\]
Using $2\pi\delta(x)=\int^{\infty}_{-\infty}d \alpha e^{ix\alpha}$, we can do
the angular integral, giving us a Bessel function:
\[
   \Gplus{x}{p}=4\pi\theta(x_0)\theta(x^2)\int_{-1}^{1} d \alpha 
      e^{i\alpha \eta}J_0(\xi\sqrt{1-\alpha^2}).
\]
Here $\eta=2(p_0 |\vec{x}|-x_0 \hat{x}\cdot\vec{p})$ and 
$\xi=2\sqrt{x^2(\vec{p}^2-(\vec{p}\cdot\hat{x})^2)}$.  This integral is in any
standard integral table \cite{gra80}.  After a bit of simplification, 
one  gets the
result (\ref{eqn:Gretdef}).  This result can be checked by performing the
Wigner transforms in momentum space, but the contour integrals needed for this
calculation are quite tedious.

\section{Detail: the Feynman Propagator}
\label{sec:feynprop}

The phase-space Feynman propagator naturally arose in our discussion of 
the exclusive reaction probabilities in Chapter \ref{chap:pips}.  There 
we choose situations where we can avoid using the phase-space 
Feynman propagator.  Here we should 
discuss it anyway.  In this section we will describe 
how the scalar propagator works and how we derive the analytic expression
for it at lowest order.  

\subsection{How it Works}
\label{sec:feynWorks}

While the Feynman propagator propagates a
particle with a given momentum (say $p_\mu=(p_0,p_L,\vec{0}_T)$) across a
space-time displacement $\Delta x_\mu=(\Delta x_0, \Delta x_L, \Delta
\vec{x}_T)$, it does so in a manner very different from the retarded
propagator.  Looking at the definition in \eqref{eqn:Gcausdef},
we see that the combination of the sign functions in the square brackets 
can be rewritten in a more transparent form:
\[ 
    [\ldots]=\left\{ 
       \begin{array}{ll}
          4& \mbox{ if } \Delta x\cdot p,\: p^2, \:\Delta x^2 >0\\
          -4& \mbox{ if } \Delta x\cdot p,\: p^2, \:\Delta x^2 <0\\
          2\:{\rm sgn}(\Delta x\cdot p)& \mbox{ if } p^2,\: x^2 \mbox{ have 
          opposite sign}
       \end{array}
    \right.
\]
Thus, particles with time-like momentum tend to travel forward
in time and inside the light-cone and particles with space-like momentum tend
to travel backwards in time outside the light cone.  
Anti-particles with time-like momentum tend 
to travel backwards in time inside the light-cone
and anti-particles with space-like momentum tend 
to travel forwards in time outside the light-cone.

The rest of the interesting features of the Feynman propagator are tied up
in the dependence on the Lorentz invariant $\lambda^2=(\Delta x\cdot p)^2 -
\Delta x^2 p^2$.   
As with the retarded propagator in Subsection~\ref{sec:classEPD}, we will 
study the $p^2>0$, $p^2<0$, and $p^2=0$ cases separately.  

To study the $p^2>0$ case, we boost to the frame where $p'_\mu=(p'_0,\vec{0})$.
In this frame, $\lambda^2={p'}^2_0 |\Delta \vec{x}'|^2 \ge 0$, so only the sine
term contributes.  The sine term is greatest for $\sqrt{\lambda^2}\lesssim 1$
so we have the following limit on the spatial propagation distance:
\begin{equation}
\daveseqn
        |\Delta \vec{x}'| \lesssim \frac{1}{|p'_0|}.\davetag{a}
\end{equation}
As with the retarded propagator, we can compute the total ``probability'' to
propagate to certain time.  This calculation gives us the following limit on
the temporal propagation distance:
\begin{equation}
        |\Delta x'_0| \lesssim \frac{1}{|p'_0|}.\davetag{b}
\end{equation}
Boosting the space-time region defined by these constraints back to the frame
with $p_\mu=(p_0,p_L,\vec{0}_T)$, we find the following constraints:
\begin{align}
\daveseqn
        |\Delta\vec{x}_T|\lesssim& R_\perp=\frac{1}{\sqrt{|p^2|}}\davetag{a}\\
        |\Delta x_L|\lesssim& R_\|=\displaystyle\frac{1}{|p_L|}\davetag{b}\\
        |\Delta x_0|\lesssim& R_0=\displaystyle\frac{1}{|p_0|}\davetag{c}
\end{align}
These limits are exactly the same as the ones we found for the retarded
propagator in Subsection~\ref{sec:classEPD}.

To study the $p<0$ case, we boost to the $p'_\mu=(0,p'_L,\vec{0}_T)$ frame.
In this frame, $\lambda^2={p'}_L^2(\Delta {x'}_0^2-\Delta {x'}_T^2)$.    
Inside the light-cone, the exponential term disappears and we get a
constraint on $\lambda^2$:
\[
        0\geq \lambda^2={p'}_L^2(\Delta {x'}_0^2-\Delta {x'}_T^2)\lesssim 1.
\]
We can integrate to find the total ``probability'' to
propagate to a certain time, giving us a limit on $\Delta x'_0$:
\begin{equation}
        |\Delta x'_0|\lesssim\frac{1}{|p'_L|}.
\end{equation}
Using the $\lambda^2$ and light-cone constraints, we find similar limits on 
$\Delta \vec{x}'_T$ and $\Delta x'_L$:
\begin{align}
\daveseqn
        |\Delta x'_L|\lesssim&\displaystyle\frac{1}{|p'_L|}\davetag{a} \\
        |\Delta \vec{x}'_T|\lesssim&\displaystyle\frac{1}{|p'_L|}.\davetag{b}
\end{align}
Boosting back to the $p_\mu=(p_0,p_L,\vec{0}_T)$ frame, we find
\begin{align}
\daveseqn
        |\Delta\vec{x}_T|\lesssim& R_\perp=\frac{1}{\sqrt{|p^2|}}
		\label{eqn:feynlimitsa}\davetag{a}\\
        |\Delta x_L|\lesssim& R_\|=\displaystyle\frac{1}{|p_0|}
		\label{eqn:feynlimitsb}\davetag{b}\\
        |\Delta x_0|\lesssim& R_0=\displaystyle\frac{1}{|p_L|},
		\label{eqn:feynlimitsc}\davetag{c}
\end{align}
which is what we found for the retarded propagator.
Now, outside of the light-cone the situation is more complicated and we must
integrate the propagator in the various directions to find limits.
We find:
\begin{align}
\daveseqn
        |\Delta x'_0|\lesssim&\displaystyle\frac{1}{|p'_L|}\davetag{a} \\
        |\Delta x'_L|\lesssim&\displaystyle\frac{1}{|p'_L|}\davetag{b} \\
        |\Delta \vec{x}'_T|\lesssim&\displaystyle\frac{1}{|p'_L|}.\davetag{c}
\end{align}
When we boost back to the frame with
$p_\mu=(p_0,p_l,\vec{0}_T)$, we find the result in 
Equations~\eqref{eqn:feynlimitsa}-\eqref{eqn:feynlimitsc}.

Finally, we investigate the $p^2=0$ case.  With $p^2=0$, $\lambda^2$ becomes
\begin{equation}
        0\le\lambda^2=|\Delta x\cdot p|
        =|p_0||\Delta\vec{x}\cdot\hat{p}-\Delta x_0|\lesssim 1
\label{eqn:alsofollowsclasspath}
\end{equation}
because the exponential term does not contribute on the light cone.
On other words the Feynman propagator functions exactly like the retarded
propagator: high energy particles tend to follow their classical path 
while low energy particles can deviate from their classical path.  
Expression (\ref{eqn:alsofollowsclasspath})
then gives a measure of the deviation from the classical path.

We find that, despite the different boundary conditions on the
two propagators, both the Feynman and retarded propagators send particles the
same distances.  This is probably no surprise since a calculation done using 
Feynman's formulation of perturbation theory for the S-matrix must give the 
same results as a calculation done using time-ordered non-equilibrium 
perturbation theory.

\subsection{Derivation}
\label{sec:feynDerivation}

The simplest derivation of $\Gcaus{x}{p}$ is far more complicated than 
the derivation of $\Gplus{x}{p}$.  We start by finding the transport-like 
equation of motion for the Wigner propagator.\footnote{The
constraint-like equation could also be used, but $G^c$ is easier to derive
using the transport-like equation.}  
The derivation is simple and very similar to the derivation for the retarded 
equation of motion in Section \ref{sec:transport}.  So  we only state 
the result:
\[
   p\cdot\partial \:\Gcaus{x}{p}=\frac{1}{\pi^2}
      \left[\pi\delta(x^2) \sin{(2x\cdot p)}-
      {\cal P}\frac{1}{x^2} \cos{(2x\cdot p)}\right].
\]
Now we define a projector onto the space perpendicular to the particle's
momentum, $g_{\perp\mu\nu}=g_{\mu\nu}-p_\mu p_\nu/p^2$.  This allows us to
change variables to $x_{\perp\mu}=g_{\perp\mu\nu} x^{\nu}$ and $\tau=x\cdot
p/\sqrt{|p^2|}\mbox{ sgn}({p^2})$.  In terms of these variables, we find
$\lambda^2=-p^2x_{\perp}^2$ and the equation of motion becomes
\[
   \partial_\tau G^{c} (\tau,x_{\perp},p)=
      \frac{\mbox{ sgn}(p^2)\sqrt{|p^2|}}{\pi^2}
      \left[\pi\delta(|k^2|\tau^2-\lambda^2)\sin{(2\sqrt{|k^2|}\tau)}
      -{\cal P}\frac{\cos{(2\sqrt{|k^2|}\tau)}}{|k^2|\tau^2-\lambda^2}\right].
\]
So, instead of doing the Wigner transform directly, we only have to solve this
ordinary differential equation.

We find the solution by integrating this differential equation.  The delta
function integral is simple and the principle value integral can be done by
contour integration.  We find 
\[\begin{split}
   G^{c} (\tau,x_{\perp},p)= &G^{c} (\infty,x_{\perp},p)\\
	& -\displaystyle\frac{1}{\pi} \left\{ 
      	\theta(\lambda^2)\frac{\sin{(2\sqrt{\lambda^2})}}{\sqrt{\lambda^2}}
      	\left[\frac{1}{2}(\theta(p^2)-\theta(x^2))+
      	\mbox{ sgn}(p^2)\theta(-\tau)\right]\right.\\
      	& \displaystyle
      	\left. -\mbox{ sgn}(p^2)\theta(-\tau)
      	\frac{\exp{(-2\sqrt{-\lambda^2})}}{\sqrt{-\lambda^2}}\right\}.
\end{split}\]
We must now divine the boundary condition at $\tau\rightarrow\infty$.

To find the boundary condition, we actually have to go back to the Wigner
transform of the propagator starting from momentum-space version of equation
(\ref{eqn:propdefa}).  We again change variable from $x$ to $\tau$ and 
$x_\perp$.  We also change from $p'$ to $p'_\perp=g_{\perp\mu\nu}{p'}^\nu$ and
$p\cdot p'=\mbox{ sgn}(p^2)\sqrt{|p^2|} k$.  With this, we perform the $k$
contour integral.  The integral is straight forward, but tedious.  However, when
we take the limit as $\tau\rightarrow\infty$, the result simplifies
dramatically:
\[
   G^{c} (\infty,x_{\perp},p)=\frac{1}{\pi^2\sqrt{|p^2|}}\int\dn{3}{p_\perp}
      \cos{(2x_\perp\cdot p_\perp)}\delta(p^2+p_\perp^2).
\]
The delta function integral is trivial and the last pair of integrals requires
integral tables, but in the end we find:
\[
   G^{c}(\infty,x_{\perp},p)=\frac{1}{\pi}\left\{
      \theta(\lambda^2)\mbox{ sgn}(p^2)
      \frac{\sin{(2\sqrt{\lambda^2})}}{\sqrt{\lambda^2}}
      +\theta(-\lambda^2)\theta(-p^2)
      \frac{\exp{(-2\sqrt{-\lambda^2})}}{\sqrt{-\lambda^2}}\right\}.
\]
Plugging this into the solution of our differential equation, we find
Equation \eqref{eqn:Gcausdef}.  This result can be checked by performing a
series of contour integrals in momentum or coordinate space.

\section{Propagating Particles with Nonzero Mass}

The issue of propagating particles with nonzero is an important one if one is 
either to dress the particles with an in-medium mass or to test conventional 
transport theory as applied to nucleons.  In this section, we will discuss the 
beginnings of work aimed at getting analytic expressions for the propagators 
for particles with nonzero mass.  There are two ways that we know for getting 
the propagator for particles with nonzero mass.  The first is a transform
we that adds mass to the massless propagators.  We demonstrate
it on the retarded propagator.  The second is an approximation developed by 
Remler \cite{tran:rem90} which we will outline and state the results he 
obtained for the Feynman propagator.

\subsection{The Retarded Propagator}

According to \eqref{eqn:propdefa}, a phase-space propagator is the Wigner 
transform of two propagators.  To find the propagator for particles with
nonzero mass, we begin by writing out the Wigner transform of the two 
retarded propagators in the 
coordinates where the transfered momentum is $p_\mu=(p_0,p_L,\vec{0}_T)$:
\begin{equation}
\begin{split}
G^{+}(x,p)=&\int\dnpi{4}{p'} e^{-x\cdot p'} G^+(p+p'/2) G^-(p-p'/2)\\
 	=&\int^\infty_{-\infty}\frac{d {p'_0}}{2\pi}
	\int^\infty_{-\infty}\frac{d {p'_L}}{2\pi}
	\int^\infty_{0}\frac{d {{p'_T}}^2}{2(2\pi)^2}
	\int^{2\pi}_{0}\frac{d \theta'}{2\pi}
	e^{-i(x_0 {p'_0} - x_L {p'_L} - x_T {p'_T} \cos \theta')}\\
	&\times \left( (p_0+{p'_0}/2)^2-(p_L+{p'_L}/2)^2-
	({\vec{p'}_T}/4+m^2) + i\varepsilon 
	(p_0+{p'_0}/2)\right)^{-1}\\
	&\times \left( (p_0-{p'_0}/2)^2-(p_L-{p'_L}/2)^2-
	({\vec{p'}_T}/4+m^2) + i\varepsilon 
	(p_0-{p'_0}/2)\right)^{-1}
\label{eqn:WholeSchmeer}
\end{split}
\end{equation}

Now, we notice that the denominator is independent of $\theta'$, so we can 
perform the $\theta'$ integral using the relation
\[
\int_0^{2\pi}d\theta'\:e^{i x_Tp'_T\cos{\theta'}}=2\pi\BesselJ{0}{x_T p'_T}.
\]
This turns Equation \eqref{eqn:WholeSchmeer} into a Fourier-Bessel transform.  We 
can invert the $p'_T$ integral to get
\begin{equation}
\begin{split}
2\pi\int_0^{2\pi} dx_T x_T &\BesselJ{0}{x_T p'_T} G^{+}(x,p)=
	\int^\infty_{-\infty}\frac{d {p'_0}}{2\pi}
	\int^\infty_{-\infty}\frac{d {p'_L}}{2\pi}
	e^{-i(x_0 {p'_0} - x_L {p'_L})}\\
	&\times \left( (p_0+{p'_0}/2)^2-(p_L+{p'_L}/2)^2-
	({\vec{p'}_T}/4+m^2) + i\varepsilon 
	(p_0+{p'_0}/2)\right)^{-1}\\
	&\times \left( (p_0-{p'_0}/2)^2-(p_L-{p'_L}/2)^2-
	({\vec{p'}_T}/4+m^2) + i\varepsilon 
	(p_0-{p'_0}/2)\right)^{-1}
\label{eqn:BessTransProp}
\end{split}
\end{equation}
Now here is the trick:  if we rename $p'_T\rightarrow\sqrt{{p'_T}^2-4 m^2}$, 
then we get the $m\rightarrow 0$ limit in the right hand side 
of~\eqref{eqn:BessTransProp}.  We can use that fact to our advantage and obtain
the result that the propagator with mass and the propagator without mass 
are related though a pair of Fourier-Bessel transforms:
\begin{equation}
2\pi\int_0^{2\pi} dx_T\: x_T\: \BesselJ{0}{x_T p'_T} G^{+}(x,p)=
2\pi\int_0^{2\pi} dx_T\: x_T\: \BesselJ{0}{x_T\sqrt{{p'_T}^2+4 m^2}}G_0^{+}(x,p).
\label{eqn:WierdResult}
\end{equation}
Here $G_0^+$ is shorthand for the massless limit of the retarded propagator.

Equation \eqref{eqn:WierdResult} can be inverted to give the propagator with mass
as a Fourier-Bessel transform of the propagator without mass!  Using
\begin{equation}
\int_0^\infty dp\: p\: \BesselJ{0}{p x_T} \BesselJ{0}{\sqrt{p^2+4 m^2} x'_T}=
\begin{cases}
	0 & \text{if $x_T>x'_T>0$}\\
	\deltaftn{}{x'_T-x_T}/x_T&\text{if $x_T=x'_T=0$}\\
	-2 m \frac{\BesselJ{1}{2m\sqrt{{x'_T}^2-{x_T}^2}}}
		{\sqrt{{x'_T}^2-{x_T}^2}} & \text{if $x'_T>x_T>0$}
\end{cases}
\label{eqn:ineedthisone}
\end{equation}
and the known result for the massless propagator, Equation \eqref{eqn:WierdResult}
can be rewritten as
\begin{equation}
G^+(x,p)=G^+_0(x,p)-2m\frac{1}{\pi}\thetaftn{x_0}\thetaftn{x^2}
\int_0^{\sqrt{x^2}} d\xi \:\BesselJ{1}{2 m \xi}\frac{\sin{\left(
2\sqrt{\lambda^2+\xi^2 p^2}\right)}}{\sqrt{\lambda^2+\xi^2 p^2}}.
\label{eqn:GetThisOne}
\end{equation}
Here, $\lambda^2$ is the Lorentz invariant defined earlier.
Notice that, as with the propagator for massless particles, massive particles 
must propagate forward in time and they must propagate inside the light-cone.

\begin{figure}
\begin{center}\includegraphics[totalheight=14cm,clip]{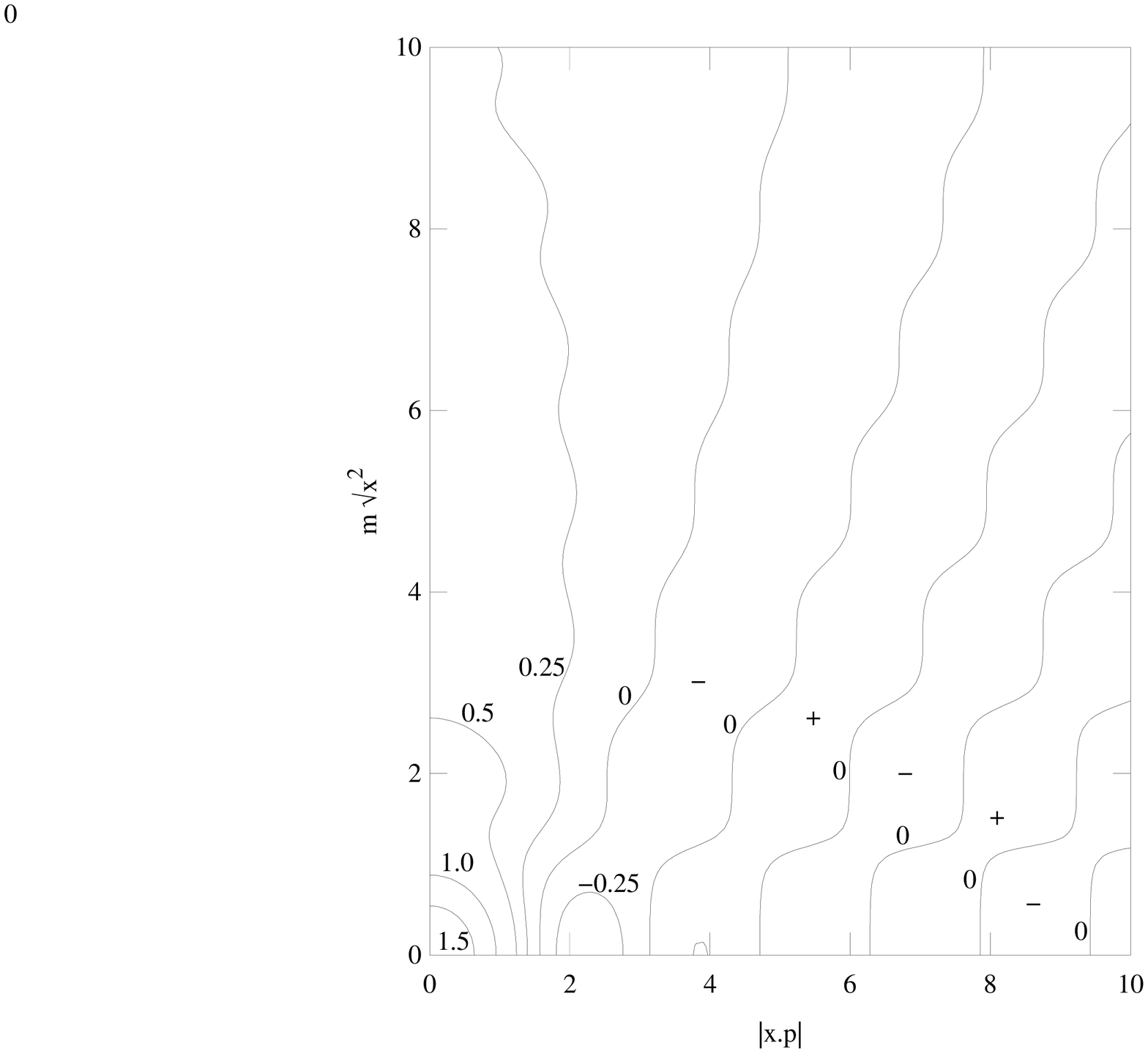}\end{center}
\caption[Plot of the phase-space retarded propagator for a particle 
	with non\-zero mass.]
	{Plot of the phase-space retarded propagator for a particle 
	with non\-zero mass.}
\label{fig:PlotofProp}
\end{figure}

Now, we do not know how to perform the integral in \eqref{eqn:GetThisOne} 
analytically, but we can evaluate it numerically.  A 
contour plot of $G^+(x,p)$ (without the $\theta$ functions) for on-mass-shell
particles is shown in Figure \ref{fig:PlotofProp}.  
One should notice the central peak at 
$|x\cdot p|, m\sqrt{x^2}\lesssim 1$.  Because this peak gives the dominant 
contribution in any integral over all of phase-space, we can estimate that
$|x\cdot p| =\sqrt{m^2+\vec{p}^2} |x_0-\vec{v}\cdot\vec{x}|\lesssim 1$ for
these on-shell particles.  For non-relativistic particles, $m\gg |\vec{p}|$
and $\vec{v}\approx\vec{p}/m$ so 
\begin{equation}
|x_0-\vec{v}\cdot\vec{x}|\lesssim 1/m.
\end{equation}
In other words, on-shell non-relativistic particles deviate from their 
classical paths to distances no more than $\hbar c/m$ -- in other words the 
Compton wavelength, $\lambda_C$.  
This estimate 
tells us that a nonrelativistic particle is localized to a region of radius 
$\lambda_C$ and follows its classical trajectory.  Given that the Compton 
wavelength of a nucleon is $\sim 0.2$~fm, this explains why it is 
reasonable to treat the nucleons as point-like particles that 
follow straight line classical trajectories in the Quasi-Particle Approximation.
For pions though, it is not so clear that they can be treated as point-like 
particles in this manner.  Their Compton  
wavelength is $\sim 1.4$~fm, a distance comparable to the effective range
of the nuclear force.

Even though we cannot evaluate Equation~\eqref{eqn:GetThisOne} in general, we 
can evaluate it in the low mass limit.  For $m^2\ll|p^2|$, we find
\begin{equation}
G^+(x,p)=G^+_0(x,p)-\frac{1}{\pi}\thetaftn{x_0}\thetaftn{x^2}\frac{m^2}{p^2}
\left[ \cos{(2\sqrt{\lambda^2})}-\cos{(2|x\cdot p|)} + 
\Order{\frac{m^4}{p^4}}\right].
\end{equation}
This result may prove useful for describing propagation of highly virtual
particles in a transport approach.

\subsection{The Feynman Propagator}

Remler \cite{tran:rem90} has found the Wigner transform of the  
Feynman propagator for particles with nonzero mass.  
Remler begins with the full Wigner transform in Equation~\eqref{eqn:propdefa}
and makes the integrals into simple contour integrals by approximating 
${p'}^2\approx(p\cdot p')^2/p^2$, where $p$ is the average momentum and 
$p'$ is the relative momentum.  We state Remler's result here:
\begin{equation}
\label{eqn:remlerprop}
\Gcaus{x}{p}=\left\{
\begin{array}{ll}
   \begin{array}{ll}
      {\displaystyle\int^{\infty}_{-\infty}} & \dn{}{\tau} 
      \deltaftn{4}{x-\frac{p}{\sqrt{-p^2}}\tau}e^{-2m|\tau|}
      \frac{1}{2m\sqrt{-p^2}(m^2-p^2)} \\
      & \times\left\{\sqrt{-p^2}\cos{(2\tau\sqrt{-p^2})}
      +m\sin{(2|\tau|\sqrt{-p^2})}\right\}
   \end{array}
   &{\rm for } \; p^2<0\\
   &\vspace*{.10cm}\\
   \begin{array}{ll}
      {\displaystyle\int^{\infty}_0} \dn{}{\tau}
      \frac{1}{2m\sqrt{p^2}} & \left\{
      \frac{\sin{(2\tau(\sqrt{p^2}-m))}}{(\sqrt{p^2}-m)}
      \deltaftn{4}{x-\frac{p}{\sqrt{p^2}}\tau}\right. \\
      & \left.-\frac{\sin{(2\tau(\sqrt{p^2}+m))}}{(\sqrt{p^2}+m)}
      \deltaftn{4}{x+\frac{p}{\sqrt{p^2}}\tau}
      \right\}
   \end{array}
   &{\rm for } \; p^2>0\\
\end{array}
\right.
\end{equation}
Now because of this approximation, the propagator is
oversmoothed in the direction transverse to the particle's momentum and we
expect these propagators to be accurate only on length scales much larger than
the smoothing scale, i.e. on lengths $>1/m$.  
Note also that the sine and exponential functions in the two terms in 
\eqref{eqn:remlerprop} become proportional
to $\deltaftn{}{p^2-m^2}$ as $\tau\rightarrow\infty$.  Thus, this
propagator reduces to the classical propagator \cite{tran:rem90}.  Finally, we 
note that the $\delta$-functions constrain the particle to move along
its classical trajectory, even though its four-momentum (and hence
its four-velocity) is being modulated by the sine and exponential functions.

\chapter{MEASURABLES OF INTEREST}
\label{append:spectra}

We cannot measure the phase-space densities directly, but 
we can measure related quantities such as inclusive and exclusive N-particle
spectra.  For example, the ratio of two-particle inclusive spectrum to 
two one-particle spectra define the correlation function which, in turn, is 
inverted to get the source function in nuclear imaging.  It turns out that 
inclusive spectra are intimately related to the particle densities -- the 
single particle inclusive spectra {\em is} the momentum space particle 
density.  The inclusive distribution is a sum over all exit channels containing
the particles of interest.  If we concentrate on one exit channel, then we have
the N-particle exclusive spectra.  Furthermore, if the reaction is dominated 
by one channel, the full inclusive distribution and the exclusive 
distributions are approximately the same.  The exclusive distribution is 
directly related to the S-matrix for the reaction and thus can be 
calculated perturbatively using the Feynman rules tabulated in many field 
theory books 
\cite{gFT:akh65,gFT:itz80,gFT:lur68,gFT:ste93,gFT:bog79}.
Because these Feynman rules are familiar and simple, we used them to 
discuss the photon and electron phase-space distributions in 
Chapter~\ref{chap:pips}.  

Let us briefly outline this appendix.  First we define the 
S-matrix.  Second, we describe exclusive and inclusive N-particle spectra
and write them in terms of the S-matrix.  Finally, we discuss how
the correlation function is written in terms of inclusive spectra and 
outline the derivation of the Pratt-Koonin equation.  The Pratt-Koonin 
equation is the starting point of the discussion in Chapter~\ref{chap:HBT}.

\section{The S-Matrix}

We begin by writing the transition amplitude from an arbitrary incoming  state 
in the Heisenberg picture $\ket{i(\text{in})}$ with an arbitrary outgoing 
state $\ket{f(\text{out})}$:
\begin{equation}
S_{i\rightarrow f}=\OverLap{f(\text{out})}{i(\text{in})}.
\end{equation}
This is commonly defined as the S-matrix and it is the main building block
of the inclusive and exclusive spectra to follow.

In practice, we are often interested in the transition amplitude to a state
that contains a known object.  For example, consider the following arbitrary
N-particle state, indexed by quantum numbers $\nu$ (here $\nu$ might 
correspond to the momentum of the particles in the state, e.g. $\vec{p}_1,
\ldots,\vec{p}_A$ of the individual particles or a center of mass momentum
$\vec{P}$ for a cluster):
\begin{equation}\begin{split}
\ket{\nu}&=\hat{\Psi}^\dagger_N(\nu)\ket{0}
\\&=\frac{1}{\sqrt{N!}}\int \dn{3}{x_1}
\ldots\dn{3}{x_N}e^{-iE_\nu t}\Phi^{(-)}_\nu (\vec{x}_1,\ldots,\vec{x}_N)
\hat{\psi}^\dagger(\vec{x}_1,t)\ldots\hat{\psi}^\dagger(\vec{x}_N,t)\ket{0}.
\end{split}
\label{eqn:ArbitraryState}
\end{equation}
This state can be anything from a single nucleon (for N=1) to an intermediate 
mass fragment as in \cite{Wig:rem86}.  Here $\hat{\psi}^\dagger(\vec{x},t)$ 
creates a particle at point $\vec{x}$ at time $t$.  Spinor, isospin, and 
other indices have been suppressed in this equation.  The exponential factor 
removes the time dependence from these 
creation operators rendering the entire state time-independent (as it must 
be for a state in the Heisenberg picture).  The total energy of the state is 
$E_\nu$.  The many particle wavefunction
$\Phi^{(-)}_\nu (\vec{x}_1,\ldots\vec{x}_N)$ is simply an arbitrary properly 
symmetrized wavefunction with outgoing boundary conditions encoding the state. 
The action of the composite
raising operator $\hat{\Psi}^\dagger_N(\nu)$ creates the N-particle state 
with the quantum number $\nu$ out of the vacuum.  With this operator, we 
can write down a state that contains this N-particle object plus anything 
else:
\[\ket{f+\nu(\text{out})}=\hat{\Psi}^\dagger_N(\nu)\ket{f(\text{out})}.\]

So, we can now write the transition amplitude from an incoming state $i$ to
an arbitrary final state $f$ plus some known N-particle object $\nu$ as an
S-matrix amplitude:
\begin{equation}
S_{i\rightarrow f+\nu}=\OverLap{f+\nu (\text{out})}{i(\text{in})}=
\ME{f (\text{out})}{\hat{\Psi}_N(\nu)}{i(\text{in})}.
\end{equation}
At first glance, this result is both trivial and apparently not all that 
useful.   This is not the case.  Any S-matrix involving free particles in
both the initial and final states can be written in the interaction picture:
\begin{equation}
S_{i\rightarrow f+\nu}=\OverLap{f+\nu (\text{out})}{i(\text{in})}
={\OverLap{f+\nu}{i}}_{\text{int}}.
\end{equation}
The Feynman rules for the S-matrix\footnote{These Feynman rules are what we 
refer to as Feynman's formulation of perturbation theory.} in the interaction 
picture are written in many field theory books (e.g 
\cite{gFT:akh65,gFT:itz80,gFT:lur68,gFT:ste93,gFT:bog79} to name a few).  
So, if we write a process in terms of
an in-out transition amplitude, we have a ready made calculation scheme.
This is the basis for the perturbation calculations in Chapter \ref{chap:pips}.

\section{Exclusive and Inclusive N-Particle Distributions}

Experimentalists often measure the momentum space particle distribution for 
one type of particle (or several types of particles).  One can do this two 
ways: one can look
for specific exit channels (i.e. ``I want particles A and B only'') or 
one can sum over channels (i.e. ``I want particle A and I don't care what
else comes out'').  The first possibility, where we select a specific channel,
gives exclusive distributions.  The second, where we include all channels,
gives the inclusive distribution.  In both cases, the distribution is
related to the cross section by
\begin{equation}
\frac{dN_\nu}{d\nu}=\frac{1}{\sigma}\frac{d \sigma_\nu}{d\nu}
\end{equation}
where $\sigma$ is the total reaction cross section.  As before, $\nu$ is the
quantum numbers of the state of interest.

\subsection{Exclusive Spectra}

The exclusive particle distribution for an arbitrary N-particle object in the
final state and an initial state specified by a density matrix is
\begin{equation}
\frac{dN_{\nu+f}}{d\nu}=\sum_{n,m}\rho_{n m}
\OverLap{f+\nu (\text{out})}{n(\text{in})}
\OverLap{m(\text{in})}{f+\nu(\text{out})}
=\sum_{n,m}\rho_{n m} S_{n\rightarrow f+\nu}S^*_{m\rightarrow f+\nu}.
\end{equation}
If the density matrix is diagonal, then this takes on a more familiar form:
\begin{equation}
\frac{dN_{\nu+f}}{d\nu}=\sum_{n}\rho_{n n} |S_{n\rightarrow f+\nu}|^2
\label{eqn:DiagonalDensityMatrix}
\end{equation}
So, given that this can be written in terms of the S-matrix,  
we have a ready made computation scheme.  In general, the density 
matrix can be arbitrarily complicated, so it may be that while a perturbative 
solution exists, it is impractical to perform the calculations.

In this thesis, we found it useful
to localize particles using wavepackets in the initial state.  From 
\eqref{eqn:DiagonalDensityMatrix} it should be clear that these wavepackets
are just another way to write the density matrix.  Also, we mention that 
Equation~\eqref{eqn:DiagonalDensityMatrix} can be rewritten in terms of 
phase-space quantities in the manner outlined 
in Appendix~\ref{sec:crosssection}.

\subsection{Inclusive Spectra}
The N-particle spectra of into a specific mode $\nu$ is simply the sum over 
exclusive spectra:
\begin{equation}
\frac{d N_\nu}{d\nu}=\sum_f \frac{dN_{f+\nu}}{d\nu}
=\sum_f \sum_{n,m} \rho_{n m}
\ME{n (\text{in})}{\hat{\Psi}^\dagger_N(\nu)}{f(\text{out})}
\ME{f(\text{out})}{\hat{\Psi}_N(\nu)}{m (\text{in})}
\end{equation}
The $f$ states form a complete basis so
\begin{equation}
\frac{d N_\nu}{d\nu}=\sum_{n,m} \rho_{n m}
\ME{n (\text{in})}{\hat{\Psi}^\dagger_N(\nu)\hat{\Psi}_N(\nu)}{m (\text{in})}
=\expectval{\hat{\Psi}^\dagger_N(\nu)\hat{\Psi}_N(\nu)}.
\label{eqn:PickaName}
\end{equation}
We recognize the spectrum then as the N-particle analog of 
the particle density $G^<$ in~\eqref{eqn:DefofDensity}.  Indeed, if we set 
$\Psi$ to be a single particle operator and fix $\nu$ to be the momentum of 
that operator, then \eqref{eqn:PickaName} gives the momentum space density 
of particles.

We can write the spectrum in the form written out by Danielewicz and 
Schuck \cite{HBT:dan92} by inserting the definition of 
$\hat{\Psi}^\dagger_N(\nu)$ in Equation~\eqref{eqn:ArbitraryState}:
\begin{equation}
\begin{split}
\displaystyle\frac{dN_\nu}{d\nu}=&\displaystyle\lim_{T,T'\rightarrow\infty}
\frac{1}{A!}\frac{1}{TT'}\int^{2T}_{T}dt\int\dn{3}{x_1}\ldots\dn{3}{x_A}
\int^{2T'}_{T'}dt'\int\dn{3}{x'_1}\ldots\dn{3}{x'_A}\\
&\displaystyle\times\exp [iE_\nu (t-t')]
\Phi^{(-)*}_\nu (\vec{x}_1,\ldots,\vec{x}_A)
\Phi^{(-)}_\nu (\vec{x}'_1,\ldots,\vec{x}'_A) \\
&\displaystyle\times\expectval
{\hat{\psi}^\dagger(\vec{x}'_A,t')\ldots\hat{\psi}^\dagger(\vec{x}'_1,t')
\hat{\psi}(\vec{x}_1,t)\ldots\hat{\psi}(\vec{x}_A,t)}
\end{split}
\label{eqn:NBodySpectra}
\end{equation}
We will use this form in all subsequent calculations in this appendix.
The $< \ldots >$ is a trace with respect to the density matrix of the system.
Again $\nu$ are the quantum numbers associated with the states. 
The additional action of the two limits 
$\lim_{T\rightarrow\infty}\frac{1}{T}\int^{2T}_{T} dt$
actually serves to project out unwanted contributions from 
heavier nuclei\footnote{such as the contribution to nucleons or 
deuterons from $\alpha$ particles in the final state} \cite{HBT:dan92}. 
These limits work by averaging away all contributions with a total energy
higher than the energy of the states of interest. 

Danielewicz and Schuck go on to show that \eqref{eqn:NBodySpectra} can be 
rewritten as 
\begin{equation}
\begin{split}
\displaystyle\frac{dN_\nu}{d\nu}=&\displaystyle
\frac{1}{A!}\int dt\int\dn{3}{x_1}\ldots\dn{3}{x_A}
\int dt'\int\dn{3}{x'_1}\ldots\dn{3}{x'_A}\exp [iE_\nu (t-t')]\\
&\displaystyle\times
\Phi^{(-)*}_\nu (\vec{x}_1,\ldots,\vec{x}_A)
\Phi^{(-)}_\nu (\vec{x}'_1,\ldots,\vec{x}'_A)\\
&\displaystyle\times\expectval{\hat{J}^\dagger_A(\vec{x}'_1,\ldots,
\vec{x}'_A,t')\hat{J}_A(\vec{x}_1,\ldots,\vec{x}_A,t)}.
\end{split}
\label{eqn:NBSpecCurrents}
\end{equation}
Here $\hat{J}_A$ is the A-particle current operator.
Their derivation assumes that the system is nonrelativistic and obeys a
Schr{\"o}dinger-type equation of motion for each particle.  However one can 
easily see from their derivation that, by replacing their equation of motion 
with a Klein-Gordon equation of motion and replacing their current with a
classical pion current, one arrives at the same equation, but for mesons.
This is the form we will use to derive the Pratt-Koonin equation.

\section{The Correlation Function and Interferometry}

In the remainder of this appendix, we define the two-particle 
correlation function and outline the derivation of the Pratt-Koonin equation
needed for Chapter \ref{chap:HBT}.
To do this, we need to derive the single particle and two-particle spectra.  


\subsection{Single and Two Particle Spectra}

For the single particle spectra, 
\begin{equation}
\begin{split}
\frac{dN_{1\nu}}{d\nu}&=\displaystyle\int dt\int\dn{3}{x}\int dt'\int\dn{3}{x'}
\exp [iE_\nu (t-t')]\\
&\times\displaystyle\Phi^{(-)*}_\nu (\vec{x})\Phi^{(-)}_\nu (\vec{x}')
\expectval{\hat{J}^\dagger_1(\vec{x}',t')
\hat{J}_1(\vec{x},t)}
\end{split}
\label{eqn:1BSpectra}
\end{equation}
Assume that the particle is in free wave state $\Phi^{(-)}_\nu (\vec{x})=
\Phi^{(-)}_{\vec{p}} (\vec{x})=
\exp (i \vec{p}\cdot\vec{x})$, then \eqref{eqn:1BSpectra} becomes the 
Fourier transform of the current-current correlator
\begin{equation}
\frac{dN_{1}}{\dn{3}{p}}=\expectval{\hat{J}^\dagger_1(\vec{p},E)
\hat{J}_1(\vec{p},E)}
\end{equation}

Similarly for the two particle spectra,
\begin{equation}
\begin{split}
\displaystyle\frac{dN_{2\nu}}{d\nu}=&\displaystyle
\frac{1}{2}\int dt\int\dn{3}{x_1}
\dn{3}{x_2}\int dt'\int\dn{3}{x'_1}\dn{3}{x'_2}\exp [iE_\nu (t-t')]\\
&\displaystyle\times\Phi^{(-)*}_\nu (\vec{x}_1,\vec{x}_2)
\Phi^{(-)}_\nu (\vec{x}'_1,\vec{x}'_2)\\
&\displaystyle\times\expectval{\hat{J}^\dagger_2(\vec{x}'_1,\vec{x}'_2,t')
\hat{J}_2(\vec{x}_1,\vec{x}_2,t)}
\end{split}
\label{eqn:2BSpectra}
\end{equation}
Assume that the two particles interact, but that the pair as a whole do not
interact with the nuclear remnants.  Then the wavefunction can be written
as a product of the center-of-mass wavefunction and the relative 
wavefunction.
\begin{equation}
\Phi^{(-)}_\nu (\vec{x}_1,\vec{x}_2)\simeq \exp[i \vec{P}_\nu\cdot(\vec{x}_1+
\vec{x}_2)/2]\phi^{(-)}_\nu (\vec{x}_1-\vec{x}_2)
\label{eqn:2partWF}
\end{equation}
This idea of separating the wavefunction into the CM part and relative part
is being used again, in a more sophisticated form, to do two-particle
correlations in the presence of a third body interacting with the other two by 
Coulomb forces in \cite{HBT:danUnPub}.

Inserting the two particle wavefunction~\eqref{eqn:2partWF} 
into~\eqref{eqn:2BSpectra} and introducing relative and average positions, 
$\vec{r}=(\vec{x}_1-\vec{x}_2)$ and $\vec{R}=\frac{1}{2}(\vec{x}_1+\vec{x}_2)$,
we find
\begin{equation}
\begin{split}
\displaystyle\frac{dN_{2\nu}}{d\nu}=&\displaystyle\frac{1}{2}\int dt  
\int\dn{3}{R} \dn{3}{r}\int dt'\int\dn{3}{R'}\dn{3}{r'}\\
&\displaystyle\times
\exp [iE_\nu (t-t')-i\vec{P}_\nu\cdot\vec{R}+i\vec{P}_\nu\cdot\vec{R'}]
\phi^{(-)*}_\nu (\vec{r})\phi^{(-)}_\nu (\vec{r'})\\
&\displaystyle\times
\expectval{\hat{J}^\dagger_2(\vec{R'}+\vec{r'}/2,\vec{R'}-\vec{r'}/2,t')
\hat{J}_2(\vec{R}+\vec{r}/2,\vec{R}-\vec{r}/2,t)}.
\end{split}
\end{equation} 
So, defining the Fourier transformed current as follows,
\[
\hat{J}_2(\vec{r},\vec{P}_\nu,E_\nu)\equiv\int \dn{}{t} \dn{3}{R}
\exp [iE_\nu t-i\vec{P}_\nu\cdot\vec{R}]
\hat{J}^\dagger_2(\vec{R}+\vec{r}/2,\vec{R}-\vec{r}/2,t)
\]
the two particle spectra becomes
\begin{equation}
\displaystyle\frac{dN_{2\nu}}{d\nu}=\displaystyle\frac{1}{2}
\int\dn{3}{r}\dn{3}{r'}\phi^{(-)*}_\nu (\vec{r})\phi^{(-)}_\nu (\vec{r'})
\expectval{\hat{J}^\dagger_2(\vec{r'},\vec{P}_\nu,E_\nu)
\hat{J}_2(\vec{r},\vec{P}_\nu,E_\nu)}.
\end{equation} 

\subsection{The Correlation Function}

The two particle correlation function is defined in terms of the one and two 
particle spectra as follows:
\begin{equation}
\frac{dN_{2}}{\dn{3}{P}\dn{3}{q}}\equiv\frac{dN_{1}}{\dn{3}{p_1}}
\frac{dN_{1}}{\dn{3}{p_2}}C_{\vec{P}}(\vec{q})
\end{equation}
where $\vec{P}=\vec{p}_1+\vec{p}_2$ is total momentum of the pair and 
$\vec{q}=\frac{1}{2}(\vec{p}_1-\vec{p}_2)$ is the relative momentum of one of 
the pair.\footnote{Experimentalists occasionally use 
$\vec{Q}_{inv}=(\vec{p}_1-\vec{p}_2)$ here.  It is likely they do this to 
confuse us.}
So, assuming no interaction between the emitted pair and the nuclear source,
we find the correlation function in terms of the one and two particle 
currents:
\begin{equation}
\begin{split}
C_{\vec{P}}(\vec{q})&=\displaystyle\left.\frac{dN_{2}}{\dn{3}{P}\dn{3}{q}}
\right/\frac{dN_{1}}{\dn{3}{p_1}}\frac{dN_{1}}{\dn{3}{p_2}}\\
&=\displaystyle\int \dn{3}{r}\dn{3}{r'}
\phi^{(-)*}_{\vec{q}} (\vec{r})\phi^{(-)}_{\vec{q}} (\vec{r'})
\frac{\frac{1}{2}\expectval{\hat{J}^\dagger_2(\vec{r'},\vec{P}_\nu,E_\nu)
\hat{J}_2(\vec{r},\vec{P}_\nu,E_\nu)}}
{\expectval{\hat{J}^\dagger_1(p_1)\hat{J}_1(p_1)}
\expectval{\hat{J}^\dagger_1(p_2)\hat{J}_1(p_2)}}.
\end{split}
\end{equation}
This result is not quite the Pratt-Koonin equation in 
Equation \eqref{eqn:PrattKoonin} in Chapter \ref{chap:HBT}.
We will make the connection in the next section.

\subsection{The Pratt-Koonin Equation}

Now, at freeze-out both the Quasi-Particle and Quasi-Classical approximations
should be valid.  At this point the particles should be fully decoupled from 
the system, so they should be both on-shell and should no longer interact.  
Under these conditions,\footnote{Note that, when 
pair emission is significantly distorted in the vicinity of the source, 
e.g.\ due to the Coulomb interaction with the source, we can either 
account for the distortions in the wavefunction in the kernel or we can 
absorb them into the definition of the source.} 
the system should act as an ensemble of incoherent particle sources.
With these assumptions, Pratt, Cs{\"o}rg\H{o} and Zim\'{a}nyi 
\cite{HBT:pra90} have shown that 
\begin{equation}
\frac{1}{2}\frac{\expectval{\hat{J}^\dagger_2(\vec{r'},\vec{P}_\nu,E_\nu)
\hat{J}_2(\vec{r},\vec{P}_\nu,E_\nu)}}
{\expectval{\hat{J}^\dagger_1(p_1)\hat{J}_1(p_1)}
\expectval{\hat{J}^\dagger_1(p_2)\hat{J}_1(p_2)}}=
\deltaftn{3}{\vec{r}-\vec{r'}}S_{\vec{P}}(\vec{r})
\end{equation}
giving 
\begin{equation}
C_{\vec{P}}(\vec{q})=\int\dn{3}{r}\left|\phi^{(-)}_{\vec{q}}(\vec{r'})\right|^2
S_{\vec{P}}(\vec{r})
\label{eqn:PK1}
\end{equation}
The source function,
$S_{\vec{P}}(\vec{r})$, is identified as 
\begin{equation}
S_{\vec{P}}(\vec{r})=\int\dn{3}{R}d t_1 dt_2 
	D(\vec{P},\vec{R}+\vec{r}/2,t_1)D(\vec{P},\vec{R}-\vec{r}/2,t_1).
\label{eqn:PK2}
\end{equation}
where $D$ is the normalized single particle source.  $D$ can also be identified
with the distribution of last collision points in space, time and momentum 
of the quasi-particles.   Equations \eqref{eqn:PK1} and \eqref{eqn:PK2} 
constitute the Pratt-Koonin equation and \eqref{eqn:PK1} is the starting 
point for our imaging work in Chapter~\ref{chap:HBT}.

For Klein-Gordon fields, with 
$(\partial^2 + m_\pi^2)\phi(x) = -j(x)$, $D$~may be written, in terms
of single-particle self-energies, as
\begin{equation}\begin{split}
D(\vec{p},\vec{r},t) =& {i \over 2 E_p} \, \Pi^< (\vec{p},
E_p, \vec{r}, t)\\  &\times\exp{\left[ -{1 \over 2 E_p}
\int_t^\infty dt' \, (-2) {\rm Im} \, \Pi^+ \left( \vec{p},E_p,
\vec{r} + \vec{v}_p (t' - t), t' \right) \right]} \, ,
\end{split}
\label{DprK}
\end{equation}
where $i \Pi^<(x,x') = \langle j(x') \, j(x) \rangle_{\rm
irred}$, and $(-2) {\rm Im} \, \Pi^+ (x,x') = \langle [j(x) ,
j(x')] \rangle_{\rm irred}$.
For the
Schr\"odinger fields, with $\left(i{\partial \over \partial t}
+ {\nabla^2 \over 2 m} \right) \Psi(x) = j(x)$, the~analogous
result is
\begin{equation}
D(\vec{p},\vec{r},t) = \mp i \Sigma^<(\vec{p}, E_p,
\vec{r}, t)  \exp{\left[ - \int_t^\infty dt' \, \Gamma
\left( \vec{p}, E_p, \vec{r} + \vec{v}_p (t' - t),t' \right) \right]} \, ,
\label{DprS}
\end{equation}
where $\mp i \Sigma^<$ is
the single-particle production rate, $\mp i
\Sigma^< (x,x') = \langle j(x') \, j(x) \rangle_{\rm
irred}$, and $\Gamma$ is the damping rate.

In a transport approach, the particles (at least the nucleons and pions) are 
good quasi-particles throughout most of the reaction.  Thus, $D$ can be 
extracted directly from a model and both source and correlation functions can 
be constructed.  This is actually the procedure Pratt uses in his correlation
code \cite{HBT:pra90}.

\chapter{THE CROSS SECTION AND PHASE-SPACE DENSITIES}
\label{sec:crosssection}

In this appendix, we discuss writing the cross sections in 
Chapter~\ref{chap:pips} in terms of phase-space 
quantities.  Since the cross section is measured by scattering a beam of 
particles off a target, we take a different approach than in 
Appendix~\ref{append:spectra} and define the cross section in terms of the
projectile/target reaction rate density and the projectile flux.  The beam is 
uniform in the beam direction and in time on the 
scale of the projectile/target interaction.  Thus, the beam can only directly 
probe the transverse structure of the interaction region.  Even this transverse
information is washed out in the typical experiment, since the beam is usually
uniform in the transverse direction on the length scale of the interaction.  
In the limit of a transversely uniform beam, we
recover the conventional definition of the cross section.  Since we
consider only  simple scattering problems, we work in Feynman perturbation
theory where we can specify both the initial and final states of the
reactions.

The beam is a collection of single particle wavepackets distributed
throughout the transverse area $A$ of the beam.  For the sake of
illustration, we take these particles to be scalars.  The 
Wigner function of these incident wavepackets is
\begin{equation}
	f(x,p)=\frac{1}{2Vp_0}\int\dnpi{4}{p'}e^{-x\cdot p'}
	f(p+p'/2)f^{*}(p-p'/2)
\end{equation}
where the wavefunction $f(p)$ is given by\footnote{The delta function that
puts the particle on-shell is absorbed into $f(p)$}
\begin{equation}
 	\ket{i}=\int \dnpi{4}{p} f(p) \ket{\vec{p}\,}.
	\label{eqn:ket}
\end{equation}
We assume the beam to be uniform in the longitudinal direction with length
$L$ and to be turned on for macroscopic time $T$.  The quantities 
$A$, $T$, and $L$ are much
larger than the projectile/target interaction region.  

The projectile/target interaction region is characterized by a reaction rate
density ${\cal W}_{i\rightarrow f}(x)$.  We assume the reaction rate density to
be localized in both space and time.  This reflects the small spatial extent of
the target and the short interaction time compared to the beam lifetime.  The
reaction rate is trivially related to the reaction probability:
\begin{equation}
	|S_{i\rightarrow f}|^2=\int d^4 x \;{\cal W}_{i\rightarrow f}(x).
\end{equation}
Thus, the reaction rate is easily identifiable in the calculations in Sections
\ref{sec:pdist} and \ref{sec:edist}.  For example, in the process 
$\gamma B\rightarrow B'$ in Figure \ref{fig:dis}(b), the reaction rate density 
is ${\cal W}_{\gamma B\rightarrow B'}(x,q)$.  For the process 
$AB\rightarrow A'B'$ in Figure \ref{fig:dis}(a), it is
\begin{equation}
	{\cal W}_{AB\rightarrow A'B'}(x)=\int d^4 r \frac{d^4 q}{(2\pi)^4}
	J^{\mu\nu}_{A}(x+r/2) D^{c}_{\mu\nu\mu'\nu'}(r,q) J^{\mu'\nu'}_{B}(x-r/2).
\end{equation}
Note that the reaction rate density is a function of the average space-time
location of all the vertices in the process.

The cross section is the effective area of the target, so we define the
cross section as the integral over the beam face of the fraction of incident
particles that interact with the target per unit area:
\begin{equation}
	\sigma=\int_{A} d^2x_T 
	\left(\frac{\# \;\mbox{scattered particles}}{\mbox{unit area}}\right)\left/
	\left(\frac{\# \;\mbox{incident particles}}{\mbox{unit
	area}}\right).\right. 
\end{equation}
The number of incident particles per unit area crossing the target plane is the
particle flux:
\begin{equation}
	\frac{\# \;\mbox{incident particles}}{\mbox{unit area}}={\cal N}_{inc}
	\int^{L/2}_{-L/2} dx_L \;\hat{n}\cdot\vec{j}(x) 
	\equiv
	{\cal F}(\vec{x}_{T}).
\end{equation} 
Here $\hat{n}$ is a unit normal to the target plane and ${\cal N}_{inc}$ is the
number of particles in the beam.  The single particle current is given in terms
of the incident particle Wigner function by \cite{tran:car83}
\begin{equation}
	\vec{j}(\vec{x})=\int\dn{3}{p}\dn{}{p^2}\vec{v} f(x,p).
\end{equation}
We do not need to average over time because the beam is uniform on the time 
scale of the reaction.
The number of scattered particles per unit area is found by multiplying the
number of incident particles by the reaction probability per unit area:
\begin{equation}
	\frac{\# \;\mbox{scattered particles}}{\mbox{unit area}}={\cal N}_{inc}
	\int^{L/2}_{-L/2} dx_L \int^{T/2}_{-T/2} dx_0 \;{\cal W}_{i\rightarrow f}(x)
	\equiv{\cal N}_{inc} \bar{{\cal W}}_{i\rightarrow f}(\vec{x}_T).
\end{equation}
Thus, the cross section is 
\begin{equation}
	\sigma=\int_A d^2 x_{T} 
	\frac{{\cal N}_{inc} \;\bar{{\cal W}}_{i\rightarrow f}(\vec{x}_T)}
	{{\cal F}(\vec{x}_T)}.
\label{eqn:crosssection}
\end{equation}

In Equation \eqref{eqn:crosssection}, all longitudinal and temporal
structure of the interaction is washed out by the beam.  Furthermore, in 
any practical  experiment, the wavepackets are delocalized in the transverse
direction on the length scale of the
interaction region.  Thus, the transverse 
structure of ${\cal F}(\vec{x}_{T})$ is gone
and the flux reduces to ${\cal F}={\cal N}_{inc}|\vec{v}|/A$, where $|\vec{v}|$
is the mean projectile velocity.  The flux can then be pulled out of the
transverse integral in (\ref{eqn:crosssection}).  The transverse integral of
the reaction probability per unit area is 
${\cal N}_{inc} |S_{i\rightarrow f}|^2$, so the
cross section becomes 
\begin{equation}
	\sigma=\frac{A |S_{i\rightarrow f}|^2}{|\vec{v}|}.
\end{equation}
This is the conventional momentum space cross section in the choice of
normalization used in this thesis.

\chapter{COULOMB FIELD IN PHASE-SPACE}
\label{append:static}

In this appendix, we describe the $\vec{v}=0$ limit of the photon distribution 
of the point charge in Subsection~\ref{sec:classEPD}.  Since the spatial 
dependence of
the Effective Photon Distribution is controlled by the Wigner transform of the
vector potential, $A_{\mu}(x)$, we only discuss $A_{\mu\nu}(x,q)$ here.  
When $\vec{v}=0$, the photon
vector potential becomes $A_\mu(x)=(e/|\vec{x}|,\vec{0})$ so 
$A_{\mu\nu}(x,q)$ is the Wigner transform of the Coulomb potential. 

Take the point charge to be resting at the origin and emitting photons with
four-momentum $q_\mu=(q_0,\vec{q})$.  Putting $\vec{v}= 0$ in 
Equation~(\ref{eqn:Amunu}), we find
\begin{equation}
\begin{array}{rl}
   A_{00}(x,q)&=\displaystyle 32\pi^2\alpha_{em}\delta(q_0)\frac{1}{|\vec{q}|}
      {\cal A}(2|\vec{x}||\vec{q}|\cos(\theta),2|\vec{x}||\vec{q}|\sin(\theta))\\
   A_{ij}&=0
\end{array}
\end{equation}
where $\theta$ is the angle between $\vec{x}$ and $\vec{q}$ and the
dimensionless function ${\cal A}$ is given in Equation~(\ref{eqn:dimlessA}).  
Clearly the photon field is time independent and is composed entirely of zero 
energy photons.  Furthermore, by virtue of the $1/|\vec{q}|$ singularity, the 
photon field is mostly composed of low momentum photons. 

\begin{figure}
    \begin{center}
    \includegraphics[totalheight=10cm]{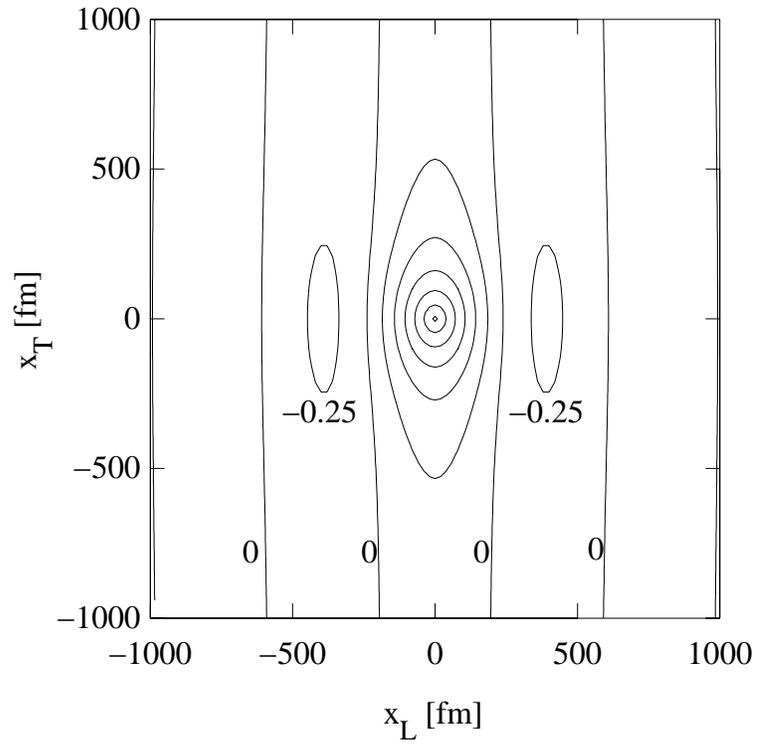}
    \end{center}
    \caption[Plot of the dimensionless function ${\cal A}$ corresponding to 
	the Wigner transform of the Coulomb field of a static point charge.  
	The 
	photons in this plot have $q_\mu=(0, 0.788, \vec{0}_T)$~MeV/c.  The 
	longitudinal axis is  defined by the photon three-momentum.]
	{Plot of the dimensionless function ${\cal A}$ corresponding to the 
	Wigner transform of the Coulomb field of a static point charge.  The 
	photons in this plot have $q_\mu=(0, 0.788, \vec{0}_T)$~MeV/c.  The 
	longitudinal axis is  defined by the photon three-momentum.}
    \label{fig:coulomb}
\end{figure}

In Figure~\ref{fig:coulomb}, we plot the dimensionless function ${\cal A}$ as a
function of $\vec{x}$ for $\vec{q}=(0, 0.788, \vec{0}_T)$ MeV/c in the plane 
defined by $\vec{x}$ and $\vec{q}$.  Note
that the central region of the distribution is circular, but becomes elliptical
as one moves away from the center.  In the transverse direction (i.e. the
direction perpendicular to the photon three-momentum), the
distribution approaches zero, but never is negative.  The width in the
transverse direction is approximately $250$~fm.  In the longitudinal
direction, the distribution drops to zero at about $x_L\approx 250$~fm and
oscillates about zero for larger distances.  These oscillations are expected
for a Wigner transformed quantity and simply reflect the fact that $x_L$ and
$q_L$ are Fourier conjugate variables.

Because the photon source is a point source,  
the shape of the Coulomb distribution comes directly from the shape of the 
the retarded propagator discussed in Appendix~\ref{append:prop}.
Thus, we can estimate the width of the photon distribution using the 
estimates of the retarded propagator in Subsection~\ref{sec:retprop}.
In the both the longitudinal and transverse directions, the propagator width 
is $\sim \hbar c/|q_L|=250$~fm, which is approximately the
width we measure from the plots.

\chapter{GAUGE ISSUES}
\label{sec:gauge}

Parton densities are supposed to be gauge invariant but the  
Effective Photon Distribution in Section \ref{sec:pdist} is gauge dependent.
In this section, we discuss how $\Afield{\mu}{\nu}{x}{q}$ transforms 
under a change of gauge and determine the gauge invariant part of 
$\Afield{\mu}{\nu}{x}{q}$.  We then state how the gauge invariant part of 
$\Afield{\mu}{\nu}{x}{q}$ is related to the phase-space Effective Photon 
Distribution.  Finally, we comment on the gauge dependence of the photon field 
of a point charge.

\section{The Gauge Independent Part}

If we gauge transform the photon field in the energy-momentum representation,
we add an arbitrary function 
in the direction of the photon momentum to the photon vector potential:
$A_{\mu}(q)\longrightarrow A_{\mu}(q) + q_{\mu} f(q)$.  
Because components of $A_{\mu}(q)$ in the direction of
$q_{\mu}$ are gauge dependent, we can write $A_{\mu}(q)$ as a sum of 
the gauge independent and dependent parts:
\[A_{\mu}(q)=A_{\mu}^{\|}(q)+A_{\mu}^{\bot}(q)\]
where
$A_{\mu}^{\|}(q)=\frac{q_{\mu}q_{\nu}}{q^2}A^{\nu}(q)$ is the gauge
dependent part and $A_{\mu}^{\bot}(q)=A_{\mu}(q)-
A_{\mu}^{\|}(q)$ is the gauge independent part.  Wigner transforming
the photon field gives us a term that is gauge independent and
terms which are gauge dependent: 
\begin{equation}
\begin{split}
   \Afield{\mu}{\nu}{x}{q} 
   & = \displaystyle\int\dnpi{4}{\tilde{q}} e^{-ix\cdot\tilde{q}}
       \left[A_{\mu}^{\|}(q+\tilde{q}/2)+A_{\mu}^{\bot}(q+\tilde{q}/2)\right]\\
   &\quad\times\left[A_{\nu}^{\|}(q-\tilde{q}/2)+A_{\nu}^{\bot}(q-\tilde{q}/2)\right]^{*}\\
   & \equiv \displaystyle \Afieldname{\mu}{\nu}{x}{q}{\bot\bot}
      +\Afieldname{\mu}{\nu}{x}{q}{\bot\|}+\Afieldname{\mu}{\nu}{x}{q}{\|\bot}
      +\Afieldname{\mu}{\nu}{x}{q}{\|\|}.
\end{split}
\label{eqn:proj1}
\end{equation}
The only gauge independent piece of $\Afield{\mu}{\nu}{x}{q}$ is
$\Afieldname{\mu}{\nu}{x}{q}{\bot\bot}$.  We do the
integrals in (\ref{eqn:proj1}) and 
identify the tensor that projects off the gauge dependent
part of $\Afield{\sigma}{\rho}{x}{q}$:
\begin{equation}
\begin{array}{ccl}
   \Afieldname{\mu}{\nu}{x}{q}{\bot\bot} 
   & = & \displaystyle(g_{\mu\sigma}-h^{+}_{\mu\sigma})
      (g_{\nu\rho}-h^{-}_{\nu\rho})\Afieldup{\sigma}{\rho}{x}{q} \\
   & \equiv & \displaystyle{\cal P}_{\mu\nu\sigma\rho}
      \Afieldup{\sigma}{\rho}{x}{q}
\end{array}
\label{eqn:proj2}
\end{equation}
where
\begin{equation}
   h^{\pm}_{\mu\nu}=\frac{(q\pm i\partial /2)_{\mu}(q\pm i\partial /2)_{\nu}}
   {(q\pm i\partial /2)^2}.
\label{eqn:htensor}
\end{equation}
This projector must be understood as a series in $q_\mu$ and $\partial_\mu$, so
can only really be used when 
$\partial_\sigma A_{\mu\nu}(x,q) < q_\sigma A_{\mu\nu}(x,q)$.  
Now, the statement of current conservation for a general 
$\Jcurrentdn{\mu}{\nu}{x}{q}$ is
\begin{equation}
   (q\pm i\partial /2)^{\mu}\Jcurrentdn{\mu}{\nu}{x}{q}=
   (q\pm i\partial /2)^{\nu}\Jcurrentdn{\mu}{\nu}{x}{q}=0.
\end{equation}
So, as expected, current conservation ensures that only the gauge independent 
part of $\Afield{\mu}{\nu}{x}{q}$ appears in the reaction probability.     

With $\Afieldname{\mu}{\nu}{x}{q}{\bot\bot}$ in hand, we can postulate the 
gauge invariant photon distribution:
\begin{equation}
   \EPdist=\sum_{\lambda=\pm}\pol{\mu}{\lambda}\polstar{\nu}
   {\lambda}\Afieldname{\mu}{\nu}{x}{q}{\bot\bot}.
\label{eqn:wwapprox3}
\end{equation}
This reduces to (\ref{eqn:wwapprox2}) if the photon field varies slowly in 
space (i.e. we neglect the gradients 
$\partial_{\sigma}\Afield{\mu}{\nu}{x}{q}\ll q_{\sigma}\Afield{\mu}{\nu}{x}{q}$),
as we now show.  Neglecting the derivatives in 
(\ref{eqn:htensor}), the projection tensor in (\ref{eqn:proj2}) reduces to
\begin{equation}
\begin{array}{rcl}
   {\cal P}_{\mu\nu\sigma\rho}&\approx& 
   \displaystyle\left(g_{\mu\sigma}-\frac{q_{\mu}q_{\sigma}}{q^2}\right)
   \left(g_{\nu\rho}-\frac{q_{\nu}q_{\rho}}{q^2}\right)\\
   &=&\displaystyle\left(\sum_{\lambda=\pm,0}\poldn{\mu}{\lambda}
   \polstardn{\sigma}{\lambda}\right)^{*}
   \left(\sum_{\lambda'=\pm,0}\poldn{\nu}{\lambda'}
   \polstardn{\rho}{\lambda'}\right).
\end{array}
\label{eqn:approxproj}
\end{equation}
Since the~polarization~vectors form a~complete~basis in~Minkowski~space, i.e.
$g_{\mu\nu}=\sum_{\lambda=\pm,0}\poldn{\mu}{\lambda}\polstardn{\nu}{\lambda}+
q_{\mu}q_{\nu}/q^2$.
Putting (\ref{eqn:approxproj}) in Equation
\eqref{eqn:wwapprox3}, we arrive back at the Effective Photon Distribution 
in~(\ref{eqn:wwapprox2}).

The tactic of projecting out the gauge dependent parts of
the photon distribution works mainly because of the simple form of the $U(1)$
gauge transformation.  Nevertheless, a variant of this technique may possibly
be applied to the gluon field.

\section{Comment on the Gauge Dependence of the Effective Photon Distribution of a Point Charge}

The Effective Photon Distribution in Equation \eqref{eqn:wwapprox2}
is observable so it is gauge invariant.  
On the other hand, the $A_{\mu\nu}(x,q)$ in Equation \eqref{eqn:Amunu}
for the classical point charge
is a gauge dependent object and so
is not observable.  Nevertheless, the features of the Effective Photon 
Distribution come directly from $A_{\mu\nu}(x,q)$.
One might ask whether the interesting features of
$A_{\mu\nu}(x,q)$ disappear under a gauge transform.  To see whether this
happens, one must insert $A_{\mu\nu}(x,q)$ into Equation \eqref{eqn:wwapprox3}.
The only things in \eqref{eqn:wwapprox3} that could significantly 
alter shape of the distribution \eqref{eqn:Amunu} are the gradients. 
Now because the photon source is extremely localized (it is a delta function),
the shape of the photon distribution comes solely from the propagator.  Since
the propagator varies significantly on length scale comparable to $1/q_\mu$,
derivatives of $A_{\mu\nu}(x,q)$ are always comparable in size to $q_\mu$ and
any expansion of the gauge projector in Equation~\eqref{eqn:proj2} will not 
converge.  So, we must conclude that our photon distribution can not be made 
gauge invariant using this technique and that we can not tell what features 
of the photon density survive a general gauge transform.
Now, had we {\em not} used a point source for our photons, 
the integration over the source could 
smooth the photon distribution so that it varies slower in space.
In that case, our distribution could be rendered gauge invariant.  
 
\chapter{WHEN FACTORIZATION FAILS}
\label{sec:probeedist}

The ``source-propagator'' picture of the phase-space particle densities, from
Chap\-ters \ref{chap:transport} and \ref{chap:pips},
and factorization of the exclusive reaction rate, from Chapter \ref{chap:pips}
and the parton model, are both conceptually 
useful concepts.  However, there are times when both fail.
In this appendix,  we discuss one such failure: lepton pair production in the 
strong field produced by two point charges.  Because the photon fields of the 
two point charges interfere, it is not possible to clearly isolate the
source or probe and we can not
factorize the square S-matrix into an electron distribution and electron/probe
interaction.  Nevertheless, we can still discuss the process in phase-space,
even though we cannot write down the electron distribution.

In this appendix, we investigate electron-positron pair production
in the strong field of two point charges.   One
might visualize this interaction as a virtual photon from one point
charge probing the virtual electron distribution of another point
charge.    Thus, 
the electron distribution would appear factorized from the virtual 
electron-virtual photon collision process.
However, we will show that this picture is incorrect because
the photon fields interfere with one another on length scales
comparable to the size of pair production region.  Of course, this also means
that our ``source-propagator'' picture fails here.  Nevertheless, we 
can still formulate the \sloppy\linebreak problem in phase space and discuss 
the interplay of the interaction length and particle production length scales.

\section{Interference of Photon Fields}

\begin{figure}
   \begin{center}
   \includegraphics{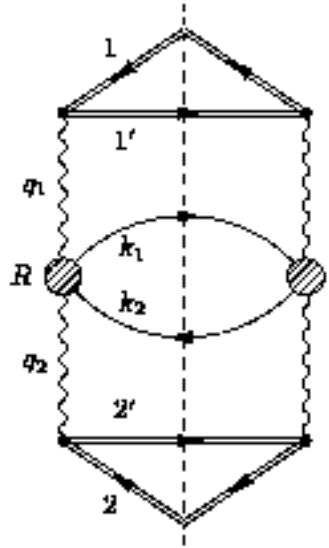}
   \end{center}
   \caption[Cut diagram for lepton pair production from a two photon 
	interaction.  $R$ is the space-time point of the center of the 
	collision region.]
	{Cut diagram for lepton pair production from a two photon 
      	interaction.  $R$ is the space-time point of the center of the 
	collision region.}
   \label{fig:2phot1}
\end{figure}

\begin{figure}
   \begin{center}
   \includegraphics{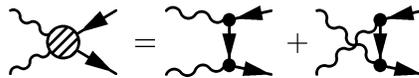}
   \end{center}
   \caption[The diagrams that contribute, at lowest order, to the 
	$\gamma\gamma\rightarrow e\bar{e}$ effective vertex.]
	{The diagrams that contribute, at lowest order, 
	to the $\gamma\gamma\rightarrow e\bar{e}$ effective vertex.}
   \label{fig:2phot2}
\end{figure}

The tree-level diagrams for pair production in a strong external field are 
shown in Figures~\ref{fig:2phot1} and \ref{fig:2phot2}.  We can write down the 
S-matrix corresponding to this using the same procedures used in
Chapter~\ref{chap:pips}.  To lowest order in the coupling strength, we obtain:
\begin{equation}
\begin{array}{ccl}
   S_{12\rightarrow 1'2'e\bar{e}} &=& \displaystyle 
   \alpha_{em}\int\dn{4}{x_1}\dn{4}{x_2}
   \dnpi{4}{k_1}\dnpi{4}{k_2}\dnpi{4}{p}\\
   &  & \displaystyle\times \frac{f^{*}(k_1,k_2)}{\sqrt{2k_{1,0}V}
   \sqrt{2k_{2,0}V}}e^{ik_1\cdot x_1+ik_2\cdot x_2+ip\cdot (x_1-x_2)}\\
   & & \displaystyle\times\Lambda_{\mu \nu}(k_1,s_1,k_2,s_2,p)
   \left\{ A_{1}^{\mu}(x_1) A_{2}^{\nu}(x_2) + 
   A_{2}^{\mu}(x_1) A_{1}^{\nu}(x_2)\right\}.
\end{array}
\label{eqn:2photSmatrix}
\end{equation}
Here $x_1$ and $x_2$ are the interaction points of the photons and 
should not be confused with the classical source particles $1$~and~$2$.
We have already separated the $\gamma\gamma e\bar{e}$ effective vertex
\[
   \Lambda_{\mu \nu}(k_1,s_1,k_2,s_2,p)
   = \bar{u}(k_1,s_1)\gamma_{\mu}iS^{c}(p)\gamma_{\nu}v(k_2,s_2).
\] 
In $\Lambda_{\mu \nu}(k_1,s_1,k_2,s_2,p)$,  $S^{c}(p)$ is the momentum-space 
Feynman electron propagator.  The final state electron-positron wavepacket is 
$f^{*}(k_1,k_2)$ and we will assume the final $e\bar{e}$ pair to be free 
and use the free wavepacket from Appendix \ref{append:current}.  The 
reader should note that we can already see the photons interfering 
in Equation~(\ref{eqn:2photSmatrix}).

As usual, we can rewrite Equation~\eqref{eqn:2photSmatrix} in terms of 
Wigner transformed \sloppy\linebreak quantities.  However, due to the photon 
fields interfering, the structure of the cross terms are complicated.  The  
$\Smatrix{12\rightarrow 1'2'e\bar{e}}$ is:
\begin{equation}
\begin{array}{ccl}
\Smatrix{12\rightarrow 1'2'e\bar{e}}&=&\alpha_{em}^2
    \displaystyle\int\dn{4}{R}\dn{4}{r}
    \dnpi{4}{k_1}\dnpi{4}{k_2}\dnpi{4}{q_1}\dnpi{4}{q_2}\dnpi{4}{p}\\
& & \displaystyle\times f(R-r/2,k_1,R+r/2,k_2)\\
& & \displaystyle\times \Lambda_{\mu\mu'\nu\nu'}(k_1,k_2,p,r)
    \twopideltaftn{4}{q_1+q_2-k_1-k_2}\\
& & \displaystyle\times\left\{ \twopideltaftn{4}{q_1-k_1+p} 
    \Afieldupname{\mu}{\mu'}{R-r/2}{q_1}{1}
    \Afieldupname{\nu}{\nu'}{R+r/2}{q_2}{2}\right.\\
& & \displaystyle+\left.  \twopideltaftn{4}{q_1-k_2-p}
    \Afieldupname{\nu}{\nu'}{R+r/2}{q_1}{1}
    \Afieldupname{\mu}{\mu'}{R-r/2}{q_2}{2} \right.\\
& & \displaystyle+\left. \int\dn{4}{\tilde{r}} \exp{[i\tilde{r}\cdot(-p+
    \frac{k_1-k_2}{2})-ir\cdot(q_1-q_2)]}\right.\\
& & \displaystyle\times\left.\Afieldupname{\nu}{\mu'}{R-\tilde{r}/4}{q_1}{1}
    \Afieldupname{\mu}{\nu'}{R+\tilde{r}/4}{q_2}{2}\right.\\
& & \displaystyle+\left.\int\dn{4}{\tilde{r}} \exp{[i\tilde{r}\cdot(-p+
    \frac{k_1-k_2}{2})+ir\cdot(q_1-q_2)]}\right.\\
& & \displaystyle\times\left.\Afieldupname{\mu}{\nu'}{R+\tilde{r}/4}{q_1}{1}
    \Afieldupname{\nu}{\mu'}{R-\tilde{r}/4}{q_2}{2}\right\}
\end{array}
\label{eqn:misceqn}
\end{equation}
This equation could look simpler if, in the interference terms, 
we Wigner transformed $A_1$ together with $A_2$.  However then we would have 
a virtual electron being emitted by some interference field and then 
reabsorbed by another interference field and the resulting equations would 
be impossible to interpret using our photon distributions.
In Equation~(\ref{eqn:misceqn}), we neglect $\tilde{k}$ relative to $k$ in the 
effective vertex and in the factors of $(2k_{0}V)$ because the final 
state wave packets are sharply peaked in momentum.  In 
Equation~(\ref{eqn:misceqn}), $R$
is the center of the interaction points $x_1$ and $x_2$ and $r$ is the 
space-time separation of these points. The final state Wigner 
density is 
\[\begin{array}{rl}
        \displaystyle f(x_1,k_1,x_2,k_2)=&      
        \displaystyle\frac{1}{(2k_{1,0}V)}\frac{1}{(2k_{2,0}V)}
        \int\dnpi{4}{\tilde{k}_1}\dnpi{4}{\tilde{k}_2}
        e^{-i\tilde{k}_1 x_1-i\tilde{k}_2 x_2} \\
        &\displaystyle\times f^{*}(k_1+\tilde{k}_1/2,k_2+\tilde{k}_2/2)
        f(k_1-\tilde{k}_1/2,k_2-\tilde{k}_2/2) 
\end{array}\]
and the Wigner transform of the effective vertex is
\[
        \Lambda_{\mu\mu'\nu\nu'}(k_1,k_2,p,r)=\int\dnpi{4}{\tilde{p}} 
        e^{i\tilde{p}\cdot r} \Lambda_{\mu\nu}(k_1,k_2,p+\tilde{p}/2)
        \Lambda^{*}_{\mu'\nu'}(k_1,k_2,p-\tilde{p}/2).
\]  
We can write the effective vertex in terms of the scalar Feynman 
propagator, 
\begin{equation}
\begin{array}{ccl}
   \Lambda_{\mu\mu'\nu\nu'}(k_1,k_2,p,r) & = & \bar{u}(k_1,s_1)\gamma_{\mu}
   (\dirslash{p}+\frac{i}{2}\dirslash{\partial}+m_e)\gamma_{\nu}v(k_2,s_2)\\
   & & \times \bar{v}(k_2,s_2)\gamma_{\nu'}(\dirslash{p}-
        \frac{i}{2}\dirslash{\partial}+m_e)
        \gamma_{\mu'}u(k_1,s_1)\Gcaus{r}{p}\\
        & \equiv & \lambda_{\mu\mu'\nu\nu'}(k_1,k_2,p,r) \Gcaus{r}{p}.
\end{array}
\end{equation}

We simplify the reaction probability by summing over the final state
electron and positron spins.  We simplify things even further by working in
the ultra-relativistic limit, namely when $v_1^2\approx v_2^2\approx 0$. 
Under these approximations,  we find
\begin{equation}
\label{eqn:noname}
\begin{split}
\Smatrix{12\rightarrow 1'2'e\bar{e}}=& \alpha_{em}^2 
    \int\dn{4}{R}\dn{4}{r}
    \dnpi{4}{k_1}\dnpi{4}{k_2}\dnpi{4}{q_1}\dnpi{4}{q_2}\dnpi{4}{p}\displaybreak[0]\\
& \times f(R-r/2,k_1,R+r/2,k_2)\displaybreak[0]\\
& \times \sum_{\rm spins}\lambda_{\mu\mu'\nu\nu'}(k_1,k_2,p,r)
    \Gcaus{r}{p}\twopideltaftn{4}{q_1+q_2-k_1-k_2}\displaybreak[0]\\
& \times\left\{ \twopideltaftn{4}{q_1-k_1+p} 
    \Afieldupname{\mu}{\mu'}{R-r/2}{q_1}{1}
    \Afieldupname{\nu}{\nu'}{R+r/2}{q_2}{2}\right.\displaybreak[0]\\
& +\left.  \twopideltaftn{4}{q_1-k_2-p}
    \Afieldupname{\nu}{\nu'}{R+r/2}{q_1}{1}
    \Afieldupname{\mu}{\mu'}{R-r/2}{q_2}{2} \right.\displaybreak[0]\\
& +\left. 2 \int\dn{4}{\tilde{r}} \cos{[\tilde{r}\cdot(-p+
    \frac{k_1-k_2}{2})-r\cdot(q_1-q_2)]}\right.\displaybreak[0]\\
& \times\left.\Afieldupname{\nu}{\mu'}{R-\tilde{r}/4}{q_1}{1}
    \Afieldupname{\mu}{\nu'}{R+\tilde{r}/4}{q_2}{2}\right\}.
\end{split}
\end{equation}
Given the relatively simple form of this equation, one would think that we
could identify the exchanged electron's phase-space density.  In fact, 
if we use free particle distributions for the 
final state electron and positron and sum over final states, we can 
identify  the virtual electron distribution 
(Equation (\ref{eqn:elecdist})) in the direct terms. 
However,  we can not make the same identification in the
interference term and factorization is not possible here.   
We might find factorization again if we had several point charges as one can
envision a situation with many photon sources screening the photons (a plasma 
for instance).  The photon field might then 
be an incoherent superposition of photon fields.  In the absence of photon
interference, we might be able to define an Effective Electron Distribution.

\section{The $e\bar{e}$ Production Region vs. the Interaction Region}

Even though Equation \eqref{eqn:noname} does not factorize or acquire the
``source-propaga\-tor'' for any of the densities (except maybe the photon 
density), we can still figure out where the electron-positron pairs are 
produced and the size of the region where the two photons interact. 
We will see that the $e\bar{e}$ production region is set by the shape and 
size of the photon distributions and that the two photon interaction region's 
size depends on the mass and virtuality of the exchanged electron.  

\begin{figure}
   \begin{center}
   \includegraphics[angle=-90,totalheight=2.5in]{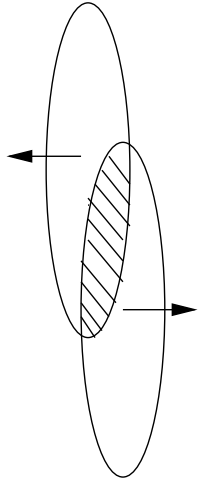}
   \end{center}
   \caption[Schematic of the pair production region.  The ellipses represent 
	the edge of the photon distributions, each  with four-momentum
        $q=(m_e,m_e/v_L,\vec{0}_T)$.  The shaded region is the geometrical
        overlap of the photon distributions and  sets the size of the
        $e\bar{e}$ production region.  The arrows point in the direction
   	of the photons' source's 3-momentum.]
	{Schematic of the pair production region.  The ellipses represent the 
	edge of the photon distributions, each  with four-momentum
        $q=(m_e,m_e/v_L,\vec{0}_T)$.  The shaded region is the geometrical
        overlap of the photon distributions and  sets the size of the
        $e\bar{e}$ production region.  The arrows point in the direction
   	of the photons' source's 3-momentum.}
    \label{fig:overlap}
\end{figure}

First, take the virtual photon distribution of the classical point charge
from Section~\ref{sec:classEPD}.   Now, the lowest energy and momentum that 
each of the interacting photons can have is\footnote{The 
distribution of photons with $q=(m_e,m_e/v_L,\vec{0}_T)$ is shown in 
Figure~\ref{fig:photondist}.} $q=(m_e,m_e/v_L,\vec{0}_T)$. 
Because the high energy or far off-shell photons are closer to the point 
charge then their lower energy and nearly on-shell cousins, photons 
with the minimum $q_\mu$ have the largest distributions.   So, the 
geometrical overlap of the high energy 
portions of the virtual photon distribution sets the size of 
the $e\bar{e}$ production region.  In Figure~\ref{fig:overlap} we illustrate
this:  the two ellipses represent the edge of the photon distribution and the
shaded region is the region where the $e\bar{e}$ pairs can be created.

Now, the size of the two photon interaction itself is determined by how 
far the exchanged electron can travel between the vertices in Figure 
\ref{fig:2phot2}.  For this, we look at the phase-space electron propagator. 
Assuming the electrons have mass,\footnote{The $m_e=0$ case is uninteresting 
because the $e\bar{e}$ production region always extends over the entire two 
photon interaction region.} we use Remler's causal propagator.  Here the 
phase-space ``probability''  for \sloppy\linebreak propagating between two 
space-time points drops like $e^{-2m_{e}\tau}$ for space-like electrons and 
like $\sin{2\tau (\sqrt{p^2}\pm m_{e})}/(\sqrt{p^2}\pm m_{e})$ for 
time-like electrons. The proper time along the electron 4-momentum is $\tau$.
In the direction transverse to the electron four-momentum, the 
``probability'' is zero.  Thus,  the interaction region has a
characteristic length scale of $\approx 1/m_{e}$.  This is 
comparable to the width of the photon distributions, so there is no scale 
separation.  Typically one requires the interaction length scale to be much
smaller than the characteristic length scale of the particle density
in order to justify the gradient expansions of the Quasi-Classical 
Approximation and allow for a transport description.  Because our approach 
does not rely on the Quasi-Classical Approximation, the transport-like 
description in Chapter \ref{chap:transport} is still be possible.

So, to summarize, we learned that the ``source-propagator'' picture of parton 
densities fails when the source particle and probe particle interact, even  
through quantum interference. 
Nevertheless, we can still discuss the process in phase-space with the
phase-space sources and propagators, even if the densities have no clear
meaning.

\chapter{WAVEPACKETS}
\label{app:wavepacket}

Throughout this paper, we use wavepackets in the initial and final states 
of a reaction (or equivalently density matrices) to provide spatial 
localization or delocalization.  In this appendix, we detail
the construction of an initial or final state wavepacket and discuss the limits
of either a completely localized or delocalized wavepacket.

\section{On--Shell Gaussian Wavepacket}

An initial (or final) state ket can be written with wavepackets:
\begin{equation}
 	\ket{i}=\int \dnpi{4}{p} f(p) \ket{\vec{p}}.
	\label{eqn:ket2}
\end{equation}
The corresponding Wigner function of the particles is
\begin{equation}
\begin{array}{rl}
	f(x,p)=&\displaystyle\int\dnpi{4}{p'}e^{-ix\cdot p'} 
		\ME{i}{\hat{\phi}^{*}(p-p'/2)\hat{\phi}(p+p'/2)}{i}\\
		=&\displaystyle\frac{1}{2Vp_0}\int\dnpi{4}{p'}e^{-x\cdot p'}
		f(p+p'/2)f^{*}(p-p'/2).
\end{array}
\label{eqn:wigpartdens}
\end{equation}
Particles in either the initial or final states are on--shell, so they can be
expanded in momentum eigenstates.  We choose the wavepacket to be a Gaussian
superposition of momentum eigenstates with a momentum spread $\sigma$:
\[
	\phi(p)={\cal N} \deltaftn{}{p^2-M^2} 
	\exp{[-(\vec{p}-\vec{p}_i)^2/2\sigma^2]}
\] 

The Wigner transform of this wavepacket can not be done analytically except 
in the limit when $|\vec{p}_i|\gg\sigma$.  In this limit, 
$\vec{p}_i\approx\vec{p}\gg\vec{p'}$
so our wavepacket is localized in momentum giving the following Wigner density
of particles:
\begin{equation}
	f(x,p)=\frac{|{\cal N}|^2}{8\pi p_0^2}\deltaftn{}{p^2-M^2}
	\exp{\left[-\frac{(\vec{p}-\vec{p}_i)^2}{2\sigma^2}\right]}
	(2\sigma\sqrt{2\pi})^3 \exp{\left[-2\sigma^2
	(\vec{v}x_0-\vec{x})^2\right]}.
\end{equation}
Here $\vec{v}=\vec{p}/p_0$ is the velocity of the wavepacket.  
Thus, the particle's Wigner function is a Gaussian in both momentum and space. 
The spread in momentum is the inverse spread in space.  The centroid of the 
Gaussian follows the particle's classical trajectory.  The magnitude of the 
energy of the packet is set by the delta function out front.  We have not
constrained the particle in energy so this
density contains both positive and negative energy contributions.  

\section{Delocalizing the Wavepacket in Space: Free Wavepacket} 

In accordance with the uncertainty principle,   the wavepacket becomes 
completely delocalized in space in the limit of complete localization in 
momentum (i.e. $\sigma\rightarrow 0$).
In this limit, the spatial Gaussian approaches unity and the momentum Gaussian
becomes a delta function.  After working out the normalization, we have
\begin{equation}
	f^{\rm free} (x,p)=\frac{1}{2Vp_0}\twopideltaftn{4}{p-p_i}.
\label{eqn:imfree}
\end{equation} 
This is no surprise since the limit $\sigma\rightarrow 0$ squeezes the state 
into a momentum eigenstate.  We use this result in Section~\ref{sec:edist}
for the final state positron's wavepacket.

\section{Localizing the Wavepacket in Space: Classical Wavepacket}
\label{append:classdens}

A classical particle is localized in both space and momentum, a seeming
violation of the uncertainty principle.  In real life, this is not a problem
since the reason classical particles appear localized is that we probe
them on length (or momentum) scales too coarse to resolve the interesting
quantum features.  In the case of our Gaussian wavepacket, this amounts to 
probing the distribution on length scales much larger than $1/\sigma$.  
In this case, the space Gaussian is too localized to resolve and we can replace
it with a delta function.  Additionally, if we assume that $\sigma$ is large, 
we can replace the momentum Gaussian with a delta function as well:  
\begin{equation}
	f^{\rm classical} (x,p) = \frac{1}{2}(2\pi)^4 
	\deltaftn{3}{\vec{p}-\vec{p}_i}
	\deltaftn{}{p^2-M^2}\deltaftn{3}{\vec{v}x_0-\vec{x}}
\label{eqn:classdens}
\end{equation}
Here we have inserted the correct normalization for the wavepacket.
This density corresponds to an on-mass-shell particle that
follows its classical trajectory $\vec{v}x_0=\vec{x}$.  Again, we left in both
positive and negative energy contributions.  We use the 
result \eqref{eqn:classdens} in the next appendix to find the current of a 
classical point charge.

\chapter{THE CURRENT DUE TO A CLASSICAL PARTICLE}
\label{append:current}

In this appendix, we derive the classical
current used in the Effective Photon Distribution calculation of 
Chapter \ref{chap:pips}.
For the sake of illustration, we take the point particle to be a
scalar particle.  The derivation takes three steps: first we define the 
Wigner current of a scalar particle, then we
derive the photon/scalar interaction vertex in phase-space, and finally we
localize the initial and final states of the scalar to give the classical
current.    

\section{Wigner Current}

We begin by restating Equation (\ref{eqn:current}):
\begin{equation}
	\Jcurrent{\mu}{\nu}{x}{q}{A}=\int\dnpi{4}{\tilde{q}}
	e^{-i\tilde{q}\cdot x}\ME{A'}{j^{\mu}(q+\tilde{q}/2)}{A}
	\ME{A}{j^{\dagger \nu}(q-\tilde{q}/2)}{A'}.
	\label{eqn:currentdef}
\end{equation}
We write the initial and final state bra's and ket's according to
Equation (\ref{eqn:ket}) and
rewrite Equation (\ref{eqn:currentdef}) in terms of initial and final Wigner
densities, 
\begin{equation}
	\Jcurrent{\mu}{\nu}{x}{q}{A}=\int\dnpi{4}{p_i}\dnpi{4}{p_f}
	\denslabel{f}{x}{p_i}{A}\denslabelstar{f}{x}{p_f}{A'}
	\twopideltaftn{4}{p_i-p_f-q}\Gamma_{\mu \nu}(q,p_i,p_f).
	\label{eqn:currwvpkt}
\end{equation}
We assume that the initial and final wavepackets are localized in momentum and
some-what delocalized in space.  Shortly, we will also assume that we probe 
this current on length scales much larger than even this delocalized space
distribution.

\section{Scalar Vertex}

$\Gamma_{\mu \nu}(q,p_i,p_f)$ is not quite the Wigner transform of the
$\gamma AA'$ vertex, although it does arise from performing the Wigner
transform in Equation (\ref{eqn:currentdef}).  It is defined by
\begin{equation}\begin{split}
	\lefteqn{\twopideltaftn{4}{p_i-p_f-q}\Gamma_{\mu \nu}(q,p_i,p_f) 
	=}\hspace*{3em}&\\
	&4V^2 p_f^0 p_i^0
	\int\dnpi{4}{\tilde{q}}\ME{\vec{p}_f}{j_{\mu}(q+\tilde{q}/2)}{\vec{p}_i}
	\ME{\vec{p}_i}{j^{\dagger}_{\nu}(q-\tilde{q}/2)}{\vec{p}_f}
\end{split}\end{equation}
Using the matrix element
\[
	\ME{\vec{p}_f}{j_{\mu}(q)}{\vec{p}_i}=
	eZ\twopideltaftn{4}{p_i-p_f-q}
	\frac{(p_i+p_f)_{\mu}}{2V\sqrt{p_f^0 p_i^0}},
\] 
We find
\begin{equation}
	\Gamma_{\mu \nu}(q,p_i,p_f)  = 
	\alpha_{em} Z^2(p_i+p_f+\frac{1}{2}(\tilde{p_i}+\tilde{p_f}))_{\mu}
	(p_i+p_f-\frac{1}{2}(\tilde{p_i}+\tilde{p_f}))_{\nu}
	\label{eqn:currstep2}
\end{equation}
The relative momenta, $\tilde{p}_i$ and $\tilde{p}_f$, become
derivatives on $x$ in the current \eqref{eqn:currwvpkt}.  
Assuming the wavepackets to be uniform the the reaction's length scales, we 
can ignore the derivatives and arrive at the phase-space scalar vertex
\begin{equation}
	\Gamma_{\mu \nu}(q,p_i,p_f)=\alpha_{em} Z^2 (p_i+p_f)_{\mu}(p_i+p_f)_{\nu}.
	\label{eqn:vertexthingee}
\end{equation}

\section{Classical Current}

We are now in a position to derive Equation (\ref{eqn:classcurrent})
for the classical current density in phase-space.  
First, we take the final state to be a momentum eigenstate and sum over it. 
Since the final state is localized in momentum around $p_f$, this is not a bad
approximation.  Second, we take the initial state to be a classical wavepacket.
In other words, we assume that the initial state is localized in momentum and
delocalized in space but that we probe it on such large length scales that we
still see a spatially localized wavepacket.  So, putting Equations
(\ref{eqn:imfree}), (\ref{eqn:classdens}) and (\ref{eqn:vertexthingee}) into 
(\ref{eqn:currwvpkt}) and summing over final states, we get 
\[
	\Jcurrent{\mu}{\nu}{x}{q}{}=2\pi \alpha_{em} Z^2\: 
	v_{\mu} v_{\nu} \:\deltaftn{3}{\vec{x}-x_0\vec{v}}
	p_{i0}\delta ((p_f+q)^2-M^2).
\]
Using $p_f^2=M^2$ and $v_\mu\approx p_{f\mu}/p_{i0}$ and assuming 
$q^2/p_{i0}\ll q\cdot v$, we get the classical current: 
\begin{equation}
	\Jcurrent{\mu}{\nu}{x}{q}{classical}=2\pi \alpha_{em} Z^2\: 
	v_{\mu} v_{\nu} \:
	\delta (q\cdot v)\deltaftn{3}{\vec{x}-x_0\vec{v}}.
\label{eqn:yeaclasscurrent}
\end{equation}
Note that this current allows for emission of both positive and negative energy
photons.  In Section \ref{sec:pdist}, we restrict emission to positive energy 
photons by inserting a factor of $\theta(q_0)$ in 
Equation \eqref{eqn:yeaclasscurrent}.  This can be justified
by suitably choosing $\vec{p}_f$ and $\vec{p}_i$ and restricting the initial 
and final states to have only positive energy.

\coversheet{REFERENCES}
\addcontentsline{toc}{chapter}{REFERENCES}
\begin{spacing}{1}
   \bibliography{QGP,neqFT,transport,HBT,imaging,plainFT,QCD,misc,Wigner}
   \bibliographystyle{alpha}
\end{spacing}
\end{document}